\documentclass[11pt,a4paper]{book} 
\usepackage[pdftex]{graphicx} 
\usepackage{url} 
\pdfoutput=1
\usepackage[bookmarks, colorlinks=false, linkcolor=black, citecolor=black, pdfborder={0 0 0}, pdftitle={Viscoelastic Interfaces Driven in Disordered Media and Applications to Friction}, pdfauthor={Francois P. Landes}, pdfsubject={Ph.D. Thesis}, pdfkeywords={quenched disorder, out-of-equilibrium, dynamical phase transition, friction, earthquakes, depinning transition, visco-elastic, Directed Percolation, multiple absorbing state transition}]{hyperref}
\usepackage{tikz}
\usetikzlibrary{snakes}
\usetikzlibrary{positioning}
\usepackage{pgfornament}
\usepackage{tikzrput}
\usepackage{graphics}
\usepackage{geometry}
\usepackage{amsfonts}
\usepackage{amsmath}
\usepackage{float}
\usepackage{color}
\usepackage{amsthm}
\usepackage{fancyhdr}
\usepackage{fancybox}
\usepackage[T1]{fontenc}
\usepackage{lmodern}
\usepackage{graphicx}
\usepackage{epsfig}
\usepackage{pdfpages}
\usepackage[english]{minitoc}

\usepackage{setspace}
\usepackage{epigraph}
\usepackage{amssymb}
\usetikzlibrary{calc}
\usetikzlibrary{intersections} 
\usepackage{booktabs}
\usepackage[titletoc]{appendix}
\usepackage{xr-hyper}
\usepackage{hyperref}
\makeatletter
\def\thickhrulefill{\leavevmode \leaders \hrule height 0.5ex \hfill \kern \z@}
\def\@makechapterhead#1{
  \vspace*{10\p@}
  {\parindent \z@ \centering \reset@font
        \thickhrulefill \quad
        \scshape $\Large{\textcolor{black}{\textsc{\@chapapp{} \thechapter}}}$
        \quad \thickhrulefill
        \par\nobreak
        \vspace*{10\p@}
        \interlinepenalty\@M
        \hrule
        \vspace*{10\p@}
        \Huge \bfseries #1\par\nobreak
        \par
        \vspace*{10\p@}
        \hrule
    \vskip 100\p@
  }}
\def\@makeschapterhead#1{
  \vspace*{10\p@}
  {\parindent \z@ \centering \reset@font
        \thickhrulefill
        \par\nobreak
        \vspace*{10\p@}
        \interlinepenalty\@M
        \hrule
        \vspace*{10\p@}
        \Huge \bfseries #1\par\nobreak
        \par
        \vspace*{10\p@}
        \hrule
  		 \vskip 40\p@
  }}
  
\DeclareTextSymbol{\degre}{T1}{6}
\DeclareTextSymbol{\degre}{OT1}{23}

\newcommand{\lp}{\left(}
\newcommand{\rp}{\right)}
\newcommand{\req}[1]{(Eq.~\ref{Eq:#1})}
\newcommand{\reqq}[1]{(\ref{Eq:#1})}
\newcommand{\pp}[1]{p.~\pageref{#1}}

\usepackage{pifont}  

\newcommand{\includefig}[3]{
\begin{figure}[]
\begin{center}
\includegraphics[width=#1]{#2}
\caption{   {\footnotesize #3}  }
\end{center}
\end{figure}}

\newcommand{\includefigtwo}[5]{
\begin{figure}[]
\begin{center}
\includegraphics[width=#1]{#2}
\includegraphics[width=#3]{#4}
\caption{   {\footnotesize #5}  }
\end{center}
\end{figure}}

\renewcommand{\d}{{\rm d}}

\begin{document}
\dominitoc
\doparttoc
\renewcommand\bibname{References}

\frontmatter
\phantomsection
\thispagestyle{empty}

\vspace*{-3.7cm}
\begin{figure}[htb!]
        \hspace{-0.4cm}\begin{minipage}[t]{5cm}
        \begin{flushleft}
					\includegraphics[height=4cm]{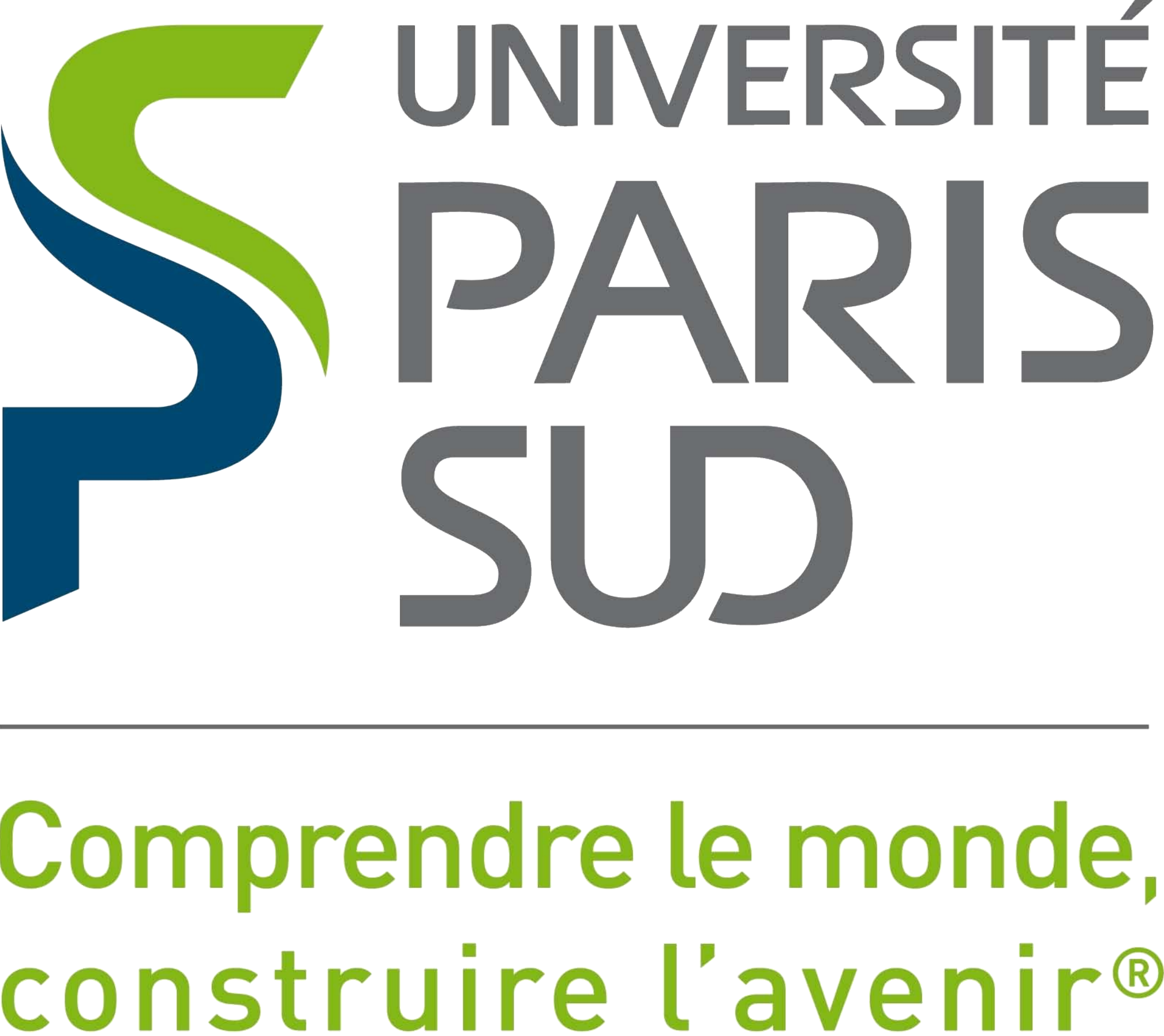}
				\end{flushleft}
				\end{minipage}
\hspace{5.6cm}
				\begin{minipage}[t]{6cm}
				\begin{flushright}
					\includegraphics[width=6cm]{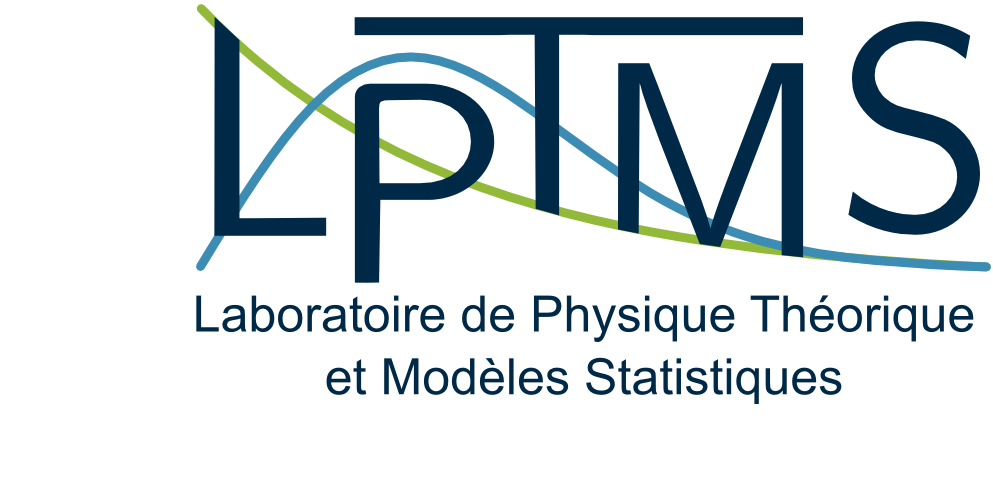} 
				\end{flushright}
				\end{minipage}
\end{figure}

\begin{center} 
\begin{tabular}{c}
		\\
		\\
    {\huge \textsc{Universit\'e Paris-Sud}} \\
		\\
		\\
    {\Large \textsc{\'Ecole Doctorale de Physique de la R\'egion Parisienne}} \\   
		{\large \textsc{Laboratoire de Physique Th\'eorique et Mod\`eles Statistiques }} \\
		\\
		{\Large \textsc{Discipline : Physique Statistique/Theorique }} \\
    \\
		\\
		\\
    \huge \textsc{Th\`ese de doctorat}\\
    \\
    \large{Soutenue le 10 Septembre 2014 par} \\
    \\
    \Huge{\textbf{{Fran\c cois P. Landes}}}\\
    \\
    \\
		\\
		\\
    \huge\bf{Viscoelastic Interfaces Driven	}\\
	\huge\bf{in Disordered Media}\\
	\huge\bf{ and }\\
	\huge\bf{Applications to Friction}\\
		
    \\
    \\
    \\
		\\
		\\
\end{tabular}\\

\begin{tabular}{p{3cm} p{3.5cm} p{5.4cm} l }
	& \footnotesize\bf{Directeur de th\`ese} : & Alberto Rosso&\\ 
	& & &\\
	& \footnotesize\bf\underline{Composition du jury :}& &\\
	& \footnotesize{Rapporteurs} : & Jean-Louis Barrat  &\\ 
  &	& Stefano Zapperi &\\ 
  & \footnotesize{Examinateurs} 	& Leticia F. Cugliandolo  &\\ 
  &	& Carmen Miguel &\\ 
  &	& Dominique Salin  &\\ 
\end{tabular}
\end{center}

\tableofcontents
\addcontentsline{toc}{section}{Table Of Contents}

\mainmatter
\chapter*{Introduction}
\addcontentsline{toc}{section}{Introduction}
\label{sec:Intro_generale_these}
 \markboth{}{Introduction}

\vspace{2cm}

There are many natural occurrences of systems that upon a continuous input of energy, react by sudden releases of the accumulated energy in the form of discrete events, that are generally called \textit{avalanches}. 
Examples are the dynamics of sand piles, magnetic domains inversions in ferromagnets, stress release on the earth crust in the form of earthquakes, and many others. 
A remarkable characteristic of most of these realizations is the fact that the size distribution of the avalanches may display power-laws, which are a manifestation of the lack of intrinsic spatial scale in these systems (similarly to what happens in \textit{continuous phase transitions} at equilibrium \cite{Landau1980, Kardar2007}, with the correlation length diverging at criticality).
The theoretical analysis is build on the features shared by these various processes, and aims at isolating the minimal set of ingredients needed to explain the common elements of phenomenology.
There are numerous models which display critical behaviour and thus power-law avalanche size distributions, however in most cases the exponents characterizing the avalanches can only take a few possible values, corresponding to the existence of a few different \textit{universality classes}.

For almost 20 years, there has been an ongoing effort to understand earthquakes in the framework of these critical and collective out-of-equilibrium phenomena.
Several theoretical models are able to reproduce a scale-free statistics similar to that present in seismic events, but miss basic observations such as the presence of aftershocks after a main earthquake or the anomalous exponent of the Gutenberg-Richter law \cite{Scholz2002}.
At a smaller and simpler scale, a general theory for the friction of solids, taking into account the heterogeneities of each surface and the collective displacements, contacts and fractures of the asperities is not yet available \cite{Persson2000, Persson1996a}.
Current theories fail to reproduce some \textit{non-stationary effects} such as the increase of static friction over time or the possibility of the decrease of kinetic friction with increasing velocity.

A first class of models displaying a single well-defined \textit{out-of-equilibrium phase transition} is that of the \textit{depinning} of an extended elastic interface\footnote{
The interface can be any manifold, i.e.~a line, a surface, a volume, etc.
} driven over a disordered (random) energetic landscape \cite{Fisher1998, Kardar1998}.
While the interface
is driven across the disordered environment, it gets alternatively stuck (\textit{pinned}) by the heterogeneities and freed (de-pinned) by the driving force. 
Despite its locally intermittent character, the overall dynamics of the interface has a stationary regime, which makes various analytical and numerical methods available.
Remarkably, one can often disregard the precise details of the microscopic dynamics when considering the large scale behaviour. 
As a result, the depinning transition successfully represents various phenomena, such as Barkhausen noise in ferromagnets \cite{alessandro1990, Zapperi1998a,Durin2000, Durin2006}, crack propagation in brittle materials \cite{Alava2006, Bonamy2008, bonamy2011failure}  or wetting fronts moving on rough substrates  \cite{Rosso2002b, Moulinet2004, LeDoussal2009a} (see \cite{barabasi1995fractal} for notions on fractals, growing surfaces and roughness).
Although the framework is also a priori well suited to describe friction and thus earthquakes, the stationary behaviour itself is the ground where major discrepancies arise between theoretical depinning results and real data: the aftershock phenomenon observed in earthquakes, for instance, is clearly not stationary \cite{Scholz2002}.

A second class of such models is that of Directed Percolation (DP) 
\cite{Odor2008, Henkel2008, Hinrichsen2006, Odor2004, Hinrichsen2000}, which models the random growth, spatial spread and death of some density of ``activity'' over time, in the manner of an avalanche.
On a lattice, each site can be either active or inactive, and at each time step, each active site tries to activate each of its neighbours, with a probability of success 
 $p$. 
When all sites become inactive, the avalanche is over and the state no longer evolves.
 This inactive state is an ``absorbing phase'' of the dynamics: the DP transition is an \textit{absorbing phase transition} \cite{Henkel2008}.
There is a critical value of the probability $p$ at which the system reaches criticality, with most stochastic observables distributed as power-laws.
Numerous birth-death-diffusion processes share the same critical exponents and scaling functions: the DP class is a wide, robust class.
We use the DP process as a toy model of avalanches with Markovian dynamics \cite{Kampen1981}. \\

In this thesis, starting from models of out-of-equilibrium phase transitions with stationary dynamics, we build and study \textit{variants} of these models which still display criticality, but in the same time have \textit{non-stationary} dynamics.\\

The physical process at the origin of most of our motivation and choices is that of solid on solid, dry friction (i.e.~in the absence of lubricants).
Actually, during this thesis our concern was initially the application of statistical physics methods to seismic events, however towards the end of the thesis we focused more on laboratory-scaled friction, as it is a much better controlled field. 
Since this subject is not a common topic in the field of disordered systems, we introduce the problem of friction in chapter \ref{chap:friction}. 
Reviewing the basic phenomenology and the well-established parts of the theory of friction, we are able to identify the main features that any friction model should include.
Two points emerge clearly.
A first is the need to account for the disordered aspect of the surfaces 
at play: asperities form a random network of contacts which constantly break and re-form, and the surfaces are heterogeneous so that the contact strengths are randomly distributed.
A second is the relevance of some slow mechanisms (plastic creep, in particular) which allow for a strengthening of the contacts over time. 
The latter point becomes especially relevant at very slow driving speeds, or when there is no motion.
We will focus on the slow driving regime, where the non-trivial frictional behaviours appear and which is crucial when considering seismic faults.

The physics of earthquakes is vast and quite complex \cite{Scholz2002}, but presents several points of interest for us.
A first is that the sliding of tectonic plates, at first approximation, may be considered as a large-scale manifestation of solid on solid, dry friction.
This ``application'' has been studied quite extensively on its own, and a large amount of data is available, so that seismic faults can be used to test the predictions of friction models. 
A second point is that due to its importance, the field of geophysics has generated numerous interesting models, which may serve as starting points to understand friction as a collective phenomenon, rather than a simple continuum mechanics problem.
This 
motivates our quick review of seismic phenomena and the related historical models, presented in chapter \ref{chap:EQs}.

The mapping of an earthquake model onto the problem of elastic depinning naturally introduces our review of the depinning transition in chapter \ref{chap:elastic}.
There, we introduce all the concepts necessary to understand our own modified depinning model, and appreciate its originality.
We explain the critical properties of this \textit{dynamical phase transition} (or depinning transition \cite{Zapperi1998a, Rosso2002b, LeDoussal2009a, Alava2006}), review the scaling relations and an original approach to the mean field. 
Even though we notice that the depinning universality class is a robust one, we are forced to acknowledge its inability to account for frictional phenomena.

With the notions presented in the previous chapters, our choice of modification of the depinning problem is quite natural.
The starting point of our analysis is to remark that conventional depinning does not allow any internal dynamical effects to take place during the inter-avalanche periods.
To address this issue, in chapter \ref{chap:visco} we introduce the model of a \textit{viscoelastic} interface driven in a disordered environment, which allows for a slow relaxation of the interface in between avalanches.
The viscoelastic interactions can be interpreted as a simple way to account for the plastic creep, mainly responsible for the peculiarities of friction at low driving velocity.
After a qualitative discussion of the novelties of the viscoelastic interface behaviour, we present a derivation of its mean field dynamics.
Extending the mean field approach that we presented for the elastic depinning 
to this new model, 
we are able to compute the behaviour of the entire system, which is found to be non-stationary, with system-size events occurring periodically.
There, we also notice that the addition of the ``visco-'' part into the elastic interactions is relevant in the macroscopic limit.
We compare the mean field dynamics at various driving velocities and find good agreement with experimental results found in fundamental friction experiments (chap.~\ref{chap:friction}) and observations on earthquakes statistics (chap.~\ref{chap:EQs}).
In two dimensions, we are limited to numerical simulations, but we are able to perform them on systems of tremendous sizes (up to $15000 \times 15000$ sites on a single CPU), which allows us to unveil some features reminiscent of the mean field behaviour.
The various outputs of our simulations (critical exponents, aftershocks patterns, etc.) compare well with the observational results from chap.~\ref{chap:EQs} (see sec.~\ref{sec:visco_conclu_2} for a more detailed summary of results).
In the comparison with models from various other contexts (amorphous plasticity, granular materials, etc.) we notice similarities in the various models construction, and a shared tendency for global, system-size events.

During this thesis, most of the work was performed on non-stationary variations on the depinning model, with a focus on the applications to seismic events. 
On the way, we studied a variation of the Directed Percolation, which has to do with non-stationarity, despite being a model completely different from those presented in chap.~\ref{chap:visco}. 
The last chapter (chap.~\ref{chap:DP}) offers the opportunity to consider the bigger picture of avalanche models.
In that chapter, we consider the celullar automaton\cite{Wolfram1983} of Directed Percolation (DP). 
We provide an intuitive 
link with the problem of interface depinning by showing 
how much one would need to modify the DP process to let it represent the avalanches of the elastic interface.
We introduce a non-Markovian variant of the DP process, in which 
 the probability to activate a site at the first try and the second one are different from those in the ulterior attempts.
This provides the system with an implicit memory, making the microscopic dynamics non-stationary.
This modified DP displays criticality with some exponents changing continuously with the first and second activation probabilities, while others do not: in particular, only one scaling relation is violated by the new dynamics, so that the new class preserves most of its parent's structure.
A long-standing challenge is to find experimental systems belonging to the DP universality class: up to now, there are no such direct examples \cite{Odor2004}.
Our new model, which includes DP as a particular case, opens the way for possible future experimental work, as we may consider universality classes larger than DP.

As a conclusion, we explain the general path that structures this thesis and draw some directions for future work (chap.~\ref{chap:Conclusion}).
\\

In each chapter of this thesis, we provide a very quick introduction, which simply details the aim of the chapter and the organization of contents.
In the chapters' conclusions, we always carefully summarize the main results, and provide the motivation for the next chapter or some directions for future work.
We sometimes refer to the Appendices for technical details or results that are not crucial to our presentation.
Although each chapter is a self-contained entity, reading the earlier ones allows to fully understand and appreciate the scope of the latter ones.
The articles published during this thesis are \cite{Jagla2014a} and \cite{Landes2012}, they essentially correspond to the chapters \ref{chap:visco} and \ref{chap:DP}, respectively.

\chapter{Introduction to Friction}
\label{chap:friction}

\vspace{-2cm}
\minitoc
\vspace{2cm}

In this chapter we aim at giving a short overview of dry friction, i.e.~frictional phenomena where the lubricants effect is negligible.
We first present the phenomenological laws derived by experimental observations, then present the rudiments of the (incomplete) theory of friction.
Excellent references on these topics are \cite{Persson2000, Persson1996a,Krim2002} and  \cite{Baumberger2006}. 
In the process, we comment on the existing literature and draw some conclusions about possible directions for future work, especially for the statistical physics community.

\section{The Phenomenological Laws of Friction}

\begin{figure}[]
\begin{small}
\begin{center}
\def\svgwidth{12cm}
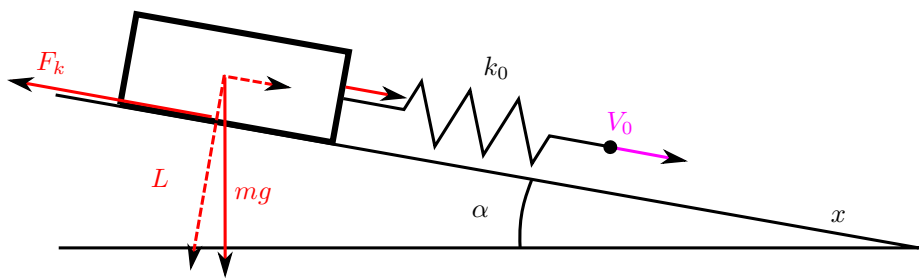
\end{center}
\end{small}
\caption{   {\footnotesize 
Solid block sliding on a solid substrate.
Solid parallelepiped sliding on an inclined plane (angle $\alpha$) at velocity $v=\dot{x}$. The weight can be decomposed in two components, one orthogonal to the surface (the load $L$), and one parallel to it (which contributes to the pulling). Additional pulling can be provided via a spring $k_0$, of which the ``free'' end may be moved at a fixed velocity $V_0$.
The kinetic friction force is denoted $F_k$
\label{Fig:bloc1} }   }
\end{figure}

Consider a solid parallelepiped  -- as depicted in Fig.\ref{Fig:bloc1} --  in contact with a large solid substrate over a surface $S$ (supposed to be flat at the macroscopic scale), with a normal load $L$ (for instance due to gravity), being pulled along the surface via a spring $k_0$, itself pulled at a fixed velocity $V_0$.
The block's velocity is denoted $v$. 
The force $F_k$ of frictional effects was\footnote{These laws were stated in the 17th century by Amontons for the first two of them, and in the 18th century by Coulomb for the third one.} claimed to follow these three laws:
\begin{itemize}
\item First law: $F_k$ is independent from the surface area $S$.
\item Second law: $F_k$ is proportional to the normal load: $F_k \propto L$.
\item Third law: $F_k$ is independent of the sliding velocity $v$. \\ 
\end{itemize}
This allows to write a phenomenological equation for the friction force:
\begin{align}
F_k = \mu_k L
\end{align}
where $\mu_k$ is the kinetic (or dynamic) friction coefficient, which depends on the nature of the surfaces in contact along with many other things, but which is here assumed to be independent from $S$, $L$ and $v$.

There is one ``exception'' to the third law which is commonly observed: for the static case ($v=0$, i.e.~when there may be pulling, but without motion) the friction coefficient takes a different value $\mu_s$, larger than the dynamical one: $\mu_s(v=0) > \mu_k(v>0)$.

\subsection{Stick-Slip Motion}
\label{sec:stickslip1}

Due to the fact that the static ($v=0$) friction force is higher than the dynamic ($v>0$) one, a mechanical instability known as ``stick and slip motion'' can occur, especially when the pulling is provided mainly in a sufficiently flexible way (small $k_0$) or at sufficiently low driving velocity $V_0$. 
As we are going to see, this is something that we experience on a daily basis.

\begin{figure}[]
\begin{small}
\begin{center}
\def\svgwidth{\textwidth}
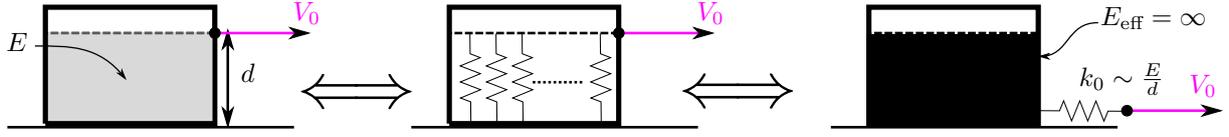
\end{center}
\end{small}
\caption{   {\footnotesize 
Effective stiffness of the driving spring. 
Left: a solid block with Young's modulus $E$ is pulled rigidly from some point at a height $d$, i.e.~this point is forced to have the velocity  $V_0$.
Middle: the solid block can be pictured as a dense network of springs, related to $ E$. 
Springs in the horizontal directions are not pictured for clarity.
Right: effective modelling by a block with infinitely rigid bulk, pulled by an effective spring $k_0 \sim E/d$. 
\label{Fig:effective_stiffness_k0}
}   }
\end{figure}

Consider the system pictured in Fig.~\ref{Fig:bloc1}, with an angle $\alpha=0$, for simplicity. The free end of the spring $k_0$ is denoted $w_0$ and is driven steadily at a velocity $V_0$. 
The spring $k_0$ can be thought of either as an actual spring through which the driving is performed, or as an effective representation for the bulk rigidity of the solid. 
As we pull the block from the side, we transmit some shear stress through its bulk. 
If the solid is  driven at constant velocity $V_0$ directly from a point on its side, the effective stiffness $k_0$ is proportional to the Young's modulus $E$ and inversely proportional to the height $d$ of the driving point (neglecting torque effects). See Fig.~\ref{Fig:effective_stiffness_k0} for a visual explanation.
In the context of a simple table-top experiment as presented here, the solid's stiffness is generally too large for stick-slip to occur, so that the use of an actual spring $k_0$ to perform driving is useful. 

Newton's equations for the center of mass of the block at position $x$ can be written in the dynamic and static cases:
\begin{align}
m \ddot{x} &= k_0(V_0 t-x) - \mu_k L   &\text{(dynamic)} \\
         0 &= k_0(V_0 t-x) - F_s       &\text{(static)} \label{Eq:StaticNewton}
\end{align}
where the static friction force $F_s$ adapts according to Newton's second law (Law of action and reaction) in order to balance the pulling force, as long as it does not exceed its threshold:  $|F_s| < \mu_s L = (F_s)_\text{max}$. 

We start with $x(t=0)=0, w_0(0)=0$, and for $t>0$ we perform the drive, $w_0=V_0 t$. 
As long as  $|F_s| < \mu_s L$, the block does not move: we are in the ``stick'' phase.

At time $t_1= \frac{\mu_s L}{k_0V_0}$, the static friction force $F_s$ reaches its maximal value $\mu_s L$ and the block starts to slide. This is the ``slip'' phase. 
Thus we have the initial condition $x(t_1)=0, \dot{x}(t_1)=0$ for the kinetic equation. The solution reads:
\begin{align}
x(t) = V_0 (t-t_1)
-\sqrt{\frac{m}{k_0}}   V_0 \sin  \left(\sqrt{\frac{k_0}{m}} (t-t_1) \right) 
+   \frac{(\mu_s-\mu_k) L }{k_0} \left( 1- \cos \left(\sqrt{\frac{k_0}{m}} (t-t_1) \right) \right).\label{Eq:x_de_t}
\end{align}
It is natural to take a look at the short-time limit of the solid's position:
\begin{align}
x(t) \underset{t\sim 0}{\sim} \frac{(\mu_s-\mu_k) L}{2m} t^2 +\frac{k_0 V_0}{6m}t^3 - \frac{(\mu_s-\mu_k) L k_0}{24 m^2}t^4 + o(t^ 4),
\end{align}
which is increasing at short time, as expected, since $\mu_s>\mu_k$.

As $x$ initially increases faster than $V_0 t$, the driving force from the spring, ($k_0(V_0t -x)$),  decreases over time, so that $\dot{x}$ may reach zero again. 
If at some point $\dot{x}=0$, the kinetic friction coefficient is replaced by the static one, and oscillations (and any form of further sliding) are prevented.
We can compute the times $t_2$ such that formally, $\dot{x}(t_2)=0$:
\begin{align}
t_2=t_1+ 2\sqrt{\frac{m}{k_0}} \left(p \pi - \arctan \left(\frac{(\mu_s-\mu_k)L}{\sqrt{m k_0}V_0} \right) \right) \label{Eq:t2}
\end{align}
where $p\in \mathbb{N}$. 
The physical solution
corresponds to the first positive time that can be obtained, i.e.~$p=1$. 
At this time, the friction force (that always opposes motion, whichever direction it goes) increases from  $\mu_k L$ to $\mu_s L$ and motion stops.
The evolution of the block is once again controlled by the static equation of motion \req{StaticNewton}, and we are in the ``stick'' phase.

\begin{figure}[]
\begin{small}
\begin{center}
\def\svgwidth{0.9\textwidth}
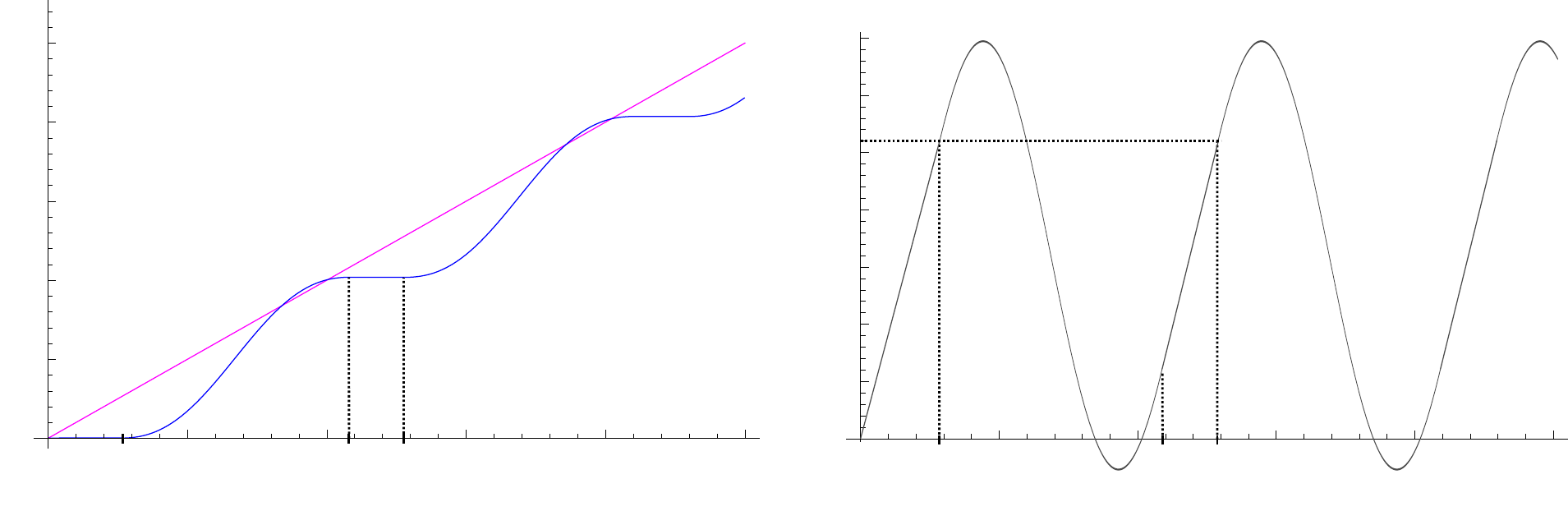
\end{center}
\end{small}
\caption{   {\footnotesize 
Stick-slip evolution of the block over time. 
Left:
Variations of the center of mass $x$ over time $t$ (solid blue) computed from \req{x_de_t}.
\label{Fig:x_de_t}
Right:Saw-tooth evolution of the stress during stick-slip motion.
Variations of the stress $\sigma= k_0(V_0 t- x)$ (solid grey line) computed from \req{x_de_t}.
\newline
The function $V_0 t$ (dashed purple) is given for reference. 
At time $t_1$, the threshold for the static force is reached and the block starts to move, with a decreased friction force $F_k$ (kinetic).
At time $t_2$, as velocity cancels, one needs to consider the static friction force. 
Loading then increases until the time $t_3$ where the threshold of static friction is once again reached.
Parameters used for the two figures are: $m=1, V_0=1, k_0=0.1, \mu_S L =0.52, (\mu_S-\mu_K) L = 0.2$.
Note that the slip phase seems long, but this is due to the parameters used: in particular, with a larger $(\mu_S-\mu_K)$ we get longer stick phases (and  -- relatively --  shorter, sharper slip phases).
Here we have a detailed view of the slip phase.
\label{Fig:sigma_de_t}
}   }
\end{figure}

The system will remain in the stick state until the time $t_3$ such that $V_0 t_3 - x(t_3) = \mu_s L/k_0$. 
Since the system has no memory (beyond $\dot{x}$), the dynamics at ulterior times is exactly periodic, as shown in Fig.~\ref{Fig:x_de_t}. 

In friction experiments, one usually measures the \textit{total shear stress} or total friction force, which is given by $\sigma = k_0(V_0 t - x)$. 
We present the evolution of $\sigma(t)$ in Fig.~\ref{Fig:sigma_de_t} (right), to be compared with experimental results, e.g.~for a mica surface pulled at constant velocity (Fig.~\ref{Fig:PerssonStickSlip}).
\includefig{10cm}{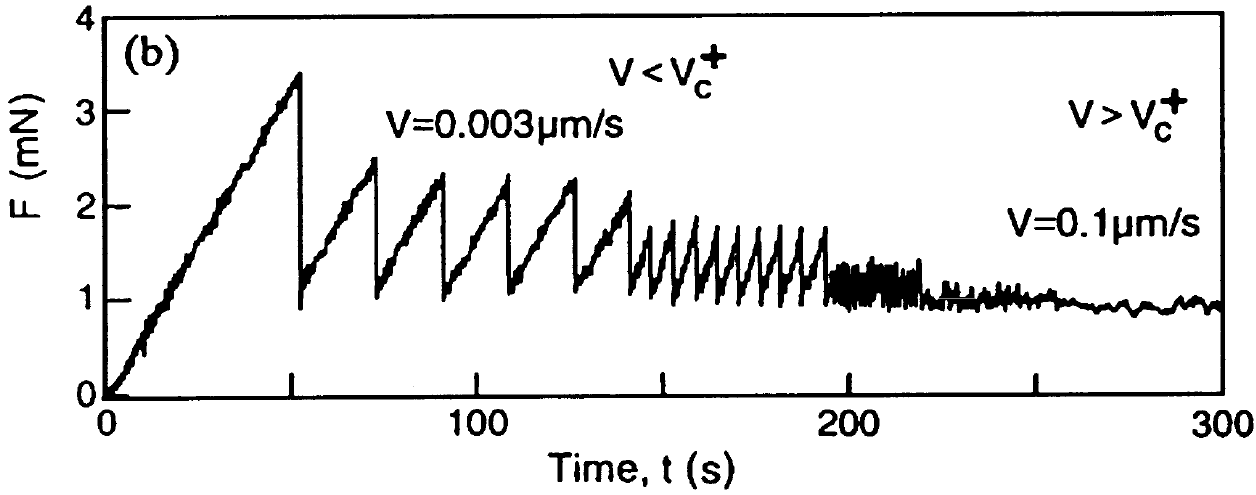}{
From \cite{Persson2000}.
Stick-slip motion of mica surfaces coated with end-grafted chain molecules (DMPE). 
The driving velocity is set to a few different values over time, while the stress or friction force (here denoted $F$, measured in $mN$) is measured.
When the spring velocity ($v$ or $V_0$) increases beyond $v_c^+$ the sliding motion becomes steady.
Here $v_c^+ \approx 0.1~\mu.s^{-1}.$
\label{Fig:PerssonStickSlip}
}

The difference between $\mu_s$ and $\mu_k$ generates a mechanical instability, in which the elastic energy provided by the driving is at times stored (static case, or ``stick'' phase) and at times released over a short\footnote{
Note that in Fig.~\ref{Fig:sigma_de_t}, the parameters chosen are such that the stick phase is rather short. 
For larger  $(\mu_S-\mu_K)$ we get longer stick phases, and  -- relatively --  shorter slip phases, since the duration of the slip phase is independent of $\mu_S$, but the loading time grows essentially linearly with it.
} period (kinetic case, or ``slip'' phase). 
This is the exact opposite of the more common situation of dissipative forces monotonously increasing with velocity so that a balance between drive and drag naturally yields stable solutions.

\subsubsection{Scope: limiting behaviours in $k_0$ and $V_0$ }

In the limits $k_0\sim \infty$ or $V_0\sim \infty$ we can derive simple analytical expressions, which allow to estimate the range of relevance stick-slip motion.

\paragraph{Duration of the Slip Phase}
The duration of the slip phase $t_2-t_1$ is obtained by developing \req{t2}:
\begin{align}
t_2-t_1 &\underset{k_0\sim \infty}{\sim}  2 \sqrt{\frac{m}{k_0}}\pi  + O(k_0^{-1})   
 \notag\\  
t_2-t_1 &\underset{V_0\sim \infty}{\sim}  2 \sqrt{\frac{m}{k_0}}\pi  + O(V_0^{-1})    
 \label{Eq:t2t1}
\end{align}
This means that the duration of the slip phase vanishes when $k_0 \sim \infty$, but remains finite  when $V_0 \sim \infty$.

\paragraph{Duration of the Stick Phase}
To fully predict how stick-slip behaviour depends on the parameters $k_0$ and $V_0$, we need to compare the durations of the slip and stick phases. 
The recurrent stick phase has duration $t_3-t_2$, which is different from $t_1$ because the initial condition we used is different from the system's state at $t=t_2$ (the spring is not extended at all at $t=0$, it is fully relaxed).
Starting from $t=t_2$ with \req{StaticNewton}, the static friction force will reach its threshold at the time $t_3$ such that $V_0 t_3 -x(t_2) = \mu_s L /k_0$ (we used $x(t_2)=x(t_3)$).
We thus have
\begin{align}
t_3-t_2 = \frac{\mu_s L}{k_0 V_0} + \frac{x(t_2)}{V_0} -t_2 \label{Eq:t3t2} .
\end{align}
It is useless to fully write down the exact value of $x(t_2)$, obtained by injecting \req{t2} in \req{x_de_t}.
Instead, we only give the relevant limits:
\begin{align}
x(t_2) \underset{k_0\sim \infty}{\sim} V_0~2 \sqrt{\frac{m}{k_0}}\pi +O(k_0^{-2}),
 \qquad~\qquad
x(t_2) \underset{V_0\sim \infty}{\sim} V_0~2 \sqrt{\frac{m}{k_0}}\pi +O(V_0^{-2}) ,
\end{align}
i.e.~the first\footnote{Actually, many higher-order terms are also equal in both developments. This is also true for the developments of $t_2-t_1$.}
 order term of both developments happens to be the same.
In this (common) term, we recognize the previous developments of \req{t2t1}:
\begin{align}
\frac{x(t_2)}{V_0 } \underset{k_0\sim \infty}{\sim} (t_2-t_1) + O(k_0^{-1}) ,
 \qquad~\qquad
\frac{x(t_2)}{V_0 } \underset{V_0\sim \infty}{\sim} (t_2-t_1) + O(V_0^{-1}) ,
\end{align}
where the dominant corrections come from \req{t2t1}.
We can inject these expressions in \req{t3t2}: $t_3-t_2  = \frac{\mu_s L}{k_0 V_0} + \frac{x(t_2)}{V_0} -t_2 = t_1 + \frac{x(t_2)}{V_0} -t_2  $:
\begin{align}
t_3-t_2  \underset{k_0\sim \infty}{\sim}  O(k_0^{-1})
\qquad~\qquad 
t_3-t_2  \underset{V_0\sim \infty}{\sim} O(V_0^{-1})  
\end{align}
This means that for a sufficiently rigid spring $k_0$ or a sufficiently high velocity $V_0$, the duration of the stick phase vanishes.

\paragraph{Existence of Stick-Slip}
More precisely, we see that in these limits, the duration of the slip phases is always large compared to the duration of the stick phase. 
For $k_0\sim \infty$, $T_\text{slip}\sim k_0^{-1/2} \gg T_\text{stick} \sim  k_0^{-1}$. 
For $V_0\sim \infty$, $T_\text{slip}\sim O(1) \gg T_\text{stick} \sim  V_0^{-1}$. 
We can conclude that in these limits, the system looses its stick-slip behaviour. 
In this very simple model, we did not include any viscous term of the form $-\eta \dot{x}$, and the friction law was assumed to be very simple. 
The addition of viscosity gives a sharper decrease of the stress in the slip phase, and smooths the displacement, which tends to suppress the stick-slip.
In more refined models, one may find a critical value of the spring stiffness, $k_0^c$ (which depends on $V_0$), as is observed in most experiments. 

The steady state can be obtained very simply by assuming a stationary behaviour. Using the kinetic equation: $0= k_0(V_0 t-x) - \mu_k L$, we get:
\begin{align}
x(t) = V_0 t + \frac{\mu_k L}{k_0}.
\end{align}

\subsubsection{Examples of Stick-Slip in Everyday Life}

There are too many examples of natural occurrences of stick-slip motion to make a comprehensive list here: we are only going to name a few.

The sound of squeaking doors originates from a motion of the hinge of stick-slip kind.
The sudden motion during each slip phase produces a sound pulse, and the periodicity of the stick-slip provides sound waves with a rather well-defined frequency. 
The fact that the phenomenon is not exactly periodic does not prevent us from classifying it as stick-slip, as the driving is still essentially monotonous.
We may notice that the computations from the previous section are validated by our everyday experience: the sound of a squeaking door can often be suppressed by opening or closing it fast enough. This is what could be expected from the fact that when $V_0\sim \infty$, the stick-slip behaviour is suppressed.

The same kind of mechanism applies to grasshoppers which produce their characteristic noise by rubbing their femur against their wings (or abdomen). The physics is essentially the same as for squeaking doors, only at different length scales. 

The bow of a violin also produces sound waves in a similar way (but it's a bit more complex, and of course the resonance of the violin's string plays an important role too).

The sudden stop of a car also involves stick-slip. 
Car brakes tend to squeal when pressed too hard: by the same mechanism as above, the gentle and rather noiseless sweep of the brake pads against the wheel (pure sliding) is then replaced by a high-pitched noise (stick-slip).
This could be expected from \req{t2}, where we see that an increase in the load $L$ is similar to a decrease in $V_0$, thus enhancing stick-slip behaviour.
The tires on the road can also (unfortunately) perform a sort of stick-slip: when the brakes are pushed so hard that they lock up the wheels (pure stick in the brake-wheels system), the tires will slide on the road (instead of rolling, i.e.~sticking to the road). 
In that case, the stick state corresponds to tires normally rolling, and the slip state corresponds to a sudden slip on the road, which can induce wear of the tires (loss of material and irreversible deformations) and ``skid marks''.
However, the intermittent behaviour (which defines stick-slip) is usually just due to an intermittent braking, so that the regularly spaced skid marks seen on roads are mostly not directly related to stick-slip, but rather are the consequence of the use of an Anti-lock Braking System.

We quickly mention a case of lubricated friction that has important implications in human health: bones articulations. 
In this system, stick-slip causes more damage than steady slip, something that can further increase the occurrence of stick-slip \cite{Lee2013}.
\\

In all of the above examples of stick-slip motion, the whole ``parallelepiped'' is considered as a single block. 
But stick-slip actually  occurs on many different length scales. 
Thus, even when the motion of the center of mass seems smooth, local ``stick-slips'' usually occur at the interface between the sliding solid and its substrate: for instance, groups of molecules or surface asperities can ``jump'' quickly in a stick-slip like fashion.
During ``steady'' sliding, these local slip events occur asynchronously, so that they essentially average out at the macroscopic level.
These local events may be probed indirectly, for instance, by studying the elastic waves emitted from the sliding interface.

We will give more details on these local events and their relevance for macroscopic friction in the following sections, but the impatient reader might jump directly to sec.~\ref{sec:asperities}.

\subsection{Ageing and Violation(s) of the Third Law}
\label{sec:ageing_violation}

\subsubsection{Observations}

\paragraph{Ageing in Static Friction}
\begin{figure}[]
\begin{small}
\begin{center}
\def\svgwidth{8cm}
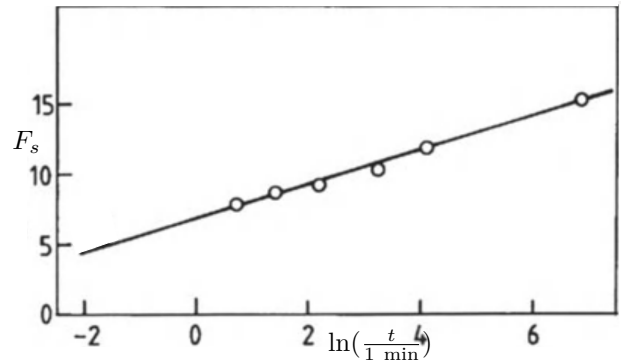
\end{center}
\end{small}
\caption{   {\footnotesize  
Static friction force versus $\ln(t)$. 
C.A. Coulomb's data (circles) is compared a simple law $A+B\ln(t)$ (solid line).
Data taken from \cite{Dowson1979}, retrieved from \cite{Persson2000}.
\label{Fig:CACoulomb_Ageing}
}   }
\end{figure}
As early as the 18th century, C.A. Coulomb measured and observed an increase of the static friction coefficient with the time of contact with the substrate. 
He found that the time dependence was rather well fit by a law $F_s = A+B t^\alpha$, with $\alpha \approx 0.2$ (see Fig.~\ref{Fig:CACoulomb_Ageing}).
However, in a more modern view one notices that his experimental data is also well fit by $F_s = A+B\ln t$, which is essentially the currently widely-accepted law for the ageing of contact in many materials\footnote{The common way to write this equation nowadays is rather $\mu_s = A+ B\ln(1+t/t_0)$, a notation that better preserves the need for homogeneity and the hatred for divergences.}. 
From these rudimentary results, we see that the strength of the contact initially increases quickly, but the time to double from $ \sim 10$ (arbitrary units) to $\sim 20$ can be extrapolated to be of $\sim 1$ hour.
More recent results about the ageing of contact at rest can be found e.g.~in \cite{Ben-David2010}. 
 
This time-dependence of the static friction with time of stationary contact is very important both in applications and conceptually. It could almost be nicknamed the 4th law of friction, due to its importance.

\paragraph{Velocity Weakening}
The third law is actually quite incorrect: how could friction be independent from the sliding velocity $v$, and at the same time, have a singularity at $v=0$? 
Upon closer inspection there is no singularity, but a smooth behaviour connecting the $v=0$ and the very small velocity regimes (as one would expect from intuition), via a friction force which decreases when the velocity increases (a rather counter-intuitive observation).
Typically, in the case of steady-state motion, the velocity-dependent friction law can be expressed in its most simplified form by:
\begin{align}
\mu_k = \mu^ * - A \ln \lp 1+ \frac{v}{V^*}\rp.
\end{align}
We further discuss the physical interpretation of this equation in the next subsection (\pp{sec:RSF1}).
For bare granite (see \cite{Kilgore1993}) parameters values range in the scales: $\mu \sim O(1)$ (typically $\mu^ *\approx 0.6$), $V^ * =1~\mu m .s^{-1} $, and $A\sim O(10^{-2})$.
These parameters can be extracted from experiments where steady-state sliding is obtained for various velocities, at different loads or other external conditions varied. 
Keeping the same setup for different velocities, one is especially interested in the relative variations of the Steady State friction coefficient $\mu_{ss} = \mu - \mu^ * =  - A \ln ( 1+ v/V^*)$, as shown in Fig.~\ref{Fig:velocity_weakening1}.
\includefig{10cm}{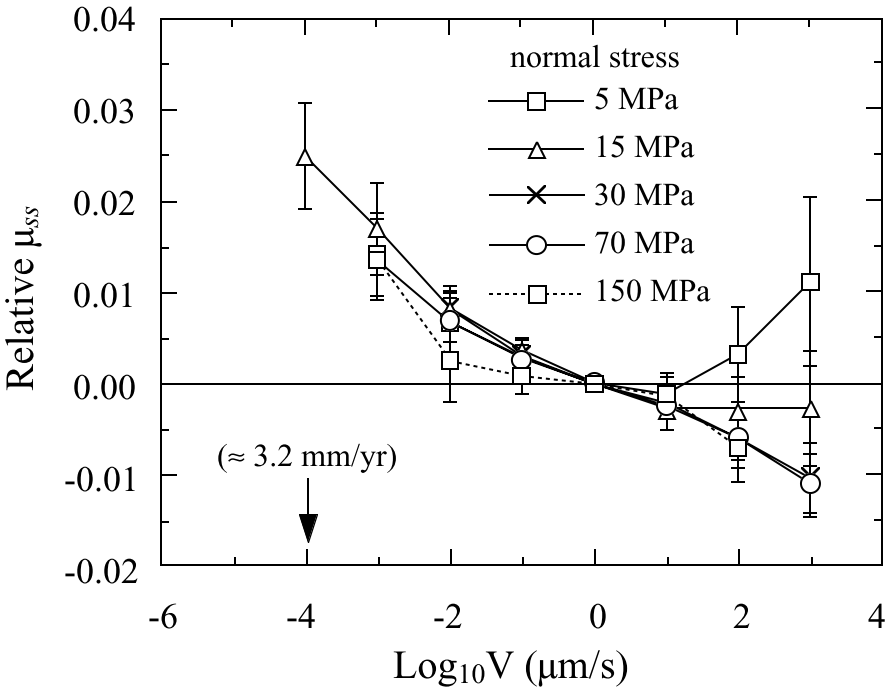}{
From \cite{Kilgore1993}. The relative variations of the steady state friction coefficient $\mu_{ss}$ at different velocities. Each curve corresponds to a normal stress (Symbols).
For loads larger than $30~\text{MPa}$, a logarithmic velocity weakening can be detected (approximately a straight line with negative slope).
For smaller loads of $5$ and $15~\text{MPa}$, there is velocity strengthening for $v>V^*$
\label{Fig:velocity_weakening1}
}

However, for most materials this continuous decrease can only be observed at very small velocities ($\sim 10 \mu m.s^{-1}$, see Fig.~\ref{Fig:velocity_weakening1}), and one needs rather good instruments to detect it in the lab. 
This also explains why it was not detected earlier.
An example of the crucial role of this weakening of friction with increasing driving velocity is found at the level of Earth's tectonic plates: as the imposed driving $\sim V_0$ is very small, plates perform stick-slip motion, with the slip phases corresponding to earthquakes.
The fact that friction is decreasing up to a limit velocity means that any initial motion of the plate triggers an instability which drives it up to this limiting velocity.
Understanding this instability of the statics is an important aspect of geophysics.
In the geophysicists' community, this decrease of friction with velocity is known as the \textit{velocity-weakening} effect.

\paragraph{Velocity Strengthening}
Let's mention also the \textit{velocity strengthening} regime (where friction increases with velocity) which is expected to occur at a higher velocity (which depends on other parameters as the load): see the right part of Fig.~\ref{Fig:velocity_weakening1}.
It is tempting to attribute velocity strengthening to viscous or hydrodynamical effects due to lubricants.
Actually, in the presence of lubricants the hydrodynamic theory predicts a friction force going as $\sim v^2$ at high Reynolds numbers (i.e.~at high velocities).
Furthermore, velocity strengthening can appear at much lower velocities via mechanisms completely independent from hydrodynamics.
A more reasonable explanation for velocity strengthening is the wear, which increases roughly linearly with velocity.
Wear may also produce an abundance of granular materials between the surfaces, which may also dissipate more energy by increasing the contacts and the sliding-induced deformation.
In this thesis we are only interested in the small velocity regime, and it is enough to know that beyond some limiting velocity, friction starts to increase instead of decreasing.
For a presentation of additional experimental results on various materials displaying velocity strengthening and some arguments to explain its origin, see \cite{Bar-Sinai2014}.

\subsubsection{The Rate- and State-Dependent Friction Law(s)}
\label{sec:RSF1}

From these diverse observations came the need to have a single constitutive law (or empirical law) that would encompass both the observed time dependence of static friction \textit{and} the velocity dependences of kinetic friction (velocity weakening or strengthening).
We now present this general phenomenological law.

In the general case of non stationary sliding velocity $v(t)$, the friction coefficient can be expressed in terms of the so-called \textit{rate and state friction law} \cite{Dieterich1979,Ruina1983}, where ``rate'' simply refers to the time derivative ($\dot{x}=v$) of the position and ``state'' refers to an internal variable which represents the quality of the contacts between the sliding solid and its substrate, $\theta(t)$ (also sometimes denoted $\phi(t)$).
A widely used form for the evolution of the variables $\mu, \theta$ is:
\begin{align}
\mu &= \mu^ * + a \ln \lp \frac{v}{V^*}\rp + b  \ln \lp \frac{V^* \theta }{D_c}\rp  \label{Eq:RSF0}\\
\frac{\partial \theta}{\partial t} &=  1 - \frac{v \theta }{D_c} \label{Eq:RSF1}
\end{align}
where typically, $V^ * = 1~\mu m.s^ {-1}$, $\mu^ *\approx 0.5$, $D_c \sim 1 - 10~\mu m$, and $a$, $b$ are dimensionless constants that need to be fit for each particular data set, but typically range in $a,b \sim O(10^{-3})$.
This is what is often called a ``constitutive relation'' for friction. 
We may note that \req{RSF0} is undefined at $v=0$. 
This can be problematic for computations, but this is compatible with the definition of friction as the normalized shear strength of a surface: there must be some slip at some scale for it to be measured.
Anyhow, \req{RSF0} is sometimes rewritten as
\begin{align}
\mu &= \mu^ * + a \ln \lp1 +\frac{v}{V^*}\rp + b  \ln \lp 1+ \frac{V^* \theta }{D_c}\rp  \label{Eq:RSF0bis}
\end{align}
to tackle this issue.

The above law is just one of several possible rate-dependent and state-dependent friction laws (RSF laws). 
Many variations are possible for the evolution of the state variable $\theta$. 
Keeping \req{RSF0}, we can have two other RSF laws by using one of these evolution equations for $\theta$:
\begin{align}
\frac{\partial \theta}{\partial t} &=  1 - \lp \frac{v \theta }{D_c} \rp^ 2, \\
\frac{\partial \theta}{\partial t} &= - \frac{v \theta }{D_c} \ln \lp \frac{v \theta }{D_c} \rp.
\end{align}
Each of these will give different behaviours when looking in details, but some of the main features are shared:
\begin{itemize}
\item In the steady state ($\partial_t \theta=0$), we obtain  $\theta^{ss} = D_c/v$.
Injecting it into \req{RSF0}, we get the steady state friction coefficient $\mu^{ss} = \mu^ * + (a-b) \ln \lp v/V^* \rp $. 
Depending on the sign of $a-b$, we will get velocity weakening or strengthening.
\item In the case of zero velocity ($v=0$), $\theta$ is a monotonously increasing function of time.
For instance, starting from $\theta(0)=0$, \req{RSF1} gives $\theta(t) = t$. 
This allows to account for the reinforcement of static friction over time.
\end{itemize}
These two shared features exactly answer to the initial need to reconcile static and dynamic observations.

\section{The Microscopic Origin of Friction Laws}
\label{sec:FrictionFromScratch}

Up to now, we have approached friction purely phenomenologically. 
At this point, the reader should be thrilled to learn about the fundamental mechanisms of friction. 
How come the friction force is not extensive in the surface of contact? 
What is the role of the load, and how come the dependence is exactly linear?
What are the mechanisms for ageing, in the static and dynamical cases? 
Are they related?
Can we find the form of the velocity-weakening law, ``from scratch''?

We are only going to give a few clues about these questions, since definitive answers are not always available: even though it has progressed a lot in the last 30 years, tribology still has many challenging questions to be answered.
Although we only present an overview of a sub-part of tribology, we will try to explain clearly the link between length scales, and how ``elemental'' objects and phenomena emerge from smaller and more fundamental ones.
This simple yet rather accurate description of friction is in large part due to Archard \cite{Archard1957}, with important improvements being very well summarized in \cite{Persson1996a,Persson2000}.

However, we won't explore much the nano-scale aspects of friction here: for reviews on nano-scale models of friction and experimental results on nano-tribology, see \cite{Vanossi2013, Capozza2013}. 
The resource letter \cite{Krim2002} contains accessible references to the relevant literature, as references are sorted and somewhat commented.
Besides, in this thesis we are interested in dry friction as opposed to lubricated friction: we explain how we may dismiss lubrication in Appendix \ref{App:lubricants_irrelevant}.

\subsubsection{Preliminary: What is The Atomic Origin of Friction?}

Small friction forces have been observed even for contacts of very few atoms: thus, it is natural to wonder about the atomic origin of friction. 
At the quantum level, there is no equivalent of ``friction forces'' between atomic clouds.
What prevents sliding at the atomic level are all the sorts of bond-formation mechanisms: chemical bonding,  Hydrogen bonds, van der Waals forces, etc.
At a larger level\footnote{
Note that we do not identify asperities and bonds. 
Bonds can be single-atomic contacts, whereas the term asperity commonly denotes micro-scale contacts.
Some bonds (as wet contacts) can be of the $\mu m$ length scale, as asperities.
We discuss these nuances in sec.~\ref{sec:asperities}.
 }, wet contacts develop capillary bridges, which are essentially liquid bonds developing due to surface tension and geometrical constraints.

In any case, the existence of bonds between surfaces in contact is an obstacle to the relative sliding of surfaces: in order to move, these bonds may first deform and at some point, break.
For a bond to break, the local force has to reach a certain threshold,
i.e.~there is an energy barrier or activation energy needed to perform local motion.
The macroscopic friction force thus emerges from these local energy barriers that have to be overcome to allow motion, so that the friction force is proportional to the number of bonds:
\begin{align}
F \propto N_{\text{bonds}}. \label{Eq:F_propto_Nbonds}
\end{align}
The intermittent nature of bonding at a local level is sometimes seen as a sort of local stick-slip occurring at the micro or nano scale (depending on the characteristic size of the bond).
However this is just an analogy: for most surfaces the local state (bound/unbound) is far from being periodic, and it is controlled mainly by surfaces' properties (and not inertia or internal stress).

Once a bond is broken, the energy is generally not recovered:  in general, no new bond is formed right after breaking.
Various detailed dissipation mechanisms can account for this ``loss'' of energy, the main ones being excitation of electrons and creation of phonons.
The energy lost in these processes can be converted into mechanical energy (elastic and plastic deformations) or directly into heat.
The dissipative nature of macroscopic friction originates from the irreversible part of these processes (even elastic oscillations dissipate energy via phonons).

\paragraph{Conclusion: Friction is Adhesion}
Aside from the relationship $F \propto N_{\text{bonds}}$, the main point of this very short discussion is that dry friction at the atomic scale can be reduced to  \textit{adhesion} (in the broad sense).
In other words, the continuum mechanics friction force 
simply emerges from 
the adhesion properties of the particles in contact at the solid-substrate interface.

\subsubsection{Outline}

In this Section (\ref{sec:FrictionFromScratch}), we will explain the three phenomenological laws with arguments based on simple microscopic mechanisms.

Since friction orginates from the adhesion of atoms that are actually in contact, the geometry of each surface is crucial.
We start our analysis by defining the main kinds of surface profiles in subsec.~\ref{sec:Roughness}, in particular we define the notion of \textit{algebraic roughness}, and provide experimental evidence of the strong roughness of most surfaces. 
This allows to understand naturally why the friction force is independent from the \textit{apparent contact area}, as specified by the first law.

In subsec.~\ref{sec:RealContactArea}, we discuss how the \textit{real contact area} evolves, and how different mechanisms (elastic or plastic deformation) for its evolution all lead to a linear dependence in the load (second law).
We also quickly discuss the role of fracture.

In the last subsection (subsec.~\ref{sec:Ageing}) we show how the third law is actually violated in experiments, explain why it is almost correct at the human scale and present some hypothetical microscopical mechanisms explaining this violation.

\subsection{Roughness}
\label{sec:Roughness} 

In the common sense, the \textit{roughness} of a surface or texture is ``how much the height profile deviates from the average height'', and it is often taken as a binary measure: things are either smooth or rough.
However, this ``definition'' implicitly promotes human length scales as references: for a height profile with a large spectrum of wavelengths, the human senses (tactile or visual) can only perceive variations over length scales larger than some threshold. 
Additionally, large wavelength variations are often considered as irrelevant for roughness ``to the eye''.

The concept of roughness as an objective measure of the texture properties of a surface is used in various areas of science and engineering, so that depending on the subject, its definition changes.
In engineering, the variation of the profile at small enough length scales is called roughness, at larger scales it is called \textit{waviness}, and at even larger scales it is called \textit{form}.
This is in contrast with the roughness as understood in most statistical physics works, where roughness is a measure embracing all length scales (as in \textit{fractals}), i.e.~where no particular length scale is favoured.

However, all definitions of roughness share a common goal:
 to reduce the tremendous amount of information contained in any given  height profile $\{h(x,y),~(x,y)\in \mathcal{D}\}$ to a few scalar variables at most  -- ideally just one, which would then be called ``the roughness''.
 The aim is of course to retain as much information as possible in these few variables. 
Depending on the symmetries expected from the profile, some definitions will be more or less fit for this purpose.

\subsubsection{Corrugation (or the false roughness)} 
The concept of \textit{corrugation} is in the neighbourhood of roughness. 
In common terms, corrugation is either ``the process of forming wrinkles'' or ``how much wrinkling there is''  at the surface of something.
Corrugation refers to how much some profile departs from being perfectly flat (as roughness does), but it implies the idea of periodicity or pseudo-periodicity for the height function $h$. 
Typical examples of profiles where corrugation rather than roughness is relevant are:
\begin{itemize}
\item Top surface of a pack of hard spheres (e.g.~glass beads), whether they are in perfect order (hexagonal lattice) or not.
\item Surface of an atomically smooth substrate (e.g.~mica surface): the electronic potential of the atoms forms regular bumps. The shape is essentially the same as for ordered glass beads, at a different length scale.
\item Underwater sand close to the shore can form a corrugated profile with characteristic lengths of a few $cm$.
\item Rail tracks tend to from quasi-periodic corrugations when excited at certain wavelengths. This increases the wear of tracks, because the ``bumps'' are extremely work-hardened, and thus \textit{fragile}. See \cite{Persson2000}, p. 41.
\item Fingerprints, or friction ridges, are ``wrinkles'' atop the fingers, which allow for a good perception of textures. See \cite{Scheibert2009} or \cite{Wandersman2011} for more details on the role of corrugation in tactile perception.
\end{itemize}
The crucial discrepancy between the concepts of corrugation and roughness is that the latter carries the idea of randomness, whereas the former one is usually a synonym for periodic behaviour.

In the case of the contact of two atomically flat surfaces (i.e.~flat at the atomic scale, without any one-atom bump or hole), there is a small corrugation due to the crystalline lattice. 
If the two lattices have lattice parameters (the length of one cell of the lattice) $a$ and $b$ such that $a/b$ is an irrational number, they are said to be \textit{incommensurate}. 
In this case, the perfect fit of the two lattices is impossible, because locations of strong bonding due to correspondence of sites of both lattice will be rare: in this case the relative corrugation ``potential'' may play an important role. 
The locations for strong bonding will appear to be random, but are indeed determined by the relative corrugation of the two surfaces.
Many friction models use this sort of corrugation to produce seemingly disordered, or random surfaces. 
One has to be careful with this interpretation, because this chaotic behaviour due to the incommensurate nature of substrates is ``not very random''.
If the ratio of lattice parameters $a/b \in \mathbb{Q}$, then the two lattices are said to be \textit{commensurate}, and then the interaction between the two will be quite strong, since the number of strong bonding sites will be extensive with the lattices size. 
We will discuss the case of commensurate surfaces a bit later, in sec.~\ref{sec:RealContactArea}.

\paragraph{Overhangs} 
\label{sec:overhangs}
\includefig{\textwidth}{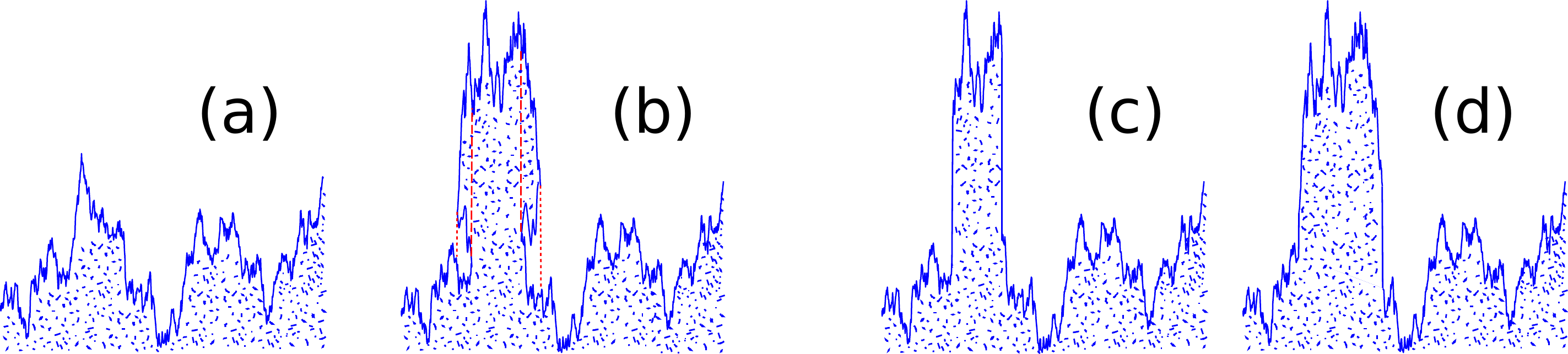}{
Various height profiles $h(x)$. 
The solid part is pictured by small dots. 
(a): ``normal profile'', without overhangs. 
(b): profile with one overhang. Two regularisations are suggested by dashed and dotted red lines.
(c): a first  regularization of profile (b), as suggested by the dashed lines.
(d): a second regularization of profile (b), as suggested by the dotted lines.
\label{Fig:overhangs}
}
The formalism used above (and below) implicitly assumes that the surfaces we consider do not have \textit{overhangs}, i.e.~for any point $(x,y)\in \mathcal{D}$ of the surface considered the function $h$ is uni-valued (not multi-valued). 
Another way to see this is to say that at any point, the local angle between the surface and the base-plane is less than or equal to $\pi/2$.
In case a surface actually has overhangs, many detection apparatus would measure a ``regularized'' surface (as shown in panel c and d of Fig.~\ref{Fig:overhangs}).

\subsubsection{Width described by a Single Scale: the Finite Roughness}

For essentially ``flat'' profiles or more generally in engineering applications (where only a certain range of length scales are relevant for friction), one may resort to simple measures of the height profile $h(x,y)$ in terms of its first moments or of some extremal values. 
The underlying assumption is that the variations of $h$ are ``finite'', i.e.~the moments of the distribution $h(x,y)$ (or even its cumulants) are finite, i.e.\footnote{This notation indicates that the function $h$ is a square-integrable function on $\mathbb{R}^2$: $\int_{\mathbb{R}^2} |h|^2 < \infty$.} $h \in \mathbf{L}^2(\mathbb{R}^2)$. 
We will see later how well this condition should be fulfilled for this sort of measures to be accurate.

Let us now precisely define a few measures of roughness. 
Consider a finite (but macroscopic) sample, defined by the domain $\mathcal{D} \subset \mathbb{R}^2$.
Suppose that the raw profile  $h$ is sufficiently regular: $h \in \mathbf{L}^2(\mathcal{D})$. 
To extract relevant variations of the height profile, we will generally subtract its average to $h$.
We use $\overline{X}$ to denote the space average of any quantity $X$: 
$\overline{h} \equiv \frac{1}{|\mathcal{D}|} \int _\mathcal{D}   h(x,y) \d x \d y$.
The most common measures of roughness are given by the following functions of $h$.
\begin{itemize}
\item The (average of the) absolute value: $R_a[h] \equiv \frac{1}{|\mathcal{D}|} \int _\mathcal{D}  \left |h(x,y)-\overline{h} \right| \d x \d y$.
\item The root mean squared $R_{\text{RMS}}$ or width: $w[h]=\sqrt{ \frac{1}{|\mathcal{D}|} \int _\mathcal{D} \left| h(x,y)-\overline{h} \right| ^2  \d x \d y  }$.
\item The maximum height of the profile: $R_t[h] = \underset{(x,y)\in\mathcal{D}}{\max}(h) -   \underset{(x,y)\in\mathcal{D}}{\min}(h) $.
\end{itemize}
Additional measures of the properties of a surface are e.g.~the skewness and the kurtosis of the profile, which come naturally as higher moments of the height function, seen as a probability distribution.

\paragraph{Relevance}

These kind of measures  -- taken as simple real values --  are well fit for engineering applications, where the roughness needs only to be assessed on a definite range of length scales, and for which the variations are usually mild in this range. 
In the case of small variations, the observables defined above are well-behaved, in particular they are essentially independent of the sample size.
However, in the more general context of the physics of friction, these measures fail to account for the rich behaviour of the surfaces we may be interested in, and more specifically, they can strongly depend on the sampling size.
Instead of looking at these functionals of $h$  as simple real variables, it is preferable to consider them as functions of the sampling length, and to extract a few relevant quantities from these functions.

In particular, in the case of numerous natural surfaces, these indicators would explode: the root mean square or width measurement for instance, $w$, would essentially diverge, if the distribution $h$ were to increase as a power-law.
We are about to see that this is indeed the case, at least in the applications we have in mind.

\subsubsection{Self-Affinity: the Algebraic Roughness}

\includefig{\textwidth}{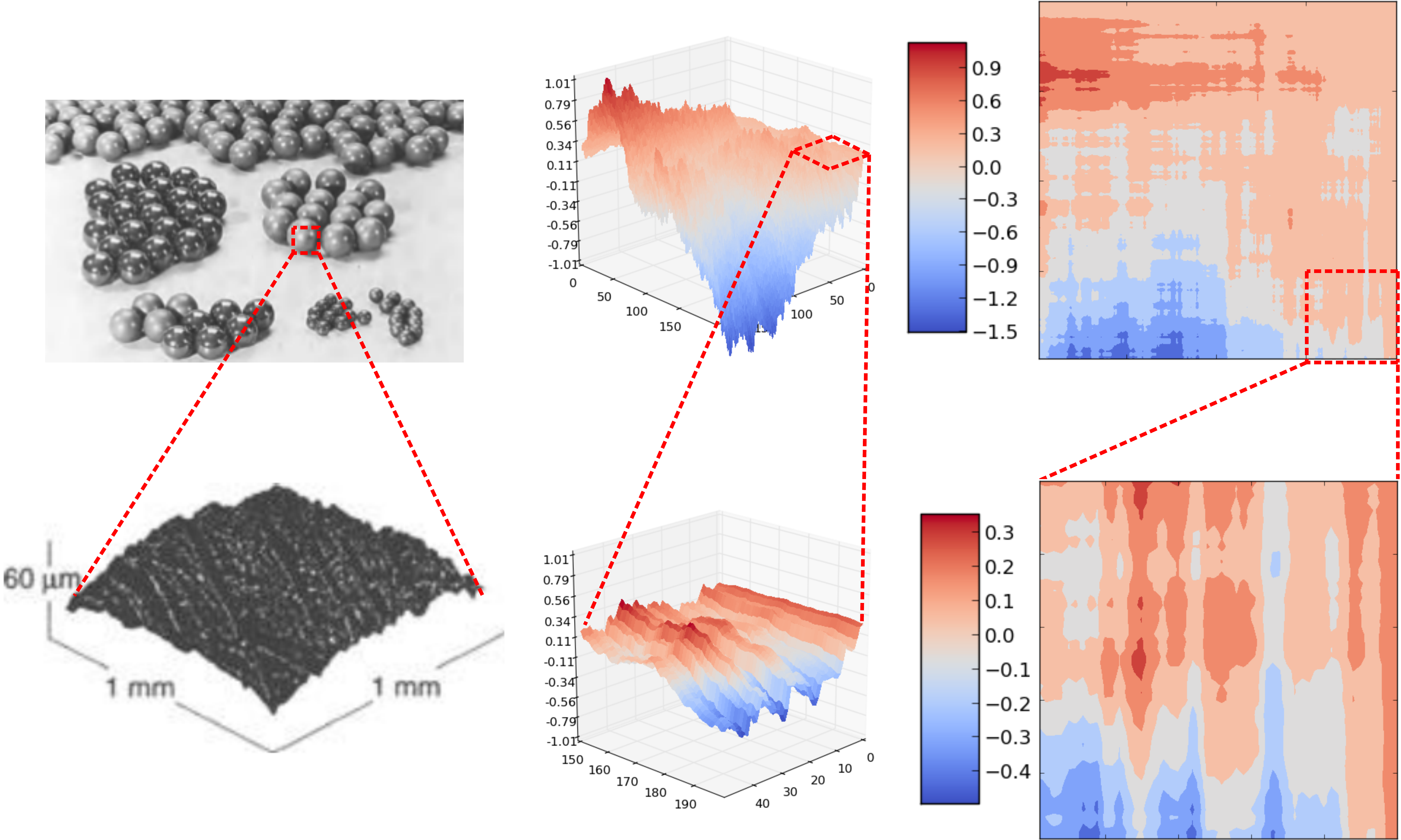}{
Left: Silicon nitride balls (used for bearings), finished (very smooth) and ``rough lapped'' (rougher). 
We zoom ($\sim \times 100$) on one of the rougher balls (below), and realize that the landscape is much rougher than it seemed, using a height resolution $\sim 10 \mu m$ 
\textit{(Images retrieved from \cite{Persson2000}, originally from \cite{Cundill1993})}.
\newline
Central (respectively right) panel: 3D view (resp.~``heat map'' colouring)  of the height profile for a toy model of surface (arbitrary units). 
We zoom ($\sim \times 3$) on a seemingly flat section, which reveals a rather irregular microscopic landscape upon closer inspection (below), similar to the large scale one.
Note that the preferred directions of our toy-surface (present at various scales) are an artefact of the generating procedure, they are not expected to be so strong for real materials.
\label{Fig:ceramic}
}
As can be observed for silicon nitride ceramic balls observed at the micrometer scale (see Fig.~\ref{Fig:ceramic}) the height profile of rather smooth objects can actually be quite irregular. 
We give a view of a rough surface from a toy model in Fig.~\ref{Fig:ceramic} (central and right panels).
This toy profile has large relative variations over a large range of length scales. 
Here we want to provide the tools for describing such kind of profiles.
Defining new tools will also allow us to characterize more precisely experimental observations. 
\\

First, we want to give clear definitions of the mathematical terms used, then see a few examples of surfaces that can be characterised using these definitions, and finally explain how we can quantitatively describe these surfaces efficiently, which will yield a natural definition of the (algebraic) roughness.

\paragraph{Self-Similarity (and related definitions)} 
Numerous objects have the property that they ``look the same'' at various length scales. 
Here we make this idea more precise by defining a few mathematical properties related to this idea.
Additional details are available in \ref{App:self_similarity}

Let us first define the property of self-similarity.
A function of two variables $g(x,y)$ is said to be self-similar if an only if (iff) it satisfies:
\begin{align}
g(x,y) = \Lambda_1 \Lambda_2 g(\Lambda_1^{-1} x, \Lambda_2^{-1} y) , \qquad \forall \Lambda_{1,2}>0,~\forall (x,y).
\end{align}
This is a \textit{re-scaling}, and it correspond intuitively (e.g.~for $\Lambda>1$) to do two things at the same time: ``zoom out'' in the $x$- and $y$-directions and to magnify (or also ``zoom in'') in the $g$-direction.
Self-similarity is a very stringent constraint, since the re-scaling in different directions has to be exactly the same.

A more general property defining objects with ``similar'' appearance at different length scales is \textit{self-affinity}.
A function of two variables $g(x,y)$ is said to be self-affine iff:  
\begin{align}
g(x,y) = \Lambda_1^{b_1} \Lambda_2^{b_2} g(\Lambda_1^{-1} x, \Lambda_2^{-1} y) , \qquad \forall \Lambda_{1,2}>0,~\forall (x,y),
\end{align}
where $b_1,b_2$ are the self-affinity or scaling exponents related to the affine transformation. 
This may be referred to as ``anisotropic'' self-affinity, but this wording is misleading, because even for $b_1=b_2 \neq1$, we already have an affine transformation (and not a similarity transformation)\footnote{
Please note that in part of the literature, these two concepts are sometimes mistaken for one another, or simply melted and seen as equivalent. 
When considering functions, it seems quite natural that the ordinate and abscissa do not share the same scaling exponent, so that considering self-affinity seems very natural. 
However, when considering geometrical objects such as self-similar or self-affine objects, the distinction becomes important. Not all \textit{fractals} are \textit{self-similar fractals}.
}.
We see that self-affinity is an anisotropic transformation which contains self-similarity as a special case ($b_1=b_2=1$).

Self-affinity is a rather general property, however it is interesting to note that it only allows to compare fully deterministic objects. 
If we are interested in a random process, we need an additional definition: \textit{statistical self-affinity}.
This is especially relevant to characterize a real surface (which is highly heterogeneous, i.e.~random).
A surface profile is said to have a roughness exponent $\zeta$ when it is statistically self-affine, i.e.~when:
\begin{align}
g(x) \overset{\text{Law}}{=} \Lambda^\zeta g(\Lambda^{-1} x), \qquad \forall \Lambda>0,~\forall x,
\end{align}
where the equality is ``in Law'' (for the random variables as distributions, not realization per realization). 

A generic example of mathematically well-defined stochastic process which is  statistically self-affine is the \textit{fractional Brownian motion} (fBm).
To give the interested reader more insight into statistcal self-affinity, we study the fBm in Appendix \ref{App:fBm}.

\subsubsection{Structure Factor}

\begin{figure}[]
\begin{small}
\begin{center}
\def\svgwidth{9cm}
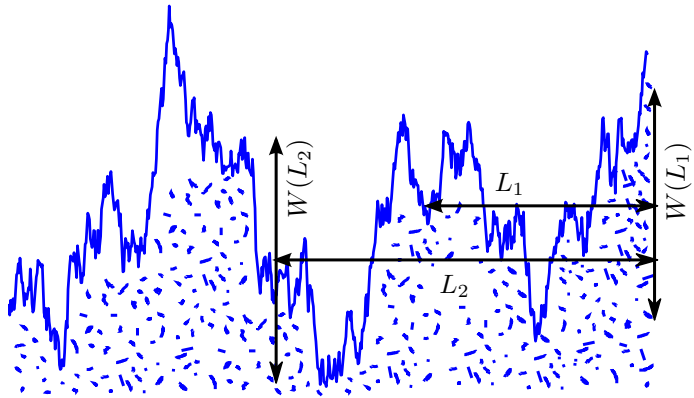
\end{center}
\end{small}
\caption{   {\footnotesize 
Illustration of the width and its dependence on the sample length $L$.
Depending on the definition of the width (or ``roughness''), the precise value of $w(L)$ will defer.
However, for an algebraically rough surface, all definitions will display a roughness exponent $\zeta$ such that $w(L)\sim L^zeta$.
\label{Fig:width_roughness}
}   }
\end{figure}

To describe height profiles with the statistical self-affinity property, 
one needs to extend the tools previously introduced.
For instance, the root mean squared $w$ (``width'') of the height profile $h(x)$ is the square root of the second moment of the distribution computed in \req{second_moment}.
For a surface being statistically self-affine at least over the range $x\in[0,L]$ with a roughness exponent $\zeta$, we thus have a width $w[h,L] = L^\zeta$ (see Fig.~\ref{Fig:width_roughness} for a concrete illustration).
This is obviously a problem, since an observable that explicitly (and much strongly) depends on the sampling size is clearly ill-defined.

The solution is to acknowledge the self-affine nature of the surface, and to use the exponent $\zeta$ to define the roughness, which is possible since
\begin{align}
\zeta \underset{L\gg 1}{\sim}  \frac{\ln(w[h,L]) }{\ln (L)}
\end{align}
does not depend on the precise value of $L$, as long as $L\gg 1$.
However, it is important to note that not all rough surfaces are exactly statistcally self-affine with a unique exponent over all length scales.
There are usually cutoffs (lower and upper) to the self-affine behaviour, and the exponent may even have two distinct values over two distinct ranges!
Thus, in order to be valid for a wider class of rough profiles, this definition of roughness needs to be extended.\\

A very general observable that helps measuring the roughness of a given height profile is the \textit{structure\footnote{Originally, the concept was used in crystallography, where \textit{structure} obviously refers to the crystalline structure. 
The idea of looking at the spectrum in Fourier space, and at the typical energy of each mode has since spread in many disciplines.} factor} $S(q)$. 
This is not a roughness, since it is not a scalar, but a function (which inherently contains more information than a single scalar).
The idea is simply to look at the energy associated to each mode in the spectrum of the height distribution. 
For a  $d$-dimensional 
profile $h(\mathbf{x})$, assuming periodic boundary conditions (for simplicity) in a system of lateral length $L$,  the averaged structure factor is defined as:
\begin{align}
S(\mathbf{q})
& \equiv \overline{ \frac{1}{|\mathcal{D}|} \left| \int_\mathcal{D} \d^ d \mathbf{x}~h(\mathbf{x})~e^ {-i\mathbf{q} \mathbf{x}} \right| ^2 }\\
&=  \int_\mathcal{D} \d^ d \mathbf{x}~\overline{h(\mathbf{x}) h(\mathbf{0}) }~e^ {-i\mathbf{q} \mathbf{x}} 
\end{align}
where $\mathbf{x}$ is the $d$-dimensional coordinate, $\mathcal{D}$ is the domain considered and where translational and rotational invariance ensure that the (spatial) frequency $S(\mathbf{q})$ only depends on $q=|\mathbf{q}|$, via  $q=2\pi n/L, n \in \mathbb{N}$. 
The average $\overline{X}$ is the average of $X$ over many samples.
For any self-affine process with exponent $b=\zeta$, we have $h(x)\sim x^\zeta $ up to a random phase  
so that we 
get: 
\begin{align}
S(q) \sim q ^{-(d+2\zeta)},
\end{align}
so that aside from finite size effects (at short and large wavelengths), it is a pure power-law (see e.g.~\cite{Kolton2009}).
The measure of the structure factor is a robust way to estimate roughness. 
A nice feature of $S(q)$ is that if the profile considered is actually not self-affine, or if it has two regimes with different exponents of self-affinity, it can be seen immediately, as for example in Fig.~\ref{Fig:paper}\\

From now on, we will be interested solely in this last sort of roughness, so that ``rough'' will refer to statistically self-affine surfaces, and $\zeta$ may be called the roughness.
Except when explicitly stated otherwise, the surfaces we will consider are rough over a large range of length scales.

We will discuss examples of rough interfaces produced by theoretical models in later sections. 
For an example of concrete use of the structure factor and some precise results on the roughness of a one-dimensional elastic line in disordered medium, see \cite{Ferrero2013}.

\subsubsection{Experimental Examples of Rough Surfaces}
\label{sec:roughness_examples}

Now that we have defined the appropriate tools, we can discuss real observations more seriously than with Fig.~\ref{Fig:ceramic}.
In Fig.~\ref{Fig:Maloy}, the roughness of some surfaces of brittle materials (close to some cracks) is observed.
\includefig{0.5\textwidth}{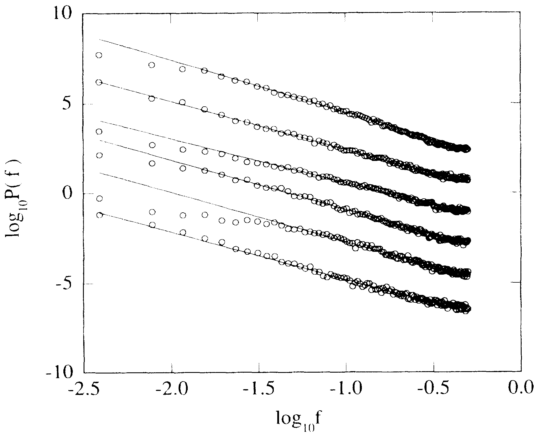}{
From \cite{Maloy1992}.
Roughness of surfaces of six different brittle materials, close to the fracture area (crack).
Measurement of the height profile along one-dimensional cuts in the direction perpendicular to the crack.	
The ``power spectrum'' $P(f)$ of the profile is exactly what we defined as the structure factor $S(q)$.
The log-log plot shows the dependence of $P(f)$ in the wavelength or space frequency $f$. The roughness $\zeta$ is extracted from the fit $P(f)\sim f^{-(1+2\zeta)}$. 
\label{Fig:Maloy}
}
If Fig.~\ref{Fig:paper}, the roughness of two-dimensional surfaces is measured for various materials, and we see how the structure factor can help to determine to what extent a surface is really self-affine.
\includefig{\textwidth}{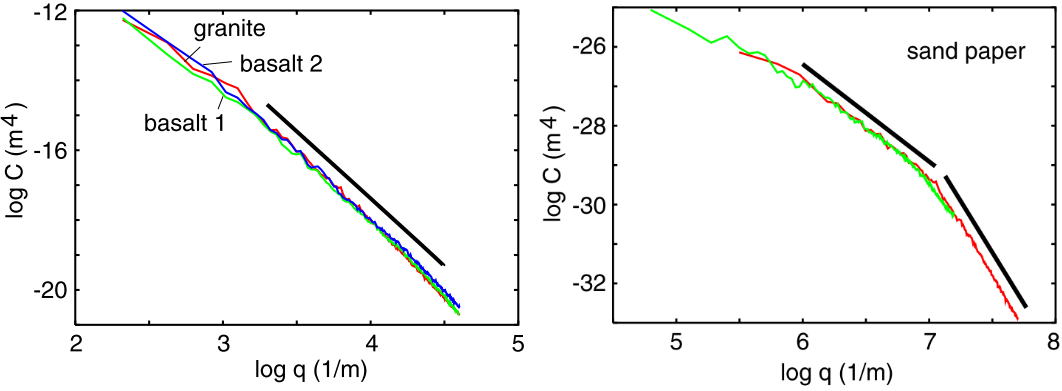}{
From a recent and excellent review on roughness, namely \cite{Persson2005} (\copyright IOP Publishing.  Reproduced by permission of IOP Publishing.  All rights reserved).
Optical measures (left panel and green curve of right panel) are combined with AFM (Atomic Force microscope) measurements (red curve of the right panel). 
The correlation function can be identified with the two-dimensional structure factor, here denoted $C(q)$. 
A fit is done to evaluate the fractal dimension, which is found to be $D\approx2$ for basalt and granit (left panel) and $D\approx 2.2$ for sandpaper at $\log q<7$ (right panel). This corresponds (for these 2D surfaces) to roughness given by $\zeta=3-D$.
Notice how there are two regimes for sandpaper, which are easily identified thanks to the use of the structure factor.
\label{Fig:paper}
}
From
 these examples of self-affine surfaces, we begin to understand why the friction force is independent from the apparent contact area: since most surfaces are very rough, they can touch each other only at few points.
If friction truly happens only where the surfaces meet, it must be proportional only to this real contact area, which we now expect to be much smaller than the apparent one.
We will explain this clearly in the following section.

\subsection{Real Contact Area} 
\label{sec:RealContactArea}

We have seen that the apparent contact area has probably little to do with the real one, and that only the latter is involved in friction. 
Here we want to compute this real contact area from macroscopic measurements.

Most people have the idea that ``smooth surfaces slide better''.
So, let's imagine the extreme case of two perfectly flat, clean and commensurate surfaces. 
What would happen if we were to put them in contact, and then apply some shear ?
The answer is that we would simply observe cold welding, i.e.~the boundary atoms would form bonds between the two surfaces. Bonds could be chemical, or just van der Waals forces\footnote{The relevance of van der Waals forces at the \textit{nano}scale has been questioned recently in \cite{Mo2009}: \textit{``friction is controlled by the short-range (chemical) interactions even in the presence of dispersive [van der Waals] forces''}.}.
If at least one of the materials has some impurities,  the shear stress necessary to obtain some strain  (deformation) would be essentially the yielding stress of the weaker of the two materials, and the shear would occur in the bulk of it, instead of occurring in the contact plane.
This simplistic example illustrates how friction would be incredibly huge, if contact was to truly occur on the complete \textit{apparent area of contact}. 
Notice that in this ideal case, ``friction'' would be proportional to the apparent contact area.
From now on when we discuss the contact area, it will be implicitly assumed that we do \textit{not} refer to this apparent area of contact.

Stepping back a little from this very extreme example, if a surface is flat except from few \textit{asperities}\footnote{
Asperities, \textit{contacts} or \textit{junctions} are all words that designate the small ``bumps'' at the top of any surface, which are responsible for the \textit{true} contact between solid and substrate.
For a rough surface, they are the top ``peaks'' of the profile.
} of approximately the same height, one may expect that the very few ``true'' contact points will allow for very low friction.
However, imagine this surface is slowly driven down towards another one with similar design (or completely flat).
As soon as the macroscopic load would be a bit more than zero, the local pressure at the asperities would quickly become enormous, since it goes as the inverse of total (true) area of contact. 
This would result on the plastic yielding of asperities, i.e.~in irreversible deformations at the atomic level, instead of reversible elastic deformation.
The ``peaks'' would be crushed, flattened, so that in the end we would have the flat solids separated by few spots of one-layer flattened asperities, resulting once again in a large contact area. 
Furthermore, if the distance between the two flat solids is indeed of only one atomic diameter, the van der Waals interactions might once again play some role by further increasing the macroscopic adhesion force.

Thus, we see that very smooth  -- nearly atomically smooth --  surfaces, contrary to popular belief, do not slide well. 
Another common idea is that very rough surfaces slide badly. 
Actually, this one is true: for a surface with macroscopic height oscillations, i.e.~``macroscopic corrugation'' or \textit{form} (or \textit{waviness}), 
the energy barriers that one needs to overcome to slide through are so high that they prevent any easy sliding. 
Even if the microscopical properties of the solids are such that the microscopic friction coefficient is small, for corrugated profiles, the surfaces will be interlocked with one another, and the macroscopic friction force will be high.
This is the case for ``roughcast'' (or for ``pebbledash''): even with a good microscopical surface treatment, two such surfaces rubbed against each other would still slide very badly.
In this sense, the engineering definitions of the waviness and form are appropriate to eliminate the large length scales contributions to friction, which can involve mechanisms other than ``small scale'' friction.

\subsubsection{Asperities at the Microscale}
\label{sec:asperities}

As it has been mentioned earlier, asperities are the  small ``bumps'' on top of a surface which are responsible for the \textit{true} contact between solid and substrate.
By definition, a \textit{contact} is the point where the two surfaces meet and where bonds can form.
The concept of \textit{junction} involves the idea of welding, which is made easier by the high pressures at the asperities.
For a rough surface, asperities are typically the top ``peaks'' of the profile.

It is important to notice at this stage that bonds and asperities are not the same thing. 
On the one hand, the notion of \textit{bond} covers length scales from the atomic size (a few \r{A}ngstr\"oms, $\sim 10^{-10}m$) to capillary bridges (up to fractions of $mm$, $\sim 10^{-4}m$). 
A bond is an elementary unit: it can get weaker or stronger due to external conditions, it can break, but it does not have relevant sub-elements.
On the other hand, the notion of asperity refers to an entity generally described by continuum mechanics: the contact between two asperities is of a size such that in the range of loading conditions studied, it can not merge with a neighbouring one. Typically, the radius of the contact area of an asperity is $\sim 10 \mu m$.

On a first approach, asperities can be seen as the building blocks of the contacts responsible for friction. 
Then, the \textit{true contact area} or \textit{asperity contact area} \footnote{What we call asperity contact area used to be consider the true contact area.} 
can be considered to be the whole area of contact between asperities, as depicted in Fig.~\ref{Fig:aperite_convex_hull}.a.
A refined approach consists in considering the inner dynamics of the contact. 
Then, the real contact area or atomic contact area is just the sum of the individual contact area of each atomic bond (See Fig.~\ref{Fig:aperite_convex_hull}.b)
The difference between these two approaches has been pointed out in \cite{Mo2009}, and opens promising avenues for a better understanding of friction, especially for nanoscale objects.
\includefig{0.85\textwidth}{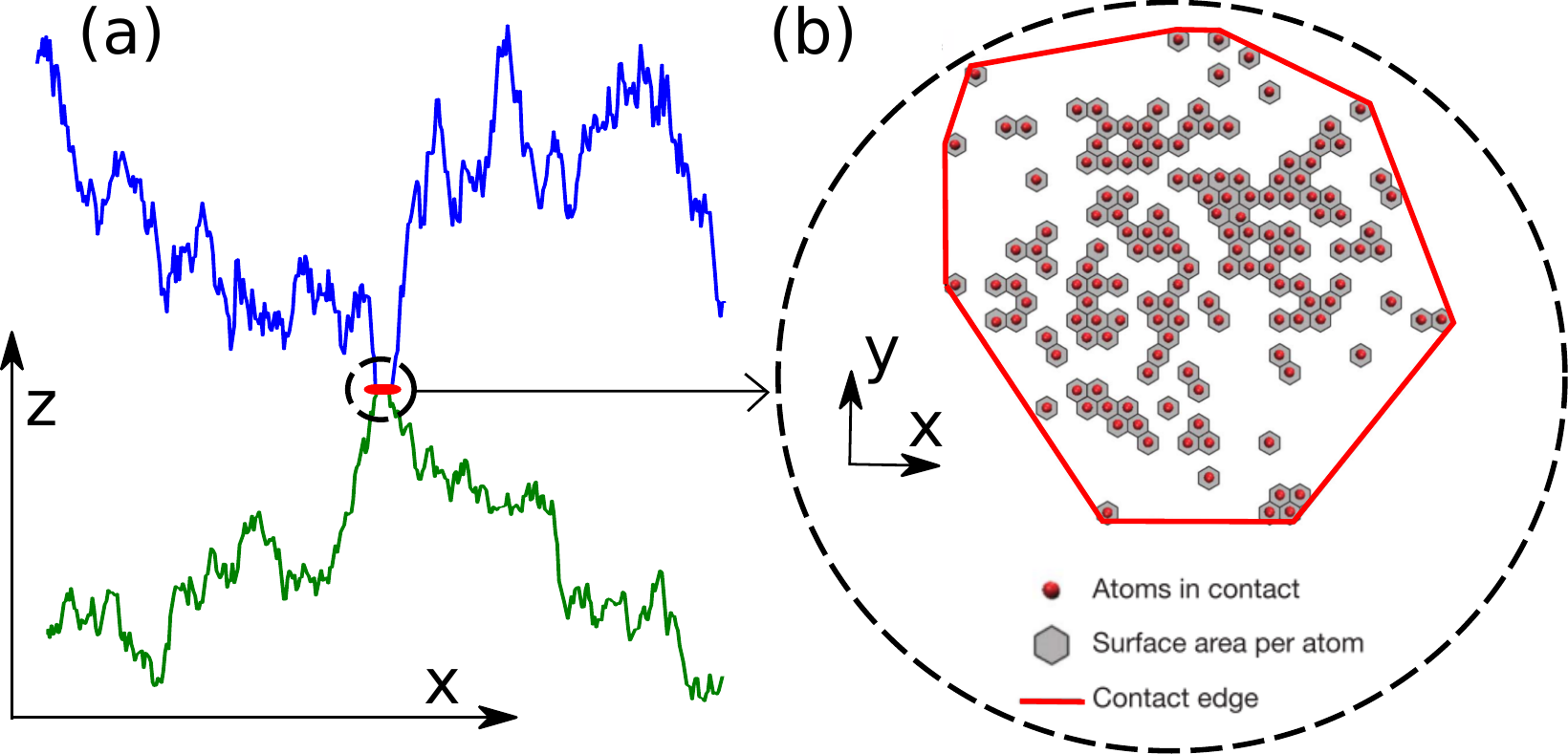}{
Left: two profiles with algebraic roughness ($\zeta=0.5$) enter in contact. 
The junction is highlighted in red.
\newline
Right: Schematic view (from \cite{Mo2009}) of the junction, from above. 
Over the area of the junction (the ``real contact area''), not all space is actually covered in bonds. 
The atomic bonds (red dots) actually cover only the grey area. 
From outside, the contact area is naturally mistaken for the contact edge (solid red line), i.e.~for the convex hull enclosing all the atomic bonds. 
In most studies, the ``real'' or ``true'' contact area implicitly refers to this convex hull, not to the grey area.
\label{Fig:aperite_convex_hull}
}

However, the notion of asperity is often not only sufficient, but more relevant than that of bond, for several reasons.
First, the fact that the real contact area is not equal to the apparent asperity area is not truly an issue, since in calculations it is (often) automatically the real contact area which is involved.
Second, asperities are the (pseudo) elementary blocks which pin the surfaces together: their scale appears as a natural length scale in many aspects of friction, and is way more practical to handle than the atomic scale. 
Consequently, it is often sufficient to study their dynamical behaviour alone (elastic and plastic deformations).
Third, asperities are large enough that one can apply most continuum mechanics to them: this is very handy.
Hence, we will mainly discuss the behaviour and dynamics of asperities in what follows. 
For a review on nanoscale models of friction and experimental results on nano-tribology, see \cite{Vanossi2013}, or the resource letters \cite{Krim2002} which contains accessible references to the literature.

\subsubsection{Role of Plastic Yielding at the Solid-Substrate Interface}

Consider a substrate upon which we set an object of which the lower surface is rough in the sense defined earlier (i.e.~it has a statistically self-affine surface).
As we approach the solid\footnote{At this point, it does not matter to know precisely the profile of the substrate: whether it is flat or rough with the same exponent as the upper solid, we can subtract the two profiles and consider the result as the effective profile for the solid, and consider the effective profile of the substrate to be flat.}
 from above, at first there is only a single asperity in contact.
At this asperity, the pressure $p_1$ over the (real) contact area $A$ is given by $p_1=L/A$,
where $L$ is the \textit{macroscopic} load. 
For a typical asperity of diameter $a\sim 10 \mu m$, we have an asperity area $A \approx 10^{-10} m^2$.
For a load given by the weight of $1kg$, $L \approx 10 N$,
 so that $p_1\approx 100\times10^{9} N/m^2$.
For reference, the yield stress\footnote{The yield stress is the stress that one needs to apply in order to obtain plastic yield. 
In the context of these estimations, the relevant quantity is the \textit{penetration hardness} or \textit{indentation hardness}.
The typical measure protocol is that of Vickers: on the sample, an indentation is performed with a tetrahedron in diamond. 
The stress needed to perform the indent is the indentation hardness.
}
 for diamond is $\sim 80 \times10^9 N/m^2$, and for steel it is between $1$ and $7\times10^9 N/m^2$ (it depends on the quality of the steel).
As the pressure in the contact area is larger than the yield stress, this single asperity must yield plastically, i.e.~it is smoothly crushed by the upper solid. 

\begin{figure}[]
\begin{small}
\begin{center}
\def\svgwidth{\textwidth}
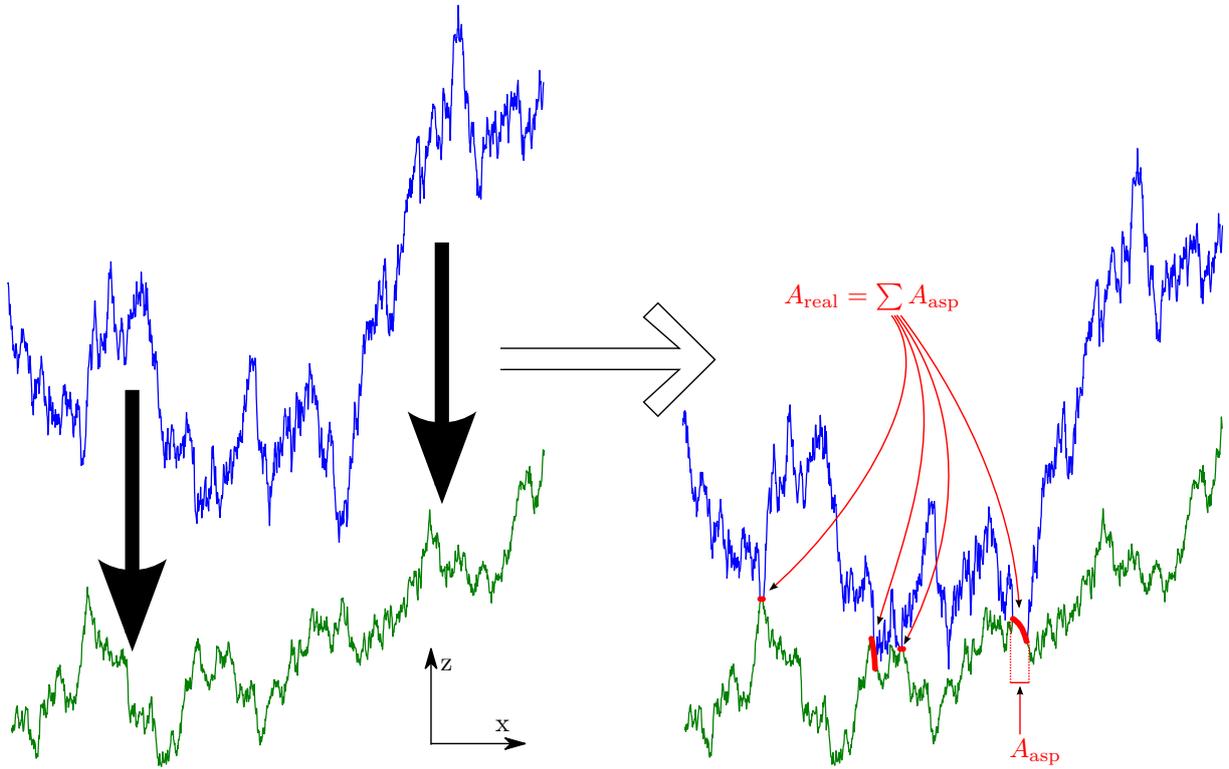
\end{center}
\end{small}
\caption{   {\footnotesize 
Schematic description of two rough surfaces (Left, $\zeta=0.5$) squeezed together.
They make more and more contacts (Right, highlighted in red) until $A_{\text{real}} = L / \sigma_c $.
The area of each asperity, $A_{\text{asp}}, $ is the projection (dotted red line) of the contact onto the $(x,y)$ plane.
This area is much smaller than the total area.
\label{Fig:plasticity}
}   }
\end{figure}

As the upper solid goes further down, it will encounter other asperities, which will increase the contact area. 
As long as the pressure remains larger than the yield stress, the solid will deform plastically.
When the contact area is large enough to strike a balance between pressure at asperities and yield stress, plastic deformation will stop. 
This gives us a natural formula for the real contact area:
\begin{align}
A_{\text{real}} = \frac{L}{\sigma_c},
\end{align}
where $\sigma_c$ is the yield stress or indentation hardness of the softer of the two materials.
To be concrete, let's continue with our mass of $1kg$, on top of a table of the same steel (or any other stronger material).
Let's assume it is made of steel with $\sigma_c=10^9N/m^2$.
The real contact area is then $A_{\text{real}} =10^{-8}m^2 $, which is completely independent from the apparent contact area. 
Surprisingly, this corresponds to only $\approx 100$ asperities of unitary area $\sim 10^{-10}m^2$.
We may compare this contact area with the apparent one $A_{\text{app}}$ by assuming the steel to be shaped as a parallelepiped, for instance with the dimensions $10cm\times 10 cm \times 1cm$ (the density of steel is $\rho \approx 10g/cm^3$). 
In this case we have $A_{\text{real}} = 10^{-4} cm^2 \ll A_{\text{app}}  = 100 cm^2$, or also $A_{\text{real}} / A_{\text{app}} = 10^{-6}$, i.e.~the real contact area is only a tiny fraction of the apparent one. 
See Fig.~\ref{Fig:plasticity} for an illustration.

When we slide a solid over a ``fresh'' area, or when the frictional wear changes the asperity landscape, and more generally as soon as some surfaces meet for the first time, the picture presented above will also be valid.
As we have seen earlier \req{F_propto_Nbonds}, $F\propto N_\text{bonds}$. The number of bonds is essentially proportional to the area of real contact, so that in the end, $F \propto N_\text{bonds} \propto A_{\text{real}}  \propto L/\sigma_c$, i.e.~we found Amonton's second law.\\

In the above cases, we have assumed that the elastic deformations of the materials are negligible. 
This is perfectly correct as long as we start from a state with few contacts: the pressure is so high that local strain is large, and most of the deformation is plastic.
Another way to put it is to say that contacts are in a state of incipient plastic flow, i.e.~that they are at their plasticity threshold (or way beyond).
When we are around the equilibrium state with $A_{\text{real}} = L / \sigma_c $, however, elastic deformations can become relevant.

\subsubsection{Role of Elastic Deformation at the Solid-Substrate Interface}

In several cases, it is elasticity rather than plasticity which controls the evolution of the surface area.
In the friction of rubber, the very low elastic modulus makes it very difficult to plastically deform the rubber, so that elastic forces prevail\footnote{
See \cite{Persson2001} for a study of this extreme case that is rubber friction. Be careful that the theory has evolved since, in particular one should consult \cite{Persson2005} for accurate results.}
For a surface that is very smooth, in the sense there are many asperities at the top with approximately equal height, one may expect the real contact area to be larger than what is expected from the plastic yield reasoning.

Another natural question is to ask 
what happens in the following ``extra-load'' experiment.
In the ``extra-load'' setup, we set our steel block onto a (hard and flat) table (the load is $L=L_1$), then press it with an extra load of $1kg \Rightarrow L_1= 10N$ (i.e.~we double the total load), then remove the extra load ($L=L_1$).
According to our reasoning, the asperities have been crushed to a point where $A_{\text{real}} =2 L_1 / \sigma_c $, 
so that we would naively expect the real contact area to be double of what is expected from the simple, current load $L_1$  (such an effect of memory of the previous loading is actually not observed, not to this extent at least).
We have just produced a ``very smooth'' surface as mentioned above, since the top asperities have all the same height.
\includefig{\textwidth}{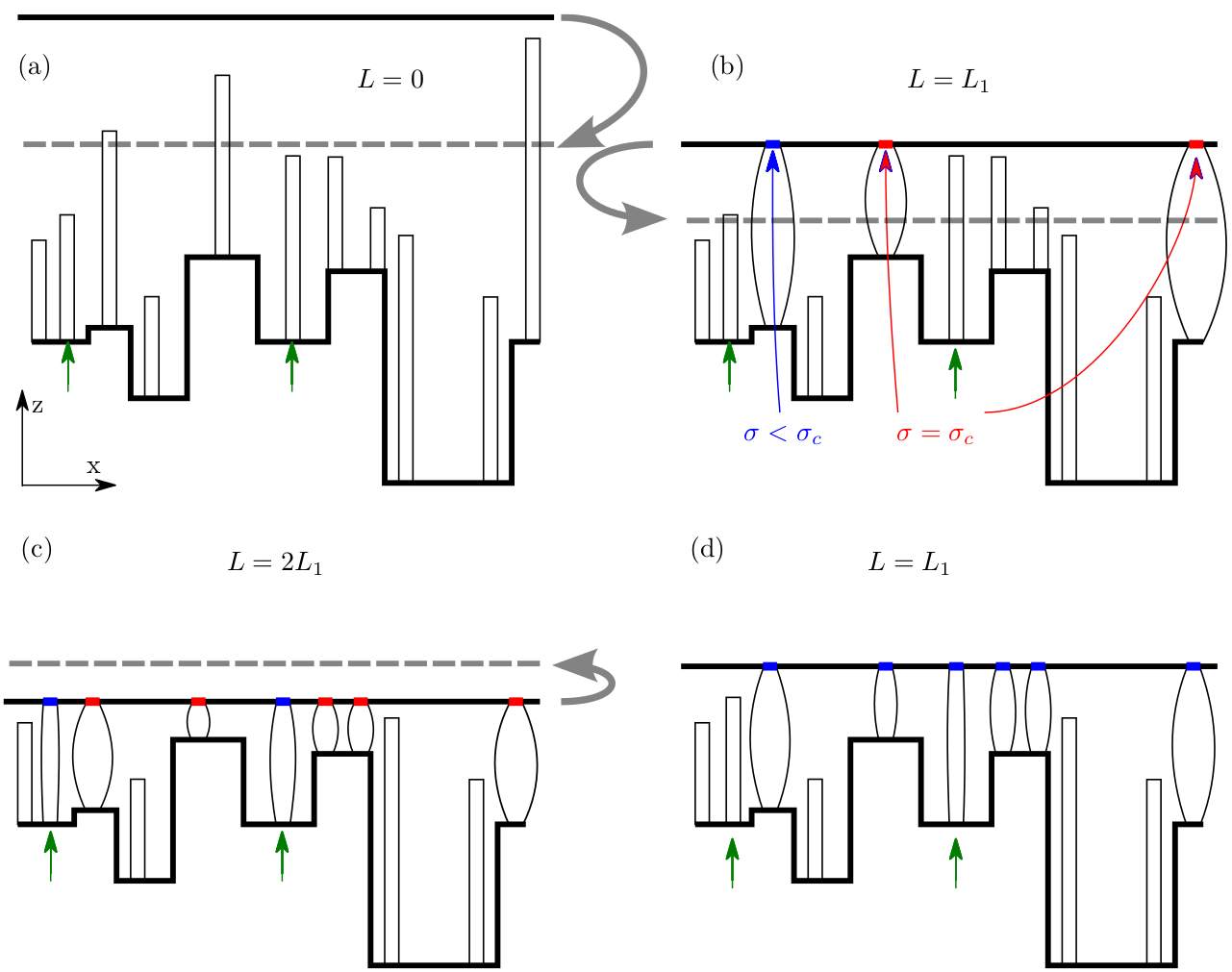}{
Schematic description of the ``extra-load'' (thought) experiment we propose.
The upper solid is considered infinitely tough compared to the lower one ($\sigma_c^{up} \gg \sigma_c^{low}$).
Two particularly interesting asperities are highlighted by green arrows along the evolution.
(a) Load is zero, all asperities are intact.
(b) Load is $L_1$, some asperities deform elastically (blue, $\sigma<\sigma_c$), others also yield plastically (red, $\sigma=\sigma_c$).
(c) Load is increased to $2 L_1$: additional elastic and plastic deformations occur.
(d) Load is decreased back to $L_1$: the upper solid is not pushed back to its initial position.
Asperities that were subject to very high stress can release a lot of it by pushing the upper solid up.
Asperities that were subject to moderate stresses go back to their original shape (left green arrow), or are only slightly compressed (right green arrow).
\label{Fig:elastic_compression}
}

In all these cases the naive analysis implies that the real area of contact is no longer proportional to the load, i.e.~that Amonton's second law is violated.
However, in all these cases the stress in the asperities can be quite high, since it is only bounded from above by the plastic yielding limit $\sigma_c$. 
With values of the local stress up to $\sigma_c$, the elastic deformations of contacts can and will play an important role.
 To compute the real area of contact and in particular its dependence on the load, we will need to consider the elastic deformations of asperities.
In this application of linear elasticity theory, we will consider adhesion forces negligible compared to the elastic tensile stresses (even though it is precisely adhesion which is responsible for friction!). 
The role of adhesion for elastic solids with rough (random) surfaces has been included in recent works as \cite{Persson2008}, where the law $F\propto L$ is still predicted. 

We now discuss the elastic response of two simple models of asperities: cylindrical asperities of which the extremity is considered flat, and spherically ended asperities (where the asperities are not elongated enough to be able to neglect the shape of the asperity extremity).

\paragraph{Model I: Cylindrical Asperities}
Here I give a schematic description of what happens in the ``extra-load'' experiment by considering the asperities as essentially cylindrical.
In this limit, the contact area at each asperity is either $0$ (no contact) or $ A_{\text{asp},1}$ (typical area of one micro-scale asperity).

When we increase the load up to $2L_1$, asperities are crushed so that $A_{\text{real}} \approx 2 L_1 / \sigma_c$ (See Fig.~\ref{Fig:elastic_compression}.c). 
When we then decrease the load to $L=L_1$, the pressure no longer overcomes the yield stress, so that plastic flow is no longer possible.
Yet, there is still some high stress concentration in the asperities: instead of having the compressive stress of the bulk, $\sigma_{zz}^ {\text{bulk}} \approx L/ A_{\text{app}}$, asperities are subject to a compressive stress 
$\sigma_{zz}^ {\text{asp}}  \in [0,\sigma_c]$, with an average:
\begin{align}
\langle \sigma_{zz}^ {\text{asp}} \rangle  \equiv p_1=\frac{L }{ A_{\text{real}}} \gg\sigma_{zz}^ {\text{bulk}}.
\end{align}
This stress corresponds to a compression of each asperity along $z$ by a compressional strain $\varepsilon$ (dimensionless variable) initially given by:
\begin{align}  
\varepsilon =\frac{d }{z_0} \propto E \sigma_{zz}^ {\text{asp}}, 
\end{align}
where $d $ is the elastic displacement of the asperity, 
$z_0$ its initial 
length\footnote{For now, we assume elongated asperities in the $z$ direction, in the sense that their contact does not depend on compression. 
Examples of such ideal shapes are cylinders or parallelepipeds, that can be modelled by a simple spring.
Examples of cases we exclude with this assumption are the spherical and cylinder-with-rounded-tips shapes. }, 
$E$ the Young's modulus of the material.
Depending on its initial length $z_0$ (length after plastic flow), each asperity is more or less compressed, as pictured in Fig.~\ref{Fig:elastic_compression}.c.

Qualitatively, when the asperities are relieved from the extra load, those which were more compressed (larger $d/z_0$) are also those which de-compress more: they rise, thus ``lifting'' the solid upwards.
Those which were less compressed (smaller $d/z_0$) ``rise'' less and can thus lose contact in the process. (See Fig.~\ref{Fig:elastic_compression}.d)

We denote $\delta d$ the``rise''  of each asperity, so that the total lift of the upper solid is equal to $\langle \delta d \rangle_s$, where the average is over the surviving\footnote{
For the asperities which lose contact (or ``die''), the variation of $d $ is even smaller than for the surviving ones (it is less than the rise of the upper solid). 
However, this smaller rise corresponds to a drop of the compressive stress from $\sigma_{zz}^ {\text{asp}}$ to zero, since contact is lost. 
This explains how some load bearing can be ``forgotten'', despite the surviving contacts being subject to an approximately constant pressure.
}
 contacts at the end of the process. 
As the rise of each surviving
 asperity is automatically equal to the macroscopic one, we have also $ \delta d = \langle \delta d \rangle_s $ (only $z_0$ is a random variable, drawn independently for each asperity).
To give a rough estimation of the dependence of the real contact area in the load, we make the assumption that the total ``rise'' of the asperities is negligible, i.e.~that  $ \delta d  \ll \langle z_0 \rangle_s$.
Thus for a surviving asperity the local change $ \delta \sigma_{zz}^ {\text{asp}} \propto \delta d / \langle z_0 \rangle_s$ in compressive stress is negligible: $\sigma_{zz}^ {\text{asp}} \propto d /z_0 \approx \text{const}$. 
For these asperities we have an average compressive stress $ p_1 = \langle \sigma_{zz}^ {\text{asp}} \rangle_s \approx \text{const}$,
i.e.~$L = p_1 A_{\text{real}} \propto  A_{\text{real}}$, i.e.~Amonton's second law is respected.
We may notice that since the asperities are cylindrical, the unitary contact area $ A_{\text{asp},1}$ is constant, and we have the more precise relation $L\propto  N_{\text{asp}}$.
We note that the approximation $ \delta d  \ll \langle z_0 \rangle_s$ is especially well respected for very rough profiles, where $z_0$ has a large distribution. 
This is clear if we consider the asperities which lose contact: if their $z_0$ is very large, a smaller rise $\delta d$ will be enough to kill contact.

The conclusion is that the main effect of decreasing (resp.~increasing) the load in the elastic regime is to remove some contacts (resp.~create new ones).

\begin{figure}[]
\begin{small}
\begin{center}
\def\svgwidth{\textwidth}
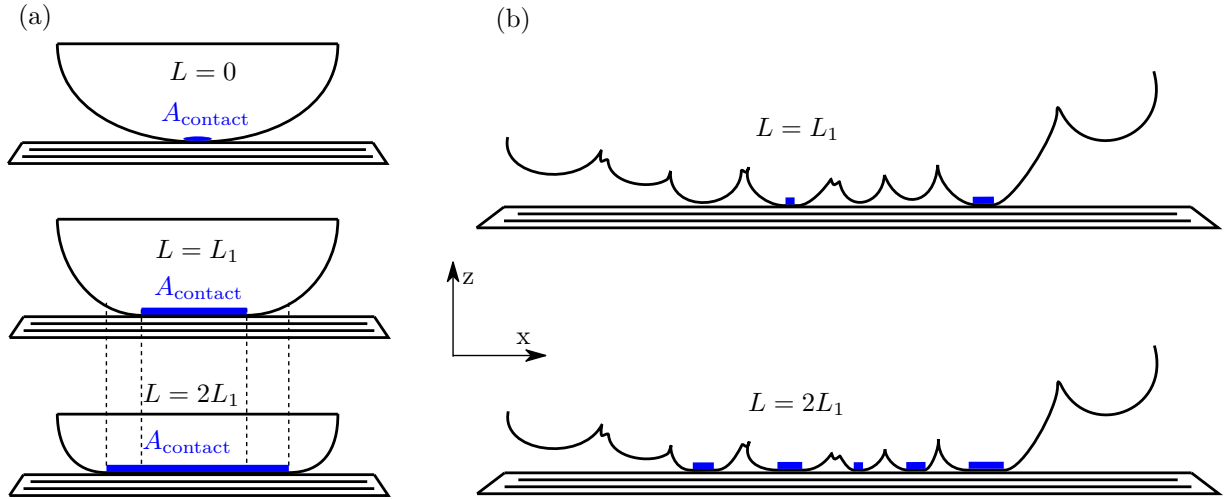
\end{center}
\end{small}
\caption{   {\footnotesize 
Schematic description of spherically shaped asperities or ``bumps'', in purely elastic compression (\textit{not} to scale).
(a): A single bump is compressed onto a rigid substrate.
At zero load, the area of contact is a single point.
At higher loads, the area of contact is elastically deformed (blue highlight) and is not proportional to the load: $A_\text{contact} \propto L^ {2/3}$.
(b): A surface is modelled by spherical bumps. 
On average, the area of contact is proportional to the load.
\label{Fig:rounded_asperities}
}   }
\end{figure}

\paragraph{Model II: Spherically Ended Asperities}
Another way to consider asperities is assimilate them as spherical bumps, as depicted in Fig.~\ref{Fig:rounded_asperities}.
Let's start with a single contact. 
Assuming purely elastic deformation and no adhesion, the Hertzian theory of contact mechanics predicts, for a sphere pressed into a half-space, a non linear dependence of the contact area with the load: $A_1 \propto L^{2/3}$.
The non linearity may seem surprising, given that we only used linear elasticity theory.

The qualitative explanation is very simple: as loading increases, the contact area increases from a point to a disk of increasing radius. 
The average pressure in the contact area is the macroscopic load divided by the contact area: it starts very large, which allows for a large indent depth $d$, but as indent increases, so does the contact area, which reduces the local pressure.
At the end of the day, even though the indentation is always proportional to local pressure, the geometry is such that the overall dynamic is non linear in the load $L$.

However at the macroscopic scale, the linear dependence is most commonly observed. 
A linear dependence emerging from the non linear law is found in a simple model of spherical asperities. 
There is a classical derivation of the area of contact and load for this model due to Greenwood, nowadays available in Chap.~2 of \cite{Persson1996a}.
We reproduce here the main line of Greenwood's argumentation. 

Consider the centres of the spherically shaped bumps (of radius $R$) as depicted in Fig.~\ref{Fig:rounded_asperities}: the centres' heights constitute a profile $\Phi(z)$ ($z=\mathcal{H}$ being the height of the flat plane onto which asperities are pressed). 
At each bump, applying Hertz theory, the bump is compressed a distance $d=z-\mathcal{H}$, leading to an (unitary) contact area $A_1=\pi R d$ and a load (borne by this single asperity) $L_1 = (4/3)E^* R^{1/2} d ^{1/2}$, where $E^*$ is the reduced Young's modulus\footnote{The reduced Young's modulus is defined as a combination of the two materials Young moduli $E_1,E_2$ and their poisson ratio $\nu_1, \nu_2$ via: 
$1/E^*=(1-\nu^2_1)/E_1+(1-\nu^2_2)/E_2$.}.
With $N$ being the number of bumps in the sample, we have the number of contacts $n$, area and load given by:
\begin{align}
n&=N \int_\mathcal{H}^\infty \Phi(z) \d z\\
A_{\text{real}} &=   N \pi R \int_\mathcal{H}^\infty \Phi(z) (z-\mathcal{H}) \d z\\
L &=  (4/3)E^* R^{1/2} \int_\mathcal{H}^\infty \Phi(z) (z-\mathcal{H})^{3/2} \d z
\end{align}
We assume a rapid decay of the function $\Phi(z)$, which seems reasonable for ``flat'' solids.
We can take any decay, e.g.~$\Phi$ can be Gaussian or more simply, $\Phi(z) \simeq e^{-\lambda z}$ is fine.
This allows to compute:
\begin{align}
n&= \frac{N}{\lambda} e^{-\lambda \mathcal{H}} 								\\
A_{\text{real}} &=   \frac{ N \pi R}{\lambda^2} e^{-\lambda \mathcal{H}}		\\
L &= \frac{E^* R^{1/2}  \pi^{1/2}}{\lambda^{5/2}}  e^{-\lambda \mathcal{H}} ,
\end{align}
from which Greenwood concludes that $A \propto L$. 
The problem with this reasoning is that it assumes a rapidly decaying profile $\Phi$: for very flat surfaces, it is well acceptable.
However, for surfaces with roughness at several length scales, the relevance of this model has been questioned, for instance in \cite{Persson2008}. 
Indeed, this model only accounts for roughness at a single length scale: the elastic deformation of larger regions (e.g.~made of several bumps) is implicitly considered to be zero, because this larger length scale is implicitly ignored.

The Hertzian theory of contact applied on spherical asperities has played an important historical role, and is still valid for ``Gaussian'' or flat surfaces. 
This non linearity in the response of spherical contacts is also interesting for the study of granular materials.
In regimes where the ``balls'' merely touch each other, it can be crucial to account for the non linear response (see for example \cite{Gomez2012} or the review \cite{Aranson2006} for more details). 
\\

There has also been some observations of sub-linear dependence of the friction in the load, in some particular contexts \cite{Butt2006,Persson2000}. 
This kind of non linear dependence at the macroscopic scale is typically obtained when the unitary contact area depends on the local load, i.e.~in all sorts of rounded or triangular shapes, but also when additional forces (e.g.~van der Waals or capillary forces) produce geometrical arrangements which strongly depend on the load, at rather large length scales.
The idea of a single-asperity with rounded extremity is also sometimes used as a rudimentary model of AFM tip, in this case the single-asperity tip is the macroscopic system.
The relevance of this sort of behaviour was summarized very early by Archard, whom asserted that \cite{Archard1957}:
\begin{quote}
{\footnotesize
``If the primary result of increasing the load is to cause existing contact areas to grow, then the area of real contact will not be proportional to the load. But if the primary result is to form new areas of contact, then the area and load will be proportional.''
}
\end{quote}

\subsubsection{The Question of Fracture}
\label{sec:question_of_fracture}

The breaking of junctions is fundamentally a fracture process: as we have said earlier, at the asperity scale, the high pressures result in cold welding, so that the separation of the two surfaces occurs through rupture.
The Fineberg's group recently developed a real-time visualization method of the real area of contact during the sliding of the blocks 
\cite{Rubinstein2004,Rubinstein2007,Ben-David2010,Svetlizky2014}, see also the review \cite{Vanossi2013}.
This method shows that the transition from static to kinetic friction is controlled by the collective behavior (and fracture) of  the ensemble of asperities that form the interface betweeen the two solids.
In particular they identify three different kinds of coherent crack-like fronts that govern the onset of slip \cite{Rubinstein2004}.
In a recent study \cite{Svetlizky2014}, it was shown that the slowest of these three fronts indeed governs the rupture, under certain conditions: 
at driving velocities such that the rupture velocity is lower than the Rayleigh wave speed, the predictions from Linear Elastic Fracture Mechanics are in quantitative agreement with experiments.

In what follows we will discuss the question of the relevance of brittle fracture on domains much larger than a single junction, or which involve some loss of material (wear), which is a different question from that discussed by Fineberg and collaborators.
As asperities and the surrounding domains are subject to high stresses and various geometrical constraints, one may naively expect mesoscopic fracture to be commonplace, especially during sliding.
We are going to see some reasons for why fracture is not so common at the scale of micro asperities, but also how it can still be relevant in some cases.

In the static case (with no driving being performed), the asperities are subject to very high compressive stresses, which tend to decrease the probability of brittle fracture\footnote{
The term brittles refers to ``pure'' fracture (without plastic deformation) as opposed to ductile fracture. 
We have already considered the plastic behaviours previously. 
}.
This is because the ductility (essentially the maximal plastic deformation possible before fracture) generally increases with the (hydrostatic) pressure.
An intuitive but hand-waving argument for this is that high pressure tends to close the micro-cracks, vacancies and other voids generated by the plastic flow in the bulk of the solid. 
As these defaults are responsible for fracture (which always starts from the largest crack in the region under stress), their relative closing by pressure tends to diminish the occurrence of fracture.
\includefig{\textwidth}{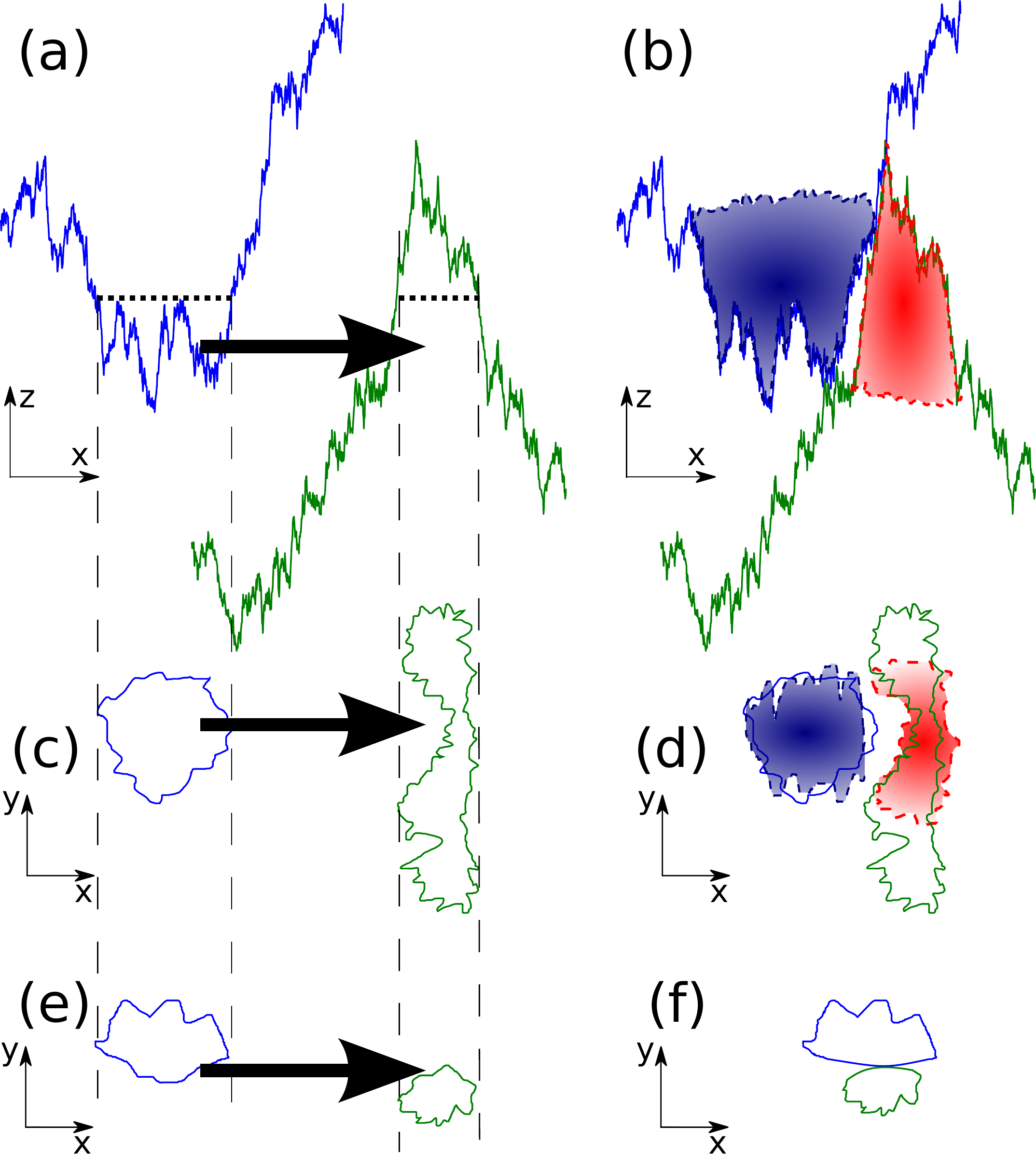}{
Schematic picture of the situation of interlocking, fracture and elastic deformation scenarios.
(a, b): sectional view of two asperities meeting under external driving (solid arrows).
(c, d, e, f): view from above of different scenarios.
(c): situation of interlocking. 
Two scenarios of fracture are suggested (b,d), with the broken zone shaded in blue (left) or red (right).
(e): situation of weak interlocking: elastic deformations can be enough to let asperities go through. 
(f): with some elastic (and a bit of plastic) deformation, asperities stay in their way.
\label{Fig:interlocking}
}

During sliding, the shear stress at the contact points can become enormous (as for the compressive stress, this is due to the small contact area). 
At the level of a single contact, as presented in Fig.~\ref{Fig:interlocking}.d, the response will be to simply break the smallest possible cross-section of the welded asperities, which we identify as a simple junction breaking, which is not the point discussed here. 
However, in the configuration of ``interlocking'' (see Fig.~\ref{Fig:interlocking}), an asperity is subject to a high shear stress in a direction orthogonal to $z$ (the main compressive stress).
Thus one may expect the small asperities to easily break by brittle fracture, rather than deform elastically or plastically (we described these two mechanisms above).
This is only half true.

On the one hand, Griffith's criterion shows that the critical stress for brittle fracture is directly controlled by the size of the largest micro-crack in the sample.
So for very small samples, this threshold stress will typically be very high: the smaller is the sample, the smaller its largest crack\footnote{This is very intuitive, and the interested reader may try to make this reasoning more quantitative by using the branch of probability theory called Extreme Value Statistics (EVS).}. 

On the other hand, we have to remember that roughness is expected at all scales.
At large scales, all sorts of geometries can create locally high stresses on the ``asperities'' domains, which are always much larger than the typical unitary asperity contact area initially blocked (see Fig.~\ref{Fig:interlocking}.d).
In the context of tectonic plates for instance, the interlocking of ``asperities'' can involve locked areas over lengths ranging from the centimetres up to several meters (or more), with widths in the same range. 
For such large domains, the size of the largest crack available can become quite large, so that the elastic stresses will easily trigger macroscopic fracture.
This threshold force needed for fracture contributes to the friction force\footnote{
How does this force scales with the real contact area as determined in the previous section (from the elastic deformations)? 
Interlocking happens only where contacts are made, otherwise the large ``bumps'' would simply go by. 
Because of that, the density of number of interlocked domains is still proportional to the real contact area (itself proportional to the load). 
Then, the area of domains (which is roughly proportional to the fracture energy) depends on the real contact area in a rather intricate way, through the roughness exponent $\zeta$. 
This is beyond the scope of this thesis.
}.

In both cases, it is interesting to note that a principle of selection is at play. 
The domains with the largest cracks (low fracture threshold) break first, and as slip occurs, only the hardest domains remain, so that during a slip phase, the prevalence of fracture typically decreases after a certain slip length $D$. We will come back to this in more detail in the next section, \ref{sec:Ageing}.

In the context of geophysics, it has been noticed that rocks are usually much less ductile than the materials commonly considered in tribology (metals, etc.), so that at equal external conditions, they break much earlier.
Thus interlocking and the associated fracture process is expected to be quite important.
An attempt at explaining friction as a process controlled mainly by the fracture of asperities was made in 1967 \cite{Byerlee1967},
 but the application of this theory has been limited to geophysics, where the presence of wear particles in large proportions makes such an hypothesis more likely.

To conclude, mesoscopic fracture plays a minor role in the dynamics of sliding\footnote{And of course, fracture plays an even smaller  role in the dynamics of static friction.} friction, regarding most applications.
However when the system is either large, made of rocks or a combination of both, fracture can become equally relevant as adhesion in explaining friction.
In geophysical applications, a comprehensive model for the sliding of plates would necessarily acknowledge the role of fracture.
Let us recall that at the level of a single asperity, fracture is omnipresent, regardless of the nature of the material and of the external conditions. This fact is the basis for numerous works on friction \cite{Rubinstein2004,Rubinstein2006,Rubinstein2007,Ben-David2010,Ben-David2011,Svetlizky2014}.

\subsubsection{Wear}

In the context of friction, \textit{wear} is usually defined quantitatively as the volume of particles which separates from one of the two surfaces \textit{during sliding}.
The separated particles may wander freely between the two surfaces (in this case, we may call them \textit{debris}) or re-attach to the other surface. 
In both cases, wear corresponds to a change in the surfaces in contact (for engineering applications, it is sometimes only the net amount of debris which is relevant).
In a first approximation, wear is proportional to the work performed by the friction force, hence it is proportional to the sliding length and friction force (but not directly to the velocity).
Interlocking and the subsequent fracture obviously causes some wear. 
Let's quickly discuss a few other mechanisms which enter in the definition of ``wear''.

A mechanism which is slightly different from plain fracture and also causes wear is that of \textit{adhesion wear}.
When two asperities enter in contact and form a junction, depending on the micro-structure of each asperity close to the junction plane, the breaking of the bond may occur elsewhere than in the welding plane, so that one of the asperity keeps a piece of the other one.
This part can either stay in place or get quickly separated from the asperity (due to the weakness of the joint): in both cases, we have some wear.
This is a possible mechanism of wear, which has much to do with adhesion, hence the name.
Note than the debris created in this way, or which are already present, can also re-attach to one of the two surfaces, thus ``regenerating'' the surface profile.
\includefig{\textwidth}{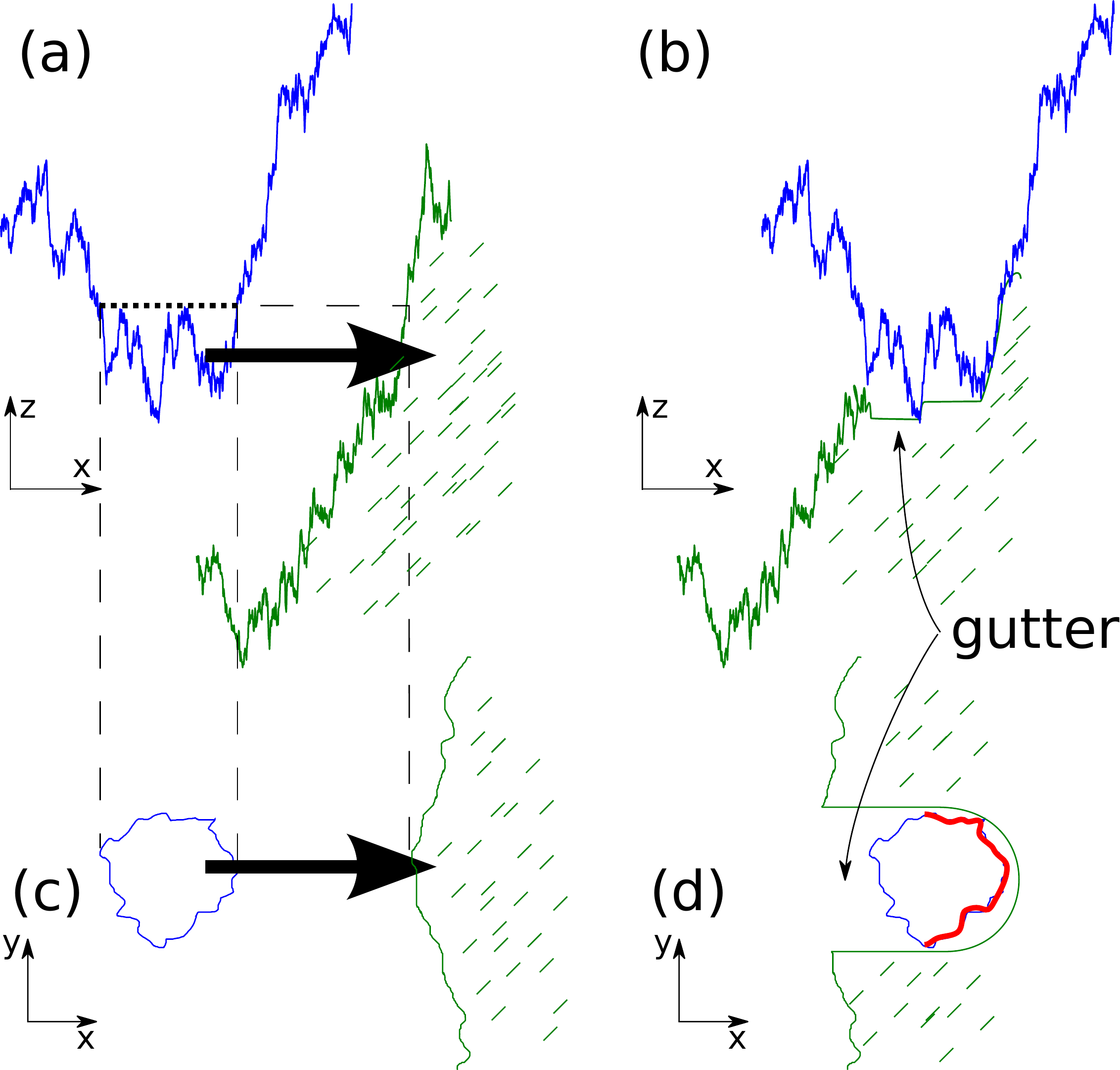}{
Very schematic picture of the ploughing scenario. 
(a, b): sectional view of two asperities meeting under external driving (solid arrows).
(c, d): view from above of the ploughing scenario.
If the upper (blue) material is much harder (larger yield stress $\sigma_c$) than the lower one (green), an asperity of the former may plough a gutter into the latter. 
The zone of the upper solid subject to the highest stresses is highlighted in red (d).
\label{Fig:ploughing}
}

The term \textit{abrasive wear} is used when one of the two surfaces is much harder than the other, in terms of plastic yielding stress $\sigma_c$.
In this case, when an asperity of the harder material indents the other, it can plough a gutter (see Fig.~\ref{Fig:ploughing}) into it (instead of being deformed or break by brittle fracture).
In a sense, ploughing is essentially plastic yielding along the surface plane, except that it can happen locally even when we are no longer in the plastic yield stress regime, macroscopically.

\subsubsection{Conclusion}

During sliding, the elastic response of asperities is twofold: for a part, asperities deform similarly as in the static case (see Fig.~\ref{Fig:interlocking}.f), and for the other part they interlock. 
The interlocking of asperities can be overcome elastically if the height of the asperity involved is small enough.
Fracture naturally appears as a limit of the elastic behaviour.
And again, during sliding, the plastic response of asperities is twofold: to some extent, asperities deform similarly as in the static case by yielding against each other one at a time, but they may also plough long gutters into the opposite surface.

\subsection{Ageing of Contact and its Consequences} 
\label{sec:Ageing}

We have explained the first two laws up to here: because of high roughness the apparent area has little to do with the real one which is truly responsible for friction, and for various reasons this area generally ends up being proportional to the load.  
However regarding the third law and its corrected version (the Rate- and State dependent Friction laws),
we have given no clue about the possible mechanisms yet.

The fact that RSF laws work very well (see sec.~\ref{sec:RSF1}) is not so surprising: with at least three fitting parameters ($\mu^*, a, b$) and somewhat five (counting $D_c$ and $V^*$), it is rather easy to ``fit the data''.
This picture becomes more satisfying when several of these free parameters can be bound to some underlying physical mechanisms.
We are about to interpret $\theta$ more precisely than just a ``state'' variable, and $D_c$ much more precisely than a simple fitting/normalisation parameter. 
Other parameters can also be interpreted, but with some caution.

\subsubsection{Microscopic Origins}

\paragraph{Ageing: definition}

Here, we are going to see that static friction and more precisely microscopical contacts display \textit{ageing}, and we will give the link with the macroscopic RSF laws.
Let's start with definitions.

We define the notion of ageing as the opposite of \textit{stationary}: a system which displays ageing has some of its properties which change over time (i.e.~they are not stationary).
A process with ageing necessarily has some long-term memory (typically a power-law decay of the autocorrelation function over time).

A corollary is that with perfect knowledge of the microscopic dynamics of a process with ageing, one can typically estimate the ``age'' of a sample from a snapshot observation at some given time (i.e.~a measure at a single time); the age being the time spent evolving from some default (known) starting configuration, to the observed one.
The most common example of materials displaying ageing are glasses (see the course 7 of \cite{LesHouches2002} or \cite{Biroli2005} for an introduction on the topic, and \cite{Bouchaud1998} for a discussion of an experimental example).
 
\paragraph{Creep}
In our study of the formation mechanisms of the real contact area 
 we have omitted the aspect of temporal evolution.
At first order, the processes we discussed are instantaneous: plastic yield, elastic response or brittle fracture all appear to happen very shortly after the appropriate constraints are applied. 
However, many time-dependent secondary processes interact with these main three, such as dislocation creep, desorption of protective films, formation of additional chemical bonds in the junction, cyclic fatigue, surface corrosion and wear of fresh surfaces, viscoelastic response in the bulk of asperities, elastic waves generated by ruptures, melting, etc.

The process usually recognized to be mainly responsible for variable behaviour over time is \textit{plastic creep}.
For crystalline materials, we may speak of dislocation creep \cite{Heslot1994,Baumberger1994,Putelat2011}. 
Dislocations are produced through plastic events, and their slow thermally-activated displacement (this is what is called \textit{creep}) may in return affect not only the plastic behaviour but also the condition for fracture.
After a quick plastic yield occurs at the time of formation of a new contact, some new dislocations are formed (they are newborns, their age is zero). 
Since they have been freshly generated, they have not reach any equilibrium, nor even a stationary state. 
Thus they slowly diffuse and possibly trigger new (typically smaller) plastic events, which can in turn generate new dislocations.
This dynamics progressively decelerates but never really reaches a stationary state, due to the infrequent bursts producing new dislocations: this is ageing \cite{Putelat2011}.
The mechanism for plastic creep in amorphous materials is a bit different (see \cite{Barrat2011} for a review and additional references).
All in all, from our understanding of creep (or even the observation of macroscopic materials), one may expect asperities to age after contact and to slowly spread around the initial junction area.

Indeed, this effect has been observed directly in experiments. 
In 1994
\cite{Dieterich1994}, the diffraction of light through transparent samples allowed to directly observe the evolution of the true contact area over time.
We reproduce these impressive results in Fig.~\ref{Fig:Dieterich1994}.
\includefig{9cm}{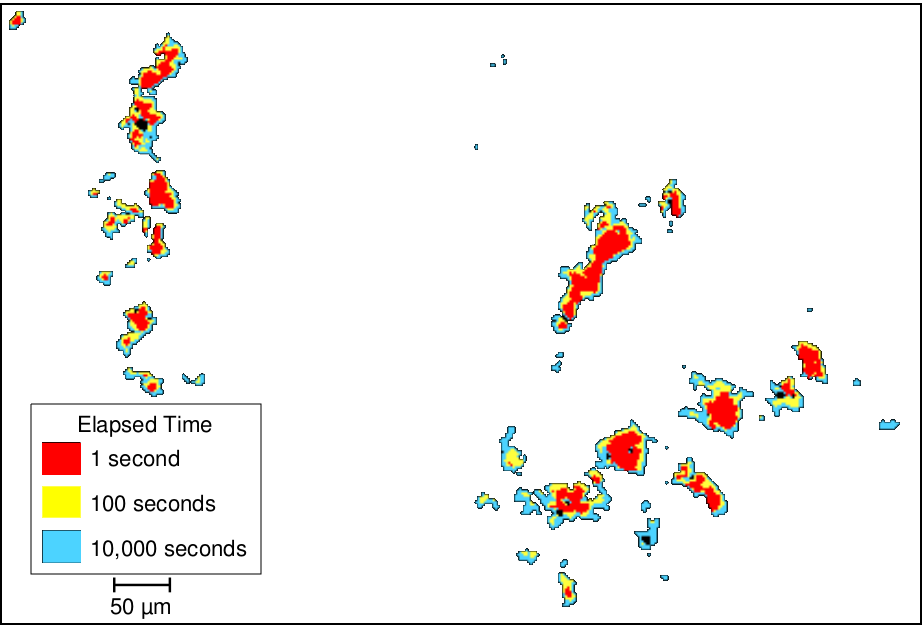}{
From \cite{Dieterich1994}. 
Using transparent samples, the contact area is made apparent thanks to the fact that the light diffracts everywhere but at the contacts. 
This allows to see the contacts directly, and most specifically to see their evolution over time.
\label{Fig:Dieterich1994}
}
Conventional techniques of contact analysis prior to these works used to be \textit{post-mortem}, i.e.~after the surfaces had been in contact, one could analyse them to sort out the properties of the last contact zones.
These post-mortem studies were of course unable to study the time evolution of contacts in such a way.
From Fig.~\ref{Fig:Dieterich1994} it is clear that despite the constant load, the true contact area slowly increases over time, i.e.~the contacts display some ageing.

In the following paragraph we show how to incorporate this ageing into an effective law for friction as the RSF laws given above.

\paragraph{Interpretation of $\theta$ and $D_c$.}
Let's assume that $\theta$ represents the additional (or ``bonus'') contact strength due to ageing, and see how we can fit this idea into our observations.

On the one hand, at rest we expect a logarithmic increase of friction over time: if $\theta(v=0)\sim t$, then the third term $b \ln (V^* \theta /D_c)$ goes like $ \sim \ln (t) +\text{const}$. 
Still at rest, $\mu^ *$ appears as the ``instantaneous'' friction, i.e.~the friction obtained immediately after contact, due to the fast processes. 
The fact that $b\ll \mu^*$ corresponds well to the fact that creep is a secondary process, which only gives corrections to the main processes.

On the other hand at finite velocity the contacts ``do not have time'' to age: since the solid constantly slips, new contacts are constantly formed, and ``old'' ones broken. 
The crucial question is to estimate the contacts typical lifetime.
Assuming a constant sliding velocity for the sake of simplicity, we may call $D_c$ the ``critical slip distance'', i.e.~the amount of slip (of the center of mass of the sliding block) necessary to break a newly formed junction. 
It takes a time $\theta_c = D_c/v$ for the bulk solid to slide over a distance $D_c$. 
Thus, the typical lifetime of a contact in the steady state is $\theta_c$, so that the average or typical ``bonus resistance'' goes like $\sim \ln(\theta_c )$.
This explains why in all RSF laws the evolution of $\theta$ must be chosen such that $\theta^{ss} =  D_c/v$.

Similarly, the values of $D_c$ can be interpreted straightforwardly. 
If asperities are sharp, in the sense that they resemble elongated needles, they may deform elastically and maintain contact over slip distances equal to several times the contact diameter $D_a$.
On the contrary, if asperities are more like flat bumps with small heights, they will break contact as soon as the slip is a fraction of their contact diameter.
In any case, for stronger bonds (larger contact diameter $D_a$), asperities will deform more before breaking, i.e.~$D_c$ increases with $D_a$.
All in all, the contact-breaking slip distance is typically of the same order of magnitude as the asperities diameter, hence $D_c \sim 1 - 10~\mu m$.

Now, in between $v=0$ and $v=\text{const}>0$, there is a world of possibilities, and each RSF law (in particular the choice for the evolution of $\theta(t)$) will react differently to different experiments, as experiments with step-like variations of the driving velocity $V_0$, slip-hold-slip experiments, etc.
The way in which each law reacts more or less realistically to each kind of input has been discussed in the reviews of reference, e.g.~in \cite{Marone1998} where experiments are discussed, or more recently in \cite{Kawamura2012} where the bibliography is abundant.
However, we are not much interested in the details of each law's pros and cons: it is enough to note that no definitive consensus has been reached yet, and that a detailed microscopic analysis from which RSF laws would emerge is still missing.
Thus, the problem in terms of fundamental physics is still largely open.

\paragraph{Interpretation of Other Variables}

The velocity $V^*$ is merely a homogeneity constant: for any choice of units, it can be absorbed into $\mu^ *$. 
Thus, the value $V^*=1~\mu m.s^{-1}$ is simply a convenient choice, since relevant velocities are usually of this order.

For $\mu^*$, the interpretation seems quite simple: it is the default friction, corresponding to the fast processes we initially described (up to the absorption of constants as $V^*$ and normalization expected at $t=0$, depending on the exact form of the RSF law, \req{RSF0} or \req{RSF0bis}).
In principle, $\mu^*$ can be estimated quantitatively: assuming a purely plastic formation of the true area of contact, we have $A_{\text{real}} = L/\sigma_c$, a number of bonds $N_\text{bonds} = A_{\text{real}} / A_{1 \text{bond}}$, and a threshold breaking force per bond $ f_1$.
Denoting $F_\parallel$ the macroscopic shear force (tangential) and $L$ the load (normal), this gives
\begin{align}
\mu^ *\approx \frac{F_\parallel}{L} = \frac{f_1}{\sigma_c  A_{1 \text{bond}}},
\end{align}
where the yield strength $\sigma_c$ is easy to measure, but the ratio $f_1 / \sigma_c  A_{1 \text{bond}}$ is very hard to get.

The interpretation of $a$ and $b$ is usually directly related to creep \cite{Heslot1994,Baumberger1994}. 
In a recent work \cite{Putelat2011}, the activation volume is defined in relation with the activation energy $E^*$ ($\Omega^*=E^*/\sigma_c$) and the parameters $a$, $b$ are predicted to be
\begin{align}
a=\frac{k_B T}{\Omega^* \sigma_c}, \qquad  b=\frac{k_B T}{\Omega^{'} \sigma_c},
\end{align}
where $\Omega^{'}$ is some other activation volume.
Unfortunately, direct access to these activation volumes and activation energies is difficult, so that these expressions for the RSF laws parameters are seldom used\footnote{
Furthermore, this interpretation of $b$ is quite new and to be taken with caution.
The interpretation of $a$ is more commonly accepted \cite{Kawamura2012}, though it should still be taken with caution.}.

Furthermore, the position of creep as dominant mechanism for ageing has been recently questioned in \cite{Li2011} where it was suggested that the strengthening of chemical bonds at junctions could be a more realistic explanation for the ageing of frictional contacts than creep.
This casts doubts upon the trust we may put into old or current interpretations of the RSF laws in terms of plastic creep.

\subsubsection{Stick-Slip Motion (with RSF laws)}
\label{sec:stickslip3}

With a friction law that continuously depends on the sliding velocity $v$ (velocity weakening), and possibly on some ageing ``state'' variable $\theta$ (increase of static friction over time), the dynamics of stick-slip becomes a bit more complex than what we forecasted in sec.~\ref{sec:stickslip1}. 
However the main results are maintained: the existence of stick-slip in general and its disappearance at large velocity ($V_0$) or hard driving spring ($k_0$).

A thorough study of RSF laws applied to a single degree of freedom (a simple rigid block) was performed early in \cite{Gu1984}.
A more concise study of this problem was performed in \cite{Rice1986}, where the differences with the Amontons-Coulomb laws were emphasized. 
There, the main difference with the more simple law of friction is the emergence of two time scales or velocities (instead of one). For velocities below a first threshold, the motion is essentially described by the quasi-static picture (which neglects the velocity dependence). 
For velocities above this threshold but below the second one, the dynamical effects cannot be neglected. Above the second threshold, stick-slip disappears (similarly to what we found in our simpler model, sec.~\ref{sec:stickslip1}).

Another complete, yet concise study of stick-slip motion was performed in \cite{Baumberger1995}.
They compare experimental results for paper on paper stick-slip with analytical computations (weakly non linear analysis around the Hopf bifurcation) and numerical integrations using the most common Rate-and-state friction law, \req{RSF0} \& \req{RSF1}.

Of course, different dynamics of stick-slip can be obtained when using various rate-and-state laws.
However, the main features we are interested in remain the same: as sliding velocity increases, stick-slip motion shifts from very regular to rather chaotic, to non-existent. 
This rich behaviour has been the playground for intensive studies in Geophysics, as we will see in chap.\ref{chap:EQs}.

The RSF law is particularly useful in geophysics in order to study the dynamics of stick-slip, which involves the static friction coefficient and where the departure from zero to finite velocity is especially relevant.
This is what we explain in the next section.

\section{Conclusion: Friction Involves Randomness and Viscoelasticity} 
\label{sec:conclu_chap1}

We have presented the basic phenomenology of Friction.
The three historical laws have been amended to account correctly for the dependence on the sliding velocity, a crucial point in the study of the dynamical stability of frictional systems (stick-slip instability).
Intuition about the physical mechanisms behind these laws has been supported with direct observations (microscopic measurements of surface profiles, contact area and its time evolution).
Simple models of (implicitly static) contact have been presented, outlining the role of elasticity, plasticity and some secondary mechanisms (fracture, plastic creep).
\\

Microscopic models of friction should take into account the presence of randomness.
A first source of randomness are the thermal fluctuations, responsible for the plastic creep \cite{Putelat2011} which plays a key role in the ageing of contacts. 
A second source of randomness is the presence of ``quenched disorder'' induced by the heterogeneities and the roughness of the surfaces. 
The idea that surface self-affinity is crucial to the friction properties is now well-established  \cite{Persson2001,Persson2005}, in particular it naturally explains the second friction law of Amontons.
However most of the phenomenological models (e.g.~\cite{Rundle1991, Persson2005}) deal with the average properties induced by the disorder and neglect the fluctuations of the dynamics.
As we will see in chap.~\ref{chap:elastic}, the validity of this assumption is a matter of scale \cite{Persson1996a, Caroli1998}.
At moderate scales (such as in laboratory experiments), the motion can be described by deterministic effective equations such as the Rate-and-State  equations.
At much larger scales, the motion is actually stochastic and displays a very complex avalanche dynamics.
This is in particular the case for fault dynamics, which is characterized by random bursts of activity (earthquakes) that are random in magnitude,  temporal and spatial location.

There have been a few tentative friction models including real randomness, but they have found rather limited echo until now: \cite{Rundle1991} is an example that received unfairly small attention.
Excellent reviews on this topic are \cite{Kawamura2012, Vanossi2013}, but we will come back to this at length later.
The problem with all other attempts is that they fail either at correctly account for randomness, or they overlook the role of microscopical ageing  which is crucial in producing the RSF laws.
All in all, no definitive consensus has been reached to this day on the foundations of the RSF law(s), even when resorting to such models:
the search for a convincing yet simple microscopical model reproducing a realistic RSF law is still an open problem.
\\

To summarize  -- crudely --  there are two main issues that must be addressed in order to properly deal with friction. 
The first is the fluctuating, heterogeneous nature of the contacts involved: one must use a stochastic approach.
The second is the ageing inherent to the microscopic mechanisms of contact. 
To deal with that, considering the natural field or degree of freedom (usually the stress field or the location of the current contacts) characterizing the instantaneous state of the system is not enough. 
One must include some additional degree of freedom atop the natural one, i.e.~consider the dynamics of the displacement field to be non-Markovian.

\chapter{Application of Friction  to Seismic Faults}
\label{chap:EQs}

\vspace{-2cm}
\minitoc
\vspace{2cm}

Here our aim is not to give a broad overview of the physics of earthquakes, which has to do with many branches of the natural sciences, from chemistry to planetary science. 
Instead, we try to give a few clues about the most commonly accepted results of geophysics and seismology, keeping in mind that one is interested in the statistical physics point of view.
Good reviews following this angle are \cite{Rundle2003, Ben-Zion2008}.
The classic reference in geophysics is \cite{Scholz2002}.
 
\section{Phenomenology of Faulting and Earthquakes}
\label{sec:EQs}

\subsection{Faults}
\subsubsection{What is a Fault?}

\begin{figure}[]
\begin{small}
\begin{center}
\def\svgwidth{\textwidth}
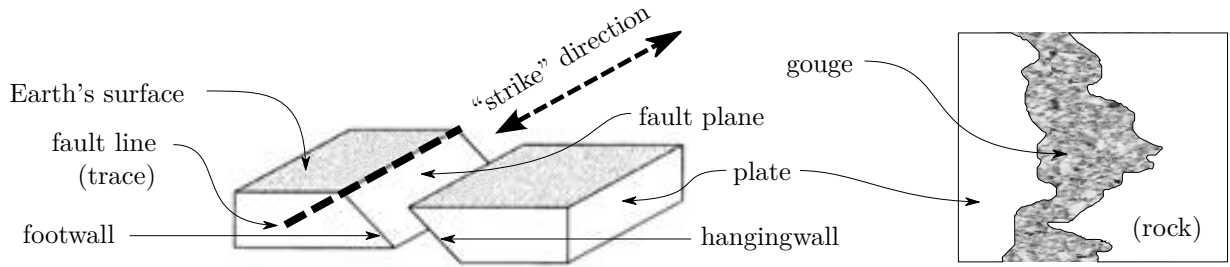
\end{center}
\end{small}
\caption{   {\footnotesize 
Left: Schematic picture of a fault and its fault plane.
If one was standing inside the fault plane, they could put a foot onto the footwall and the hangingwall would be hanging above them, hence the names.
Right: Cross-sectional view of the fault plane.
The fault plane is actually not a mathematically well-defined plane, rather it is a zone with variable width. 
Between the two plates there is the fault gouge layer (essentially thinly sheared, broken rocks which form a mix of granular materials).
\label{Fig:faults-plane}
}   }
\end{figure}

A fault is a fracture in Earth's crust along which there has been significant displacements of the two surrounding rock slabs. 
Its depth is that of the corresponding fractured tectonic plate, i.e.~the fracture extends from Earth's surface (or the ocean floor) into the \textit{schizosphere} (literally, the broken part) and stops at the \textit{plastosphere} (literally, the moldable part), in which rocks become extremely ductile and the notion of fracture is irrelevant \cite{Scholz2002}. 
We define the basics of the terminology of faulting in Fig.~\ref{Fig:faults-plane}.
The length of faults (in the direction of the trace, along Earth's surface) is widely distributed: it ranges from a few to several hundreds of kilometres. 
The boundaries between plates (representing perimeters of thousands of $km$) thus consist in a \textit{fault system}, a network of inter-plate\footnote{
There are also intra-plate faults, which form into the bulk of the tectonic plates and are responsible for a much smaller fraction of earthquakes.
}
 faults that accommodate the constraints coming from the bulk of the plates, from the magma currents in the (liquid) mantle, and from the neighbouring faults of the system itself.
These adjustments of the faults occur via sudden slips, which correspond to earthquakes in the schizosphere
(generally at depths of less than twenty kilometres) and to an overdamped plastic deformation in the plastoshpere.
When the existing faults cannot accommodate these constraints, a new fault may form, with a fault plane essentially aligned with the sum of the forces.
In some areas where a dense network of faults makes it difficult to clearly identify a single fracture plane (the fault plane), the term \textit{fault zone} is preferred to the notion of fault line.
Note that some faults slowly slip without producing earthquakes: this is why some plate boundaries are seismically inactive.

An important feature of faults is the presence of \textit{fault gouge}, a layer of thin granular materials (broken rocks) which fills the inter-plate space (see right panel of Fig.~\ref{Fig:faults-plane}).
The gouge layer can easily flow \cite{Anthony2005}, compared to the rocks of the crust which are formed by the cooling of magma and consist in large solid slabs: in this sense, the gouge layer lubricates friction between the solid plates\footnote{
The analogy between fault gouge and lubricants is limited: a reason is that granular materials do not adhere in the way liquid lubricants do.
}.
The thickness of the gouge layer fluctuates along the fault plane, and can span between a few millimetres to several hundreds of meters, depending on the history of the fault.
Too understand better the possible roles of fault gouge and a review on models accounting for the flow in granular or more generally amorphous materials, see the recent review \cite{Daub2010}.

\paragraph{Slip Geometries}

\begin{figure}[]
\begin{small}
\begin{center}
\def\svgwidth{\textwidth}
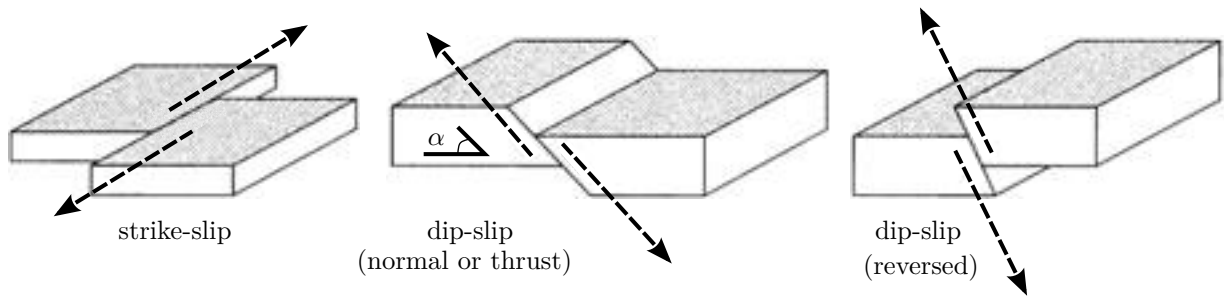
\end{center}
\end{small}
\caption{   {\footnotesize 
The two elementary kinds of slip, strike-slip and dip-slip.
When the dip (angle $\alpha$ between the fault plane and Earth's surface) is small, ``normal slip'' is also called ``thrust''.
When the footwall is going down instead of up, the dip-slip is said to be reversed.
Of course all faults are always part strike-slip, part dip-slip, but there is usually one component of the displacement that strongly dominates the other.
When both components are important, we have an \textit{oblique} fault, as that pictured in Fig.~\ref{Fig:faults-plane}.
\label{Fig:faults}
}   }
\end{figure}

Depending on the forces applied on a fault, it may remain locked (i.e.~plates do not move) or slip in various ways: see Fig.~\ref{Fig:faults} for a description of a few basic scenarios.
If the angle between Earth's surface and the fault plane (the ``dip'') is smaller than $45$\degre, the normal dip-slip fault is referred to as a \textit{thrust} fault, a particularly interesting case.
In the subduction zones (where oceanic tectonic plates sink into the mantle) this angle can actually be zero, i.e.~the two plates may lay horizontally on top of each other. 
In this sense, subduction zones are a special class of thrusts, which correspond to the largest faults on Earth, and give rise to the largest earthquakes\footnote{
Nine out of the ten largest earthquakes that occurred in the $20^ {th}$ century were subduction zone events. This includes the 1960 Great Chilean Earthquake, which at a Magnitude of $9.5$ was the largest ever recorded, the 2004 Indian Ocean earthquake and tsunami, and the 2011 Tohoku earthquake and tsunami.
}.
For these reasons, thrust faults are also the typical case of study for physicists.

\subsubsection{The Stick-Slip Instability}

In a fault system or inside a single fault, there are regions with velocity-weakening friction (in which the slip accelerates once it starts) and regions with velocity-strengthening friction (in which further slip is inhibted by the increase of friction). 
The regions of velocity weakening accumulate energy until the static friction force threshold is met. 
Once this is the case, the friction force decreases with increasing slip velocity, so that the velocity can increase up to the value at which velocity strengthening starts\footnote{
The regions of velocity strengthening also have a static friction force higher than zero, but as the friction force increases with velocity, there is no such instability, thus no earthquakes.
}.
This stick-slip instability is at the origin of earthquakes (see the section on stick-slip, sec.~\ref{sec:stickslip1}, \pp{sec:stickslip1} and the RSF laws, sec.~\ref{sec:RSF1}).
The neighbouring regions stop the propagation of slip by remaining locked, either by absorbing the stress (if their local stress is far enough from the static friction force threshold) or because they display velocity-strengthening (in which case the slip velocity sets to a very slow value and does not correspond to an earthquake).
As it is very ductile, the plastosphere is in the velocity-strengthening regime even at the smallest velocities, and thus absorbs the slip without sudden motion, i.e.~without quakes.
We note that in this context, the ageing of contacts at rest (corresponding to an increase of the static friction force during stationary contact) can play an important role in faults, as the time between two earthquakes in a given region can be very large.
This is indeed the case, and this effect is referred to as \textit{fault healing} in the geophysics community.

Studying a single fault with a well defined fault plane, regardless of its orientation (strike-slip fault on Earth's surface or thrust fault in a subduction zone), we may consider it as a simple tow-body system.
Applying the laws of friction to this system should help us to get some understanding of the mechanisms for earthquakes. 
In this simple description of a sliding fault, we see how important the RSF laws can be for seismology.
We present a few models based on these considerations later, in sec.~\ref{sec:afewEQs_models}.

We now focus on the phenomenological description of earthquakes: first their individual properties, then their statistical analysis.

\subsection{Earthquakes: Individual Characteristics}

\subsubsection{Geometrical Definitions}
\includefig{12cm}{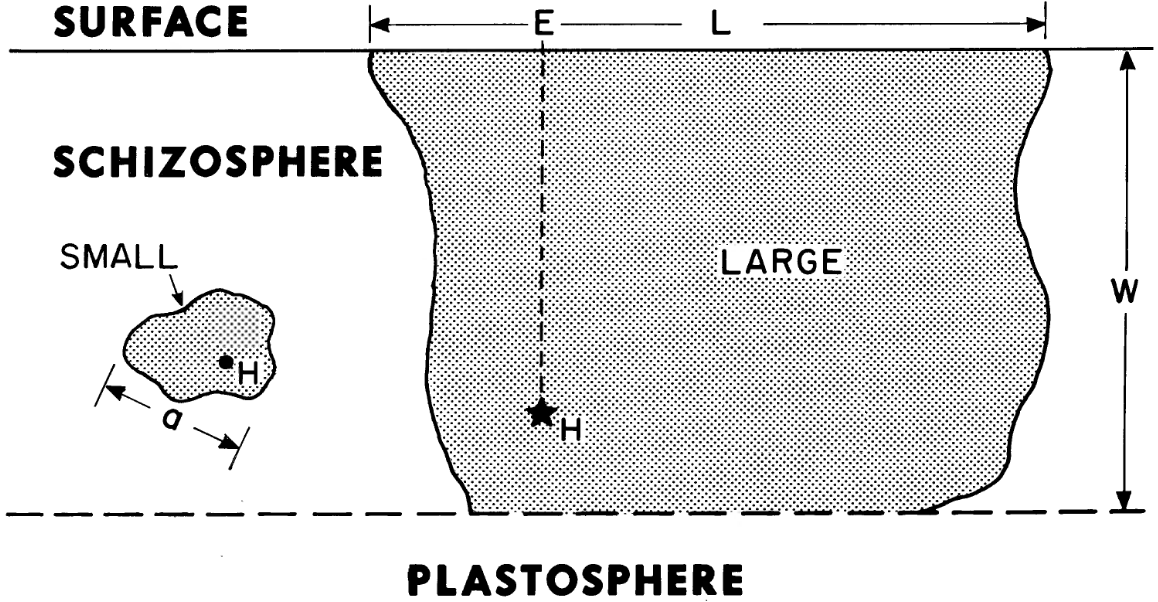}{
From \cite{Scholz2002}.
Schematic cross-sectional view of an earthquake rupture area (grey).
Left: for a \textit{small} earthquake, the rupture area can be characterized by its diameter $a$.
\newline
Right: for a \textit{large} earthquake, the rupture area reaches the surface and the plastosphere and is characterized by the schizosphere width $W$ and the along-strike length $L$.
In both cases, the \textit{hypocenter} is the point where rupture was initiated and is just below the \textit{epicentre}, defined as the projection of the hypocenter on Earth surface.
\label{Fig:hypocenter}
}
We define
 the most important characteristics of an earthquake in figure \ref{Fig:hypocenter}.
The zone that slipped, the \textit{rupture} area $A$ (grey) is especially important.
We note than independently of the magnitude of the earthquake,
it is physically meaningful to distinguish two classes of earthquakes, depending on whether the rupture area reaches both ends of the schizosphere (\textit{large}) or not (\textit{small}).
The rupture area scales either as $A\sim W L$ (large events) or $A\sim a^ 2$ (small events).

\subsubsection{Seismic Moment and Magnitude}

As a first approximation, the energy of an earthquake with average slip $\overline{\Delta u}$ over an area $A$ reads:
\begin{align}
E_S \approx \frac{1}{2} \Delta \sigma \overline{\Delta u} A,
\label{Eq:Energy_S}
\end{align}
where $\Delta \sigma$ is the average stress drop, generally assumed to be the difference between the initial stress and the threshold stress for sliding.
As the stress and stress drop are actually difficult to define unequivocally and to measure, this definition is difficult to relate to field observations.
The scalar value of the seismic moment, $M_0 = 2 \mu E_S / \Delta \sigma$, can sometimes be estimated from direct measurements:
\begin{align}
M_0 \equiv \mu \overline{\Delta u} A,
\end{align}
where $\mu$ is the shear strength in the fault. 
The full seismic moment is the core measure that we should refer to when discussing earthquake size.
Note that the definition does not depend on the nature (small or large) of the earthquake.

\paragraph{Link with the (Famous) Richter Scale}
With the recent improvements of observational tools, the estimation of the average slip has become increasingly precise, making a direct estimation of $M_0$ possible.
Historically, geophysicists had to rely mainly on the observation of seismic waves for a quantitative description of earthquakes.
Precisely, they defined the magnitude $M_S$ (the notation is a bit misleading, but firmly established) of an earthquake as the logarithm of the amplitude of a specified seismic wave measured at a particular frequency, with the distance from the hypocenter appropriately accounted for.
Seismic waves are defined by their direction (angle with the surface plane), their nature (longitudinal or transversal) and their amplitude: we do not detail the mechanisms for dissipation through radiation here, but simply remark that they could be measured quite early.
By considering the full spectrum of seismic waves, one can derive a magnitude-moment relation empirically,  as \cite{Scholz2002, Rundle2003}:
\begin{align}
\log M_0 = \frac{3}{2} M_S + 9.1,
\end{align}
where the prefactor $3/2$ is well established in the literature while the constant $9.1$ is subject to fluctuations.

\subsubsection{Characteristic Earthquakes and the Seismic Cycle}
\label{sec:characteristic_EQs}

Considering a single fault and assuming that it is not perturbed by neighbouring faults activity,
one may expect a simple stick-slip dynamics. 
This is actually what happens in some seismic regions where earthquakes occur
 on a given spot, almost periodically, with an almost constant magnitude. 
Those periodic earthquakes are referred to as \textit{characteristic} earthquakes.
We give an example of such a region in Fig.~\ref{Fig:characteristic_earthquakes}
\includefig{9.3cm}{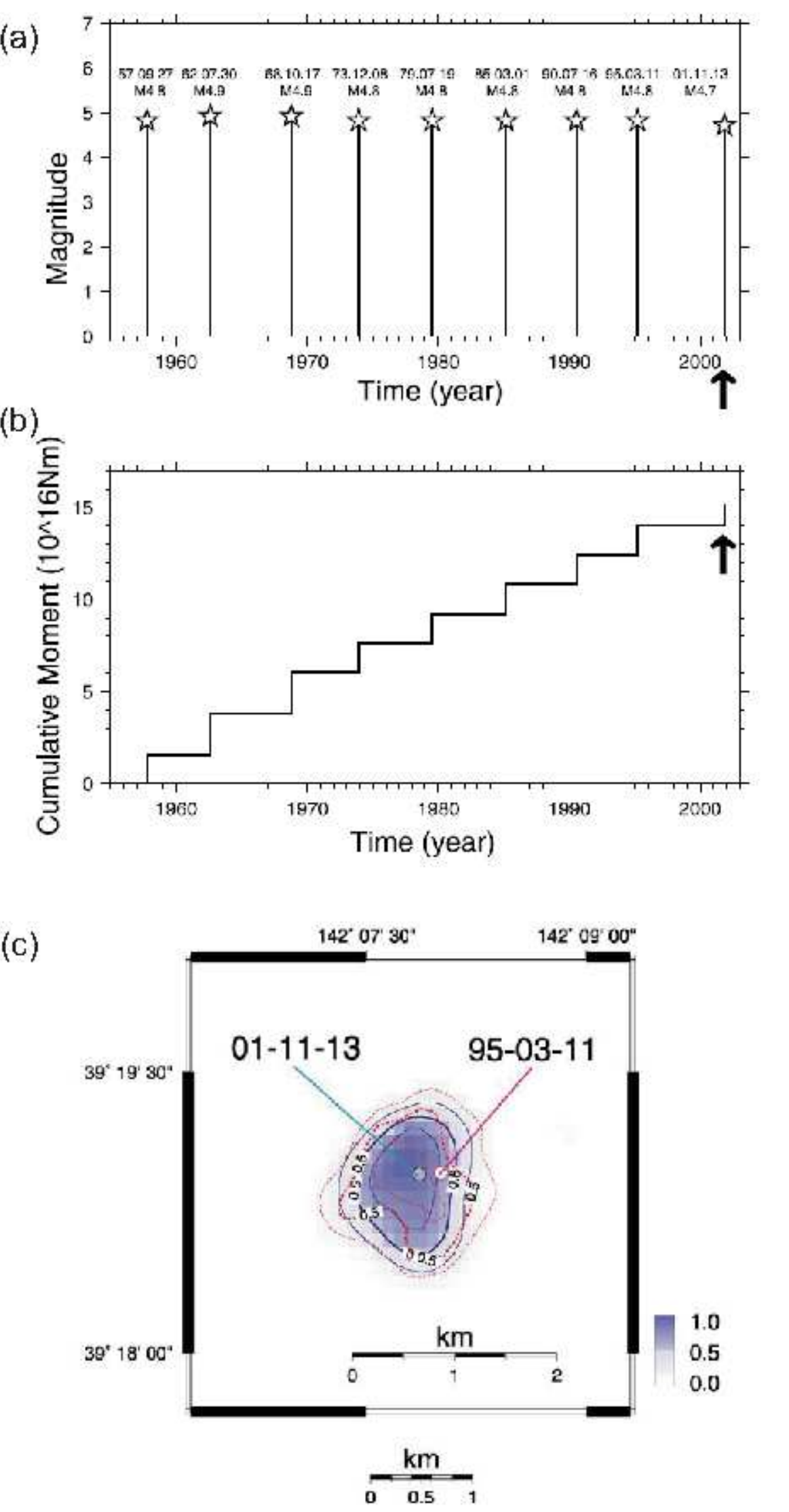}{
Originally from \cite{Okada2003}, retrieved from \cite{Kawamura2012}.
Upper panel: Large earthquakes in the Kamaishi region, with nearly the same magnitude $M\simeq4.8$ and recurrence intervals of $\sim 6$ years. 
\newline
Central panel: Cumulative seismic moment of Kamaishi earthquakes.  
\newline
Lower panel: The ``coseismic'' slip distribution of the 1995 and 2001 Kamaishi earthquakes estimated from seismic waveforms is between $0$ and $1$ meters (white to blue scale, on the right). 
Epicenters are indicated by dots linked to their date of occurrence.
The dotted purple contour line (resp.~solid blue) denotes the area of seismic slip larger than $0.5~m$ in the 1995 (resp.~2001) earthquake. 
These three panels all point towards an almost periodic behaviour, in terms of time, magnitude and location.
\label{Fig:characteristic_earthquakes}
}

The occurrence of ``characteristic'' earthquakes in a few geographical areas is echoed more generally in numerous seismic faults, for which it is argued that the inner, single-fault activity is naturally periodic.
This more general quasi-periodicity of the local seismic activity is referred to as the \textit{seismic cycle} \cite{Barbot2012, Ben-Zion2003}.
In this perspective, the non-periodic occurrence of earthquakes in most regions is interpreted as resulting from  mutual triggering of neighbouring faults between them, of which the different seismic cycles are not synchronized.
Because neighbouring faults can trigger earthquakes before the local cycle is complete, the overall seismic activity of a fault system will appear to be random \cite{Scholz2002}.
Note that this argument does not explain the non-periodic behaviour of some very large faults (as in subduction zone areas).

\subsubsection{``Constant'' Stress Drop : a Scaling Law}
\label{sec:constant_stress_drop}

We have seen that the seismic moment scales as the average stress drop and as the rupture area, which are both a priori independent random variables.
However, there is a phenomenological scaling law which seems to indicate a linear relationship between moment and rupture area, so that the stress drop seems to be fairly constant.
\begin{figure}[]
\begin{small}
\begin{center}
\def\svgwidth{\textwidth}
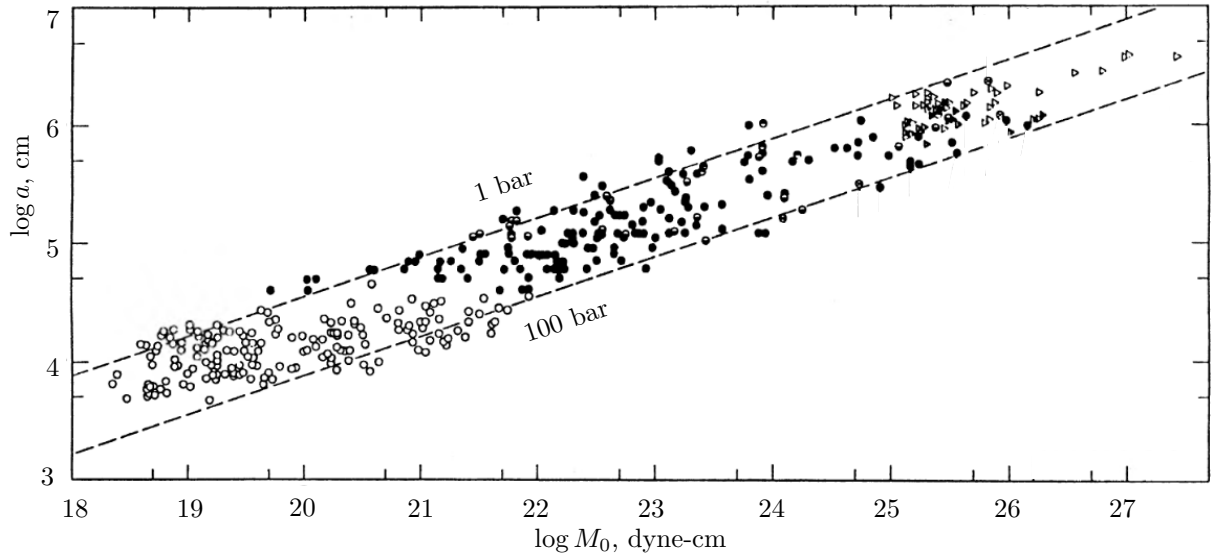
\end{center}
\end{small}
\caption{   {\footnotesize 
From \cite{Scholz2002}.
Source radius $a$ against seismic moment $M_0$ for ``small'' earthquakes.
Note the logarithmic scales: the two dashed lines indicate constant stress drops of $\Delta \sigma =1$ bar and $100$ bar.
\label{Fig:constant_stress_drop}
}   }
\end{figure}
In Fig.~\ref{Fig:constant_stress_drop}, we see that the range of stress drop seems to be controlled by the seismic moment, but the large width of this range, $\Delta\sigma \in [0.03$MPa, $30$MPa$]$ makes this ``scaling law'' a rather weak prediction.
Some models use a non-constant stress drop in order to describe fault dynamics,
 while the validity of this scaling law is also supported by recent studies \cite{Shaw2009}.
As several important points in seismology, the community has not yet reached a definitive consensus on the question of the validity of this law \cite{Scholz2002}.

\subsection{Earthquakes: Statistical Properties}

Here we provide the main two laws characterizing the statistical properties of earthquakes. 
For more details and discussion on additional scaling laws, there is the classic \cite{Scholz2002} and a comprehensive review,  \cite{Wells1994}.

\subsubsection{The Gutenberg-Richter law}
\label{sec:Gutenberg-Richter}

A very important scaling law concerns the statistical properties of earthquakes: the celebrated Gutenberg-Richter law relates the magnitude of earthquakes to their frequency.
The law states that in any region, it is found that during a given period, the number $N(M_S)$ of earthquakes with magnitude $\geq M_S$ is:
\begin{align}
\log(N(M_S)) = a - b M_S
\end{align}
where $b$ is the Gutenberg-Richter (GR) exponent and $a$ is a constant that depends on the region and time considered, which indicates the overall degree of seismicity.
\begin{figure}[]
\begin{small}
\begin{center}
\def\svgwidth{\textwidth}
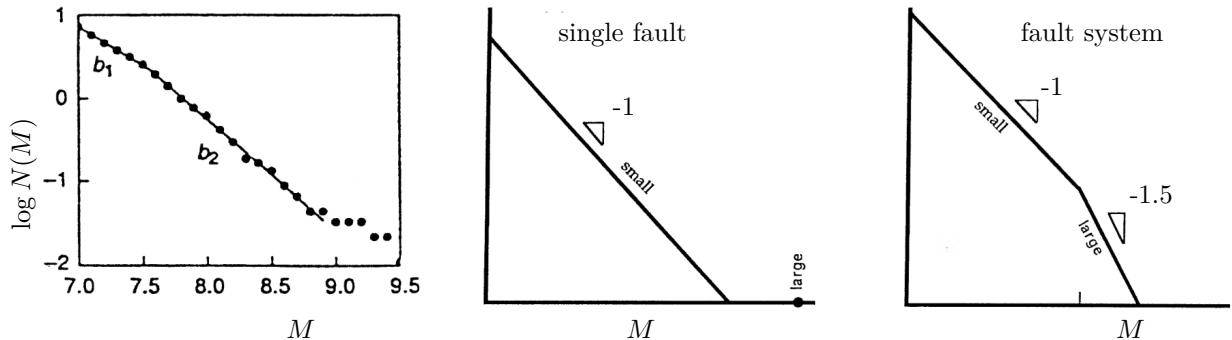
\end{center}
\end{small}
\caption{   {\footnotesize 
Adapted from \cite{Pacheco1992}.
Left: cumulative histogram $N(M)$ of the number of earthquakes with magnitude larger $M$ in some region.
Around $M\approx 7.5$ for this fault system, the $b$ exponent shifts from $b_1\approx 0.9$ to $b_2\approx 1.3$.
Central panel: for a single fault, large earthquakes occur with a typical magnitude much larger than what could be expected from interpolation of the small earthquakes regime. 
The small earthquakes are characterized by $b_1\approx 1$, typically.
Right panel: for a fault system, both small and large earthquakes have a wide distribution of magnitudes, typically characterized by exponents $b_1\approx 1$ and $b_2\approx 1.5$. 
The value of the crossover between the two regimes depends on the width $W$ of the schizosphere for the considered fault(s).
\label{Fig:b_exponents}
}   }
\end{figure}
This relation for the cumulative distribution of magnitudes $N(M_S)$ becomes a power-law for the probability distribution of the seismic moment $M_0$, with an exponent $1+B = 1+ 2b/3$:
\begin{align}
P(M_0) = \frac{1}{\mathcal{N}} M_0^ {-(1+B)},
\end{align}
where  $\mathcal{N}$ is a normalization factor.
It is often claimed that the GR exponent has a universal value of $b\approx 1$, but the situation is actually a bit more complex.

First, the small and large earthquakes (as defined above in geometrical terms) seem to have different exponents.
We report some results concerning this question in Fig.~\ref{Fig:b_exponents}.
This double scaling with a pair of exponents is rather well interpreted in terms of the geometric picture described above (Fig.~\ref{Fig:hypocenter}): it seems to be consistent with a ``finite size effect'' due to the confinement of the earthquake.
In this thesis, we will consider models for ``small'' earthquakes, which are not affected by the finite width of the schizosphere, and denote $b$ the exponent $b_1$ of ``small'' earthquakes.
This allows us to dismiss the question of the complex boundary conditions inherent to large earthquakes.

Second, there seems to be regional variations of the value of $b$: 
taking the world-wide average, one obtains $b=1$, but from one fault system to the other the value actually ranges from $0.8$ to $1.2$, i.e.~the range of values has a width of $0.4$ ($0.4$ at least: there are also claims of wider variations \cite{Hirata1989, Utsu2002}).
This question is however debated, and some of the regional variations are blamed on insufficient sampling.

\paragraph{Interpretations}

The GR law displays a robust power-law behaviour, despite its exponent being subject to fluctuations.
This intriguing scale-free property calls for an interpretation.
In the literature, we identify three common approaches that aim to explain this law.

A first approach is to consider that each fault typical produces an earthquake that scales with the fault size: since the faults' lengths are distributed as power-laws \cite{Scholz2002}, a power-law distribution for the events' sizes is natural.
This approach provides a good quantitative agreement between the different observations (earthquakes magnitudes and faults lengths) but fails to explain why the faults length have a scale-free distribution in the first place. 
In this sense, it does not explain the GR law from first principles.
The variability of events sizes in very large subduction-zone faults is also unexplained.

A slightly different approach is to argue that each fault, taken alone, would essentially follow a periodic seismic cycle with one characteristic earthquake followed by aftershocks occurring at regular intervals (this main shock does not need to entirely invade the fault plane, so that the seismic moment need not scale as the fault size, as assumed above).
If faults were independent, we would a priori obtain some distribution of earthquakes centred around the average characteristic  earthquake value.
However, as faults interact, an important earthquake in one fault can trigger an event in the neighbouring one ``before its time''.
This argument explains very well the low number of truly ``characteristic'' faults observed, however its application to a quantitative description of the GR law is subject to debate \cite{Wesnousky1994,Kagan1996,Stein2004}. 

An alternative approach is to simply assume that all faults follow a RSF friction law, are driven by the plates bulk, and possibly interact between them.
In this view, the power-law behaviour emerges from the competition between the randomness in the initial state, 
the nonlinearities of the RSF law and the driving from the plates bulk (and mantle).
A complex dynamics ensues, that some spring-blocks models somehow capture (see sec.~\ref{sec:BKmodels}).
In this approach, the heterogeneities of the crust are accounted for via the RSF laws.

A fourth angle, which is the one followed in this thesis, is to build simple models based on the fundamental features of seismic faults.
In our model, we will account for the viscoelastic interactions in the plates bulk and the heterogeneities will be represented by \textit{quenched} disorder. 
Under driving, this kind of simple model yields a rich dynamics which reproduces numerous important features of real earthquakes and RSF laws.
The important difference with the previous approach (spring-block models) is that we will account for heterogeneities and slow plastic creep directly, instead of using the effective description provided by RSF laws.

\subsubsection{The Omori Law}
\label{sec:Omori}

In seismically active regions, there is generally a \textit{background noise} of numerous very small earthquakes (with magnitude $ \lesssim 2$) that continuously occur (typically, $1$ million per year, world-wide). 
The earthquakes we discuss here are large in the sense that they are above this background noise, but they are typically ``small'' in the sense defined in Fig.~\ref{Fig:hypocenter}.
When an earthquake occurs, there are neighbouring regions in which the stress is increased: this produces secondary earthquakes, strongly correlated with the initial one (the \textit{main shock}), that are called \textit{aftershocks}.
A main shock can also be preceded by \textit{foreshocks}, i.e.~events above the background noise but much smaller than the main shock. 
Thus, the main shock is not defined as the first, but as the largest event of a correlated sequence of earthquakes\footnote{
Some alternative definitions based on some qualitative properties of the main shock formally allow the aftershocks to be larger, but this is not the most common case.
}.

A second very important law concerning earthquakes statistics is the modified Omori law (or Omori-Utsu law), which describes the aftershocks decay rate, following the main shock.
Its most widely used form is \cite{Scholz2002}:
\begin{align}
n_{as}(t)= \frac{1}{t_0(1+t/t_1)^ p},
\end{align}
where $t$ is time since the occurrence of the main shock, $n_{as}(t) \d t$ is the number of aftershocks with magnitudes greater than a specified value occurring in the time interval $[t,t+\d t]$, $t_0$ and $t_1$ are constants, and the exponent $p$ is found to be $p\approx 1$.
Note that $t_1$ is also often denoted $c$, and is generally quite small, e.g.~less than $100~s$.
The Omori law is quite important for seismic hazard estimates, as it allows to estimate the true value of the background noise and thus to identify potential foreshocks.

\section{A Few Earthquakes Models}
\label{sec:afewEQs_models}

In the first chapter we presented friction in the most common environment, i.e.~in conditions much simpler than that between plates.
In the geophysical applications, one ought to consider a plethora of secondary effects: physical peculiarities of rocks (prevalence of fracture, compared to metals), the fact that fault gouge is present and affects friction in ways very different from lubricants \cite{Bretz2006},
 high heterogeneities in the rock formations, difficulties of scaling laboratory studies up to field scales, variations in fluid pressure, rock melting at the interface, dependence of constitutive laws on pressure, temperature (which depends on the depth), etc.
Since the dynamics of seismic (and aseismic) faults is a priori quite complex, it seems reasonable to make drastic simplifications in our description, in order to sort out the relevant physical mechanisms at play.

This is the approach followed by the models we present in this section.
We won't fully review the (impressively large) literature on models of seismic faults or earthquakes propagation, but focus on the historical model (sec.~\ref{sec:BKmodels}) and a few simple variants which are connected to our work (sec.~\ref{sec:cellular_autom_OFC}).
We quickly mention the role of finite element simulations in sec.~\ref{sec:finite_elements}.

The interested reader may consult one of the two following recent works, which review the topic with a statistical-physics point of view.
We already mentioned \cite{Kawamura2012}, which emphasizes the statistical aspects of friction and of simple earthquakes models.
In the lecture notes \cite{Bhattacharyya2006}, various phenomena connected to earthquakes are reviewed, including friction, plasticity, fracture, which are treated via simple models (e.g.~fibre bundle models).
There, the emphasis is more on the mathematical treatment of the problems.

\subsection{The Burridge-Knopoff model (spring-block model)}
\label{sec:BKmodels}

\begin{figure}[]
\begin{small}
\begin{center}
\def\svgwidth{\textwidth}
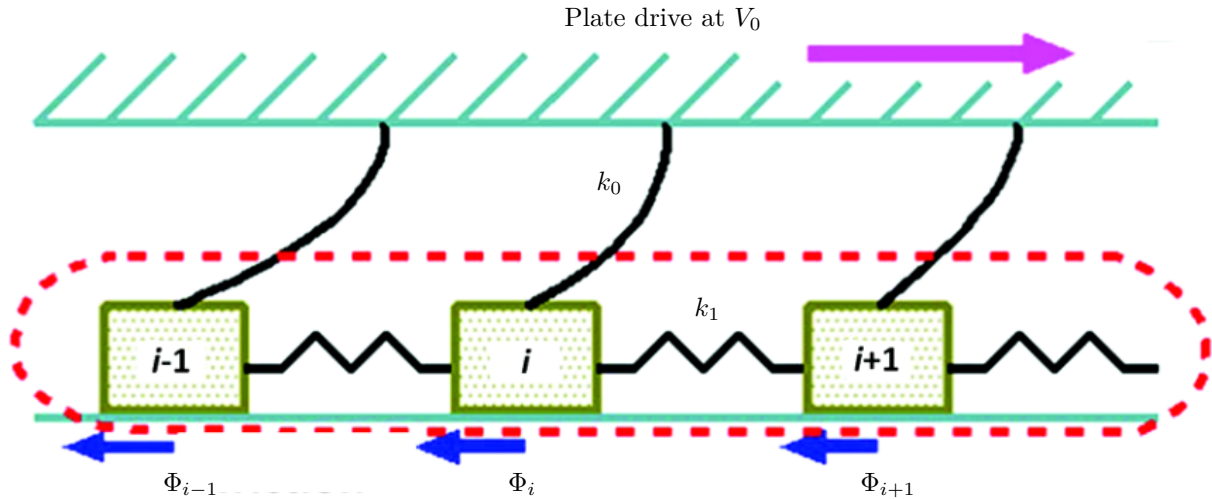
\end{center}
\end{small}
\caption{   {\footnotesize Adapted from \cite{Kawamura2012}. 
The one dimensional BK model. 
Springs $k_1$ connect the blocks together while springs $k_0$ connect them to the driving plate (pictured above). 
Some effective friction force $\Phi$ acts on each block, which are atop some rough substrate.
\label{Fig:BK_model}
}   }
\end{figure}
The Burridge-Knopoff (BK) \cite{BK1967} model (or spring-block model) is designed as a mesoscopic approach to friction in the context of seismicity:  a tectonic plate is divided in virtual blocks which are connected via elastic interactions, loading being performed via elastic interactions with a rigid plate, itself driven at some fixed velocity $V_0$ (see Fig.~\ref{Fig:BK_model}).
The core assumption of the BK model is that each block is subject to some given phenomenological RSF law.
Denoting $h_i$ the distance travelled by the block $i$ in the driving direction, $k_0$ the stiffness of the connection with the driving plate, $k_1$ the stiffness of the interactions between blocks and $m$ the mass of each block, the equation of motion reads:
\begin{align}
m \partial _t^ 2{h}_i = k_0( V_0 t -h_i) + k_1 (\nabla^ 2 h)_i -\Phi_i,
\end{align}
where $\nabla^ 2$ is a shorthand for the discrete Laplacian and $\Phi_i$ is the local friction force acting on the block $i$. 
Initially, the BK model is in one dimension, but extending to the two dimensional case is trivially done by reinterpreting the index $i$ as a couple of integers $i\equiv (x,y)$.

Of course, BK models do not intend to explain any RSF law, since the law is directly injected in the model via the function $\Phi$.
However, they provide a framework to study the collective dynamics emerging from complex friction laws, something which is expected to be relevant in individual seismic faults and fault systems.

Even in the case of the simplest friction law, defined by only two coefficients (static and kinetic) and applied to a single block, we already have an interesting stick-slip instability.
In presence of many blocks, the finite slip of a single one may pull on neighbouring blocks and trigger an avalanche of numerous one-block slips, an event that can be identified with an earthquake.
The occurrence of earthquakes in such a conceptually simple model triggered a large activity around the BK model: variations include two-dimensional blocks assemblies, models with long-range elastic interactions between blocks (which are an effective representation of the interactions via the bulk of the plate), or driving via the system boundary (train model\footnote{It is similar to a train since only one block per column of blocks is directly driven.}).
In most cases, there is no randomness in BK models: avalanches follow regular patterns, except when chaotic behaviour allows for seemingly random events.
In some variations of the BK model, the initial configuration is random, which allows power-law distribution to occur.

Using appropriate choices of RSF laws, geometry and numerical parameters, models of the BK type have been rather successful at reproducing many features of seismic dynamics \cite{Carlson1994}. 
In particular, power-law distributions of avalanches similar to the Gutenberg-Richter law and (in some occurrences) an Omori-type law for the aftershocks decay has been observed.
Variants of the original model are still studied to this day \cite{Ohmura2007, Gran2012}, especially in the geophysics community. 
See \cite{Kawamura2012} for a recent review on the results of the 1D, 2D, short-range and long-range BK models, or \cite{Ben-Zion2008} for a table summarizing the key results associated to each ingredient included into the models.

The difficulty of simulating systems with a large number of blocks (due to the nature of the equations, i.e.~coupled continuous ODEs) has pushed the statistical physics community to study simpler models in which general statistical results can be obtained, such as cellular automata representing sliding blocks.
Most importantly, the BK model \textit{assumes} a complex friction law rather than letting it emerge from simple microscopical considerations: in this sense it is simply a way to probe the collective effects of the RSF laws, not a fundamental description of frictional processes.

\subsection{Cellular Automata}
\label{sec:cellular_autom_OFC}
We define a cellular automaton simply as any system that can not be defined by a Hamiltonian or by applying the Newtonian force balance, but only via a set of \textit{rules}.
This definition includes all systems that can not be written in terms of an equation involving some time derivatives of some local state variable.

Here we focus on the Olami-Feder-Christensen (OFC) model, closely related to ``sandpiles'' models, and its connection to the problem of elastic depinning.
Other models of cellular automata (e.g.~forest fire models) and their connections to earthquake phenomena are reviewed 
 in  \cite{Rundle2003}, with an emphasis on the accurate description of actual earthquakes.
For references on sandpiles themselves, there is the classic \cite{Dhar1995} and the more recent \cite{paoletti2013}.

\subsubsection{The Olami-Feder-Christensen model}

A simple cellular automaton model that has had a large success in the Olami-Feder-Christensen model \cite{Olami1992}.
This model is equivalent to a quasistatic (infinitely small $V_0$) two-dimensional version of the Burridge-Knopoff model, using the simplest friction law possible. 
The key simplification of the OFC model is that in the quasistatic limit, one may disregard the continuous time nature of motion and replace the Newton equation for the block position (involving a numerically difficult second order time derivative) with a simple set of rules for the local stress\footnote{The stress is a tensorial quantity, but one may consider only the scalar stress resulting from the projection onto the sliding direction.}.

The friction law stipulates that when the stress of a block reaches a threshold (the static friction force), the block will slide until the pulling force (or stress) acting on it becomes zero (i.e.~the kinetic friction force is chosen to be zero and inertia is neglected).
This translates into simple rules for the stress $\sigma_i$ acting on block $i$, which were derived in \cite{Olami1992}.
Consider a square lattice of $L\times L$ sites. The system state $\{ \sigma_i, i \in[1,L^ 2]\}$ is initialized with \textit{random} values $\sigma_i\in [0,1]$.
We then have the steps:
\begin{itemize}
\item[(1)] All the $\sigma_i$'s are uniformly increased at a constant rate $k_0 V_0$ until a block has $\sigma_i=1$.
\item[(2)] Any block that has $\sigma_i \geq 1$ slips: 
the $\sigma_i$ is set to zero and all neighbouring blocks each receive an additional stress $\alpha$.
This is done in parallel for all blocks.	
\item[(3)] Repeat Step (2) until $\sigma_i<1, \forall i$. When this is the case, the avalanche is over and we may repeat Step (1).
\end{itemize}
The parameter $\alpha$ represents how much the system is conservative: on a square lattice each site has $4$ neighbours and the system is exactly conservative for $\alpha=1/4$.
This latter case corresponds to the dynamical rule of the BTW model \cite{Bak1987} (or ``Abelian Sandpile'' model), which actually inspired the OFC model.
For all values $\alpha < 0.25$, the system dissipates a stress $1-4\alpha$ at each slip, which allows avalanches to be finite even when using periodic boundary conditions\footnote{In 
the BTW model 
 dissipation occurs only at the boundaries, thus the avalanches cutoff is controlled by the system size, and it is impossible to use periodic boundary conditions.
In the OFC model the dissipation occurs in the bulk, thus the cutoff is controlled by this dissipation rate, which allows for open or periodic boundary conditions.
}.

However, the OFC model still strongly depends on boundary conditions: power-laws (and more generally Self-Organized Criticality features) are obtained only using open or free boundary conditions, which allow for additional dissipation at the boundaries \cite{Olami1992}.
This peculiarity, shared with  -- conservative --  Abelian sandpiles, is a symptom of the deterministic nature of the system.
Despite displaying seemingly random avalanches events, the OFC model randomness lies only in its initial condition, 
so that it is ``less random'' than models with \textit{quenched} randomness.

\subsubsection{Quenched Disorder and The OFC* model}

\paragraph{Quenched Disorders and Early Results}

The reason for studying ``quenched'' randomness is contained in its name: for some systems as metallic alloys, the thermal noise, fluctuating in time (corresponding to fluctuations of density, charge, etc.) can be frozen by a quench, i.e.~by sudden cooling of the hot metal (e.g.~by dipping it into cool water).
For many real systems  -- as Earth's crust --  this kind of mechanism is at the origin of heterogeneities, also called ``disorder''.
Formally a noise is said to be quenched when it is an explicit function of space but not of time.
The evolution of a $d$-dimensional system inside a $d+1$ dimensional space allows it to explore new values of the disorder over time, so that the system continually explores new values of the disorder along its evolution.

Variants of the OFC model with quenched disorder have been studied soon after the original paper: with heterogeneous redistribution coefficients $\alpha_i$ at the different sites \cite{Ceva1995, Mousseau1996} or  with heterogeneous stress thresholds \cite{Janosi1993} (which are renewed upon slip).
The first implementation of disorder is weaker than the second in the sense that the randomness is set once and for all for each $\alpha$, whereas in \cite{Janosi1993} new values of the thresholds are continuously drawn at random.
In all cases, various power-law distributions of the avalanche sizes can be obtained at least by an appropriate selection of parameters, sometimes as a robust feature (as in the second case).

Some of these early results should be taken with caution, since finite size effects may be mislead for universal properties, due to the limited system sizes available at the time.
For instance, some transition between regimes \cite{Mousseau1996} have later been shown to be simple crossovers \cite{Bach2008}. 
A persisting feature in the case of variables $\alpha_i$'s is the observation of a full synchronisation of the bulk sites (producing system-sized events, reminiscent of ``characteristic'' earthquakes), over a given parameter range \cite{Mousseau1996,Bach2008}. 
Although being very interesting, this feature is limited to the tuning of some parameters into a given range, i.e.~it is not universal.
More generally, the values of the power-laws exponents depend on the amplitude of variation of the random $\alpha_i$'s

Conversely, the features specific to the second kind of quenched disorder are very general and robust to parameter changes.
We now detail this second kind of disorder, which was introduced in \cite{Janosi1993} and studied at multiple occurrences \cite{Ramos2006,Yamamoto2010,Jagla2010}.
We will refer to this model as the OFC* model, in reference to \cite{Jagla2010}.

\paragraph{The OFC* model}
\label{sec:OFC*}
Consider a square lattice of $L \times L$ sites.
The system state $\{ \sigma_i, i \in[1,L^ 2]\}$ is initialized with an homogeneous state $\sigma_i=0,\forall i$, with each block $i$ having a random threshold $f^\text{th}_i$ drawn from some square-integrable distribution $\rho$.
The rules defining the dynamics of the OFC* model are the following: 
\begin{itemize}
\item[(1)] All the $\sigma_i$'s are uniformly increased at a constant rate $k_0 V_0$ until a block has $\sigma_i=f^\text{th}_i$.
\item[(2)] Any block that has $\sigma_i \geq f^\text{th}_i$ slips:
the $\sigma_i$ is set to zero and 
all neighbouring blocks each receive an additional stress $\alpha \sigma_i$.
A new threshold $f^\text{th}_i$ is drawn from $\rho$.
 This is done in parallel for all blocks.
\item[(3)] Repeat Step (2) until $\sigma_i<f^\text{th}_i, \forall i$. When this is the case, the avalanche is over and we may repeat Step (1).
\end{itemize}

We do not give any details about the phenomenology of this model, instead we map it to the model of the elastic interface embedded in random media, of which the behaviour is  detailed in the next chapter.

\paragraph{Mapping with Elastic Interfaces}

We consider the set of blocks $i$ and their positions $h_i$ as defined in the original BK model, and want to  translate the OFC* dynamics for the stress variable $\sigma_i$ back into an evolution equation for the positions $h_i$.
Actually, the work is already done since by definition, the stress (projected onto the driving direction) is defined as the sum of forces on the block $i$:
\begin{align}
\sigma_i \equiv k_1 \nabla^ 2 h_i + k_0 (V_0 t - h_i).
\end{align}

Considering the set of blocks as a single object, a flexible membrane or an \textit{elastic interface}, the equation of motion for this object can be written, in the overdamped limit (see sec.~\ref{sec:Equations_elastic_driving}):
\begin{align}
\partial_t h_i \propto  k_1 \nabla^ 2 h_i + k_0 (V_0 t - h_i) - f^\text{dis}_i(h_i)  ,
\end{align}
where $f^\text{th}_i$ corresponds to the local static friction force threshold, or simply to the ``disorder'' in which the interface is embedded.
This continuous equation of motion can be shown (again, see sec.~\ref{sec:Equations_elastic_driving}) to be equivalent to the rules:
\begin{itemize}
\item[(1)] Time $t$ increases until the total force acting on some site is larger than zero, i.e.~until $k_1 \nabla^ 2 h_i + k_0 (V_0 t - h_i) - f^\text{th}_i >0$ for some site $i$.
\item[(2)] Any block fulfilling this condition slips: in a discrete setup, its $h_i$ is increased by $1$.
This increases the force on each neighbour by $k_1$, while the force on $i$ decreases by $4 k_1+k_0$.
The random force $f^\text{dis}_i(h_i)$ takes a new value, $f^\text{dis}_i(h_i+1)$.
\item[(3)] Repeat Step (2) until the total force $\sigma_i- f^\text{th}_i  $ on each site is smaller than or equal to zero. 
When this is the case, the avalanche is over and we may repeat Step (1).
\end{itemize}
This corresponds to the OFC* model iff $\alpha = k_1$ and $k_0 + 4k_1=1$, i.e.~iff 
\begin{align}
k_0= 1-4k_1=1-4\alpha.
\end{align}

We see that in the case where the RSF law is replaced by a random static friction force threshold and a kinetic friction force of zero, the Burridge-Knopoff model maps onto the well studied problem of the depinning of an elastic interface in a random environment.
In this thesis, we will extend this kind of relationship to more complex models, in the same spirit.

\subsection{Finite Elements}
\label{sec:finite_elements}

\includefig{\textwidth}{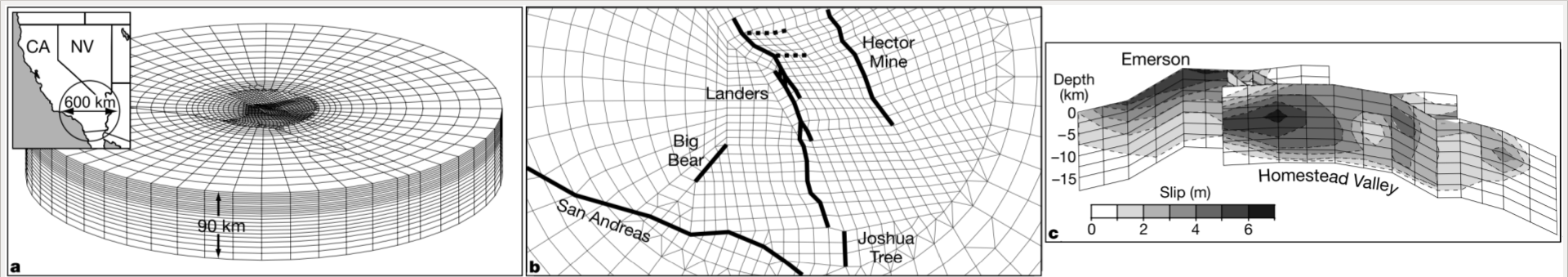}{
From \cite{Freed2001}.
Finite elements simulations of the tectonic plate in the California region.
(a): Global view of the volume simulated and geographic situation.
(b): Close up of the central region, in which the mesh is more refined.
A few important fault lines associated to large earthquakes are highlighted with bold lines.
(c): Cross sectional view of the region of slip during the \textit{Landers} earthquake.
\label{Fig:Freed2001}
}

It would be unfair to conclude an overview of earthquakes and faults models without mentioning the approach of finite elements, in which realistic continuum mechanics stress-strain laws are used to predict or retro-predict the evolution of Earth's crust.
A pioneering paper \cite{Graves1996} initially introduced the idea of using 3D finite elements methods to simulate the propagation of seismic waves into the crust.
In the same line of thought, a promising model for the evolution of the fault (and neighbouring crust) after a large earthquake was presented in \cite{Freed2001}, where a whole region of plate was simulated by finite elements (see \ref{Fig:Freed2001}).
Using a three-dimensional viscoelastic model, they simulate the stress transfer in a large region of the plate, during the 7 years following the \textit{Landers} earthquake, which allows to discuss several evolution scenarios and to find good agreement of the (retro-)predictions with observations.

Such finite element methods, with an output which is difficult to interpret intuitively, are essentially unable to predict general laws.
However, they may be used as an efficient way of probing which constitutive laws and physical effects are necessary to obtain a realistic evolution of the faults.
Reciprocally, the general results of statistical physics (e.g.~the relevance of disorder in fault systems) could be included in finite elements simulations, thus helping to improve their predictive power, something useful for producing precise seismic hazard estimations.

\section{Conclusion: Earthquakes As Test Cases}

The dynamics of seismic faults is much more complex than that of a simple large-scale manifestation of friction.
However frictional forces play a central role in faults dynamics, and geophysics can be used as a playground or test case for friction models, which can, reciprocally, help us to develop some intuition about the microscopic mechanisms at play in seismic faults.
In this respect, one should acknowledge the role of geophysics as a strong driving force for understanding the detailed mechanisms of friction.
At the same time, numerous fault models simply incorporate RSF laws as an explicit ingredient.

In this thesis, we are interested in models with a micro- or meso-scopical foundation, as our main concern is to understand how non-trivial friction laws can emerge from simple, well-understood microscopic interactions.

\chapter{Elastic Interfaces Driven in Disordered Media}
\label{chap:elastic}

\vspace{-2cm}
\minitoc
\vspace{2cm}

In this chapter, we present a model for extended disordered systems driven out of equilibrium by external forces, which displays an out-of-equilibrium phase transition or \textit{dynamical phase transition}.
Starting from a few simple microscopic rules, one obtains a non-trivial critical behaviour with avalanches that are somewhat reminiscent of earthquakes.

We first define the model of an elastic interface uniformly driven in a disordered medium.
We then introduce another type of driving, more appropriate for the study of avalanches. 
After explaining the transition, its exponents and scaling relations in finite dimensions, we present two techniques for solving the mean field problem, the second one being more flexible (and useful in the next chapter).
Finally we present a few conventional extensions of the problem, thus showing how broad this framework is.
Despite a strong robustness of the model to various microscopical changes and a large range of applicability, we unveil several flaws in its relevance for friction or seismic faults dynamics.

\section{The Elastic Interface in a Disordered Medium }

The model of an elastic interface in a disordered medium we are about to present provides a good description of the interfaces between magnetization domains appearing in disordered ferromagnetic materials ($3$ dimensional case, two-dimensional interface) or in thin magnetic films (two dimensional medium, one dimensional elastic line).
In particular, this model succeeds \cite{Durin2000} at explaining the so-called Barkhausen noise \cite{alessandro1990, Zapperi1998a, Durin2006} measured in ferromagnetic materials, i.e.~the surprising observation that magnetization domains can move via large jumps or avalanches which follow power-law distributions over a large range of length scales.
This critical phenomenon is captured by the so-called \textit{depinning transition}.
There are numerous excellent reviews on the depinning transition, such as the historical ones \cite{Fisher1998, Kardar1998}, or more recently \cite{Giamarchi2006, Giamarchi2008}.

There are several other successful applications of the depinning framework to real systems, such as 
crack propagation in brittle materials \cite{Gao1989,Alava2006, Bonamy2008, bonamy2011failure},
contact lines in wetting \cite{Joanny1984, Moulinet2004}, 
in particular wetting fronts moving on rough substrates  \cite{Rosso2002b, Moulinet2004, LeDoussal2009a} or
 wetting fronts in porous media \cite{Saha2013,Atis2013,BouMalham2010},
 dislocation assemblies (i.e.~crystal deformation) \cite{Miguel2006,MiguelBook2006,Zapperi2006,Miguel2008},
 in particular in vortex polycrystals in type II superconductors \cite{Moretti2004,Moretti2004a,Moretti2005,Moretti2005a}.
 In this last case, the connection between  (poly)crystalline and amorphous vortex matter was also studied, within the depinning framework \cite{Moretti2009}.
However we will refer to the historical setup of the original, ferromagnetic case as the physical reference in the following discussion.

\subsection{Construction of the Model}

\subsubsection{Continuous Equation of Motion}

Consider a $d$-dimensional manifold embedded in a $(d+1)$-dimensional space.
We may denote $(\mathbf{x},z)$ or $(x,z)$ the $(d+1)$-dimensional coordinates of any point and $h(x,t)$ the scalar function describing the position (or ``height'') of the manifold along the last coordinate (i.e.~$z$).
The function $h$ is univalued, i.e.~it has no overhangs (see sec.~\ref{sec:overhangs}).
For numerical simulations, we discretize the $\mathbf{x}$ space on a lattice with $L^d$ sites, numbered by an index $i$. 
The $z,t$ coordinates still vary continuously (up to numerical precision), i.e.~$h(i,t) \in \mathcal{L} (L^d\times \mathbb{R},  \mathbb{R})$.

\paragraph{Elasticity}
The elastic interactions within the interface tend to minimize the local curvature, $\nabla^2 h$. 
The elastic energy of the line can be written:
\begin{align}
E_\text{elastic} =  \int \frac{k_1}{2} (\nabla_\mathbf{x}  h)^2 \d^d \mathbf{x},
\end{align}
where $k_1$ is some effective stiffness constant (homogeneous to a spring stiffness or membrane elasticity).
This corresponds to a local force $F_\text{elastic}(x) = -\partial E_\text{elastic}/\partial h= k_1 \nabla_x^2 h$.

\paragraph{Driving}
As the magnetic field $F$ is increased in the material, the average magnetization increases at the interface, thus pulling (or pushing) it accordingly: this is modelled by the driving term with coupling energy  
\begin{align}
E_\text{drive} = - F. h(x,t),
\end{align}
 where $F$ accounts for the intensity of the applied magnetic field. 
 This gives a driving force $ F_\text{drive}(t) \equiv -\partial E_\text{drive}/\partial h = F = \text{const.}$, similar to a simple drift term.

\paragraph{Disorder}
Space is filled with quenched disorder, i.e.~we have a random force $\eta(z,x)$ which does \textit{not} evolve over time. 
Its statistical properties are determined by its first two moments, the higher ones being essentially irrelevant (one generally assumes a Gaussian statistics):
\begin{align}
\overline{\eta(z,x)} &=0\\
\overline{\eta(z_1,x_1) \eta(z_2,x_2)} &= \delta_x(x_1-x_2) \Delta_z(z_1-z_2),
\end{align}
where $\delta_x, \Delta_z$ are some short-range functions (typically a Dirac for $\delta_x$, but a function with some range $r_f$ for $\Delta_z$).
The notation $\overline{X}$ stands for average over realizations of the random variable $X$.
We can always choose the average to be zero because a non-zero average can be absorbed into the driving force (a simple shift of $F$).
The corresponding force reads $F_\text{disorder}= f^\text{dis} \eta[h(x,t),x]$, where $f^\text{dis}$ is the typical strength of the disorder.
Note that this kind of action of the disorder is called \textit{Random Field}, since it is the local fields which are random\footnote{An 
alternative kind of disorder is the \textit{Random Bond}, which is discussed at the end of this chapter, along with other variations on this precise kind of elastic interface.}.

For simulations, a typical choice is to take the $\eta(z,i)$'s to be a set of $L^d$ independent Gaussian noises with short-range correlations in the $z$ direction. 
To obtain a range $r_f \approx 1$, one may draw each $\eta(\lfloor z \rfloor,i)$ from a zero mean, unit variance normal distribution and interpolate between nearest neighbours for the non-integer values of $z$.
This allows for an easy and rather efficient numerical implementation, however there is a better strategy that we discuss in sec.~\ref{sec:narrow_wells}.

\paragraph{Conclusion: the Equation}
In the overdamped limit, denoting $\eta_0$ the effective viscosity for the interface, we may apply Newton's equation to each point $h(x,t)$ and obtain the Langevin equation:
\begin{align}
\eta_0 \partial_t h(x,t) = F + k_1 (\nabla_x^2 h)(x,t) - f^\text{dis} \eta[h(x,t),x], \label{Eq:ForceDepinning}
\end{align}
where the brackets $[..]$ highlight the functional dependence.
Note that the equation is non-linear due to the last term: since $\eta$ is a random distribution it is definitely not linear in $z$. 
The key difficulty of the depinning problem is to deal with this non-linearity.

We just want to add that the equation without disorder (but with thermal noise) is historically referred to as the \textit{Edwards-Wilkinson} equation, so that the Langevin equation for the depinning is sometimes nicknamed \textit{quenched Edwards Wilkinson}, as for ``Edwards Wilkinson equation with \textit{quenched} noise''.

\subsubsection{Dynamics}

\paragraph{Continuous Dynamics}
\label{sec:cellullar_auto1}

\begin{figure}[]
\begin{small}
\begin{center}
\def\svgwidth{0.9\textwidth}
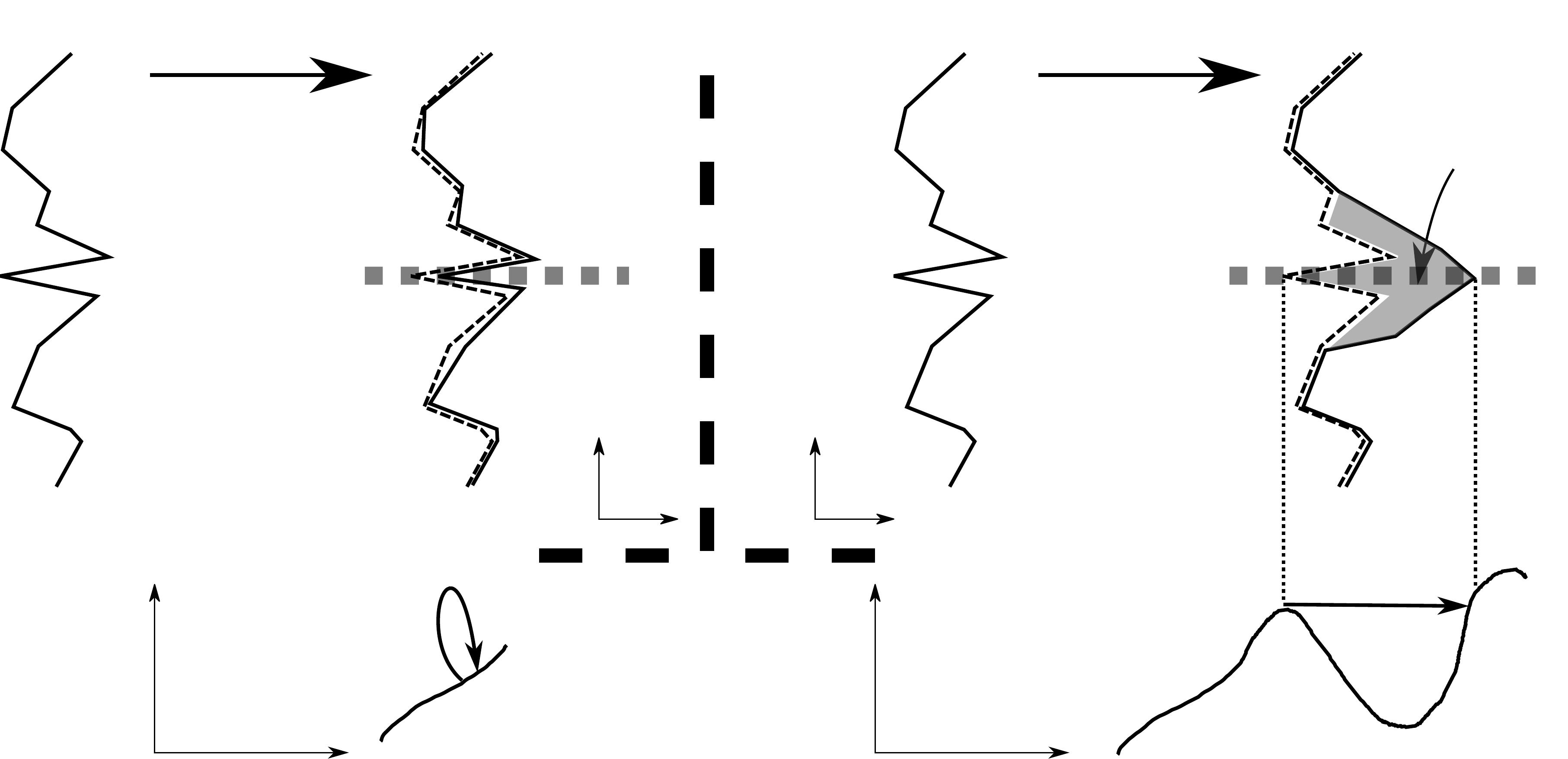
\end{center}
\end{small}
\caption{   {\footnotesize 
Left: infinitesimal advance of the interface upon an infinitesimal increase of the force by $\delta F$.
The disorder force is locally increasing at all points.
\newline
Right: avalanche triggered by an infinitesimal increase of the force by $\delta F$.
The disorder force is locally decreasing in $x=x_1$: this point will move forward until either the disorder force $\eta[h(x_1),x_1]$ or the elastic interactions stop it.
\label{Fig:Interface_continue}
}   }
\end{figure}
The dynamics of motion is twofold, depending on the local slope of the disorder function $ \eta[h(x),x]$ at each point $x$.
\begin{itemize}
\item[(i)]
If $\eta[h(x),x]$ is increasing everywhere, an infinitesimal increase in $F$ will result in an infinitesimal advance of the interface and to a corresponding infinitesimal adjustment in  $ \eta[h(x),x]$. See the left part of Fig.~\ref{Fig:Interface_continue}. 
\item[(ii)]
If $\eta[h(x),x]$ is decreasing at a point $x$, an infinitesimal increase in $F$ will result in an advance of the interface which will stop only when the forces acting on the interface cancel once again.
See the right part of Fig.~\ref{Fig:Interface_continue}.
\end{itemize}
This point of view is especially adapted to treat sets of successively pinned (motionless) configurations.
In a dynamical regime, \req{ForceDepinning} would be more suitable.

This kind of dynamics is 
impractical because upon a slight increase in the drive the interface may adapt smoothly, resulting in infinitely many infinitesimal ``avalanches'' that need to be discarded by some small-size cutoff in the avalanches definition.
This is worrisome for analytical arguments, but also implies many fruitless computations in terms of numerics.

\paragraph{The Narrow Wells ``Approximation''}
\label{sec:narrow_wells}

We are interested in the universal properties of the large and discontinuous avalanches.
To get rid of the  numerous infinitesimal avalanches, we propose to replace the continuous disorder function $f^\text{dis}\eta[z,\mathbf{x}]$ with a function being zero everywhere except for countably many, randomly located positions where it has random values, similarly with the strategy adopted in \cite{Fisher1998}.
\begin{figure}[]
\begin{small}
\begin{center}
\def\svgwidth{13cm}
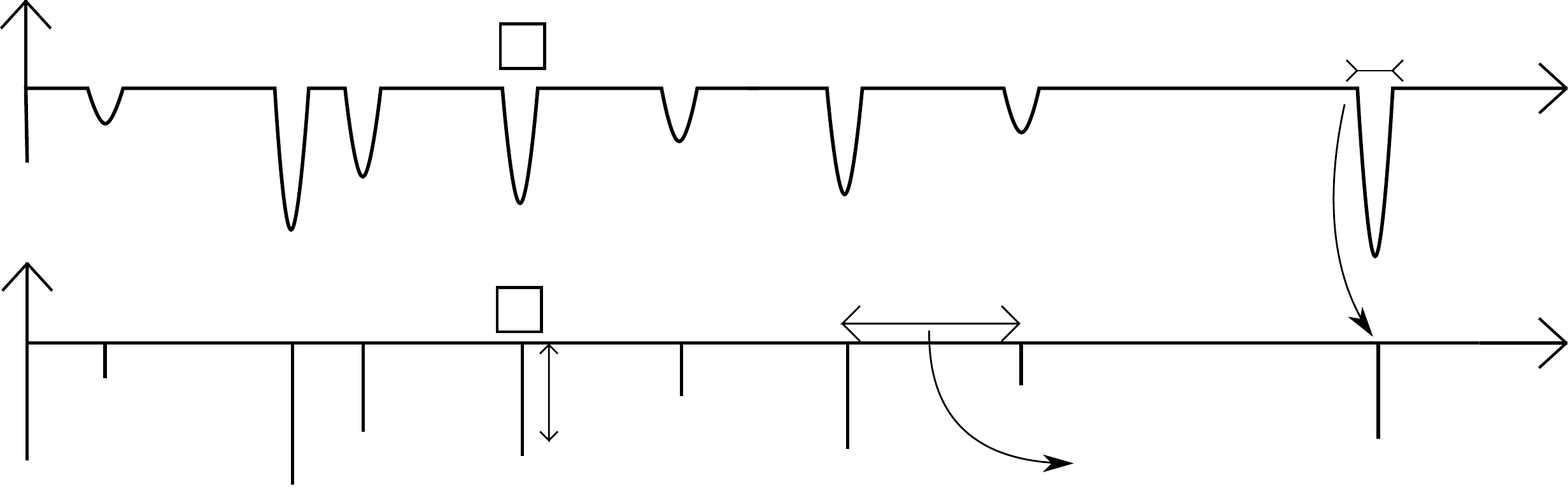
\end{center}
\end{small}
\caption{   {\footnotesize 
Up: Physical picture of the narrow wells: the disordered energy potential consists in wells with some finite width.
\newline
Down: idealized picture. 
The width is taken to be zero, so that each block's position $h_i$ can only take values in a countable set of positions.
The narrow wells are now characterized only by the threshold force $f_i^{\text{th}}$ needed to exit and their spacings $z$.
\label{Fig:narrow_wells}
}   }
\end{figure}

See the upper part of  Fig.~\ref{Fig:narrow_wells}.
Physically, the disorder energy landscape is seen as a collection of narrow wells representing impurities.
Along the $h$ direction, the narrow wells are separated by random intervals (spacings) $z$ with distribution $g(z)$ and mean length $\overline{z}= \int_0^\infty z g(z) \d z$. 
A natural choice for $g(z)$ is the exponential law, which corresponds to the case where impurities are uncorrelated in space\footnote{The law of the spacings between points uniformly drawn on a line is the exponential law. 
The average spacing is easily obtained from the linear density of the uniform distribution.
}.
The value of the disorder force in a well depends on its shape, essentially defined by the width along the $h$ direction and the depth. 
We will assume that the spacings are not too large compared to the well's depth\footnote{To do it properly, we need to compare $\sim k_1 \overline{z}$ with the disorder force, which is essentially a ratio of the depth to the width of the well.}, so that any time a site escapes from a well, it will directly jump to the next one, never staying in between two wells (if $ k_1 \overline{z}$  is small enough).

We will also make the crucial assumption that the wells are narrow, i.e.~their widths are negligible compared to $\overline z$, so that the displacement of a point trapped in a well is negligible compared to the jumps between wells (see the lower part of Fig.~\ref{Fig:narrow_wells}).
To exit from a well, a block will need to be pulled by a force larger than some threshold $f^{\text{th}}_i$ related to the well's shape. 
With a given (infinitesimal) width and some random distribution of depths, we obtain some stochastic distribution for the threshold forces $f^{\text{th}}_i$ (there is one set of those for each site $i$). 

To summarize, using infinitely narrow wells of finite depth, with randomly distributed spacing lengths, each block is always located in one of the (discrete) wells and its coordinate $h_i$ evolves only via finite jumps $z$ with a distribution $g(z)$.

Under these assumptions, the continuous dynamics of the blocks becomes fully discrete and the issue of having infinitely many infinitesimal avalanches disappears. 
The dynamics is straightforward: we only have the possibility (ii) of the previous dynamics.
As long as each site fulfils the stability condition:
\begin{align}
F+ k_1 (\nabla_x^2 h)(x) < f^ \text{th}_i, \qquad \forall x\in L^d,  \label{Eq:NWstability1}
\end{align}
the interface does not move at all.
When the increase of the force is enough to violate \req{NWstability1} in one point $i$, the interface locally jumps forward to the next well, i.e.~$h_i$ increases by $z$ (drawn from $g(z)$), and a new threshold force $f^{\text{th}}_i$ is drawn at random.
This process is iterated for all unstable sites until  \req{NWstability1} is valid again.

We want to stress that the ``approximation'' of narrow wells does not limit the generality of our presentation: the narrow wells disorder can be reduced to a  Gaussian white noise, using $g(z)\approx \delta^{Dirac}(z-1)$ and $f^{\text{th}}_i$ drawn from a zero mean, unit variance normal law.

\subsection{The Depinning Transition (Constant Force Driving)}

\paragraph{The Critical Force}
If we start from a very small force $F$, the interface will easily be pinned.
As we increase the force, at some point an impurity (sticking less strongly to the interface) will finally let go.
This may cause the neighbouring impurities to also detach right after the next one, and then their own neighbours, and so on, almost instantaneously (on a time scale $\eta_0^{-1}$).
This chain reaction or \textit{avalanche} stops when 
the interface is finally pinned down and the local velocity is zero everywhere: $\partial _t h = 0$.

If we again increase the force by an infinitesimal amount $\delta F$, a new avalanche may be triggered.
Keeping the  perturbation $\delta F$ constant, for larger forces $F$, the interface will need to find stronger impurities in order to stop, something that will become more rare: the avalanches will get bigger with increasing $F$ (and constant $\delta F$).

\begin{figure}[]
\begin{small}
\begin{center}
\def\svgwidth{8cm}
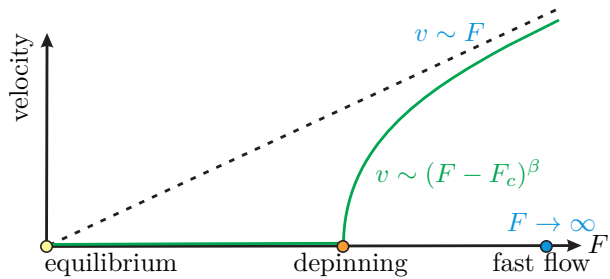
\end{center}
\end{small}
\caption{   {\footnotesize 
Adapted from \cite{Ferrero2013a}: the ``phase diagram'' of the depinning transition at zero temperature.
For $F<F_c$, we are in the pinned regime, $v=0$.
For $F \gtrsim F_c$ we are in the depinned regime, $v\sim (F-F_c)^\beta$.
See Fig.~\ref{Fig:v_de_F_Creep} for the case with temperature $T>0$.
\label{Fig:v_de_F_T=0}
}   }
\end{figure}

Above a certain force, the occurrence of impurities strong enough to pin the entire line will switch from rare to non-existent, so that the center of mass will never rest: we say that the interface is \textit{depinned}, and we have: 
\begin{align} 
v(t) &>0, \qquad \forall t  \\
\text{with }\quad v(t) = \langle  \partial_t  h(x,t) \rangle &\equiv \frac{1}{L^d} \int \partial_t h(x,t) \d ^d x .
\end{align}
This threshold force above which the velocity is positive is called the \textit{critical force} and is often denoted $F_c$. 
Note that above it, some pieces of interface may be at rest sometimes, i.e.~we may have locally $\partial_t h(x,t) =0$.
The precise relationship between the instantaneous average interface velocity and the force $F$ is shown in Fig.~\ref{Fig:v_de_F_T=0}, where we see that the critical force clearly plays a role analogous to that of a \textit{critical point}, separating a \textit{pinned phase} from a \textit{depinned phase}.

\paragraph{The Critical Force and the Larkin Length}
\label{Larkin}

The prediction of a finite pinning force is due to Larkin \cite{Larkin1979}, in the context of vortices depinning, dislocations in solids \cite{Labusch1970} or domain walls in ferromagnets \cite{Hilzinger1975}.

At small length scales the non-linearity of the disorder can be neglected (namely $f^\text{dis} \eta[h(x,t),x]$ can be replaced with $f^\text{dis} \eta[x]$) and the system reaches a stationary state, in which the interface moves rigidly, without internal rearrangements.
At the length scales where the interface deformation 
 is of the same order as the microscopic disorder correlation length, this approximation does not hold and the non linearities of the disorder should be accounted for.
 This length is called the Larkin length $L_c$ (or ``correlation length'' \cite{Muser2004, Persson1999}) and separates two regimes: 
 \begin{itemize}
 \item
For systems sizes smaller than $L_c$, we observe a deterministic dynamics with a pinning force which depends on the system size.
\item 
 For larger systems ($L>L_c$), the critical force becomes independent of the system size and the dynamics is strongly intermittent.
 \end{itemize}
The Larkin length can be computed analytically, and is shown to be very large in systems with long-range interactions.
The absence of intermittent dynamics (avalanches) in friction experiments at the laboratory scale can be interpreted as a finite size effect (system size smaller than the Larkin length) \cite{Caroli1998, Persson1999, Vanossi2013}.

In what follows we deal with systems of infinite size, for which the depinning transition displays an intermittent dynamics characterized by universal scaling laws and exponents, defined below.

\paragraph{Critical Exponents}
Considering $F$ to be a control parameter and $v$ to play the role of the order parameter, we see that the system undergoes what we may call a \textit{dynamical phase transition}, in analogy with equilibrium phase transitions\footnote{
Note that phase transitions only happen in the macroscopic limit (infinite system size).
This is no exception and for finite systems (as in numerics) one observes a dynamical crossover instead of a transition. 
A careful analysis of the size effects reveals that we have truly a transition.
We do not discuss that here.
}.
As the velocity near $F_c$ is given by
\begin{align}
v&= 0, \qquad \text{ for } F<F_c \\
v&\sim  |F-F_c|^{\beta}, \qquad \text{ for } F>F_c,
\end{align}
 we have a second-order phase transition\footnote{A phase transition is of second order if only the derivative of the order parameter is discontinuous at the transition. If the order parameter is discontinuous, it is a first-order phase transition. If only the second-order derivative was discontinuous, it would be a third-order phase transition. 
}.
Furthermore, close to criticality the interface develops roughness, i.e.~it becomes self-affine and is characterized by an exponent $\zeta$:
\begin{align}
\sqrt{\langle [h(x)-h(0)]^ 2 \rangle } \sim x^\zeta.
\end{align}
We will precise the domain of validity of this relation soon.

The huge difference with equilibrium phase transitions is that here, we are not at equilibrium: since the impurities pin the line, the system does not fully explore the phase space, and there is no way for it to find equilibrium.
The transition is called \textit{dynamical} because the ``phases'' correspond to different dynamical states: pinned (not moving) or de-pinned (moving)\footnote{In this sense, it's the phases that are dynamical (or not), not the transition.}.

However, the analogy goes further than a simple qualitative change upon variation of a parameter.
At $F\sim F_c$ the competition between disorder and elasticity prevails and a critical state emerges, in which all the quantities of interest are distributed as power-law distributions of the distance from criticality, $\Delta_c= |F-F_c|$.
We qualitatively explain this critical behaviour below.\\

\begin{figure}[]
\begin{small}
\begin{center}
\def\svgwidth{\textwidth}
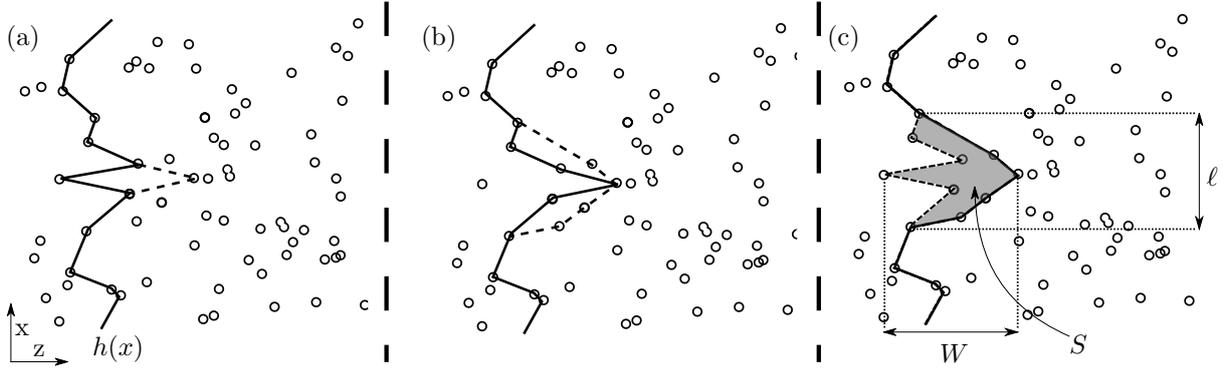
\end{center}
\end{small}
\caption{   {\footnotesize 
Avalanches characteristic properties.
Schematic view of a one-dimensional interface embedded in a random (disordered) medium. 
(a): Initial state (solid line). The line $h(x)$ is depinned at a single point (dashed line).
(b): Intermediate state. The neighbours of this point are also depinned, which allows the interface to move even further.
(c): After some additional local de-pinnings of the interface, it is once again pinned in the final state (solid line). For comparison, we also recall the initial state (dashed line). 
The avalanche lasted for a very short time, it is characterized by three spatial measures:
the lateral length $\ell$, the width $W$ and the size $S$ (the area swept over by the interface during motion, highlighted in grey), which is $S \sim \ell^d W$.
\label{Fig:interface_schema_simpliste}
}   }
\end{figure}

\paragraph{Avalanche Lateral Length}

Far below $F_c$ (very small forces), an initially flat interface will remain essentially flat, and its correlations will be those of the initial configuration. 
As the force increases towards $F_c$, the three contributions start to balance each other, increasing the instability of the interface: each point is in a metastable state, on the verge of jumping ahead.
Hence each growing avalanche can easily destabilize the neighbouring (metastable) areas, thus increasing its size: upon infinitesimal increases $\delta F$ of the force, larger and larger avalanches are triggered.  
As all forces are present equally, it also takes a long time for an unstable part to find a locally stable configuration and stop. 
At the transition, the interface becomes depinned, i.e.~it is in an infinite avalanche. 
This divergence to infinity of the avalanches maximal lateral length $\ell^\text{max}$ reads:
\begin{align}
\ell^\text{max}_{\{F<F_c\}} \sim (F_c-F)^{-\nu}.
\end{align}
We note that configurations verifying \req{NWstability1} close to the transition should be called \textit{metastable} rather than stable, because an infinitesimal perturbation ($\delta F$) may trigger a large avalanche instead of a return to the same configuration.

Far above $F_c$ (very large forces) the disorder force is swept over very fast, resulting in an effective thermalization (or annealing) of the noise. 
This results\footnote{
Anticipating on the following definitions of the critical exponents, we can prove this.
In the reference frame of the center of mass of the moving interface, the disorder effectively felt is $\eta[h(x)+vt,x]$. 
Using $h(x)\sim x^ \zeta$, $v(x) \sim x^ z (F-F_c)^ \beta$ and the fact that $z>\zeta$, we can see that $\xi \to 0 $ when $F\gg F_c$.
} in different pieces of the interface moving relatively independently, thus $\ell^\text{max} \to 0$ when $F\gg F_c$.
At smaller forces ($F \gtrsim F_c$), when the three terms compete equally, the interface motion is continuous (it never stops everywhere at once) but essentially consists in numerous (almost) individual avalanches acting in parallel.
Thus the avalanches still have a typical length scale $\ell^\text{max}_{\{F>F_c\}} = (F-F_c)^{-\nu}$, with the very same exponent $\nu$, an a priori surprising ``coincidence'', that is predicted by analytical works and verified numerically.

We now link the avalanche lateral length and the correlation length.
Below or above the threshold force, a point locally stops exactly when the local forces reach a balance, so that the local correlations at rest are large.
Thus, at the end of an avalanche, the region that moved is strongly correlated within itself, and we can identify the correlation length of the interface at rest $\xi$ with the maximal avalanche lateral length $\ell^\text{max}$:
$\xi_{\{F<F_c\}} \sim (F_c-F)^{-\nu}$,
where we understand that the name $\nu$ was chosen in analogy with equilibrium phase transitions.
Above threshold, there is also a correlation length, even though the interface is never fully at rest: it represents the correlations of the moving interface (which can have some parts at rest).
The common origin of the correlation length above and below threshold allows us to write the general form:
\begin{align}
\xi = |F_c-F|^{-\nu}.
\end{align}
We note that in this out-of-equilibrium transition, a static observable ($\xi$) derives from a dynamical one ($\ell^\text{max}$).

\paragraph{Avalanche Width and Size}

We have only mentioned the lateral length (along the $\mathbf{x}$ plane) up to now.
The size $S$ of an avalanche is the total area or volume swept over by the interface during an avalanche:
\begin{align}
S \equiv \int \d^d x \lp h_\text{after}(x) - h_\text{before}(x)\rp,
\end{align}
where the integration spans over the entire system (or the zone affected by the avalanche, it is the same) and the labels are explicit.
By extension, the avalanche of size $S$ is sometimes called $S$, and $\{S\}$ may denote the sites involved in an avalanche.

The width of an avalanche can be defined as the maximal gap between the height $h(x)$ of any two points $x$ that where involved in an avalanche: 
\begin{align}
W= \underset{x\in \{S\}}{\max} (h_\text{after}(x)) - \underset{x\in \{S\}}{\min} (h_\text{before}(x)),
\end{align}
a definition that is best understood via Fig.~\ref{Fig:interface_schema_simpliste}

The same line of argumentation as for the avalanche lateral length shows that the avalanche width must diverge as some power of the distance to criticality, i.e.
\begin{align}
W^\text{max} \sim |F_c-F|^{-\nu \zeta}
\end{align}
which can be more practically written $W\sim \xi^\zeta$.

\label{depinning_roughness}

\begin{figure}[]
\begin{small}
\begin{center}
\def\svgwidth{\textwidth}
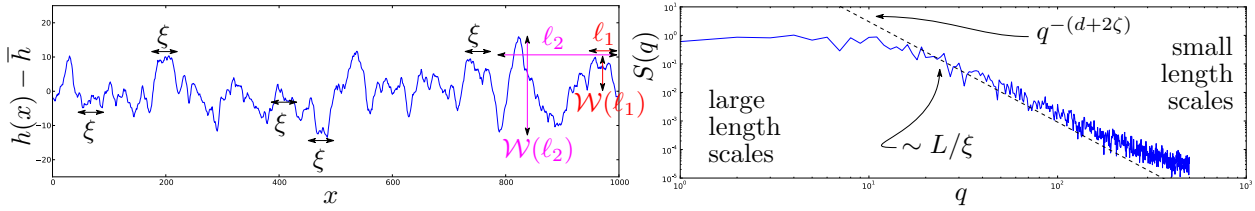
\end{center}
\end{small}
\caption{   {\footnotesize 
Illustration of the notion of interface roughness.
Left: Profile or height elastic interface in random medium in $d=1$.
Here the interface is discretized over $L=1000$ sites. 
We may notice that the interface is strongly correlated over distances $\leq \xi \approx 40$.
Right: Average structure factor associated to a few realizations of the interface shown on the left.
For large length scales $\ell>\xi$, the structure factor takes large values, i.e.~the interface is not correlated over large distances (no more than at $\xi$).
For small length scales, the interface is strongly correlated. 
Its correlations are cahracterized by a self-affinity exponent $\zeta\approx 1.26$.
\label{Fig:roughness_simple_zeta}
}   }
\end{figure}

\paragraph{Interface Roughness}
The interface roughness or typical width is defined as: 
\begin{align}
\mathcal{W}(\ell) &\equiv \sqrt{\langle [h(\ell)-h(0) ]^2  \rangle}.
\end{align}
For the same reasons that explain the correspondence of $\ell^\text{max}$ and $\xi$, the maximal avalanche width (or depth) $W^\text{max}$ and the interface typical extension $\mathcal{W}(\ell) $ are of the same order of magnitude.
Thus, the metastable states and the moving interface are both characterized by a self-affine profile $h(x)$ with roughness exponent $\zeta$ and a large-size cutoff given by the correlation length $\xi$:
\begin{align}
\mathcal{W}(\ell) &\sim \ell^ \zeta, \quad \forall \ell \leq \xi \\
\mathcal{W}(\ell) &\sim \xi^\zeta , \quad \forall \ell \geq \xi
\end{align}
Note that the second line denotes a flat interface at large scales, since for $\ell \gg \xi$, $ \xi^\zeta \ll \ell^\zeta$.
An intuitive definition of $\mathcal{W}(\ell)$ is given in Fig.~\ref{Fig:roughness_simple_zeta}.

\paragraph{Avalanche Duration}

We define the avalanche duration as the time $T$ between the start and end of the avalanche.
During an avalanche, as the interface gets locally re-pinned and de-pinned again, its local velocity may vary widely.
When many sites are de-pinned together, they may move much faster collectively than if they had to struggle individually. 
However, the extension of the moving region into the pinned region is restrained by the impurities.
This collective effects are such that $T$ does not scale linearly with the avalanche spatial extension:
\begin{align}
T \sim |F_c-F|^{-\nu z} = \xi ^z,
\end{align}
where some unimportant prefactor is controlled by the characteristic time $\eta_0$.
Due to the pinning from the impurities, the growth of the moving region is slower than linear in time, i.e.~we have $z>1$ (a smaller $z$ means a faster growth of the avalanche over time).

\subsection{Scaling Relations}

We have defined some relevant observables up to here ($v, \xi, W,S, T $), and some critical exponents ($\beta, \nu, \zeta, z$) which link them to the distance to criticality $|F_c-F|$.
We expect that a few simple heuristic arguments should show how these observables scale with one another, thus providing us with \textit{scaling relations} between the exponents.

\paragraph{The Statistical Tilt Symmetry (STS)}
\label{sec:STS}
Originally, this symmetry and the ensuing scaling relation was discussed in \cite{Narayan1993}.
Following this seminal work, the STS was also studied in terms of field theory and renormalization group. 
A precise derivation of the STS is possible using the Martin-Siggia-Rose formalism \cite{Chauve2000, Martin1973}. 
However here we want to give a simple argument for it.

Consider the average susceptibility of the interface,
\begin{align}
\chi = \frac{\text{response}}{\text{perturbation}},
\end{align}
where we consider the \textit{average} response.
Since an increase of the local stress by $\delta F$ can either do nothing or trigger a large avalanche, the susceptibility is expected to diverge at the critical point.

On the one hand, consider a ``tilt'' $\delta f(x)$ of the driving force with zero mean (and constant over time). 
The response of the interface is formally denoted $\delta h(x)$. 
Noting $G_{el}$ the most general linear elastic kernel (the short-range elastic interaction we used until now is $G_{el}\equiv k_1 \nabla^ 2$), we can write the equation of motion for the interface with the tilt, $h(x)$, then make a change of variable $h(x) \equiv \widetilde{h}(x)+ G_{el}^ {-1}(\delta f)$:
\begin{alignat}{4}
\eta_0 \partial_t h &=& 
F & + &\delta f +& G_{el} (h) - f^\text{dis} \eta[h,x] \label{Eq:justeapres} \\
\eta_0 \partial_t \widetilde{h} &=&  
F & + &         & G_{el} (\widetilde{h})   - f^\text{dis} \eta[\widetilde{h} + G_{el}^ {-1}(\delta f) ,x]  
 \label{Eq:htilde}
\end{alignat}
We note that the equation for the new field $\widetilde{h}$ is that of an un-tilted interface, but with a different disorder.
The key to deriving the exact STS relation is to note that on average (over realizations) the disorder does not actually change under the tilt:
\begin{align}
\overline{ \eta[\widetilde{h}_1+    G_{el}^ {-1}(\delta f(x_1))   ,x_1] ~\eta[\widetilde{h}_2+   G_{el}^ {-1}(\delta f(x_2)) ,x_2] }
&=\delta(x_2-x_1) \Delta[\widetilde{h}_2-\widetilde{h}_1 + G_{el}^ {-1}(\delta f(x_2)) - G_{el}^ {-1}(\delta f(x_1)) ]  \notag \\
&=\delta(x_2-x_1) \Delta[\widetilde{h}_2-\widetilde{h}_1 ]  \notag \\
&= \overline{\eta[ \widetilde{h}_1,x_1]~ \eta[ \widetilde{h}_2 ,x_2] } ,
\end{align}
where we went from the first to the second line thanks to the product with the Dirac $\delta(x_2-x_1)$.
The effect of the tilt thus disappears from the two-point correlation function of the disorder 
and assuming a Gaussian disorder (or showing similar relations for higher moments) we get:
\begin{align}
\eta[\widetilde{h}_1+    G_{el}^ {-1}(\delta f) ,x ] \overset{law}{=}  \eta[\widetilde{h},x]. \label{Eq:disorder_equal}
\end{align}
Thus the average susceptibility of the auxiliary interface $\widetilde{h}$  to the tilt is zero: $\overline{  \delta \widetilde{h} / \delta f }   =  0$.
The average susceptibility of $h(x)$ is given by the response $\delta h = \delta \widetilde{h} + G_{el}^{-1}(\delta f) $ to the tilt:
\begin{align}
\chi = \overline{  \frac{\delta h}{\delta f} } = \overline{  \frac{\delta \widetilde{h} +G_{el}^{-1}(\delta f) }{\delta f} }  = \frac{G_{el}^{-1}(\delta f)}{\delta f} \sim G_{el}^{-1} \sim  \xi^{\alpha},
\end{align}
where the last equivalence comes from the dimensional analysis\footnote{
We denote $[X]$ the dimension of the variable $X$ in what follows.
} 
of the interaction kernel. 
For instance, for the short-range elastic kernel $G_{el}\equiv \nabla^2$  we have $[G_{el}^{-1}] =1/[\nabla^2]= [x]^{2}$, i.e.~$G_{el}^{-1}$ is homogeneous to a length squared, and $\alpha =2$.
For elastic interactions with longer range, we may have $\alpha \leq 2$.

On the other hand, the average local response to a local perturbation $\delta f=\delta F$ in the driving is given by the average and maximal avalanche width $W \sim  \xi^\zeta $. 
Since the perturbation $\delta F$ is homogeneous to $|F_c-F|$, we consider that $\delta F$ needs to be some finite fraction of it, and we have 
\begin{align}
\chi =\frac{\delta h_\text{local}}{\delta F_\text{local}}=\frac{ \xi^\zeta }{ |F_c-F|} \sim \xi^{\frac{1}{\nu} + \zeta} \label{Eq:chi=1nu+zeta}
\end{align}

To conclude, we have 
\begin{align}
 \xi^{\alpha} = \xi^{\frac{1}{\nu} + \zeta} 
 \quad  \Longrightarrow  \quad 
 \nu = \frac{1}{\alpha - \zeta},
 \label{Eq:STS_alpha}
\end{align}
where in our case of short-range interactions $\alpha=2$.
This general relation is often called the STS relation and allows us to reduce the number of fundamental exponents by one: we no longer need to measure or report $\nu$ since it is given by this relation. 
In numerical simulations of the depinning of the elastic interface, this relation is well verified.

\paragraph{Scaling Relation for the Velocity}
Concerning the average velocity (over the whole system) $v(t)$, assuming that we reached some steady state $v(t)=v$, we can derive a simple scaling relation for the depinning regime ($v>0$).
The average velocity is simply given by the maximal avalanches, which displace the interface locally on a distance $W\sim \xi^\zeta$ over a time $T\sim \xi^z$. 
Thus, 
\begin{align}
v\sim \frac{W}{T} \sim \xi^{\zeta -z}.
\end{align}
By definition, $v\sim (F-F_c)^\beta = (\xi ^{-\frac{1}{\nu}})^ \beta$. 
Thus,:
\begin{align}
\xi^{\zeta -z} \sim \xi ^{-\frac{\beta}{\nu}} 
 \quad  \Longrightarrow  \quad 
 \beta = \nu (z-\zeta).
\end{align}
We note that $z>\zeta$ ensures that $\beta>0$ i.e.~that the average velocity vanishes at the critical point.
This scaling relation was also found in \cite{Narayan1993}, from renormalization arguments (change of scale, field dimensions).
Combined with the STS, it gives $\beta  = \frac{z-\zeta}{\alpha - \zeta}$.

\paragraph{Conclusion}
We have reduced the number of ``fundamental'' exponents from $4$ to $2$ thanks to two scaling relations.
The exponents that we choose to be ``fundamental'' are $z$ and $\zeta$.
Up to now we have discussed the properties of the maximal avalanches and of the interface itself both below and above threshold, but not the \textit{statistical properties of the avalanches} which are also expected to display critical behaviour at the transition.

Above the threshold (depinned regime), the dynamics consists essentially in numerous \textit{almost} independent avalanches.
However when a point is almost stopped (just before the end of an avalanche) it may keep on moving by participating in a new one: because the motion truly consists in a single very large, never-ending avalanche, these are not really independent.
This makes it difficult to properly define finite avalanche events, above the threshold.

Below the threshold (pinned regime, $F<F_c$), an infinitesimal increase $\delta F$ of the force may trigger avalanches.
By taking  $\delta F$ small enough, one may hope to ensure that exactly zero or one avalanche will be triggered.
In this way, one can a priori trigger a large number of avalanches at fixed $F$, given that $\delta F \ll |F_c-F|$. 
However numerically it may prove difficult to keep $F$ constant while increasing it by $\delta F$ numerous times.

\section{Avalanche Statistics at The Depinning Transition}

As we have just seen, the depinning problem with constant force driving is not the most appropriate way to study avalanche statistics.
Here we introduce another way to drive the system, which is more relevant for frictional or seismical applications and allows for unlimited avalanche statistics while staying below the critical force $F_c$ (and very close to it).

\subsection{Origin of the Elastic Drive}
\label{sec:origin_elastic}

In the context of magnetization domains, in the previous approach we neglected the effect of demagnetizing fields, which are actually relevant in most geometries \cite{Zapperi1998a}.
Essentially, the demagnetizing fields  are due to some boundary effects which generate a field proportional and opposed to the magnetization, so that in the equation of motion we should add a force $-k h(x,t)$, where $k$ is the demagnetization factor:
\begin{align}
\eta_0 \partial_t h(x,t) = F -k h(x,t) + k_1 (\nabla_x^2 h)(x,t) - f^\text{dis} \eta[h(x,t),x].
\label{Eq:demagnetization_field}
\end{align}
This seemingly small variation is actually crucial.
Suppose that $F>F_c$ and $h\approx 0$: as an avalanche unfolds the local height $h(x,t)$ grows, and the effective local driving force $F_\text{drive}\equiv F-k h(x,t)$ decreases. 
At some point, the ``demagnetization'' from the term $-k h$ will be enough to have  $F_\text{drive}<F_c$ in the avalanche region, and the avalanche will be able to stop.

Thus for any initial value of $F$ the system will end up in the pinned phase, precisely around the transition: $F_\text{drive}^\text{final}= F-k h^\text{final}  \lessapprox F_c$.
A fruitful approach is to choose $F$ to be a time-dependent force, slowly increasing over time. 
On the time scale of an avalanche ($\sim \eta_0$), the external force $F$ is constant and the final value of the average driving force will be $F_\text{drive}^\text{final}  \lessapprox F_c$.
After an avalanche the slow increase in $F$ will eventually trigger a new avalanche, approximately when $F_\text{drive} \approx F_c$.
In this sense we obtain a stationary dynamics, since avalanches properties are expected to be controlled by the initial value of $|F_c-F_\text{drive}|$.
Using this setup we may obtain as many identically distributed avalanches as we need by simply waiting long enough.
Furthermore, for a large enough system each new avalanche occurs on a location far away from the previous one, and  avalanches will be nearly independent. 

After this brief link with the previous case of the constant force, we present the problem more formally in appropriate notations, in a self-contained way.

\subsection{Construction of the Equation}
\label{sec:Equations_elastic_driving}

\subsubsection{Langevin Equation}

Instead of driving the system with a constant force $F$ equal in all points of the interface and independent of its progression, we may want to pull it elastically via springs (one per site) attached to a common surface (set in the plane $z=w$) with an externally imposed velocity $V_0$ (i.e.~$w=V_0 t$), as we did in the Burridge-Knopoff or OFC* models.
This is equivalent to a coupling with an energy parabola (potential energy well) of which the minimum or center $w$ moves at velocity $V_0$.
The related energy contribution reads:
\begin{align}
E_\text{drive} =\int  \frac{k_0}{2	}  (w- h( x))^2 \d^d  x ,
\end{align}
where $k_0$ is the coupling constant between the external field and the interface, and corresponds to the stiffness of the springs aforementioned. 
It corresponds to a local driving force $F_\text{drive}(x) = k_0 (w  - h)$. 
We can identify with the context of domain walls: $k_0 w \equiv F$ and $k_0 h \equiv k h$.

In general the driving function $w(t)$ could be  anything.
However, the drive is usually taken to be a monotonously increasing function of time (there are good reasons for that \cite{Dobrinevski2013}), but interestingly the non-monotonous case has also been considered\footnote{We will come back to this only in the next chapter.} recently \cite{Dobrinevski2013a}.
The role of non-stationarity was  also studied in \cite{Durin2006}.

In this thesis we use $w=V_0 t$ with small $V_0$, i.e.~$V_0 \ll r_f / \eta_0$ (where $r_f$ is the disorder correlation length, i.e.~a small characteristic length along the $z$ direction).
We write the evolution equation in a self-contained way:
\begin{align}
\eta_0 \partial_t h(x,t) = k_0( V_0 t - h(x,t)) + k_1 (\nabla_x^2 h)(x,t) - f^\text{dis} \eta[h(x,t),x], \label{Eq:myDepinning}\\
\overline{\eta(z,x)}=0, 
\qquad 
\overline{\eta(z_1,x_1)\eta(z_2,x_2)} = \delta^D(x_2-x_1) e^{-(z_2 -z_1)^2/2 r_f^2},
\end{align}
where $\delta^D$ is the Dirac distribution, and the correlation function of $\eta(z,x)$ along $z$ may be any short-range function with range $\sim r_f$ (we just give an example here).

\begin{figure}[]
\begin{small}
\begin{center}
\def\svgwidth{8cm}
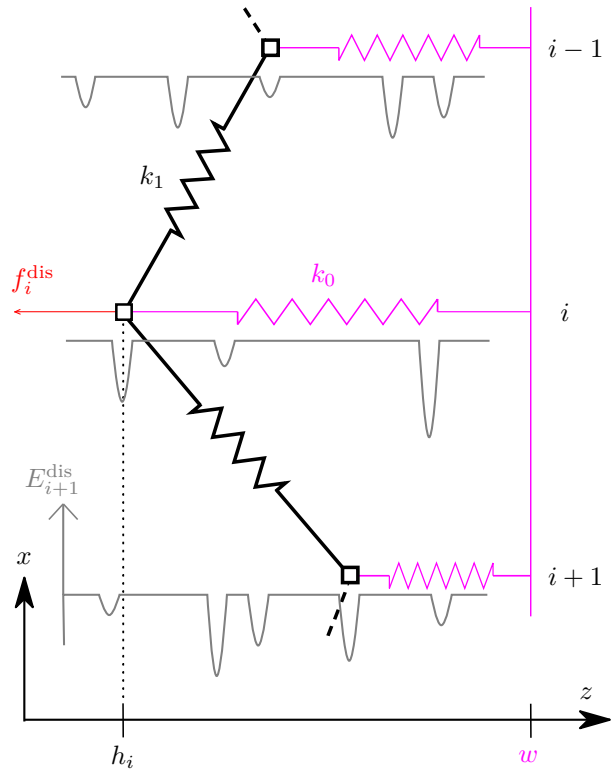
\end{center}
\end{small}
\caption{   {\footnotesize 
Mechanical ``circuit'' or sketch of the one-dimensional elastic interface model.
The interface itself (bold black line) consists in blocks (empty squares) located at discrete sites $i-1,i,i+1$ along the $x$ axis and are bound together via springs $k_1$ (the interface  elasticity is thus $k_1$).
The driving is performed via springs $k_0$ linked to a common position $w$ (thin purple lines).
The disorder force $f^\text{dis}_i$ (red) for the site $i$ derives from a disordered energy potential $E^\text{dis}_i$, which is here simplified as a series of narrow wells separated by random spacings.
The damping (proportional to $\eta_0$) is not pictured.
\label{Fig:depinning1}
}   }
\end{figure}

In Fig.~\ref{Fig:depinning1} we represent the one dimensional system in terms of a mechanical ``circuit'' (in analogy with an electrical circuit) consisting in blocks connected by springs.
This kind of representation will prove especially useful in the next chapter, but it can already give us an intuitive view of the problem.

\subsubsection{Quasi-Static Dynamics}

By definition, a metastable state $\{w,h(x)\}$ of the system fulfils the stability condition:
\begin{align}
k_0(w-h(x)) + k_1 (\nabla_x^2 h)(x) < f^ \text{dis} \eta[h(x),x], \qquad \forall x. \label{Eq:wstability}
\end{align}
The quasi-static ($V_0=0^ +$) dynamics can be summarized very simply, as an infinitesimal increase of $w=V_0 t$ has only two possible effects.
Either the infinitesimal advance of the interface is such that \req{wstability} is still fulfilled and the motion cannot be called an avalanche.
Or one site becomes unstable (i.e.~it violates \req{wstability} locally), triggering an avalanche (or ``event'') that unfolds until \req{wstability} is verified again.
As the avalanche duration is $\sim \eta_0 \ll r_f/V_0$, we may consider $w$ constant during the event.
Algorithmically, $w$ is kept constant during the avalanches.
Once the system is in a new metastable state, $w$ increases again.

This formulation of the dynamics in terms of well-defined avalanches (or \textit{shocks}) between \textit{static states} allows for a clearer understanding of the problem and an easy numerical implementation of the system as a cellular automaton.
As in the constant force setup, we can use the narrow wells ``approximation'' (\ref{sec:narrow_wells}) to avoid the issue of the infinitesimal avalanches. 
In that case, the dynamics is exactly that of the OFC* model (see sec.~\ref{sec:OFC*}).
There is a smart numerical scheme due to Grassberger \cite{Grassberger1994a} which allows to simulate this kind of dynamics very efficiently. 
The core reason for the efficiency of this method is the disappearance of the infinitesimal avalanches in the algorithm, but there are also some purely technical ``tricks'' that are very useful, which we present in  Appendix \ref{App:Grassberger_efficient_method_two_cases}.

\subsection{The Depinning Transition (Elastic Driving)}

\subsubsection{A Different Protocol}

\paragraph{Tuning the Interface Velocity}
Since the driving force is no longer constant, it cannot be the control parameter any more: however we do control the driving velocity $V_0$ (and the stiffness $k_0$).
We want to relate the parameter $V_0$ to the observable $v$.

Consider some macroscopic increase in the driving force $\Delta F = k_0 V_0 \Delta t \gg 1$. 
The variations of the elastic and disorder terms (as measured on metastable states) are necessarily bounded.
Thus the only way to balance this change $\Delta F$ in \req{wstability} is to move the interface by a distance $\Delta h \sim V_0 \Delta t$ during this time interval $\Delta t$, so as to get a stationary driving term $k_0(V_0 t-h)$.
Hence the time- and space-average velocity of the interface is given by
\begin{align}
\langle v \rangle 
&\equiv \frac{1}{\Delta t}\int_0^{\Delta t} \d t~v(t) \\
&\equiv \frac{1}{\Delta t}\int_0^{\Delta t} \d t \frac{1}{L^d}\int \d^d x~\partial_t h(x,t)\\
&= V_0,
\end{align}
where the $\langle .\rangle$ stands for time average.
This means that what used to be similar to an order parameter is now a controlled quantity.
For an infinite system there will always be a point of the interface moving (i.e.~some avalanche occurring) and the time average on $v$ does not need to be taken over a long interval $\Delta t$, i.e.~we have an instantaneous space-averaged velocity $v(t) = V_0$. 
However this is obtained by averaging over space the very fast, large, localized jumps of the interface, typically separated by large distances. 
In other words, we have huge spatio-temporal fluctuations of the ``order parameter'' $v(x,t)$ (as expected in a phase transition).

\paragraph{Measured Observable: the Stress or Force}
The stress or force $\sigma \equiv k_0 (w-h) \equiv F_\text{drive}$ is now an observable that we measure rather than control.
It has huge fluctuations $\sigma(x,t)$, but it is a response function (unlike $v(x,t)$ which is simply equal to $V_0$ on average).
We find that the space- and time-average value $\sigma(V_0)$ follows the exact same law as  $F(v)$ did in the other protocol, the constant force setup (see Fig.~\ref{Fig:v_de_F_T=0}), i.e.~we have the same phase diagram as previously.
We understand the behaviour at large velocities by considering the interface as simply following the drive $w$, with the average stress representing how much the interface lags behind. 

The behaviour at small (vanishing) velocities $V_0 = 0^+$ simply corresponds to the limit $v\sim |F_c-F|^\beta \sim 0^+$, i.e.~to $F \approx F_c$.
Thus, the elastic driving method (in its stationary regime) does not allow to explore the whole region $F<F_c$ of the phase diagram. 
Instead, it automatically drives us to the critical point, which is much more interesting.

\paragraph{Where is the Transition?}
A natural question is to ask ourselves: where is the critical point? 
More precisely, what is the critical velocity?
At $V_0=0$, after a possible short transient, nothing happens and we have $F-kh <F_c$ everywhere: we are below criticality.
At $V_0=0^+$ i.e.~in the quasi-static regime, the system evolves via discrete and well-defined avalanches.
At the end of each avalanche, the system is stable and $F-kh<F_c$ everywhere, but this never lasts: the system oscillates around the critical point.
At any finite velocity $V_0>0$, the infinite system is always in motion and we are above the critical point.
Thus, the ``critical velocity'' is $V_0^c=0^+$.

\subsection{A new Scaling Relation}
\subsubsection{Scaling of the Correlation Length}
In what follows we are in the quasi-static regime ($V_0=0^ +$).
A naive approach would conclude that since we are automatically driven towards the critical point, all quantities of interest, in particular the correlation length $\xi$ will diverge to infinity: since $V_0 = v\sim |F_c-F|^\beta \sim 0^+$, we might expect $\xi\sim V_0^{-\nu/\beta}\sim \infty$.
However, during an avalanche the term $-k_0 h$ actually takes us a bit below criticality.
We now need to characterize quantitatively how far we typically are from criticality, depending on $k_0$ (keeping $V_0=0^+$).

The elastic driving term corresponds to a quadratic energy potential $E_\text{drive} = k_0(w-h)^2$.
It acts as a confining potential for the interface, for which the energetic cost of large excursions from $w$ grows quadratically.
Calling $\ell$ some unspecified length in the $x$ direction, we perform the dimensional analysis of the driving and elastic energies
 over a patch $\ell ^d$:
\begin{alignat}{3}
E_\text{drive}(\ell) &= \frac12 \int_{\ell^d} \d ^d x~k_0 (w-h)^2&   &\sim k_0 \ell^d [h^2]\\
E_\text{el}(\ell)    &= \frac12 \int_{\ell^d} \d ^d x~k_1 (\nabla h)^2& &\sim k_1 \ell^{d-2} [h^2].
\end{alignat}
The role of the disorder is a priori more complex. 
However, the argument used to derive the STS in the constant force setup can be extended to the elastic driving case.
In particular, we find that the response $\delta h = (k_1 \nabla^2 +k_0)^{-1} \delta f$ does not affect the disorder\footnote{To be precise, the correlator of the disorder with or without the tilt is the same (in law).}, which implies that $(k_1 \nabla^2h +k_0 h)$ is not affected (or ``renormalized'') by the disorder.
We define $\xi$ as the length scale where the elastic and driving contributions have equal weights:
\begin{align}
E_\text{el}(\xi)  &\sim E_\text{drive}(\xi)   \notag \\
\xi &\sim  \lp \frac{k_1}{k_0} \rp^{1/2}
\end{align}
For $\ell \gg \xi$, the drive contribution outmatches the elastic one and the interface energy is dominated by the driving term.
The interface shape is thus controlled by the confining potential, i.e.~it is flat (at this length scale).
For $\ell \ll \xi$, the competition between disorder and elasticity prevails, and the interface will be self-affine (rough) with a roughness exponent $\zeta$, as pictured in Fig.~\ref{Fig:roughness_simple_zeta}.
We see that this length $\xi$ actually behaves as the correlation length, thus we identify it with the one in the previous section (which was defined as $\xi = |F_c-F|^{-\nu}$).

Up to the dismissal of the disorder this argument is a very classical one: its most common form is in the field theory of a field $\phi$ with mass $m$, where the action reads $S=\int \d^d x [ (\nabla \phi)^2+m^2 \phi^2]$. In that case the ``correlation length'' is $\xi_\phi = 1/m$.
The addition of the disorder term is expected to induce non-trivial effects, that may a priori disturb this scaling.
It is found, however, that the argument still holds in the presence of disorder, due to the invariance of the relative weights of the elastic and driving terms under renormalization (a rather non trivial result).

The most important thing to remember is that the correlation length is controlled by the parameter $k_0$ via $\xi \sim k_0^{-1/2}$.

\subsubsection{Self-Organized Criticality (SOC)}
The restoring force $-k_0 h(x,t)$ decreases the driving force when an avalanche unfolds, allowing to automatically set ourselves \textit{at} the depinning transition critical point (we already explained it in the introduction, sec.~\ref{sec:origin_elastic}).
Since we do not need to tune any parameter to go there (except for $V_0\to 0, k_0\to 0$), we may recognize this as an example of Self-Organized Criticality (SOC).

In contrast with original models of SOC, here the dissipation occurs in the bulk of the system. 
For an avalanche of size $S$, the corresponding  total (system-wide) decrease of the driving force is $-k_0 S$.
As the avalanche size increases with applied force and still increases above the threshold $F_c$, the dissipation $-k_0 S$ becomes extremely powerful when we reach the threshold, thus preventing us from going beyond.
The explanation for the ``self organization'' simply lies in the continuous driving toward criticality of a system that strongly dissipates ``energy'' above the critical threshold.
To comment on the SOC nature of the system, we cannot hope to put it better than Fisher \cite{Fisher1998}:
\begin{quote}
Whether critical behavior is considered ``self-organized'' or not is somewhat a matter of taste: if the systems we are considering are driven at very slow velocity, then they will be very close to critical. In another well known situation, when a fluid is stirred on large scales, turbulence exists on a wide range of length scales extending down to the scale at which viscous dissipation occurs. In both of these and in many other contexts the parameter which is ``tuned'' to get a wide range of scales is the ratio of some basic ``microscopic'' scale to the scale at which the system is driven.
\end{quote}
For us the microscopic scale occurs at the most local scale with a dissipation parameter $k_0$ and the driving scale is that of the system.
We need to set the velocity $V_0 $ to a very low value in order to obtain (self-organized) criticality.

\subsection{Statistical Distributions}

\subsubsection{Distribution of the Avalanches Sizes}

At the transition ($V_0=0^+, k_0 \ll k_1$) we expect to have many avalanches, with a typical size $\xi$ diverging as $k_0^{-1/2}$.
However, the avalanches are random events whose sizes span the entire range from microscopic to $\xi$-wide.
The distribution of avalanches $P(S)$ follows a power-law distribution with a cutoff (fast decay) for sizes larger than some characteristic (or ``maximal'') size $S_m$:
\begin{align}
P(S) = S^{-\tau} g(S/S_m),
\end{align}
where $g(s)$ is a scaling function which decays very fast for $x>1$.
An avalanche with lateral length $\ell$ has size $S=\ell^{d+\zeta}$, so that 
\begin{align}
S_m = \xi^{d+\zeta}.
\end{align}
We want to relate the new exponent $\tau$ to those previously introduced.

\begin{figure}[]
\begin{small}
\begin{center}
\def\svgwidth{\textwidth}
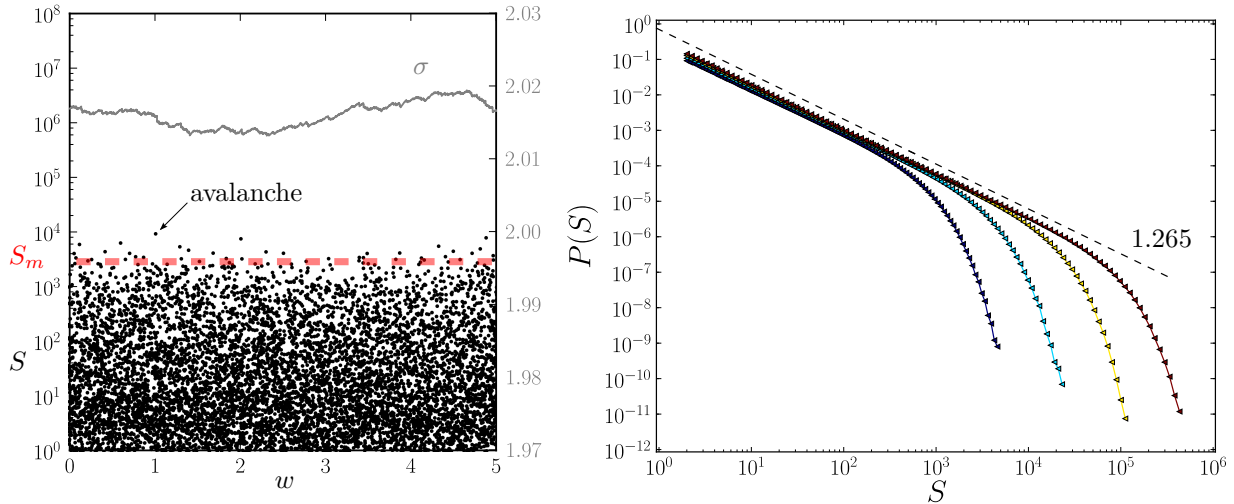
\end{center}
\end{small}
\caption{   {\footnotesize 
\textit{Avalanche size distributions.}
Left: Avalanches sizes $S$ (indicated by dots) are evenly distributed over time, with only short-range correlations in time and space. 
The corresponding stress $\sigma $ has small fluctuations (solid grey line) over time ($w=V_0 t$).
Parameters are $k_1=1, k_0=0.03$. 
\newline
Right: Probability distribution $P(S)$ of the avalanches for $k_1=1$ and $k_0$ decreasing from left to right: $k_0=0.03,0.01, 0.003,0.001$.
The dashed line indicates the pure power-law $\sim S^{-1.265}$ as guide to the eye.
The curves are shifted for an easy comparison and events with $S<2$ have been cut off; the number of remaining events is $10^ 7$.
\label{Fig:scatter_elastic+tau126}
}   }
\end{figure}

\paragraph{Derivation of the Exponent $\tau$}
Adapting an argument from \cite{Zapperi1998a}, we can directly derive the exponent $\tau$.
The average response (total displacement $\Delta h \equiv \int \Delta h(x) \d^d x$) to an increase of the force $\Delta F \sim |F_c-F|$ corresponds to the sum of avalanche sizes under many infinitesimal increases $\delta F = \Delta F/N$:
\begin{align}
\langle S \rangle  \sim \frac{ \Delta h}{ \Delta F }, \label{Eq:Smean_suscept}
\end{align}
since each avalanche $S$ is the response $\Delta h$ to an increase $\delta F$.

Besides, similarly to the argument that led us to $v=V_0$, we see that the total drive $\Delta F = k_0 V_0 \delta t$, must be compensated by a displacement $k_0 \Delta h$. 
Hence, $\Delta F \sim k_0 \Delta h$.

Combining these two arguments we get $\langle S \rangle  = 1/k_0 \sim \xi ^{2}$.
By definition we also have $\langle S \rangle= \int S P(S) \d S \sim S_m^{2-\tau} = (\xi^{d+\zeta})^{2-\tau}$.
By identification we get
\begin{align}
\tau = 2 - \frac{2}{d+\zeta}       \label{Eq:tau1}
\end{align}
Another approach is to identify $\langle S \rangle $ in \req{Smean_suscept} with the average of the susceptibility $\chi $ computed in \req{chi=1nu+zeta}, since on average the displacement under an increase $\delta F$ in the force is $\langle S \rangle = \overline{\chi} $. 
We then have $\langle S \rangle = \chi = \xi^{1/\nu +\zeta}$.
Since $\langle S \rangle \sim(\xi^{d+\zeta})^{2-\tau}$, this yields the more general relation:
\begin{align}
\tau = 2 - \frac{\frac{1}{\nu}+\zeta}{d+\zeta}, 
\label{Eq:tau2}
\end{align}
which collapses to \req{tau1} since $1/\nu +\zeta = \alpha =2$ in the case of short-range elasticity (we used the STS relation).
In Fig.~\ref{Fig:scatter_elastic+tau126}, we find $\tau=1.265\pm 0.005$, in agreement with numerical results found in the literature for $d=2$ (see \cite{Durin2000, Rosso2009}).

\subsubsection{Other Distributions}

\paragraph{Distribution of the Avalanches Lateral Length}

There are several ways to compute the avalanches lateral length $\ell$. 
The simplest one is to measure the maximal length $\ell_X$ in the X direction (or $\ell_Y$ in the Y direction).
This gives us the power-law with $\tau_\ell\simeq1.75$, although this is not the smoothest result.
Another way is to compute this length as $\sqrt{\ell_X  \ell_Y}$.
Yet another way is to compute it as $\sqrt{n}$, where $n$ is the actual number of sites affected by the avalanche (which is always smaller than the product $\ell_X \ell_Y$).
In any case, we find the power-law $P(\ell)\sim \ell^ {-\tau_\ell}$ with $\tau_\ell\simeq1.75$.

Denoting $\ell$ the lateral length (along the $\mathbf{x}$ plane) of an avalanche, we may call $P(\ell)$ the avalanches lateral length distribution, $\tau_\ell$ the associated exponent and $\ell_m = \xi$ its cutoff.
Since an avalanche with length $\ell$ has size $S=\ell^{d+\zeta}$, we have:
\begin{align}
P(\ell) \d \ell 
&= \ell^{-\tau_\ell} g_\ell(\ell/\xi) \d \ell 
= P(S) \d S \\
&\sim S^{-\tau} \d S , &\quad \forall S \ll S_m\\
&\sim \ell^{-(d+\zeta)\tau}  \d (\ell^{d+\zeta}) , &\quad \forall \ell \ll \xi \\
&\sim \ell^{-(d +\zeta) (\tau-1)-1 } \d \ell , &\quad \forall \ell \ll \xi.
\end{align}
So that we identify $\tau_\ell \equiv (d +\zeta) (\tau-1)+1 = d+\zeta-1$ using \req{tau1} or $\tau_\ell = d+1-1/\nu$ using the more general \req{tau2}.
In Fig.~\ref{Fig:tau_ell_sqrt_k2=0}, we observe that $\tau_\ell\simeq 1.75$, in agreement with the measurement of $\zeta\approx 0.75 $ in $d=2$.

\includefig{8cm}{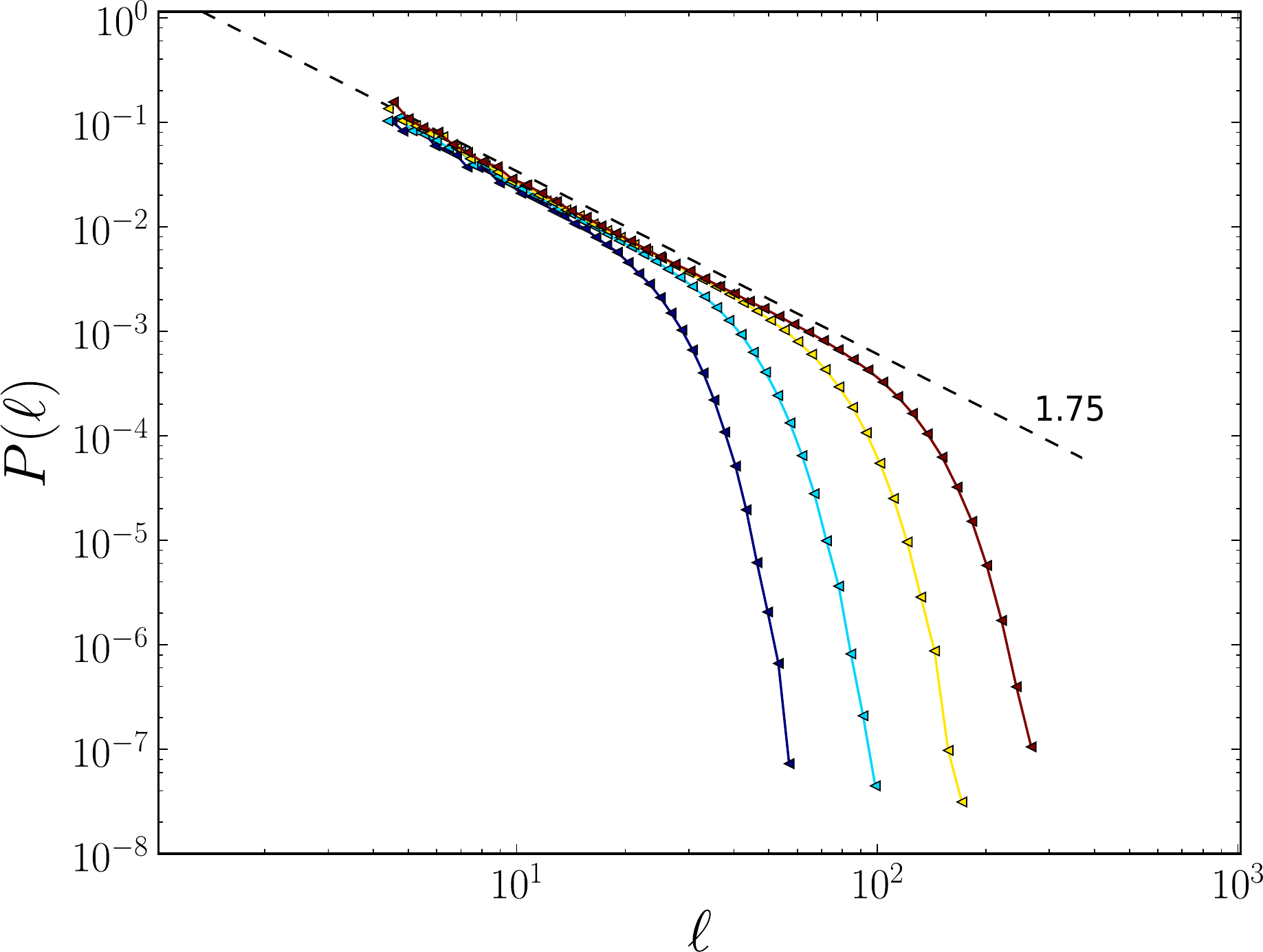}{
Avalanches lateral length distribution $P(\ell)$.
The length is computed as the square root of the number of sites affected by the avalanche.
The dashed line indicates the pure power-law $\sim \ell^{-1.75}$ as guide to the eye.
Note that all events are much smaller than the system size $L=5000$, so that we do not have any finite-size effect.
We used $k_1=1$ and $k_0$ decreasing from left to right: $k_0=0.03,0.01, 0.003,0.001$.
The curves are shifted for an easy comparison and events with $S<2$ have been cut off; the number of remaining events is $10^ 7$.
\label{Fig:tau_ell_sqrt_k2=0}
}

\paragraph{Distribution of the Avalanche Areas}
Similarly the avalanche areas, i.e.~the number $n \sim \ell^ d$ of sites affected by an avalanche has a power-law distribution. The associated exponent $\tau_n$ and cutoff $n_m=\xi^d$ can be derived from the identity  $n=\ell^d$. 
We find $\tau_n =  1+ (\tau_\ell -1)/d = 1 + (d+\zeta)(\tau-1)/d$.
Using the STS relation \reqq{tau1}, this simplifies into $\tau_n = 1+ (d+\zeta-2)/d$.
The correspondence with the exponent measured in the simulations can be found in Fig.~\ref{Fig:tau_enn_k2=0}.
\includefig{8cm}{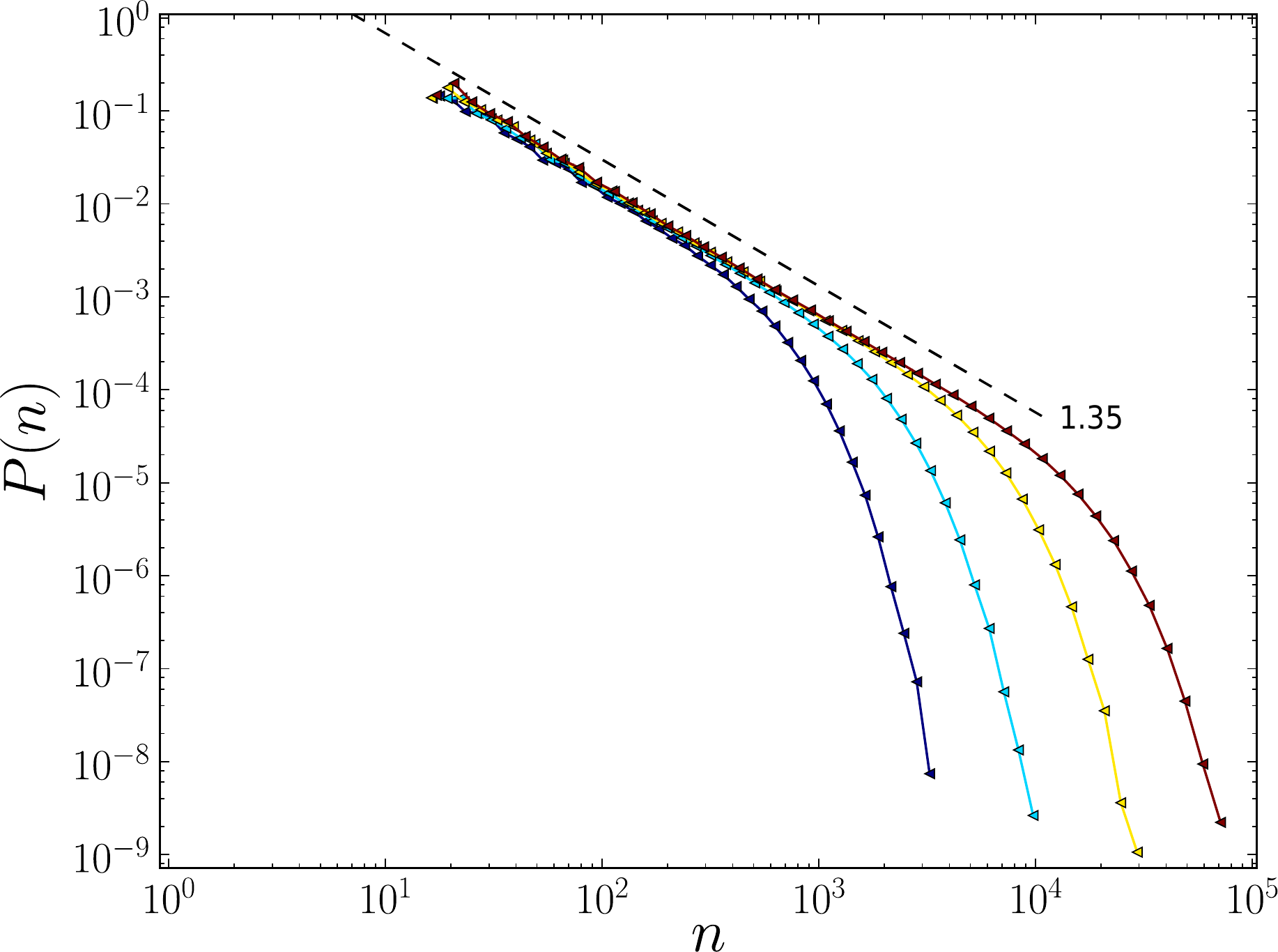}{
Avalanches area distribution $P(n)$, in terms of the number of sites $n$ affected.
The length is computed as the square root of the number of sites affected by the avalanche.
The dashed line indicates the pure power-law $\sim \ell^{-1.75}$ as guide to the eye.
Note that all events are much smaller than the system size $L=5000$, so that we do not have any finite-size effect.
We used $k_1=1$ and $k_0$ decreasing from left to right: $k_0=0.03,0.01, 0.003,0.001$.
The curves are shifted for an easy comparison and events with $S<2$ have been cut off; the number of remaining events is $10^ 7$.
\label{Fig:tau_enn_k2=0}
}

\paragraph{Distribution of the Avalanches Duration}
Similarly the avalanches duration, i.e.~the number of waves of updates necessary for the avalanche to completely unfold has a power-law distribution. 
The associated exponent $\tau_T$ and cutoff $T_m=\xi^z$ can be derived from the identity  $T=\ell^z$. 
We find $\tau_T =  1+ (\tau_\ell -1)/z = 1 + (d+\zeta)(\tau-1)/z$.
Using the STS relation \reqq{tau1}, this simplifies into $\tau_T = 1+ (d+\zeta-2)/z$.

\subsubsection{Conclusion}

Up to now, we have presented only simulation results and scaling arguments.
We have explained how the dynamical phase transition (depinning transition) can occur via qualitative arguments, and gave links between the exponents, thanks to the scaling relations.
However, the evidence of a critical state with diverging length scales, interface roughness and power-law distributed avalanches comes from numerics only. 
The same is true for the values of the ``fundamental'' exponents $z$ and $\zeta$, which we obtain solely from numerics.

In the following section we present two \textit{mean field} approaches, which allow to solve the problem of the depinning transition analytically in the cases of infinite-range interactions or very high dimensions.
The analytical approach allows a better understanding of the inner mechanisms of the depinning transition and is a very convincing argument for the case of a critical phenomenon in low dimensions.
An alternative and more general way to deal with the depinning problem analytically is to use the Functional Renormalization Group (FRG) approach. 
For an introduction to this subject, see \cite{Polonyi2003}, for an historical review see \cite{Fisher1985}, for applications to the problem of depinning see \cite{LeDoussal2002,Rosso2007}.

\section{Mean Field Approaches}
\label{sec:MF_depinning}

\subsection{Introduction}

\paragraph{Defining the ``mean field''}
The term \textit{mean field} encompasses numerous microscopic models.
The common idea between all mean field approaches is that if the interactions within the system are dense enough, then the dominant contribution to the interaction of any point with the rest of the system (represented by some field) will be given by the interaction with the average field, i.e.~the fluctuations are neglected.
In the extreme limit, the mean field can be studied via a fully-connected model, where each site interacts equally strongly with all the others.
This limiting case is very simple since we go from $N$ degrees of freedom to a single one (or a couple of them at most).

A common and seminal example of such an approach is that of the Ising transition, where the Curie-Weiss model assumes that each degree of freedom (the local value of the spin $s_i$) interacts with the macroscopic magnetization (which is just the average $M=\langle s \rangle$).
A way to build this interaction from a microscopic model is to assume that each spin interacts equally with all the others, regardless of their distance: this is an example of \textit{fully connected} model.

\paragraph{High Dimensionality}
A natural occurrence of mean field behaviour is when the system considered is of high spatial dimension.
Formally, when $d=\infty$ every site interacts with infinitely many other sites and we obtain the same mean field as in the fully connected case.
As is, this is not really useful since reality does not have infinitely many dimensions.
However there is generally some \textit{upper critical dimension} $d_{uc}$ beyond which the model behaves as in mean field: for any $d\geq d_{uc}$, we have mean field exponents due to the high connectivity of each site.
Most models have a finite and rather low upper critical dimension $d_{uc}$: for instance, the elastic depinning with short-range interactions has $d_{uc}=4$.

Because of this correspondence, the mean field case is often presented as complementary to the ``finite dimensional'' case (understand ``not mean field case'').

\subsubsection{Infinite Range: The Fully Connected Model}
\label{sec:fully_connnected_definition}
\paragraph{Construction of the Interaction Term}
For the sake of simplicity, we will use the discrete system where $h(x)$ is modelled by a lattice of sites with values $h_i$, $i\in L^ {d}$. 
The continuum equations can be derived straightforwardly from the discrete ones.
For the same reason, we will use $d=1$ in some derivations to decrease the volume of terms.

Originally, the elastic interaction is written as the sum of contributions from the neighbours of the current site $i$:
\begin{align}
F_{el,i} = k_1 (h_{i+1}-h_i) + k_1 (h_{i-1}-h_i) \equiv k_1 (\Delta h)_i,
\end{align}
where $(\Delta h)_i$ is the discrete Laplacian of $h$ evaluated at $i$.
Here in one dimension each site has $2d=2$ neighbours.

With infinite range, the elastic interaction for $i$ reads:
\begin{align}
F_{el,i} =\frac{k_1}{N} \sum_{\forall j\neq i}(h_{j}-h_i) ,
\end{align}
where we must divide by the total number of sites $N$ to keep the system's energy extensive in $N$ (and not growing as $N^2$). 
The sum over all neighbours can be rewritten: 
\begin{align}
\sum_{\forall j\neq i}(h_{j}-h_i) 
&= \lp\sum_{\forall j}(h_{j}-h_i)\rp - (h_i-h_i) \\
&= \sum_{\forall j} h_j - \sum_{\forall j} h_i \\
&= N(\overline{h}-h_i)
\end{align}
To conclude, in order to get the fully connected Langevin equation for depinning, we just need to replace $k_1 \nabla^ 2 h$ with $k_1 (\overline{h}-h)$.
The equation of motion for a single site $i$ in the fully connected model then reads:
\begin{align}
\eta_0 \partial_t h_i = k_0  (w-h_i) + k_1 (\overline{h} - h_i) - \eta_i[h], \label{Eq:MF0}
\end{align}
where the individual contribution of each site to the interaction term (second term of the r.h.s.) was in $k_1/N$.

\paragraph{Remark on Roughness}
A crucial feature of the fully connected model is the irrelevance of geometry. 
Since all sites interact equally with each other, the notion of space becomes irrelevant. 
Consequently, the notion of spatial correlation and of roughness becomes meaningless: spatial fluctuations with a well-defined shape are forbidden, since each site is the neighbour of all the others.
Thus in mean field the interface is ``flat'' in the sense that its spatial fluctuations are bounded, and the roughness exponent is:
\begin{align}
\zeta=0.
\end{align}

\subsection{The ABBM Solution}

The dynamics of the center of mass of the interface in the fully connected model can be mapped to the study of a single particle in some effective disordered potential \cite{Zapperi1998a}, the ABBM model.
It was initially Alessandro, Beatrice, Bertotti, Montorsi who proposed the problem of a single particle driven in a Brownian force landscape \cite{alessandro1990}.

We first derive the mapping between the two problems, then we derive the exponent $\tau$ from stochastic calculs arguments.

\subsubsection{Mapping of the Fully Connected Model to a Single Particle Model}

\paragraph{Equation of Motion for the Center of Mass}
We sum \req{MF0} over $i$, divide by the number of sites $N$ and obtain the equation of motion for the center of mass of the interface:
\begin{align}
\eta_0 \partial_t \overline{h} = k_0  (w-\overline{h}) - \frac{1}{N} \sum_{\forall i} \eta_i[h_i],
\end{align}
where the interaction terms cancelled each other.
The key point is to understand the statistical properties of the disorder felt by the center of mass of the interface $\overline{\eta}[h] \equiv  \frac{1}{N} \sum_{\forall i} \eta_i[h_i]$, which currently explicitly depends on the $h_i$'s.

An avalanche of size $S\sim \ell^{d+ \zeta} = \ell^d$ involves a number $n\sim \ell ^d$ of sites and corresponds to a shift of the center of mass by $\sim n/N$.
For an avalanche involving $n$ sites, the mean force of the disorder $\overline{\eta}[h]$ changes by an amount $\mathcal{N}(0, 2 n \sigma_\mu^ 2 / N^ 2) \sim \sqrt{2 n} \sigma_\mu / N$ because $n$ random numbers $\eta_i$  are replaced with new ones. 
Taking this into account, the random force acting on the center of mass can be rewritten:
\begin{align}
\overline{\eta}[h] = \frac{1}{\sqrt{N}} \sum_{j=1}^{j=\lfloor \overline{h} \rfloor} \mathcal{N}\lp 0, 2  \sigma_\mu^ 2\rp,
\end{align}
i.e.~it depends on $h$ only through the average $\overline{h}$.
That is, the center of mass of the fully connected model behaves as a single particle of which the position may be denoted $\overline{h}$, and which follows
 the equation:
\begin{align}
\eta_0 \partial_t \overline{h}(t) = k_0  (w-\overline{h}(t)) - \frac{1}{\sqrt{N}} BM(\overline{h}(t)), \label{Eq:effectiveSingle}
\end{align}
where $BM(t)$ is a Brownian Motion process with finite variance, $\sqrt{2}\sigma_\mu$.
Note that this equation is still non linear, because of the last term.

\subsubsection{Statistics from the ABBM Picture}

The dynamics of \req{effectiveSingle} is very simple because it relates to a well-known problem of stochastic calculus, that is the problem of the first crossing of a random walk with a line.
The single particle in a Brownian potential has been studied in depth by Sinai \cite{Sinai1983}.
Under a drive like ours, \req{effectiveSingle} can easily be mapped onto that of a single particle in a \textit{tilted} Brownian potential.

\begin{figure}[]
\begin{small}
\begin{center}
\def\svgwidth{12.5cm}
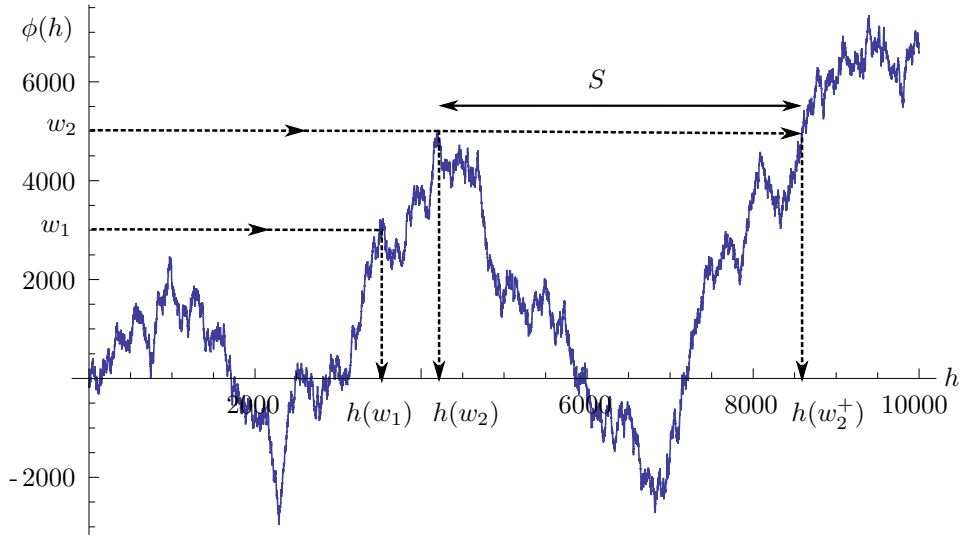
\end{center}
\end{small}
\caption{   {\footnotesize 
Maxwell Construction: plot of the function $\phi(h) = h + BM(h)/k_0$.
For $w=w_1$, we find the corresponding $h(w_1)$ via a simple geometrical construction.
For $w=w_2$, there are two solutions. The avalanche size is the difference: $S=h(w_2^ +)-h(w_2)$.
\label{Fig:Maxwell_construction}
}   }
\end{figure}

\paragraph{Reformulation in Terms of First Crossing}
In the quasistatic limit ($V_0=0^ +$), \req{effectiveSingle} can be summarized by a simple rule.
If the particle verifies the condition
$k_0  (w -\overline{h}) < BM(\overline{h})$,
it does not move
 (the prefactor $	\sqrt{N}$ has been absorbed in $k_0$).
When $w$ is increased, as soon as the equality is fulfilled,
the particle becomes unstable and it moves forward as long as the distance $s$ it has moved is such that $k_0  (w -\overline{h} -s) >  BM(\overline{h}+s)$.
As the Brownian Motion is continuous, the avalanche actually stops as soon as this equality is verified.
This allows for a simple geometrical construction of the solution: for any $w$, the position $h(w)$ is always the smallest $h$ verifying $w =h + BM(h)/k_0$. 
Introducing $\phi(h) \equiv h + BM(h)/k_0 $, we have the so-called Maxwell construction presented in Fig.~\ref{Fig:Maxwell_construction}.

Assuming that the avalanche happens infinitely fast, we have $t=t_1$ at the end of the avalanche.
Under an appropriate change of reference frame,  the avalanche stopping condition reduces to:
\begin{align}
 BM(s) = - k_0 s  , \qquad \text{with } BM(0)=0,
\end{align}
meaning that the avalanche ``size'' $s$ (or length) is distributed as the time of first crossing of a Brownian Motion with the line of slope $- k_0$.

\paragraph{Probability of First Crossing}

In the limit $k_0\to 0$, the problem reduces to that of the return at the origin for a Brownian walker starting in zero.
This distribution decays as:
\begin{align}
P(s) \propto s^ {-3/2},
\end{align}
so that the average size diverges: $\langle s \rangle = \int P(s) s \d s = \infty$.

For finite $k_0$, the distribution has a cutoff at large length scales.
The cutoff can be found intuitively\footnote{
A full computation should account for the new correlations of the walker induced by the condition that it survived up to time $t$ (necessary to be able to cross for the \textit{first time}).
}  by comparing the typical extension of the ``killing wall'' $-k_0 s$ with the typical extension\footnote{
Another (equivalent) way is to compute the probability density (or propagator) to be at coordinate $X=-k_0 s$ at ``time'' $s$. 
This is $P(X=-k_0 s, s) = (2\pi s)^ {-1/2} \exp(-(-k_0s)^ 2/2s) \sim \exp(-k_0^ 2 s/2) $.
The exponential term is of order one iff $s\sim k_0^{-2}$ and quickly decays for larger $s$: we get the cutoff $s_m\sim k_0^{-2}$.
}
 of the free Brownian Motion (i.e.~without killing wall), $\sim s^ {1/2}$.
The typical size at which these two intersect is $s_m\sim k_0^{-2}$. 
For larger values of $s$ the linear term prevails and the probability quickly decays: the cutoff is thus $s_m$:
\begin{align}
P(s) \sim s^ {-3/2} \exp \lp - s/4 s_m \rp.
\end{align}
The average size is then given by $\langle s \rangle \sim s_m^{1/2}$.
An exact computation is provided in \cite{Doussal2008}.

\paragraph{Consistency with Scaling in Finite Dimensions}
This result is perfectly consistent with the scaling relations we derived in the case of finite dimensions.
The key to link the two approaches is to consider that the dimension corresponding to the mean field is the upper critical one, i.e.~to inject $d=d_{uc}=4$ in the scaling relations found above. 
Keeping in mind that $\zeta=0$ for $d\geq d_{uc}$, we find again $\tau=2-2/d=3/2$, $s_m\sim \xi^4\sim \lp k_0^{-1/2} \rp^4 $,  $\langle s \rangle \sim 1/k_0$.
We note that we only computed the statistics of the sizes of the avalanches, but much more is available \cite{LeDoussal2009a, Dobrinevski2012}.

\subsection{The Fokker-Planck Approach}

In the ABBM all the degrees of freedom are reduced to a single one, and the only reminiscence of the spatial extension of the system is contained in the correlations of the disorder.
This will become an issue in the following chapter, where the number of degrees of freedom  is two per site.

To address this issue, we present another method to derive the fully connected behaviour which relies on the reduction of the complete system state $\{w, h_i, f^\text{dis}_i[h_i], (\forall i)\}$ to the probability distribution $P(\delta)$ of a single, simpler and local variable $\delta_i=\mathcal{F}(w, h_i, \overline{h})$.
After recalling the dynamics in terms of $\delta$'s, and introducing an important simplification, we derive and integrate a simple equation for $P(\delta)$, in the spirit of the Fokker-Planck approach\footnote{
We may also call it a Master Equation, but here the state is described by a continuous variable $\delta$.}.

\subsubsection{Definition of The Local Variable $\delta$ and  Narrow Wells}

Let us define explicitly the variable $\delta$, which represents the amount of additional stress that a site can hold before becoming unstable (its ``remaining stability range''):
$\delta_i \equiv f^\text{dis}\eta[h_i,i] - k_0(w-h_i) - k_1 (\nabla^2 h)_i$.
As is, this expression is impractical because upon a slight increase in the drive the interface may adapt smoothly, resulting in infinitely many infinitesimal avalanches (as we discussed earlier, see sec.~\ref{sec:cellullar_auto1}).
This shortcoming can be addressed using the narrow wells ``approximation'' (see later in sec.~\ref{sec:narrow_wells}), for which the variable $\delta_i$ reads:
\begin{align}
\delta_i \equiv f_i^{\text{th}} - k_0 (w-h_i)  - k_1 (\nabla^2 h)_i,
\label{Eq:deltai_introduction}
\end{align} 
This variable $\delta_i$  plays the same role as the quantity $f_i^{\text{th}} - \Sigma_i$ of the OFC* model, and follows the exact same dynamical rules if we choose $g(z)=\delta^ {Dirac}(z-1)$.

\paragraph{Identical Wells  --  Constant Thresholds}
For the mean field treatment, we make an additional simplification: we consider the depth of the narrow wells to be the same for all wells.
In terms of threshold forces, this means that $f_i^{\text{th}} = \text{const.} \equiv f^{\text{th}}$, which simplifies the expression for the $\delta$'s:
\begin{align}
\delta_i = f^{\text{th}} - k_0 (w-h_i)  - k_1 (\nabla^2 h)_i.
\label{Eq:deltas_depinning_dD}
\end{align} 
The dynamics now only depends on the distribution $g(z)$ of the spacings between narrow wells.
However, this is a rather minor change in the physics: we still have quenched randomness, so the universal properties of this particular choice of disorder are expected to be the same as for a more general one.

\paragraph{In Mean Field}
As we have seen in sec.~\ref{sec:fully_connnected_definition}, in the fully connected model all blocks are linked via springs of stiffness $k_1/N$, which results in a simple replacement of $ k_1 (\nabla^2 h)_i$ with $ k_1 (\overline{h}-h_i)$:
\begin{align}
\delta_i \equiv f^{\text{th}} - k_0 (w-h_i)  - k_1 (\overline{h}-h_i),
\label{Eq:deltas_depinning_MF}
\end{align} 
where $\overline{h} = (1/N)\sum h_i$ is the average height and $N$ is the number of sites in the system.
Thanks to the fully connected graph of interactions, every site has the same expression of $\delta_i$, involving only the average $\overline{h}$ and $h_i$ itself: the notions of space and neighbours have disappeared.
The dynamics of the $\delta_i$ variable is quite simple.
\begin{itemize}
\item[1.] Upon an increase in the load $w$, all the $\delta$'s decrease uniformly until a block becomes unstable ($\delta_i \leq 0$)
\item[2.] Unstable sites ($\delta_i \leq 0$) each move to their next pinning wells:   $\delta_i \mapsto \delta_i+ z(k_0+k_1)$ with a different value of $z$ drawn from $g(z)$.
For each jump $z$ there is a drop in \textit{all} the $\delta$'s: $\delta_j \mapsto \delta_j - z k_1/N, \forall j$. 
\item[3.] If $\exists i / \delta_i \leq 0$, perform Step 2.
Else ($\delta_i >0), \forall i$, perform Step 1. 
\end{itemize}

\subsubsection{The Distribution $P(\delta)$}

\paragraph{Infinite Size Limit}
As we have just seen, all the sites are equivalent and the $\delta_i$'s are independent identically distributed (i.i.d.)~variables characterized by their probability distribution $P_w(\delta)$, which in general depends on the initial condition $P_0(\delta)$ and on the current value of $w$.
For a finite system with $N$ sites, the typical configuration $\{\delta_i, i \in [|1,...,N |] \}$ will correspond to a set of $N$ i.i.d.~random variables drawn from $P(\delta)$.

In the thermodynamic limit $N\to \infty$, the fluctuations vanish and the description of the system via the sole distribution $P(\delta)$ becomes exact.
Our aim is now to write down the evolution equation for $P_w(\delta)$ when $w$ increases.

An interesting observable is the average force applied on the system (or stress), defined as $F\equiv \sigma \equiv \overline{k_0(w-h)}$. 
In our case, the stress simply reads:
\begin{align}
\sigma = f^{\text{th}}  - \overline{\delta},
\end{align}
thanks to the cancellation of the interaction term, on average.

\paragraph{Dynamics}
\label{sec:elastic_depinning_dynamics_Mf_hardcore}

When the external driving is increased by an infinitesimal quantity $\d w$, the distribution evolves from its initial shape $P_w(\delta)$ to a new shape $P_{w+\d w}(\delta)$. 
In order to compute the latter, it is useful to artificially decompose the dynamical evolution in different steps. 

In a first step, the center of the parabolic potential moves from $w$ to $w + \d w$ and all $\delta_i$'s decrease by $\Delta \delta_{{\rm step} 0} = k_0 \d w$: $P(\delta) \d \delta$ is increased by $ (\partial P_w / \partial \delta) \d \delta  k_0 \d w $.
Still in this first step, a fraction  $P_w(0) k_0  \d w $  of sites\footnote{Since $\d w$ is infinitesimal and $P$ is continuous, we have  $P_w(0) \approx P(k_0  \d w) $, and the fraction of unstable blocks can also be written $ P(k_0  \d w) k_0 \d w $.}
 becomes unstable and moves to the next wells: $P(\delta) \d \delta$ is increased by $P_w(0) k_0  \d w g_1(\delta) \d \delta$, where $g_1(\delta) \d \delta$ is the probability for a block to fall in the range $[\delta, \delta +\d \delta]$ after a jump\footnote{By definition, $g_1(\delta) \d \delta = g(z) \d z$.}. 
The new $\delta_i$'s are given by $z (k_1+k_0)$,  with $z$'s drawn from the distribution $g(z)$. 
This  writes:
\begin{align}
\frac{P_{{\rm step} 1} (\delta) - P_w(\delta)}{ k_0  \d w }
&=  \frac{\partial P_w}{\partial \delta} (\delta)  +  P_w(0) \frac{ g\left( \frac{\delta}{k_0+k_1}\right) }{ k_0+k_1}.  \label{Eq:Pdelta}
\end{align}
In this expression we have not accounted for the increase of $\overline{h}$ due to the numerous jumps.
This increase is given by the fraction of jumping sites multiplied by their average jumping distance, i.e.~it is worth\footnote{
The average jump size of any finite number of jumps is not $\overline{z}$, so this expression should be puzzling.
However, we work with $P(\delta)$, i.e.~we work in the infinite system size limit. 
In this limit an infinitesimal fraction of sites that jump corresponds to infinitely many sites, so that the average jump is exactly $\overline{z}$.
}
 $\overline{z} P_w(0) k_0 \d w$.
The corresponding change in the $\delta$'s is a uniform decrease by $\overline{z} k_1 P_w(0) k_0 \d w $ (see \req{deltas_depinning_MF}).

This shift in the $\delta$'s is accounted for in a second step, which acts on $P_{{\rm step} 1}(\delta)$ exactly as the first did on $P_w(\delta)$, but with an initial drive given by the shift $\Delta \delta_{{\rm step} 1}= \overline{z} k_1 P_w (0) k_0 \d w $:
\begin{align}
\frac{P_{{\rm step} 2} (\delta) - P_{{\rm step} 1}(\delta)}{\Delta \delta_{{\rm step} 1}}
&=  \frac{\partial P_{{\rm step} 1}}{\partial \delta} (\delta)  +  P_{{\rm step} 1}(0) \frac{ g\left( \frac{\delta}{k_0+k_1}\right) }{ k_0+k_1}. \label{Eq:Pstep2}
\end{align}
In turn, this second step does not account for the increase of $\overline{h}$ due to the ``driving'' by $\Delta \delta_{{\rm step} 1}$: this is accounted for in a third step, and so on.

As these steps go on, the drive from the increase in $\overline{h}$ is given by the geometrical series:
\begin{align}
\Delta \delta _{{\rm step} k} = k_0 \d w \prod_{j=0}^{k-1} (\overline{z}k_1 P_{{\rm step} j} (0) ) , 
\label{Eq:Deltadeltak}
\end{align}
where we identify $P_{{\rm step} 0} \equiv P_w$.
The convergence of the series to zero is guaranteed if $P_{{\rm step} j} (0) < 1/(\overline{z}k_1), \forall j $.
At this point it is enough to assume that this condition is fulfilled at all times. 
In the next chapter we study another model for which this condition may be violated at some times, there, we discuss this issue.

The general set of equations for the $P_{{\rm step} k}$'s is 
a closed form since  $P_{{\rm step} k}$ only depends on the previous $P_{{\rm step} j}, (j<k)$.
Denoting $s\equiv \text{step} k$ the internal time of the avalanche in terms of steps, we can write the evolution as:
\begin{align}
\frac{ \partial P_s}{\partial s} \frac{1}{\Delta \delta _{{\rm step} k} }
&=  \frac{\partial P_s}{\partial \delta} (\delta)  +  P_s(0) \frac{ g\left( \frac{\delta}{k_0+k_1}\right) }{ k_0+k_1}. \label{Eq:Pstepk=s}
\end{align}
This evolution stops either when $P_{{\rm step} k}(0)=0$ (hence $\Delta \delta_{{\rm step} k+1}=0$), or when the r.h.s of \req{Pstepk=s} is zero.
If $P_{{\rm step} k}(0)=0$, 
 some additional driving (increase in $w$) will eventually lead to $P_w(0)>0$ at ulterior times. 
Upon successive increases of $w$, \req{Pstepk=s} will be iterated again and again, each time with a renewed initial drive $k_0 \d w$: we will see that this lets the distribution $P(\delta)$ flow to its fixed point, where the r.h.s of \req{Pstepk=s} cancels.
We now study this case, i.e.~the cancellation of the r.h.s of \req{Pstepk=s}.

\paragraph{Integration of the Dynamics}

As we explained, \req{Pstepk=s} has a fixed point $P_*(\delta)$ that is found when:
\begin{align}
 \frac{\partial P_*}{\partial \delta} (\delta)  +  P_*(0) \frac{ g\left( \frac{\delta}{k_0+k_1}\right) }{ k_0+k_1} = 0 .
 \label{Eq:ilFaut_une_IPP}
 \end{align} 
 This equation can easily be integrated, with $P_*(0)$ computed from the normalization condition\footnote{To let the term $\int\d \delta P(\delta)$ appear, one should multiply \reqq{ilFaut_une_IPP} by $\delta$ and integrate by parts.
 }
 $\int\d \delta P(\delta) =1$.
This gives:
\begin{align}
P_* (\delta)   = \frac{  1 - G\lp \frac{\delta}{k_0+k_1}  \rp }{\overline{z} (k_0+k_1)}   ,   \label{Eq:pdelta_depinning}
\end{align}
where $G(z)\equiv\int_0^z d z' g(z')$.
A simple stability analysis shows that the fixed point is attractive, so that any initial condition converges to it.
Moreover, it is possible to prove that for any given initial condition, there exists a finite $w_*$ at which the distribution reaches the fixed point and remains there for $w>w_*$.
This indicates that the large time dynamics is stationary, as in the 2D case.
The expression for the average stress can be computed explicitly with an integration by parts:
\begin{align}
\sigma 
&= f^ \text{th} - \int_0^ \infty \d \delta~  \delta ~\frac{1-\int_0^ {\delta/ (k_0+k_1)} g(z) \d z}{\overline{z} (k_0+k_1)}\\
&= f^ {th} - \frac{(k_0+k_1) \int_0^ \infty \d z g(z) z^ 2}{2\overline{z}}\\
&= f^ {th} - \frac{(k_0+k_1) \overline{z^ 2}}{2\overline{z}}.
\label{Eq:sigma_mean_general1}
\end{align}
It is worth to note that the average stress only depends on the first tow moments of $g(z)$.
As the elastic driving with $V_0=0^ +$ takes us exactly at the point of depinning transition, this expression is actually the explicit expression for the critical force $F_c$ that we defined in the constant force setup.
We see very well that it is a non universal quantity. 
For the reasonable example of an exponentially distributed $z$ (i.e.~for $g(z)=e^{-z/\overline{z}}/\overline{z}$), we have for instance  $\sigma =  f^ \text{th} - (k_0+k_1)\overline{z}$, so that the critical force (with $k_0\to 0$) is $F_c= f^ \text{th} -  k_1 \overline{z} $.

\subsubsection{Statistics}

\paragraph{Mapping to the Problem of First Crossing}
For any distribution $P(\delta)$, we can compute the probability distribution of the avalanche sizes $N(S)$, for finite values of the parameters $k_0, k_1, \bar{z}$. 
To be concrete, we first consider the case where $g(z)=\delta(z-\overline{z})$.
For a finite system with $N$ sites, the typical configuration $\{\delta_i\}$ corresponds to a set of $N$ independent and identically distributed random variables drawn  from $P(\delta)$.
Let us sort the set: $ \delta_0 < \delta_1 < \dots <\delta_{N-1}$. 
When the system becomes unstable we have by definition $\delta_0=0$.
This site jumps to the next well at distance $\overline{z}$, so that all $\delta_i$'s are decreased by $\overline{z} k_1/N$. This will produce at least another jump if $\delta_1<  \overline{z} k_1/N$. 
More generally, the avalanche size $S$ corresponds to the first time  that the relation:
\begin{align}
\delta_{S-1} \leq   \frac{\overline{z} k_1}{N} S < \delta_S  \label{Eq:ava_stop}
\end{align}
is fulfilled. 

It is thus important to study the statistics of the  $\delta_i$ with $i \ll N$. 
Let us observe that when  $N$ is very large, all these $\delta_i$'s are close to zero, and their distribution can be approximated with a uniform distribution: $\delta \ll 1 \Rightarrow P(\delta) \approx P(0) = \text{const}$.
Within this approximation, the spacings $X_i=\delta_{i+1}-\delta_i$ are independent exponential variables of mean $1/P(0)N$ and variance $1/	(P(0)N)^ 2$.
We conclude that the sequence $\delta_0, \dots, \delta_i$ is a random walk with diffusion constant $1/(P(0)N)^ 2$ and drift $1/(P(0)N)$.
When it crosses the line of slope $\overline{z} k_1/N$, the avalanche is over (see \req{ava_stop}).

\paragraph{Probability of First Crossing}
The statistics of $S$ thus corresponds to the problem of first crossing with $0$ of  a random walk with diffusion constant $D=1/(P(0)N)^ 2$ and drift $d =  \frac{\bar{z}k_1}{N}- \frac{1}{P(0)N}$. For a positive drift, there is a finite probability that this random walk never crosses $0$, which corresponds to an infinite avalanche.
For a negative drift, the time of zero crossing  is always finite, and has been computed for the Brownian motion in  \cite{Majumdar2002}.
The distribution of the avalanche sizes thus reads:
\begin{align}
N(S) &\sim  S^ {-3/2} e^ {-S/2S_\text{max}}  \notag\\
\text{with } &S_\text{max} = \frac{D}{d^ 2}  
= (1 - P(0)\overline{z}k_1 )^ {-2} \label{Eq:Smax}
\end{align}
where for simplicity we have neglected the short-scale regularization in the expression of $N(S)$. 
If now we replace the choice $g(z)=\delta(z-\overline{z})$ with a broader function $g(z)$, only the diffusion constant changes, thus $S_\text{max}$ is the same up to a constant. 

We note that the possibility of divergence for the avalanches re-appears in the expression of  $S_\text{max}$, since if $P(0) = 1/ \overline{z}k_1$, it formally diverges.
In the stationary regime found in \req{pdelta_depinning}, we have $P_*(0)=1/(\overline{z}(k_0+k_1))$ for \textit{any} distribution $g$, and thus $S_\text{max} \propto ((k_0+k_1)/k_0)^2 $ (for any $g$): this illustrates the universality of the scaling relations.

Finally let us remark that the results we obtain here by focusing on $\delta_i$ coincide with the results obtained using the mapping to the ABBM model.

\subsubsection{ Numerical Integration of the FP Equations}

In this simple depinning model, we have the complete analytical solution $P_*$ by direct integration.
However if one is interested in the transient dynamics, i.e.~in how any initial configuration leads to the stationary one, the integration of the equations ``by hand'' proves very hard.
Furthermore, in the model we propose in the next chapter, the dynamics does not lead to a simple stationary solution, and one is interested in the complete evolution over time.
For these reasons, we present here a scheme for the numerical integration of the system  \req{Deltadeltak}-\req{Pstepk=s} under quasi-static increase of $w$.

\includefig{8cm}{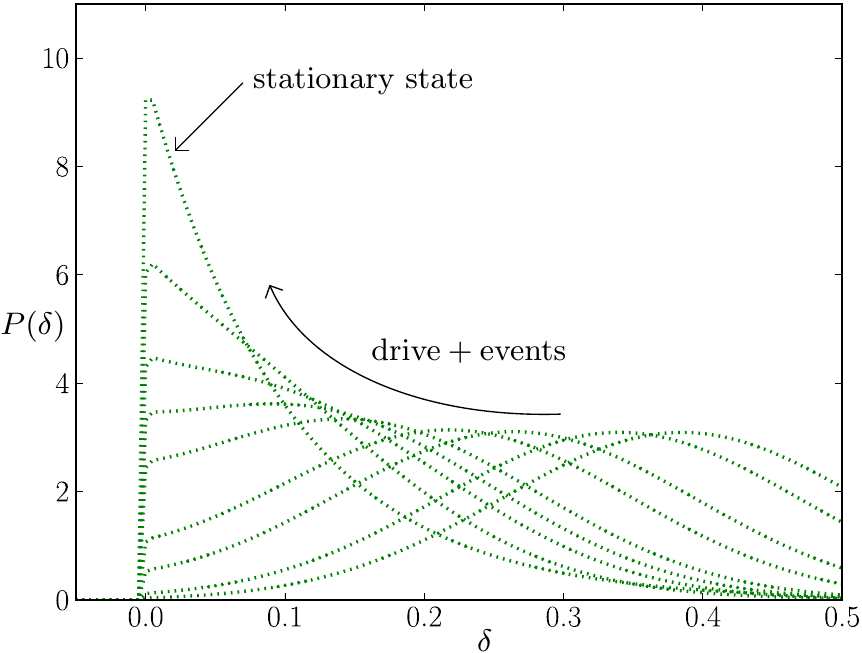}{
The evolution of $P_w(\delta)$ for the depinning model when $w$ is increased. 
The initial distribution is a Gaussian centred in $\delta=0.4$, with standard deviation $0.15$, and the weight at the left of $\delta=0$ cut. 
$ P(\delta)$ quickly reaches its stationary form.
\label{convergence}}
Let us discretize $P(\delta)$ with a bin of size $\varepsilon$.
The distribution probability is then a vector $P_i$ (related to $P(\delta)$ by $P_i = P(\delta = \varepsilon i)$) which evolves with the following rules:
\begin{itemize}
\item {\em Driving process}:
We shift  $P_i$ of one bin: $P_i \leftarrow P_{i+1} $ (physically, $\varepsilon \equiv k_0 \d w$).
\item {\em Instability check}:
We compute the weight of unstable sites: $$P_{\text{inst}} = \varepsilon \sum_{i<0} P_i $$ 
If $P_{\text{inst}} > 0 $, we perform the {\em Avalanche process}. \\
Else we go back to the {\em Driving process}.
\item {\em Avalanche process}: it is composed by a ``jumping sites'' and a ``driving step''.
\begin{itemize}
\item Jumping sites: 
\begin{align}
P_{i\geq0}&\leftarrow P_{i}+  P_{\text{inst}} \frac{g\left(\varepsilon i/(k_0+k_1)\right)}{k_0+k_1}  \nonumber\\
P_{i<0}&\leftarrow 0  \nonumber
\end{align}
\item Driving step ($\Delta \delta_{{\rm step} }$): we shift $P_i$ of $n_{\text{shift}} = \text{Int} [\frac{\overline{z} k_1 P_{\text{inst}}}{\varepsilon}]$ bins.
\begin{align}
P_i \leftarrow P_{i+n_{\text{shift}}}  \nonumber
\end{align}
\end{itemize}
\noindent Then we perform the {\em Instability check}.
\end{itemize}
This algorithm converges very quickly from any initial configuration to $P_*(\delta)$ for any choice of $g(z)$, as is shown in the example of Fig.~\ref{convergence}.

\section{Is the Depinning Framework relevant to Friction?}
\label{sec:criticizing_depinning_friction}

We have described several techniques and results dealing with the depinning transition in the constant force and elastic driving setups. 
We gave a short motivation for the problem, mentioning the context of magnetic domain walls (Barkhausen noise), yet the relationship between the depinning and friction or seismic faults has been eluded.

In this section we first present a few variations of the depinning problem, that received significant attention in the literature.
This overview (sec.~\ref{sec:largeUC}) outlines the broad spectrum of situations covered by the depinning transition and the relative robustness of this universality class.
We then discuss how some characteristic results from this universality class do not compare well with friction experiments or earthquakes dynamics (sec.~\ref{sec:depinning_output_not_friction}).
Despite the large spectrum of physics covered by the depinning universality class, we are forced to acknowledge that friction (and a fortiori seismic phenomena) cannot be adequately captured by any of the depinning instances presented.

\subsection{Depinning: a Robust Universality Class}
\label{sec:largeUC}

We have already presented two variants of the depinning problem via the different forms of driving (constant force and elastic driving), leading to different facets of the same problem.
Here we present other variations that are also in the depinning universality class.
This presentation is not intended to be exhaustive: our aim is to give an idea of the generality of the depinning framework and define the vocabulary for the curious reader. 
Along the lines, we try to show the relationship of the models with experimental works.

Up to now, our default depinning equation was \req{ForceDepinning}:
\begin{align*}
\eta_0 \partial_t h(x,t) = F + k_1 (\nabla_x^2 h)(x,t) - f^\text{dis} \eta[h(x,t),x].
\end{align*}
The most general equation for the depinning problem reads:
\begin{align}
 \eta_0 \partial_t h(\mathbf{x},t) = \mathcal{F}_\text{drive}[h(x,t),t] + \mathcal{F}_\text{elastic}[h(x,t)]  + \mathcal{F}[\eta[h(x,t),x]],
 \label{Eq:depinning_VERY_GENERAL}
\end{align}
where $\mathcal{F}_\text{drive}, \mathcal{F}_\text{elastic}, \mathcal{F}$ are some general functionals. 
We now want to discuss each of these terms.

\subsubsection{Random Bond versus Random Field}
\label{sec:random_Bond}

There are two universality classes of disorder: ``Random Bond'' (RB) and ``Random Field'' (RF) (the names come from magnetic realizations of the depinning problem).
The RB kind of disorder corresponds to impurities that directly attract or repel the interface, while in RF the pinning energy
of the interface depends on all the impurities that the interface has swept over see Fig.~\ref{Fig:two_disorders}, \cite{Giamarchi2006}.
\includefig{\textwidth}{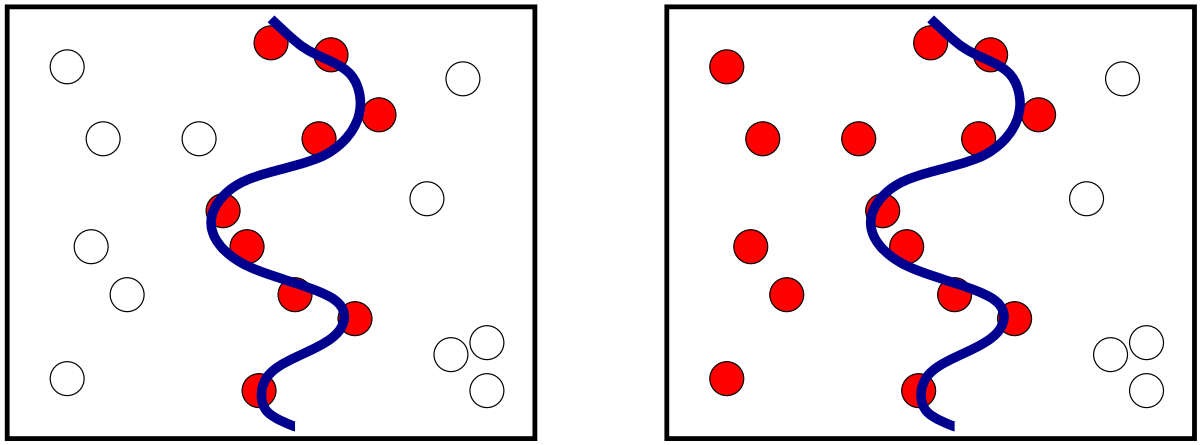}{
From \cite{Giamarchi2006}.
Interface (solid line) and impurities (circles). Filled circles indicate impurities which contribute to the interface energy.
Left: Random Bond, the interface energy only depends on the impurities around it.
\newline
Right:  Random Field, the interface energy depends on all the impurities that have been swept over.
\label{Fig:two_disorders}
}
Denoting $\eta(z,x)$ the microscopic random energy potential associated to the impurities, the corresponding energy terms (and forces) read:
\begin{align}
E_\text{RB}  &\equiv \eta[h(x,t),x] &\Longrightarrow & \qquad \mathcal{F}_\text{RB}  \equiv \frac{\partial}{\partial h}  \eta[h(x,t),x] &\\  
E_\text{RF} &\equiv \int_{-\infty}^{h(x,t)} \d z~\eta(z,x)  &\Longrightarrow & \qquad \mathcal{F}_\text{RF}  \equiv  \eta[h(x,t),x].&
\end{align}
The impurities at the origin of both kinds generally have positions uncorrelated in space (typically they follow the uniform law) and correspond to values of the force distributed over a range of finite width, i.e.~it is reasonable to assume that $\eta(z,x)$ is a white noise (and in particular that it has short-range correlations).
The crucial distinction is on the way that these impurities affect the interface, which can lead to either RB or RF kind of correlations for the disorder force.
The functional forms of the disorder energy correlator $R$ (and the force correlator  $\Delta=-R''$ \cite{Chauve2000}) are reviewed in \cite{Giamarchi2006, Giamarchi2008}.

Up to now we have used the RF kind of disorder, i.e.~the correlations of the random \textit{force} were short-ranged. 
This kind of disorder is appropriate e.g.~when applying the depinning transition to the problem of fracture in the tensile mode (Mode I, \cite{bonamy2011failure}), where the impurities in the fractured plane still contribute to the total energy of the system after they broke \cite{Alava2006, Bonamy2008}.
More generally, the random field is appropriate whenever there is an asymmetry between the half-space that has been visited (broken, in the context of fracture) and the other one.

In the static case, i.e.~in the absence of an external driving ($F=0$), the equilibrium properties of the two classes are largely different.
However, the dynamics ($F>0$) of the two classes have been proven \cite{Narayan1993, Chauve2000} to be in the same universality class, with a unique set of exponents characterizing the depinning transition of these two kinds of disorder.
Numerics continue to verify this prediction with increasing precision, up to recent works \cite{Ferrero2013}.
In this thesis we will be interested only in random field disorder, i.e.~in a short-range correlated disorder function for the pinning \textit{force}.

\subsubsection{Short-Range versus Long-Range Elasticity}

\paragraph{Short-Range}
Elasticity is fundamentally a short-range process:
as atoms form bonds, each one of them is in the minimum of the energetic potential generated by its neighbours.
At first order, the deviations from this minima are quadratic\footnote{
The first non zero development of any function around a local minimum is always quadratic. 
A similar argument explains the null hypothesis stating that ``fluctuations are Gaussian''.
}, i.e.~the elastic energy is $E_{el} \sim k (\delta x)^2$, where $\delta x$ is the deviation from the minimum.
This gives an elastic force linear in the deviation from elastic equilibrium and with a short    range.
When we put together several atoms on a chain, the interaction of each one of them with its closest neighbours naturally builds the discrete Laplacian,  $\nabla^2 h_i = h_{i-1} -2 h_{i} + h_{i+1}$.
In the continuum limit, we can write the elastic force as  $F_{el} = k_1 \nabla^2 h = k_1 \partial_x ^2 h$ (in higher dimensions the force is still given by a Laplacian).
For many systems, elasticity will naturally be accounted for via this term, and we may speak of \textit{short-range elasticity}.

\paragraph{Long-Range}
However, there are some situations where the cohesion forces are better accounted for via long-range interactions. 
In the context of fracture \cite{Gao1989}, the propagation of the crack front (a rough line) depends on the elastic interactions over the whole crack plane\footnote{Precisely, the interactions tend to minimize the overall post-mortem surface.} when considering the dynamics of this elastic line, one should use a long-range elastic interaction as an effective description of the interactions mediated by the surroundings of the crack front.
Similarly, in the wetting of a (disordered or rough) surface \cite{Joanny1984, Rosso2002b, Moulinet2004, LeDoussal2009a}, the contact line of a liquid meniscus is affected by the 
forces from its surroundings,
 so that the effective elastic interactions for the line alone are long-ranged.
In the seismic context, long-range elastic interactions within the 2D fault would be an effective representation of the elastic interactions mediated by the bulk of the half-space (3D structure).
In some magnetic systems, there are also dipolar interactions \cite{Nattermann1983} which are naturally long-ranged\footnote{
Note that even in these systems, the fundamental interaction is mediated by short-range processes (exchange of photons, etc.), which are (much) more simply accounted for via a long-range interaction kernel.
In all cases, we consider systems where the propagation of the interaction (speed of light or sound) is much faster than the system's evolution (avalanche velocity).
}.

Let us define precisely the notion of long-range: considering the interface displacements in Fourier space, the total elastic energy of the system is defined as:
\begin{align}
E_\text{elastic} 
= \frac{k_1}{2}  \int \frac{ \d^d \mathbf{q} }{(2\pi)^d}~|\mathbf{q}|^ \alpha~\widehat{h}(\mathbf{q})\widehat{h}(\mathbf{-q}), 
\end{align}
where $\widehat{h}(\mathbf{q}) $ is the Fourier transform of $h(\mathbf{x})$.
The crucial point here is that the spectrum of the kernel scales as $\sim k_1 |q|^\alpha$: for $\alpha=2$, we recover the short-range elastic kernel, and for any $\alpha <2$, we have a long-range elasticity.
We note that a smaller exponent $\alpha$ gives a larger importance to the smaller $q$'s in the kernel's power spectrum, i.e.~to the larger length scales, as expected. 
In the direct space, the elastic force applied on the point $x$ can be written (in one dimension):
\begin{align}
F_{elastic}(x)= \frac{\partial E_{elastic}}{\partial h}  =- k_1 \int \d x' \frac{h(x)-h(x')}{\vert x-x' \vert ^ {1+\alpha} } ,
\end{align}
with a possible alternative being the use of fractional derivatives \cite{Zoia2007b}.
With long-range interactions, the exponents of the depinning transition change and depend continuously on $\alpha$ \cite{Tanguy1998}. 
However the elastic kernel is still convex in the variable $h$, so that the scaling relations hold and the depinning framework is still appropriate. 
We recall the general scaling relations here:
\begin{align}
\nu = \frac{1}{\alpha-\zeta}; \qquad \beta= \nu (z-\zeta) ; \qquad \tau=2-\frac{\zeta+1/\nu}{d+\zeta},
\end{align}
where $\alpha, z$ and $\zeta$ are our ``fundamental'' exponents.
These exponents do change, while these scaling relations remain the same. 
The extreme case of infinite range ($\alpha=0$) corresponds to the mean field, fully connected  approximation, that we discussed in sec.~\ref{sec:MF_depinning} (remember that in this case we also have $\zeta=0$).

We note that the addition of non-linear (non harmonic) terms breaks the STS relation and take us into a different universality class (for the dynamics, $F>0$).
The example of a quartic term such as $E_{elastic}=\int \d ^ d x k_1' (\partial_x h)^ 4 $ yields the quenched KPZ universality class \cite{Rosso2003, Kolton2009} for which we have e.g.~$\zeta=0.63$ instead of $1.25$ (in $d=1$).

\paragraph{Eshelby Problem}
\label{sec:AmorphousPlastoc}

In the field of amorphous plasticity, it is well known that local plastic events redistribute the stress over long distances via an anisotropic stress propagator \cite{Barrat2011, Martens2011, Nicolas2014}.
Indeed, it has been shown from microscopic models that on average, the long-time equilibrium response yields the results predicted from the Eshelby inclusion problem \cite{Eshelby1957} (see \cite{Puosi2014} and references therein). 
In the context of seismicity this also seems inappropriate, as earthquakes seem to entail quadrupolar stress redistributions \cite{Scholz2002, Bhattacharyya2006}.

In extensions of the  depinning transition framework to amorphous plasticity, this anisotropic long-ranged kernel is taken into account \cite{Talamali2011, Budrikis2013}.
Approximating the effect of the rearrangements of the elastic interface (corresponding to local plastic events) by a force quadrupole, one expects a four-fold quadrupolar symmetry for the inhomogeneous part of the stress propagator.
The form of this propagator for an infinite two dimensional medium  reads \cite{Martens2012}:
\begin{align}
G(r,\theta) = \frac{1}{\pi r^ 2}\cos (4\theta),
\end{align}
where the stress-strain ($\sigma - \varepsilon$) relation reads:
\begin{align}
\partial_t \sigma(r,t) = \mu \dot{\gamma} + \int \d r' G(r-r') \dot{\varepsilon}^ {pl}(r',t),
\end{align}
with $\mu$ the shear modulus, $\dot{\gamma}$ the strain rate and $\varepsilon^ {pl}$ refers to the plastic part of the strain.

However, the non-convexity of this kind of elastic kernel renders several fundamental defining properties of the depinning transition  invalid: in particular, we no longer have only forward movements of the ``interface''.
The universality class thus changes, and is different from the mean field one \cite{Budrikis2013}.
However, these developments are quite recent and numerous questions remain open.

In this thesis, we focused on the microscopic modelling of the 2D surfaces with short-range elastic interactions, and the effect of including long-range elasticity (possibly anisotropic) in our models remains an open question.

\subsubsection{Zero Temperature versus Finite Temperature (Creep)}

Consider the addition of temperature, i.e.~the addition of a random force $\theta(x,t)$:
\begin{align}
\eta_0 \partial_t h(x,t) = F + k_1 (\nabla_x^2 h)(x,t) + \eta[h(x,t),x] + \theta(x,t)  
\label{Eq:FiniteTemperature} \\
\text{with } \qquad \langle \theta(x_1,t_1)\theta(x_2,t_2) \rangle = 2\eta_0 k_B T \delta^D(x_1-x_2) \delta^D(t_1-t_2),
\end{align}
where $\delta^D$ denotes the Dirac distribution, $F$ is the driving force and $T$ is the temperature.
The addition of temperature means that configurations that would normally be pinned forever can now overcome small energy barriers, thanks to thermal fluctuations.
This phenomenon of escaping local energy minima is called \textit{creep}.
On average, the thermal fluctuations push the interface in the direction of the force, 
something that can generate an avalanche.
\begin{figure}[]
\begin{small}
\begin{center}
\def\svgwidth{\textwidth}
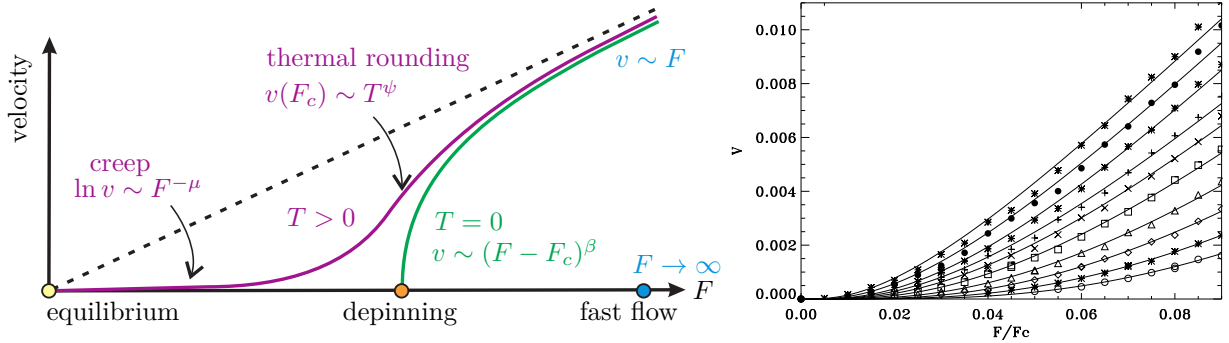
\end{center}
\end{small}
\caption{   {\footnotesize 
Effect of the temperature on the depinning transition of an elastic line in a disordered landscape.
Left: Adapted from \cite{Ferrero2013a}: the ``phase diagram'' of the depinning transition, with and without temperature.
\newline 
Right: Focus on the creep regime, from \cite{Kolton2005}. 
Temperature is increasing, from bottom to top $T=0.24, 0.26, ..., 0.42$.
Markers indicate numerical results and solid lines the fit with \reqq{creep_depinning}, with $U_c$ and $\mu$ as fitting parameters.
The calculation of the creep exponent $\mu$ can be performed via a Functional Renormalization Group \cite{Chauve1998}.
\label{Fig:v_de_F_Creep}
}   }
\end{figure}
Thus, the average velocity of the interface $v$ will be larger than zero even when $F<F_c$. 
This changes the characteristic curve $v(F)$, which can be fitted by (see Fig.~\ref{Fig:v_de_F_Creep}):
\begin{align}
v(F)\sim \exp\lp-\frac{U_c}{k_B T}  \lp\frac{F_c}{F}\rp^ {\mu} \rp,
\label{Eq:creep_depinning}
\end{align}
where $\mu$ is the creep exponent and $U_c$ some characteristic energy. 
Strictly speaking, for any temperature $T>0$ the transition is lost and we get a simple crossover \cite{Bustingorry2008} (i.e.~no sharp transition even in the macroscopic limit).
In this sense, we see that temperature is \textit{relevant}.
For further discussion on the role of temperature on interfaces with or without disorder, see \cite{Kolton2002, Cugliandolo2002}, or \cite{Iguain2009} for the effect of temperature on the ageing (relaxing) properties of these interfaces.

A discussion and a table of the exponents for the depinning models with RF or RB disorder, with or without quenched KPZ non linearities and the effect of temperature on these models is available in \cite{Kolton2009}.
However, for many systems the prefactors and the pace of avalanches are such that temperature can be neglected, in a first approach.
In this thesis, we are interested in the approximation of zero temperature.

\subsection{Depinning: a Model for Frictional Processes?}
\label{sec:depinning_output_not_friction}
There are micro- and mesoscopic arguments for expecting the depinning of an elastic interface to be related with frictional processes: the random distribution of asperities may be accounted for via a quenched disorder term, the cohesion forces within each sliding surface may be represented by elastic interactions and the driving from a side of one block may be represented by the elastic driving term $k_0(w-h)$ (see also Fig.~\ref{Fig:effective_stiffness_k0}).
These arguments can be debated.
Here, we focus on the statistical output of the models of depinning of elastic interfaces and compare them with observations in the frictional and seismical contexts.

\subsubsection{Context of Friction}
\label{sec:confusion_stick_slip}

\paragraph{Stick-Slip: a Confusion}
At the macroscopic scale, the slow driving of a solid relative to another is expected to produce periodic stick-slip  (see \ref{sec:stickslip1}), either when $V_0\sim 0$ or when $k_0 \sim 0$.
As we explained in Chap.~\ref{chap:friction}, asperities are also expected to perform ``a kind of stick-slip'' motion in the sense that they alternate between phases of contact ($\sim$static, stick) and free phases ($\sim$dynamic, slip).
Even in the seemingly steady state regime of kinetic friction, there is still this microscopic stick-slip which occurs locally, characterized by a pseudo-periodic behaviour and finite advances of the asperities during the slip phases.

\begin{figure}[]
\begin{small}
\begin{center}
\def\svgwidth{\textwidth}
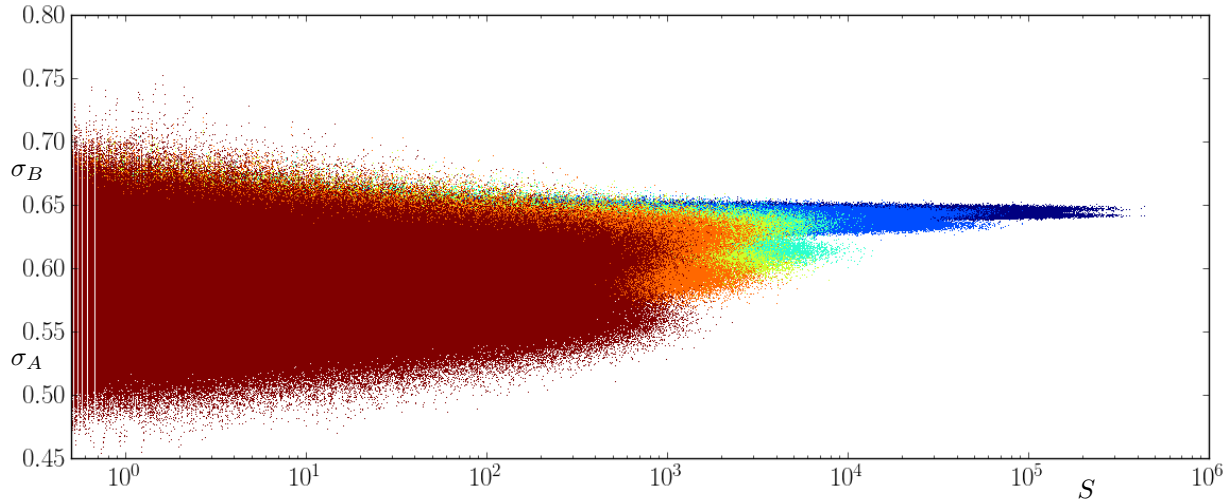
\end{center}
\end{small}
\caption{   {\footnotesize 
Scatter plot of the average stress restricted to the avalanche area against the avalanche size $S$: $\sigma_B$ (resp.~$\sigma_A$) is the stress before (reps. after) the avalanche occurred.
We used $k_1=1$ and (from left to right): $k_0=0.05, 0.025, 0.012, 0.003, 0.001$.
When $k_0 \to 0$, the two values $\sigma_B,\sigma_A$ converge to a common value: the stress drop $\Delta \sigma = \sigma_B-\sigma_A$ associated to the avalanches is thus infinitesimal (in particular for the large avalanches).
\label{Fig:scatter_plot_k2is0_1}
}   }
\end{figure}

In the depinning model, the interface rapid motion during the avalanches alternates with static phases, something which seems reminiscent of stick-slip. 
However, this local pinned-unpinned alternation does not depict the expected stick-slip behaviours observed in friction.
In particular, there are two major discrepancies with expected stick-slip.

First, in friction the duration of the stick phase for a large set of synchronized asperities following stick-slip is pseudo-periodic, whereas in the elastic model the waiting times between avalanches (at a single location) are not (the waiting times for the entire interface between two events is actually exponentially distributed).

Second, the local stress felt by a small section of the interface does not operate a saw-tooth like pattern as expected in friction.
This is due to the random occurrences of avalanches over time, but also to the fact that a small patch of the system 
cannot  -- macroscopically --   accumulate energy over time, as we now show.
The typical energy accumulated by a patch of diameter $\ell$ is given by the maximal avalanche size $S_{max}\sim \xi^{d+\zeta}$, 
so that local variations of the stress scale as $\Delta \sigma \sim k_0\xi^{d+\zeta}$ (for $\ell = \xi$).
Since $\xi \sim k_0^{-1/2}$, this corresponds to a variation per surface area $ \sim k_0\xi^{\zeta} \sim k_0 ^{1-\zeta/2}$.
 As $\zeta<2$, 
 when $k_0 \to 0$ (i.e.~at criticality), $\Delta \sigma \to 0$: the energy released locally by the largest avalanches is thus infinitesimal (instead of being macroscopic), as can be seen in Fig.~\ref{Fig:scatter_plot_k2is0_1}.

\paragraph{Velocity-Weakening}
In the depinning framework, the stress-velocity curve is monotonously increasing, and its reciprocal is also monotonous: the driving force $F$ is always increasing with velocity, i.e.~there is no room for velocity-weakening.
Note that this is also true in the finite temperature case.

\paragraph{Ageing of Contact at Rest}

In the case of the elastic line at exactly zero temperature, if the driving is stopped (constant force $F$, or $V_0$ set to zero), nothing happens to the line once the avalanche is over.
On the contrary, at finite temperature there will be some creep i.e.~the interface will move forward without any additional driving being performed.
As the interface moves forward, the stress decreases (decrease of $k_0(w-h)$) until the interface reaches the metastable states of lowest energy, i.e.~those where $h(x)\approx w$.
This stress decrease is actually qualitatively compatible with friction observations.

However, if driving is then restored ($V_0>0$), the interface stress will increase back to its steady state value, without any statistically significant overshoot.
This is unlike what happens in friction, where the stress (static friction force) overshoots compared to the steady state value.

\subsubsection{Seismic Context}

\paragraph{Correlations}
In the depinning model the time distribution of avalanches is essentially poissonian (uncorrelated), whereas in the seismic context main shocks and aftershocks are strongly correlated over time.
There can be no naturally defined aftershocks in the context of a purely elastic line (without inertia), simply because there are no characteristic times besides the avalanche inner time scale $\sim \eta_0$ and the driving time scale $\sim h_0/V_0$:
aftershocks are impossible to render using purely elastic (overdamped) models.

\paragraph{Exponents}
The GR law for the earthquakes magnitude-frequency distribution is typically considered to be a power-law with an exponent $b \simeq 1\pm 0.25$, which corresponds to $\tau =1 + 2/3 b \simeq 1.7 \pm 0.2$ (see sec.~\ref{sec:EQs} for the full definitions and historical origins).
The mean field value $\tau=3/2$ of the depinning is thus just in the limit of the acceptable range, and the 2D value $\tau=1.26$ 
is definitely out of it\footnote{
More generally, in the depinning problem we have $\tau \leq 1.5$ in all dimensions.
}.

\subsubsection{Depinning Is Not a Model for Friction}

The conclusion is very clear: although there are a few qualitative similarities between the depinning of an elastic interface and friction or earthquakes, there are crucial discrepancies which force us to discard the model of the depinning of an elastic interface as an appropriate representation of friction (and a fortiori, of seismic faults dynamics).

\section{Conclusion: Elastic Depinning is not Friction}
\label{sec:conclu_chap3}

We have introduced depinning and presented its most salient features.
With simple scaling arguments, mean field calculations and numerical simulations, we have clearly stated the defining and characteristic features of the depinning transition. 
In this dynamical phase transition between a moving and a static phase, we observe critical power-law distributions of avalanches 
 and a rough (self-affine) interface. 
These features are somewhat reminiscent of frictional processes, where the interfaces in contact are usually self-affine, and of earthquakes, where the distributions of seismic events follow power-law distributions.

However, despite its wide application spectrum and its robustness to small changes, the avalanches at the depinning transition do not satisfactorily reproduce the properties of frictional or seismical processes.
Quantitatively the power-law exponents do not match the observed ones, and important qualitative features such as stick-slip, velocity-weakening or aftershocks do not appear.

In terms of microscopic construction there are several natural correspondences with seismic faults, but also important discrepancies.
A first approximation of the elastic depinning models is the absence of local ageing mechanisms, a shortcoming which explains the absence of velocity-weakening, a crucial feature in the genesis of seismic phenomena.
We will see how it is also responsible for the other failures of the elastic depinning model at describing frictional processes. 
A second point is the approximation of overdamped dynamics, which may not be justified for friction. 
As we will see in the next chapter, when inertia is included, the depinning transition can become first-order like \cite{Marchetti2000, Prellberg2000}.
\\

We conclude that the framework of the depinning transition offers a promising basis for understanding friction and possibly earthquakes, but is a long shot from providing a definitive answer.
Despite representing a broad universality class, the depinning framework needs to be extended to account for some fundamental mechanisms relevant in friction, as the ageing of contacts.
This is precisely what we do in the next chapter\footnote{
From now on, the term ``depinning'' will refer to the main case studied in this chapter except when explicitly stated otherwise.
See Appendix \ref{App:important_remark} for a complete list of the choices we made in our model of elastic interface.
}.

\chapter{Viscoelastic Interfaces Driven in Disordered Media}
\label{chap:visco}

\vspace{-2cm}
\minitoc
\vspace{2cm}

As we have seen in the previous chapter, the driven dynamics of heterogeneous systems often proceeds by random jumps called avalanches, which display scale-free statistics. 
This critical out-of-equilibrium behaviour emerges from the competition between internal elastic interactions and interactions with heterogeneities and is understood in the framework of the depinning transition \cite{Fisher1998, Kardar1998}. 
In this description of avalanches a trivial dynamics is usually assumed in the inter-avalanche periods, characterized by a monotonous driving \cite{Fisher1998, Sethna2001}.
However, the inclusion of viscoelastic effects with their own characteristic time scales brings about novel dynamical features, which we study in this chapter.
The existence of viscoelastic interaction has drastic consequences on the macroscopic behaviour of the system, as in the context of friction, where it is linked to the increase of static friction over the time of contact \cite{Dieterich1972, Marone1998} (see also sec.~\ref{sec:ageing_violation} and sec.~\ref{sec:Ageing}).	
Here we show how these relaxation processes generically induce a novel avalanche dynamics characterized by new critical exponents and bursts of aftershocks strongly correlated in time and space.
Due to its simplicity, the model allows for analytic treatment in mean field, and for extensive numerical simulations in finite dimensions.
We compare our model with the existing literature in two times: we start with models that are strongly connected to ours, and conclude with a discussion on models from other contexts, showing that a global trend seems to emerge, showing that our model may play a role in other areas than friction and its applications.

\section{Previous Literature}

\label{sec:previous_litt}

\subsection{Viscoelastic Interfaces Driven Above the Critical Force}  
\label{sec:Marchetti}

Vortices in type-II superconductors, due to their mutual repulsion, tend to form a triangular crystal which is pinned and deformed by the presence of impurities \cite{Abrikosov1957}.
In presence of a current $I$ the Lorentz forces acting on the magnetic flux can eventually depin the vortices.
This depinning can either be \textit{elastic} or include some degree of plasticity.
In the elastic depinning the crystal can be deformed but moves collectively.
In the plastic depinning, topological defects proliferate and only a fraction of the system moves while the rest remains pinned.
This latter case is observed more frequently in experiments and numerical simulations, where array of vortices are depinned together and flow through channels.
In this regime, the assembly of vortices can not be described as an elastic solid sliding on a disordered substrate.
Instead, it seems appropriate 
to extend the elastic depinning framework to include viscous, fluid-like interactions \cite{Marchetti2000,Marchetti2002, Prellberg2000,Marchetti2005, Marchetti2006}.

In the seminal paper \cite{Marchetti2000}, the ``visco'' part is described by a memory kernel $C(t)$, so that the equation of motion for the coarse-grained displacement field $u_i(t)$ (representing deformations of regions pinned collectively by the disorder) reads:
\begin{align}
\eta_0 \dot{u}_i(t) =  \nabla^2 \left( \int_0 ^t \d s~C(t-s) \dot{u}_i(s) \right)_i  + F + f_i(u_i)
 \label{Eq:Marchetti1}
\end{align}
where $f_i(u_i)$ is the disorder function.
The first term in the r.h.s.~is the viscoelastic interaction force felt by the interface at the position $i$. For $C(t-s)=\delta(t-s)$, it reduces to a purely viscous force $\nabla^2 \dot{u}_i$, corresponding to a purely fluid-like dynamics. 
For $C(t-s)= \text{const}$, it reduces to a purely elastic force $\nabla^2 u_i(t)$, i.e.~we come back to elastic depinning. 
For $C(t-s)= \mu e^{(t-s) \mu / \eta_u}$, it corresponds to an interaction of the ``Maxwell'' type, discussed in sec.~\ref{sec:Maxwell_fluid}.

The mean field case is studied via the fully-connected limit, in which both the viscous and the elastic parts of the interactions become of infinite range.
The analytical results are obtained under the assumption of a constant average velocity $\overline{v}=\sum_i \dot{u}_i$.
Using the exponential kernel $C(t-s)= \mu e^{(t-s) \mu / \eta_u}$, with $\eta_u$ an effective viscosity (or friction  coefficient), they find a self-consistency condition relating the average velocity $\overline{v}$ to the driving force $F$.
For sufficiently large values of  $\eta_u$, there are several solutions to this condition, i.e.~several  $\overline{v}$  are compatible with a given driving force $F$. 
This is interpreted as the possibility of an hysteresis for the force-velocity curve, which is actually also observed in numerical simulations (where the constraint of constant  $\overline{v}$  is relaxed).
See Fig.~\ref{Fig:Prellberg} for a comparison of the analytical and numerical response curves $\overline{v}(F)$.
\includefig{8cm}{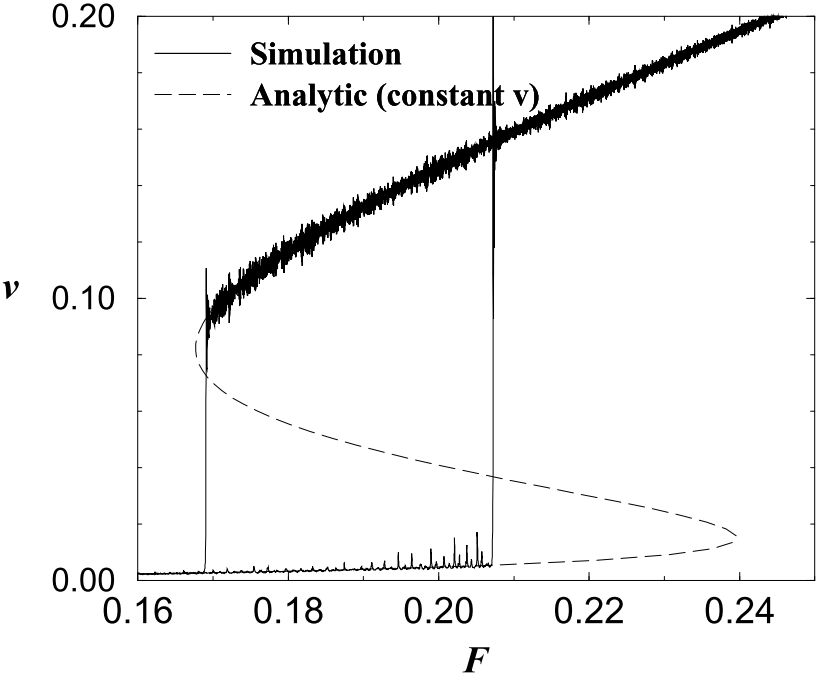}{
From \cite{Marchetti2000}: the hysteretic depinning.
In the depinned regime ($v>0$), the (spatially) averaged velocity is always positive.
In the numerics, the force $F$ is slowly increased at first, and $\overline{v}$ switches to the upper branch around $F\approx 0.205$. 
When $F$ is slowly decreased, $\overline{v}$ switches back to the lower branch around $F\approx 0.17$. 
There is thus a range of values of $F$ ($\approx [0.17,0.205]$) for which  $\overline{v}$ can take two values, depending on the system previous history: this is the hysteretic depinning.
\label{Fig:Prellberg}
}

The main conclusion that can be drawn from this study is that the depinning of elastic and viscoelastic interfaces in disordered media differ significantly in the mean field (and thus probably also in finite dimensions). 
However, the analytical studies on viscoelastic interfaces are done in the constant force setup, where the focus is generally on the velocity-force relationship instead of avalanches.
Some of the results from \cite{Marchetti2000} can be compared with those presented in \cite{Fily2010}, where molecular dynamics simulations (of two-dimensional vortex lattices) 
 are performed.

\subsection{The Relaxed Olami-Feder-Christensen Model (OFCR)}

As we have seen in sec.~\ref{sec:OFC*}, \pp{sec:OFC*}, the OFC* model (equivalent to elastic depinning) is also an honourable candidate for modelling frictional systems (e.g.~tectonic plates), except for its crucial lack of any ageing mechanism.
In order to account for the slow processes occurring between seismic events (plastic events, water flow, etc.), a ``relaxation mechanism'' is introduced, which slowly smooths the stress field $\sigma$ of the OFC* model over time.

\begin{figure}[]
\begin{small}
\begin{center}
\def\svgwidth{12cm}
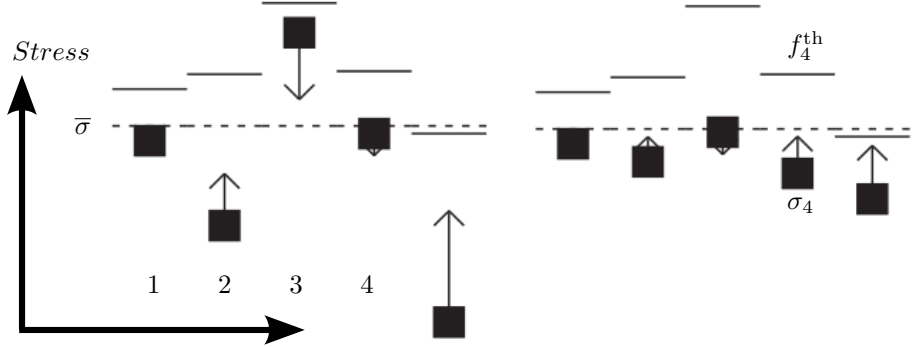
\end{center}
\end{small}
\caption{   {\footnotesize 
Schematic illustration of the stress relaxation in the OFCR model in one dimension, for five blocks $1,2,3,4,5$, with $k_0 V_0 \ll R$.
Each square indicates the stress of the corresponding site, horizontal bars indicate the thresholds $f^\text{th}_i$.
The dashed line indicates the average stress.
Left: initial situation. 
Due to the local nature of the Laplacian operator, the block $4$ does not move towards the average stress, but towards the average of its two neighbours.
\newline
Right: same system observed after it relaxed for some time.
The block $4$ went downward, now it goes upwards.
In the absence of aftershocks, the stresses converge to the average value of the stress (dashed line).
Here the block $5$ is going to meet its threshold, thus causing an aftershock.
\label{Fig:monOFCR_relaxation}
}   }
\end{figure}

It is natural to ask for a decrease of the stress, as it represents the local energy density, which can only be minimized by microscopical processes (aside from thermal fluctuations which are neglected).
This approach was followed by Jagla et.~al.~in \cite{Jagla2010a, Jagla2010}, where an effective equation for the stress variable $\sigma_i$ in the inter-avalanche periods was proposed, in the so-called OFCR model:
\begin{align}
\frac{\d \sigma_i}{\d t} = k_0 V_0 + R \nabla^2 \sigma_i. 
\label{Eq:OFCR}
\end{align}
The second term is an effective way of translating the relaxation of the stress due to microscopical processes.
We comment on the effect of this second term in Fig.~\ref{Fig:monOFCR_relaxation}.
The rules defining the dynamics of the OFCR model are the following: 
\begin{itemize}
\item[(1)] All the $\sigma_i$'s evolve according to \req{OFCR} until a block has $\sigma_i=f^\text{th}_i$. 
This can be due to an increase of the relaxing term or to the drive with rate $V_0$.
\item[(2)] Any block that has $\sigma_i=f^\text{th}_i$ slips:
all neighbouring blocks each receive an additional stress $\alpha \sigma_i$ and the $\sigma_i$ is set to zero.
A new threshold $f^\text{th}_i$ is drawn from $\rho$.
 This is done in parallel for all blocks.
\item[(3)] Repeat Step (2) until $\sigma_i<f^\text{th}_i, \forall i$. When this is the case, the avalanche is over and we may repeat Step (1).
\end{itemize}

This model and a few variants of the relaxation mechanism \reqq{OFCR} were studied in great detail in \cite{Jagla2010a, Jagla2010, Jagla2011b, Rosso2012}, via numerical simulations.
There, several features were observed, which are in good agreement with earthquakes phenomenology:
\begin{itemize}
\item The presence of aftershocks as side-effects of main shocks (they continue to happen after a main shock even when driving is stopped).
\item A Gutenberg-Richter law, with an exponent $b\approx 1$.
\item The Omori law of aftershocks decay with an exponent $p \approx 1.1$.
\item The presence of a seismic cycle (in the two dimensional system) \cite{Rosso2012}.
\end{itemize}
A model of elastic material with a somehow similar relaxation mechanism was applied to study the evolution of the contact area of solids at rest in \cite{Jagla2010b}.

There have been a study done in parallel with ours,where the effect of the relaxation of the contacts was studied analytically in mean field in \cite{Braun2013}.
There, the focus was on the limited range of applicability of the GR law (which applies only for ``small earthquakes'').

These results indicate that the general philosophy of the model contains an element essential to  the dynamics of seismic faults and maybe more generally of frictional processes.
However, there are limitations to this approach.
Due to the formulation of the problem in terms of a cellular automaton, the physical interpretation of the model is difficult: the relaxation mechanism is merely an effective way of accounting for many microscopical processes, all reduced to a single parameter $R$.
Consequently, the respective roles of the various microscopical processes at play in friction (as plastic yielding of asperities, etc.) can hardly be sorted out in the model.
Another important limitation due to the very formulation of the model is that a field theoretic treatment, or even a simple mean field description is not easily obtained for this cellular automaton describing the evolution of a single field $\sigma$.

In this thesis, we tackle these issues by studying a variant of the depinning model which includes a ``relaxation mechanism'' inspired from the OFCR model (see sec.~\ref{sec:visco_elastic_model}).
The model is defined by continuum evolution equations of a microscopically well defined viscoelastic interface, which has a natural interpretation of its own, and can be studied analytically in the mean field limit.

\subsection{Compression Experiments and the Avalanche Oscillator}
\label{sec:papanikolaou}

In a recent paper \cite{Papanikolaou2012}, a model of depinning with a relaxation mechanism reminiscent of that used in OFCR was applied to the context of crystalline plasticity.
Due to its pseudo-periodic behaviour in a certain regime, this model was named the \textit{avalanche oscillator}.

The avalanche oscillator is built 
by considering a singe slip plane (two dimensional structure), in which the motion or slip is characterized by a single scalar variable $h(x)$ denoting a component of the plastic distortion tensor.
When the locally applied stress $\sigma(x)$ (along the appropriate direction) reaches the random stress barrier to slip $\sigma_{dis}$, the system slips with a rate proportional to the excess stress $\sigma - \sigma{dis}$.
Thanks to dislocation hardening and elastic interactions via the crystalline lattice, the stress decreases during slip, which thus eventually stops.
\begin{figure}[]
\begin{small}
\begin{center}
\def\svgwidth{\textwidth}
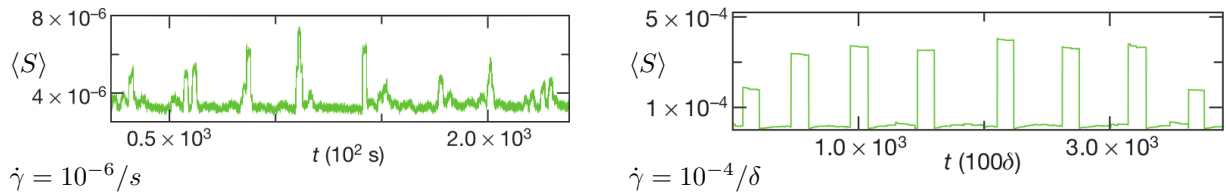
\end{center}
\end{small}
\caption{   {\footnotesize 
Adapted from \cite{Papanikolaou2012}.
Bursts of activity in the slow compression of micro-crystals.
Left: Experimental measurements. The average avalanche size $\langle S \rangle$ is strongly dependent upon the strain rate $\dot{\gamma}$ (for $\langle S \rangle $ the average is taken over small time windows of $400~s$).
At low strain rate of $\dot{\gamma} = 10^ {-6}~s^ {-1}$ the activity is mostly concentrated in short, quasi periodic bursts.
\newline
Right: results from simulations, which very well reproduce the almost periodic behaviour. $\delta$ is the time unit of the simulation.
\label{Fig:Papanikolaou2012}
}   }
\end{figure}
Adapting some of the notations from \cite{Papanikolaou2012} to ours, the evolution for the slip $h$ can be written:
\begin{align}
\eta_0\frac{\partial h }{\partial t} = \eta_0 D \lp \frac{\sigma(x)}{\mu} \rp^ n \Theta(\sigma(x))  +\frac{1}{\mu } \lp \sigma(x) - \sigma_{dis}(x) \rp  \Theta \lp \sigma(x) - \sigma_{dis}(x) \rp, \\
\text{with } \sigma(x) \equiv k_0(V_0 t-h) + k_1 G_{el}(h) 
\label{Eq:papa_phi}
\end{align}
where typically only $n=1$ is used, $\mu$ is the shear modulus, $\Theta$ the Heaviside function, and $\eta_0 D$ plays a role similar to that of the $R$ in OFCR, i.e.~it controls the rate of thermally activated processes responsible for the slow relaxation.
The first term corresponds to a slow relaxation mechanism while the second renders the fact that slip is allowed only beyond a certain stress threshold.
The local applied stress $\sigma(x)$ itself depends on $h$ via:
\begin{align}
\sigma(x) = k_0 V_0 t - k_0 h(x) + \int \d^2 x' G(x-x')h(x'),
\end{align}
where the first term comes from the externally applied stress (uniform), the second from local dislocation hardening, and the last accounts for long-range elastic interactions (see \cite{Papanikolaou2012} for more details on the kernel $G$).

The novelty lies in the relaxation term (first term in the r.h.s.~of \reqq{papa_phi}), which slowly increases the slip in the inter-avalanche periods, but only when the stress is positive\footnote{This constraint is chosen mainly for numerical purposes, it prevents the slip to decrease during relaxation. The authors show that this choice does not qualitatively affects the output of the model.}.
To implement the two competing time scales associated to relaxation (slow process) and avalanches (fast process), the formulation of the model relies on the Heaviside $\Theta$ function (of the second term), which is an elegant reformulation of the cellular automaton presentation.

The numerical integration of this model compares well with the experiments of compression of Nickel micro-crystals, also reported in \cite{Papanikolaou2012}.
In particular, at small strain rates they observe avalanches distributions with a larger exponent $\tau$, accompanied by periodic bursts of very intense activity, as measured in experiments (see Fig.~\ref{Fig:Papanikolaou2012}).
This model (and its results) was inspirational for this thesis: we will come back to it later on.

\section{A Viscoelastic Interface in Disordered Medium}
\label{sec:visco_elastic_model}

\subsection{Physical Motivations for Viscoelastic Interactions}

As we pointed out in the previous chapter, the framework of the depinning transition offers a promising basis for understanding friction and possibly earthquakes, but is far from providing a definitive answer.
Here we use the depinning terminology to discuss the inclusion of additional microscopic effects. 

In chap.~\ref{chap:friction}, we showed that the slow ageing of junctions 
is the microscopic mechanism at the origin of several effects: velocity-weakening, the increase of static friction at rest and more generally the Rate- and state-dependent friction laws (RSF).
The local increase of adherence over time (under constant external constraints) is difficult to characterize precisely, but roughly corresponds to a slow increase of the contact strength  -- at the junctions --  over time.
As this variation only applies to the asperities which are actually in contact, using a time-dependent disorder force 
is not adequate:
it would make the disorder evolve even in the areas not in contact, which may correspond to the nucleation of new contact points (which is not the main mechanism of contact ageing).

A first alternative to strengthening the disorder is to introduce a ``relaxation mechanism'' that slowly weakens the only local force competing with the disorder, i.e.~the elastic interaction.
The importance of local relaxation mechanisms in fault dynamics was pointed out in \cite{Jagla2010, Jagla2010a, Rosso2012} but in these works relaxation was introduced with ad-hoc rules, in a cellular automata fashion (OFCR model).

In this thesis we study the continuum evolution equations of a microscopically well defined mechanical model that allows for a slow relaxation of the interface. 
The local ageing is modelled via an elemental velocity-dependent term (a 	``dashpot''), which naturally accounts for the creep plasticity occurring at the contact points.

\subsubsection{Viscoelasticity: Two Techniques}

\paragraph{Dashpots and Springs}
The notion of \textit{dashpot} is at the core of our viscoelastic model.
Unlike springs which naturally represent elastic\footnote{
In the regime of interest here, the atomic interactions are just deviations from an equilibrium position and the first non-trivial term in the Taylor expansion is the second order (quadratic) term.
}
 interactions inside a solid, dashpots provide an effective representation of the various interactions that are velocity-dependent rather than position-dependent, as e.g.~in liquids.
They represent the simplest form of velocity dependence at the mesoscopic level. 
The force acting on a point of coordinates $h_i$ linked via a dashpot (resp.~a spring) to a point of coordinates $h_{i+1}$ is given by:
\begin{align}
F_\text{dashpot}(h_{i+1}\rightarrow h_i) &= \eta_u (\dot{h}_{i+1}-\dot{h}_i) \equiv \eta_u \frac{\partial}{\partial t}(h_{i+1}-h_i),\\
F_\text{spring} (h_{i+1}\rightarrow h_i)  &= k (h_{i+1}-h_i),
\end{align}
where $\eta_u$ is a constant homogeneous to a viscosity (force/velocity or $kg.s^{-1}$) and $k$ is a stiffness (force/length or $ kg.s^{-2}$).
Combined with springs, dashpots can easily provide a system with a form of memory.

\paragraph{The Memory Kernel Approach}
An alternative to using springs and dashpots is to introduce a memory kernel.
One considers the force deriving from a general kernel $C(t)$ coupling the velocities $\dot{h}\equiv \partial_t h$:
\begin{align}
F_\text{kernel} = \int_0^t C(t-s) \lp \dot{h}_{i+1}(s)-\dot{h}_i(s)  \rp  \d s.
\end{align}
This coupling is non-local in time, so that the ``memory'' aspect of dashpot models is immediately apparent.
The only constraint is that the coupling must have a finite first moment, i.e.~$\int_0^\infty \d s C(s) =\eta_u < \infty$.

One of the most simple choices is the exponential decay: $C(t)=k_2 e^{-t k_2 / \eta_u}$, which is the form used in \cite{Marchetti2000}.
More generally, most simple mechanical models built with springs and dashpots can be expressed via memory kernels $C(t)$.
The reciprocal is false: not all memory kernels can be expressed as simple mechanical models involving only a finite number of springs and dashpots.
In this thesis, we favour the use of mechanical circuits, as they represent the most minimal models possible and give a clear intuition of the microscopic physics in the system.
We presented the memory kernel approach for completeness and to improve the readability of the literature.
We give two examples of memory kernels in the following.

\subsubsection{Solids, Liquids and Viscoelastic Matter}

\paragraph{Elastic Solid}
A simple model of solid elasticity is provided by a simple spring. 
We do not detail this simple yet fundamental case here, since the harmonic oscillator has already been well studied.

\paragraph{Maxwell Fluid}
\label{sec:Maxwell_fluid}
\begin{figure}[]
\begin{small}
\begin{center}
\def\svgwidth{\textwidth}
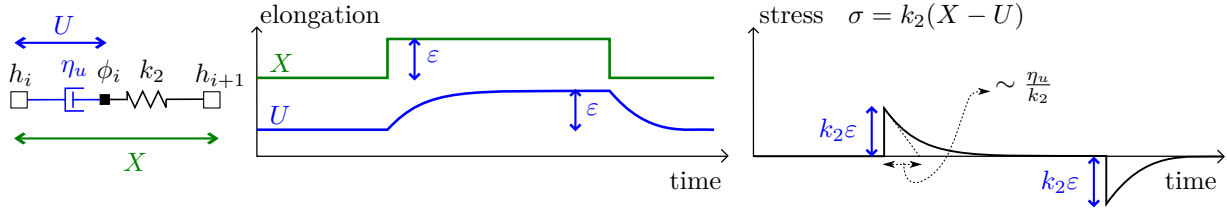
\end{center}
\end{small}
\caption{   {\footnotesize 
Maxwell model for viscoelastic material, or Maxwell fluid.
A spring $k_2$ is in series with a dashpot $\eta_u$ (left). 
Centre: keeping $h_i$ fixed, the position of $h_{i+1}$ is submitted to a step increase by $\varepsilon$, and the (ideal) spring instantly adapts (is elongated by $\varepsilon$). 
On a characteristic time scale $\sim \eta_u/k_2$, the length $U=\phi_i-h_i$ adapts to the strain and the dashpot dissipates the stress stored in the spring $k_2$ (right).
Another step of decreasing elongation $X$ is applied at a later time and results in a similar behaviour (step variation followed by an exponential decay of the stress).
\label{Fig:maxwell}
}   }
\end{figure}
Consider the Maxwell model, a simple model for viscoelastic materials, as depicted in Fig.~\ref{Fig:maxwell} (top).
Denoting $X=h_{i+1}-h_i$ the total length of some material, we study its response to a finite step in the imposed strain $\varepsilon$.
This response can be computed by solving first-order differential equations (see \reqq{NewtonOnPhi}, \reqq{viscoqEW1}, for  the method of derivation), here we provide their solution in Fig.~\ref{Fig:maxwell}.

There is no restoring force in this model: any imposed strain induces a stress which eventually relaxes to zero, with the final configuration losing all memory of its initial state.
Thus, formally, we are modelling a viscous fluid rather than a solid.
We also see in this example that dashpots react as rigid bars to high frequency constraints (frequencies higher than $\sim k_2/\eta_u$) but dissipate the low frequency inputs.
Over a short observation time scale, a Maxwell fluid with small $k_2/\eta_u$ will react essentially as a solid.
An example of material that we may model (at first approximation) using the Maxwell model is honey: the time scale of relaxation is typically $1-10~s$, and strongly depends on temperature.
Note also that a Maxwell element can be elongated indefinitely: since stress can be fully relaxed, the energy needed to impose a finite velocity to the point $h_{i+1}$ grows only linearly with time.
This is actually the case in \cite{Marchetti2000}, where the regime probed is that of finite velocity, or ``steady shear'' (see \ref{sec:Marchetti} and references therein).

The memory kernel $C(t)=k_2 e^{-t k_2 / \eta_u}$ produces Maxwellian dynamics: to prove it, we consider the same step increase of $X(t)\equiv h_{i+1}-h_i$ as in Fig.~\ref{Fig:maxwell}.
This corresponds to a Dirac distribution for the velocity: $\dot{h}_{i+1}(s)-\dot{h}_i(s) = \dot{X}(s)= \varepsilon \delta^D(s-t_0)$.
The kernel response is $F_\text{kernel}= \varepsilon k_2 e^{-(t-t_0)k_2/\eta_u}$ for $t>t_0$ and zero before.
This is exactly the response of the Maxwellian model presented earlier.
As the force $F_\text{kernel}$ is linear in the field $\dot{X}(t)$, we may consider any input $X$ has a sum of step 	functions (Heaviside functions) and the total response will be the sum of the step responses, so that this kernel and the microscopic Maxwell model introduced above yield the same physics.

\paragraph{Standard Linear Solid (SLS)}
\label{sec:plastic_creep_explains_visco}

\begin{figure}[]
\begin{small}
\begin{center}
\def\svgwidth{\textwidth}
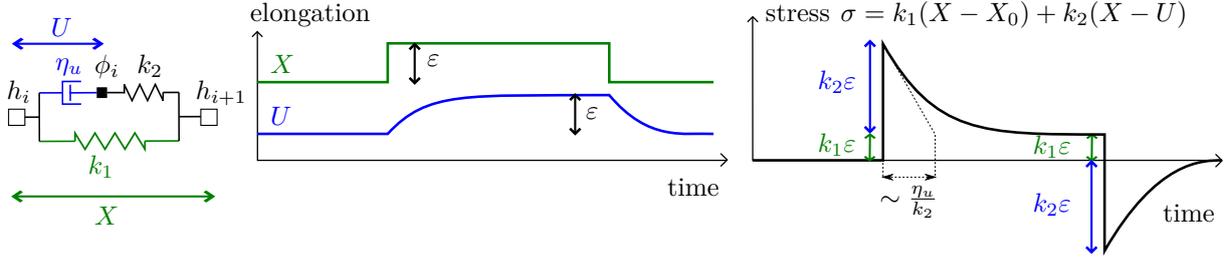
\end{center}
\end{small}
\caption{   {\footnotesize 
Standard Linear Solid model (SLS) for a viscoelastic material.
A spring $k_2$ is in series with a dashpot $\eta_u$ and in parallel with a spring $k_1$ (top).
Middle: keeping $h_i$ fixed and moving $h_{i+1}$ by $\varepsilon$, 
the (ideal) springs $k_1,k_2$ instantly adapt (are elongated by $\varepsilon$).
On a characteristic time scale $\sim \eta_u/k_2$, the length $U=\phi_i-h_i$ adapts to the strain and the dashpot dissipates the stress stored in the spring $k_2$, i.e.~$\sigma_2=k_2(X-U)$.
We note $X_0$ the equilibrium elongation for the spring $k_1$. 
The stress $\sigma_1=k_1(X-X_0)$ contained in $k_1$ cannot be dissipated: it provides the system with a memory of its initial configuration.
Another step of decreasing elongation $X$ is applied at a later time and results in a similar behaviour.
Note that the stress that was stored in $k_1$ is instantly recovered.
\label{Fig:SLS1}
}   }
\end{figure}

As we are interested in modelling a solid, we want to include a restoring force which may bring back the system closer to its original position (under a fixed stress) or which may retain some stress (at imposed strain), for any observation time scale.
This is done in the SLS model, where an additional spring $k_1$ is set in parallel to the Maxwell model: denoting $X=h_{i+1}-h_i$ the total length of an SLS element, we study its response to a finite step in the imposed strain $\varepsilon$ in Fig.~\ref{Fig:SLS1}.

Note that an SLS element always keeps some memory of its initial state: part of the strain imposed ($k_1 \varepsilon$) is never forgotten, thanks to the restoring force induced by $k_1$.
As for the Maxwell material, the system reacts rigidly to high frequency constraints (frequencies higher than $\sim k_2/\eta_u$). 
However, the low frequency inputs are not necessarily fully dissipated.
In particular, an SLS element can not be elongated indefinitely: since stress can not be fully relaxed, the energy needed to impose a finite velocity to the point $h_{i+1}$ scales as the square of the displacement (for large displacements).

To recover the SLS model from a memory kernel, we use the combination: $C(t-s)=k_1 + k_2 e^{-t k_2 / \eta_u}$.
The proof is similar to that of the Maxwell case: we decompose any input strain as a sum of step functions, and observe that their response is that of this memory kernel.
Note that in the mechanical perspective (springs and dashpots), this model is obtained by the addition of a single additional degree of freedom, which is the most minimal choice we can think of.

\paragraph{Origin of Viscoelastic Interactions: Plastic Creep}

Since we use the SLS model as the elemental internal interaction of our interface, we now give some idea about its microscopic origins.
For materials under stress that are close to their limit of elasticity, small plastic events occur, which correspond to local rearrangements of the atoms for amorphous materials, and to dislocations motion for crystalline structures. 
These rearrangements allow to locally release stress while loading the neighbouring region, with an overall decrease of stress due to the greater satisfaction of constraints in the final configuration (and some dissipation). 
Here we only mention a simple activation theory, but the study of (crystalline and amorphous) plasticity is a field in itself \cite{Barrat2011,MiguelBook2006,Miguel2006,LesHouches2002,Zapperi2006,Miguel2008}.

During a rearrangement, the stress constraints are badly satisfied, so that there is an energy barrier that prevents plastic events.
Thanks to thermal fluctuations, this barrier can be overcome, with a rate given by an Arrhenius law:
\begin{align}
Rate \propto  \exp \lp -\frac{E_a}{k_B T} \rp,
\label{Eq:Arrhenius}
\end{align}
where the height of the energy barrier is also called the activation energy, $E_a$.
This $Rate$  corresponds to the relaxation time scale $\eta_u/k_2 = \tau_u$ mentioned above.
This relation between time scales and temperature allows to model \cite{Lubliner2008} 
the increasing deformation occurring under constant stress known as \textit{creep}. 
In this sense, our model of a viscoelastic interface will depend on temperature.
Still, we will use the depinning formalism in its zero temperature limit, as the term $\theta(x,t)$ of the depinning equation does not properly account for this viscoplastic creep.
This kind of macroscopic behaviour is often referred to as viscoplasticity rather than viscoelasticity, however in what follows we will refer to our model as viscoelastic rather than viscoplastic, in order to remember that it results from a combination of viscous and elastic elements.

\subsection{Derivation of the Equations of Motion}

Inspired by some of these previous works and by my collaboration with E.A. Jagla and Alberto Rosso, I designed the model 
that we study in this chapter\footnote{
To be precise I initially designed another model and from it we defined the one presented here.
This other model (with ``Laplacian relaxation'') has a big physical advantage and a big numerical disadvantage compared to the present one, it is discussed in sec.~\ref{App:variant3global}.
}.
Intuitively, our model corresponds to replacing the purely elastic interface of depinning with a viscoelastic one (using the SLS model), thus accounting for some irreversible processes (local plastic events) occurring at the micron scale (we give more details about the microscopical interpretation of viscoelasticity in sec.~\ref{sec:plastic_creep_explains_visco}).
The model we use for viscoelasticity is a quite common and general phenomenological model called the ``Standard Linear Solid'' (SLS).
We now present a few insights about this model.

\paragraph{Continuous Equations of Motion}

\begin{figure}[]
\begin{small}
\begin{center}
\def\svgwidth{\textwidth}
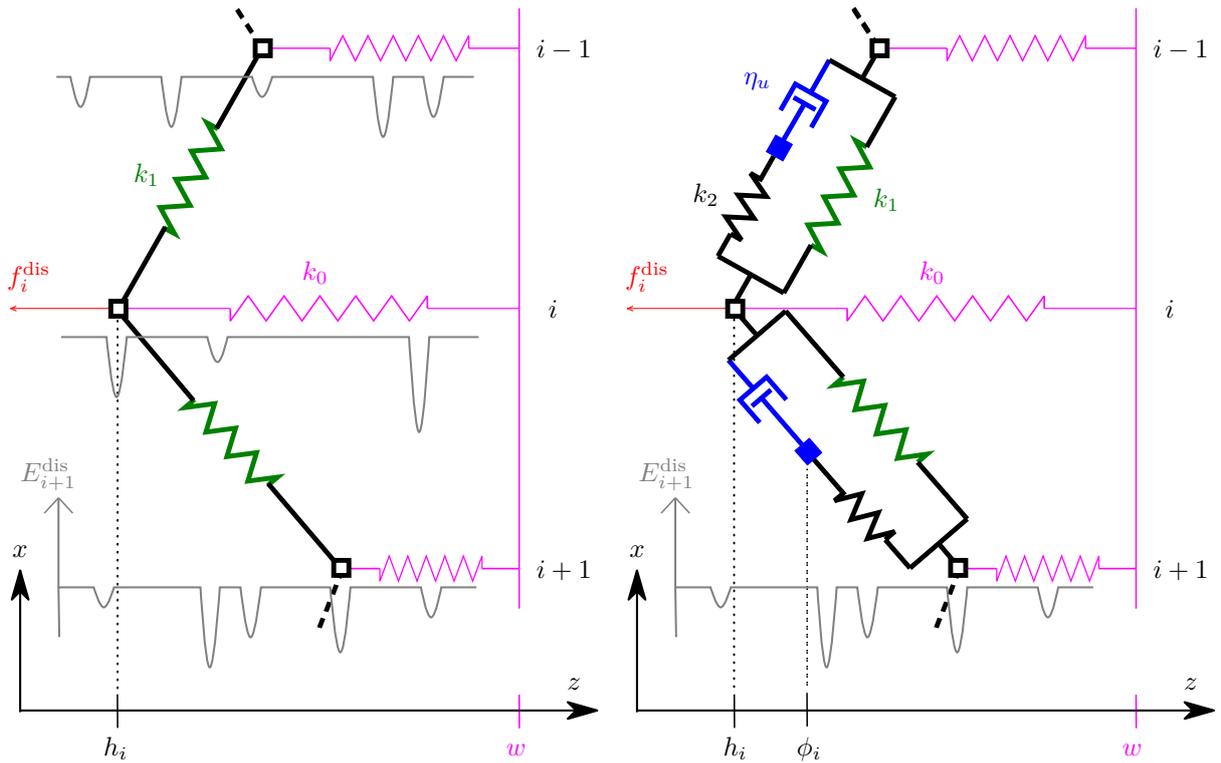
\end{center}
\end{small}
\caption{   {\footnotesize 
Mechanical ``circuit'' or sketch of the one-dimensional elastic interface model (left) and of the viscoelastic model (right).
The interface itself (bold black line) consists in blocks located at discrete sites $i,i+1, \dots$  (empty squares with location $h_i,h_{i+1},\dots $) along the $x$ axis and are bound together via springs $k_1$ in the purely elastic model and a combination of springs ($k_1,k_2$) and a dashpot ($\eta_u$) in the viscoelastic model.
In the viscoelastic model the additional (internal) degree of freedom $\phi_i$ is represented by a full square (blue).
The driving is performed via springs $k_0$ linked to a common position $w$ (thin purple lines).
The disorder force $f^\text{dis}_i$ (red) for the site $i$ derives from a disordered energy potential $E^\text{dis}_i$, which is here simplified as a series of narrow wells separated by random spacings.
The damping (proportional to $\eta_0$) is not pictured.
\label{Fig:depinning3}
}   }
\end{figure}

In the previous chapter, we presented the original depinning model (of a purely elastic interface) in terms of a \textit{mechanical circuit} 
consisting in blocks connected by springs (in Fig.~\ref{Fig:depinning1}).
This kind of definition via a sketch allows for an intuitive extension of the model.
The model for a viscoelastic interface (or ``depinning with relaxation'') we propose is defined by the mechanical circuit of Fig.~\ref{Fig:depinning3}.
We first study the one-dimensional case, as presented in the figure.

The interface is decomposed in blocks of mass $m$, labelled $i$ and moving along horizontal rails $h_i$. 
The action of the dashpot is to resist the change in $\phi_i - h_i$ via viscous friction, with a resulting force on $h_i$ given by $\eta_u  \partial_t (\phi_i-h_i)$.
The blocks move in a medium with some effective viscosity $\eta$ and we study the overdamped regime, $m \partial_t^2 h_i \ll \eta\partial_t h_i$. 
As each block is described by two degrees of freedom $h_i$ and $\phi_i$, the time evolution is governed by two equations.
We now provide a pedestrian derivation of the equations, for the sake of completeness. 
The first equation comes from the force balance on $h_i$:
\begin{align}
\eta \partial_t h_i 
=&  f^\text{dis}\eta[h_i,i] + k_0 (w-h_i) + k_1(h_{i+1} - h_i)  \notag  \\
&+ k_1 (h_{i-1} -h_i) + \eta_u \partial_t(\phi_i -h_i) + k_2 (\phi_{i-1}-h_i)
\label{Eq:viscoqEW1}
\end{align}
The second equation is derived from the force balance on $\phi_i$: 
\begin{align}
0= k_2(h_{i+1} - \phi_i) + \eta_u \partial_t(h_i - \phi_i) 
\label{Eq:NewtonOnPhi}
\end{align}
where we assume that the internal degree of freedom $\phi_i$ has no mass.
Similarly the force balance on $\phi_{i-1}$ yields:
\begin{align}
0= k_2(h_{i} - \phi_{i-1}) + \eta_u \partial_t(h_{i-1} - \phi_{i-1}) .
\label{Eq:NewtonOnPhi-1}
\end{align}
In order to let the Laplacian term $k_2 (h_{i+1}- 2h_i+h_{i-1})$ appear, we introduce the variable 
\begin{align}
u_i \equiv \phi_i - h_i + h_{i-1} - \phi_{i-1},
\end{align}
which represents the elongation of the dashpot elements connected to site $i$.
We inject \req{NewtonOnPhi} into \req{viscoqEW1} to get rid of the time derivatives, and we subtract \req{NewtonOnPhi-1} from \req{NewtonOnPhi} to obtain \req{detaille2}:
\begin{align}
\eta \partial_t h_i &= f^\text{dis}\eta[h_i,i] + k_0 (w-h_i) + (k_1+k_2)(h_{i+1} -2 h_i + h_{i-1} ) 
- k_2 u_i  
\label{Eq:detaille1}\\
\eta_u \partial_t u_i &= k_2 (h_{i+1}- 2h_i+h_{i-1}) - k_2 u_i. 
 \label{Eq:detaille2}
\end{align}
A more elegant notation using the Laplacian operator $\nabla^ 2$ is:
\begin{align}
\eta \partial_t h_i &= f^\text{dis}\eta[h_i,i] + k_0 (w-h_i) + k_1 \nabla^ 2_i h_i  +k_2 (\nabla^ 2_i h_i  -u_i) \notag \\
\eta_u \partial_t u_i &= k_2 (\nabla^ 2_i h_i  -u_i). \label{Eq:viscoDetaille3}
\end{align}

To generalize this to higher dimensions (on a square lattice), one simply has to connect each block $h_i$ to its neighbours via viscoelastic elements, using a single orientation per direction.
The equations obtained are exactly \req{viscoDetaille3} if we reinterpret the label $i$ as referring to $d$-dimensional space, the Laplacian $\nabla^ 2$ as the $d$-dimensional one, and the $u_i$ variable as:
\begin{align}
u_i = \sum_{j=1}^ {d}( \phi_j - h_j) + \sum_{j'=d+1 } ^ {2d} (h_{j'} - \phi_{j'}) \label{Eq:definitionGeneraleDe_u},
\end{align}
where indices $j$ denote the $d$ first neighbours, connected via a dashpot followed by the spring $k_2$ (and $k_1$ in parallel) and indices $j'$ denote the last $d$ neighbours, connected via the spring $k_2$ followed by a dashpot  (and $k_1$ in parallel).

Our viscoelastic interface model 
is described in full generality by the equations \reqq{viscoDetaille3}, so that from now on we forget about the intermediate variables $\phi_i$ and consider the dynamics solely in terms of the principal field $h$ and the auxiliary field $u$.

\paragraph{Narrow Wells}
Using the narrows wells representation for the disorder (see sec.~\ref{sec:narrow_wells}),  we may rewrite \reqq{viscoDetaille3} as a slightly simpler set of evolution equations:
\begin{align}
\eta_0 \partial_t h_i &= k_0 (w-h_i) - f^\text{th}_i  + k_1 \nabla^ 2 h_i  +k_2 (\nabla^ 2 h_i  -u_i)
\label{Eq:hvisco} \\
\eta_u \partial_t u_i &= k_2 (\nabla^ 2 h_i  -u_i) , 
\label{Eq:uvisco}
\end{align}
where the threshold force $f^\text{th}_i$ has some random distribution (e.g.~a Gaussian) and the narrow wells are separated by spacings $z$ with some distribution $g(z)$ with finite average $\overline{z}$.
As previously, we are interested in the case of steady driving, $w=V_0 t$.

\paragraph{Backward Motion}
\label{sec:noBackwardMotion}
In the purely elastic case, there is a Middleton theorem  \cite{middleton1992asymptotic} that guarantees that the interface moves only forward, thanks to the convexity of the Laplacian operator and the monotonicity of the driving ($w=V_0 t$ is an increasing function of time). 
In presence of viscoelastic elements, the term $-k_2u_i$ may decrease the pulling force over time and the Middleton theorem does not apply: backward movements of the interface $h$ are a priori possible. 

Let's study the possibility of backward motion in the narrow wells case.
There should be a backward jump when the sum of the forces on a block exceeds the threshold force to exit the narrow well in the decreasing $z$ direction: assuming that the wells have symmetric shape along the $z$ axis, this threshold is simply $|-f^\text{th}_i|$.
Thus, the interface moves backwards ($\partial_t h < 0 $) when 
\begin{align}
 k_0 (w-h_i) + f^\text{th}_i  + k_1 \nabla^ 2 h_i  +k_2 (\nabla^ 2 h_i  -u_i) < 0,
\label{Eq:hviscoback} 
\end{align}
where the important point is the change of sign in front of $f^\text{th}_i $.
The stability range for the site $i$ is thus given by the condition  $ k_0 (w-h_i) + k_1 \nabla^ 2 h_i  +k_2 (\nabla^ 2 h_i  -u_i) \in [-f^\text{th}_i, f^\text{th}_i]$.
We expect backward movements to be rare, since in general $w-h_i>0$.

To verify this proposition, we perform the following test, using the narrow wells disorder.
We build an algorithm which, after each increase in $w$ (or change in the $u_i$'s), sweeps over all sites twice: during the first sweep, the criterion for backward movements is checked, and backward jumps are performed (in parallel). 
In the second sweep, the criterion for forward movements is checked, and forward jumps are performed (in parallel).
Among the numerous possibilities to implement backward and forward jumps at the same time, this one is the one which favours the backward jumps the most.
Using this algorithm, in all the parameter ranges that we have explored, we have not detected a single backward jump. 
The only exception is when the parameters chosen produce a negative stress (i.e.~$w<\overline{h}$), an unphysical feature that appears in particular for large $\overline{z}$ and small $f^ \text{dis}_i$'s.
We discard this exception as it is unphysical and vanishes in the limit of small $k_0$'s.
The conclusion is that using the narrow wells disorder, accounting for the possibility of backward movements  -- or not --  does not affect the dynamics of the viscoelastic model (at all).

More generally, for any choice of disorder these movements are not frequent, thanks to the biased driving term $k_0(w-h)$: there, we have also observed numerically that the real dynamics yields the same statistical results as the dynamics that allows only forward movements.
In all of the following, we will restrain the dynamics to forward movements.

\paragraph{Discrete Dynamics (Cellular Automaton)}

Similarly to the purely elastic case, within this choice of narrow wells disorder we can reformulate the continuous time dynamics in terms of a cellular automaton with (partly) discrete behaviour.
This is especially practical in numerical simulations.
In a similar spirit as \req{deltai_introduction}, we introduce the local variable $\delta_i$ which represents the amount of additional stress that a site can hold before becoming unstable (its ``remaining stability range''):
\begin{align}
\delta_i \equiv  f^ \text{th}_i  -& k_0(w-h_i) - ( k_1+k_2 )(\nabla^2 h)_i +k_2 u_i . \label{Eq:deltai_visco_2d}
\end{align}
Restricting ourselves to forward motion\footnote{
If we allowed backward motion, it should happen when $\delta_i \geq 2 f^\text{th}_i$, however we discarded such a possibility earlier.
}, the definition of a metastable state $\{w,h_i,u_i,~\forall i\} $ (also denoted $\{ \delta_i,u_i,~\forall i\}$) is to fulfil the stability condition:
\begin{align}
&\delta_i >0 , \qquad \forall i   
\label{Eq:viscostability}
\end{align}
which is reminiscent of the conditions \req{NWstability1} or \req{wstability}.

The quasi-static ($V_0=0^ +$) dynamics is very simple.
\begin{itemize}
\item[1] Increase $w$ until $\exists i / \delta_i \leq 0$.
\item[2] For all sites $i$ with $ \delta_i \leq 0$, draw a $z$ from $g(z)$, increase $h_i$ by $z$ and draw a new threshold $f^ \text{th}_i $ for the site $i$. 
This changes $\delta_i$ and its neighbors $\delta_j$.
Repeat until \req{viscostability} is fulfilled again.
\item[3] The system is in a new metastable state, so relaxation acts via \req{uvisco}.
If at some point during relaxation we have a site $i$ with $\delta_i \leq 0$, go to Step 2.
If this does not happen, i.e.~if we reach the state $\forall i, u_i = \nabla^2 h_i$, then we are in the fully relaxed state, go to step 1.
\end{itemize}

\subsection{A model with Laplacian Relaxation}
\label{App:variant3global}

\begin{figure}[]
\begin{small}
\begin{center}
\def\svgwidth{9cm}
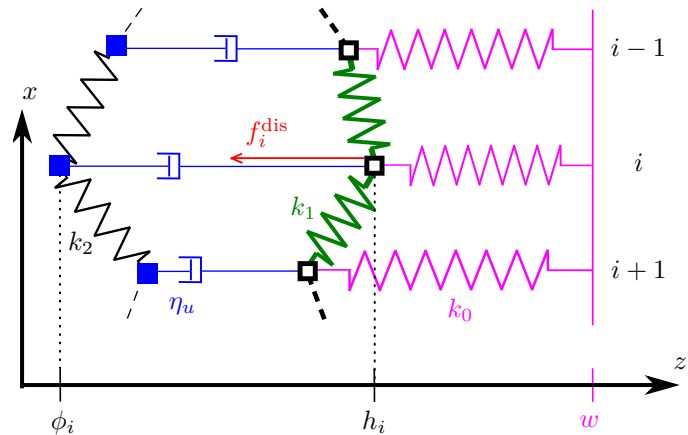
\end{center}
\end{small}
\caption{   {\footnotesize 
Mechanical circuit of the model of viscoelastic interface with Laplacian relaxation, to be compared with the elastic and local relaxation models presented in Fig.~\ref{Fig:depinning3}. 
It can be interpreted as representing the viscous interactions between two elastic surfaces, one of them being pulled and subject to disorder. 
There is no restoring force (spring in parallel to the dashpots), but the whole auxiliary field $\phi$ prevents the distance between the two parts of the ``interface'' ($\phi$ and $h$) from becoming too large, so that we have an effective viscoelasticity similar to that of the SLS model.
\label{Fig:variant3}
}   }
\end{figure}
I also studied a variant of our model with ``long-range'' relaxation (non-local relaxation), inspired by some of the relaxation mechanisms studied in \cite{Jagla2010, Jagla2010a}.
From the mechanical circuit associated to this model (see Fig.~\ref{Fig:variant3}), we derive the equation of motion, which is very much alike \req{viscoDetaille3}:
\begin{align}
\eta \partial_t h_i &= f^\text{dis}\eta[h_i,i] + k_0 (w-h_i) + k_1 \nabla^ 2_i h_i  +k_2 (\nabla^ 2_i h_i  -\nabla^ 2 u_i) \notag \\
\eta_u \partial_t u_i &= k_2 (\nabla^ 2_i h_i  -\nabla^ 2 u_i),
\label{Eq:viscoDetaille3_globalVar3}
\end{align}
where the variable $u_i$ is still the elongation of the dashpot connected to the site $i$: $u_i \equiv \phi_i-h_i$.
Thus the only difference is that the $u_i$'s relaxation now involves $u_i$ and its neighbourhood, via the Laplacian term $\nabla^2 u_i$ (which replaces the simple $u_i$ of \req{viscoDetaille3}).
Hence, we will refer to this model as that with ``Laplacian relaxation'', that we may oppose to the primary model, with ``local relaxation''.

The model with Laplacian relaxation is computationally much more demanding to simulate in finite dimensions than the local model, due to the non-local nature of the relaxation step.
Thus, in what follows we will primarily report results on the first model (with local relaxation), and refer to the second one when relevant differences arise.
In the mean field case, we will see that the two models collapse on a single one, so that all results apply indifferently to both models.

\subsection{Qualitative Dynamics of the Viscoelastic Interface}

The following description of the dynamics is valid in the general case, independently from the form of disorder chosen (narrow wells or not).
It is also qualitatively the same for the model with Laplacian relaxation. 
However it is useful to have the narrow wells approximation in mind, since some things are conceptually simpler in that case.

\paragraph{Three Time Scales}
The relaxation constant $\eta_u$ sets a new time scale:
\begin{align}
\tau_u = \frac{\eta_u}{k_2},
\end{align}
 which is characteristic of the relaxation of the dashpots.
It can be compared with two other time scales:
\begin{itemize}
\item[(i)] $\tau_D = \overline{z}/V_0$, which accounts for the slow increase of the external Drive $w$.
\item[(ii)] $\tau_0 = \eta_0/\max [k_0,k_1,k_2]$, which is the response time of the position $h$ of the blocks, i.e.~the characteristic avalanche duration.
\end{itemize}
Except when explicitly stated otherwise, in what follows we will assume that the three time scales are well separated:
\begin{align}
\tau_0 \ll \tau_u \ll \tau_D.
\end{align}

\paragraph{Avalanche Dynamics (time scale $\tau_0$)}
On the time scale $\tau_0$ of the avalanche duration, we have $ \vert \partial_t u \vert \sim \vert\partial_t h \vert \eta_0/\eta_u \sim  0$:  the dashpots are completely rigid and \reqq{hvisco} is simply the equation for an elastic interface with elasticity $k_1+k_2$, up to the term $-k_2 u_i $ which is constant in time\footnote{
On this time scale, the term $-k_2 u_i $ has the same properties as the tilt $\delta f(x)$ we introduced to prove the Statistical Tilt Symmetry (STS).
}.
We may refer to this abstract elastic interface related to our viscoelastic model as the \textit{rigid} interface.

\begin{figure}[]
\begin{small}
\begin{center}
\def\svgwidth{\textwidth}
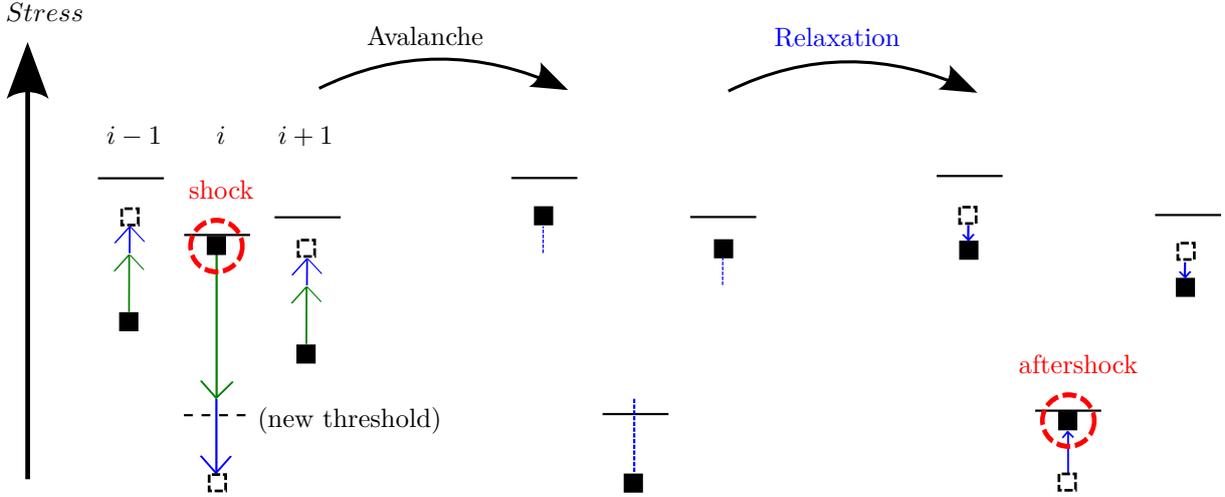
\end{center}
\end{small}
\caption{   {\footnotesize 
Schematic description of the evolution of the local stress $\sigma_i$ over time for three sites.
Left: A simple shock affecting a single site produces a stress drop on this site ($i$).
\newline
Central panel: Part of the stress drop is due to the spring $k_2$ (blue part), i.e.~it is due to the viscoelastic part of the interactions (spring $k_2$ in series with a dashpot).
\newline
Right: During relaxation this part of the stress drop is lost and we may have an aftershock.
If at some point all the  viscoelastic part of interactions (blue) is relaxed, then uniform driving resumes.
\label{Fig:3steps}
}   }
\end{figure}

\begin{figure}[]
\begin{small}
\begin{center}
\def\svgwidth{8cm}
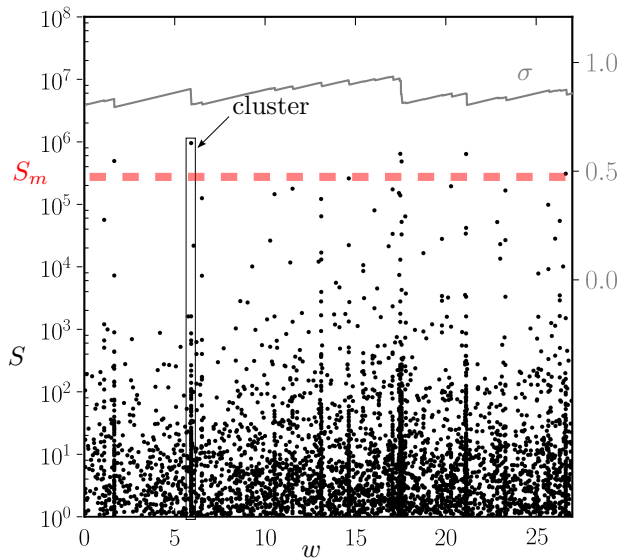
\end{center}
\end{small}
\caption{   {\footnotesize 
Avalanche in the two-dimensional viscoelastic model.
Avalanches sizes $S$ (indicated by dots) are grouped in clusters, with strong correlations over time inside each cluster.
We show the fluctuations of the corresponding stress $\sigma $ (solid grey line) over time ($w=V_0 t$).
\label{Fig:scatter_visco+tau126}
}   }
\end{figure}

\paragraph{Relaxation (time scale $\tau_u$)}
At the end of an avalanche the blocks are pinned and the $h_i$'s are almost constant in time, i.e.~they do not participate in any avalanche (with the narrow wells choice, they are exactly constant).
Thus \reqq{hvisco} cancels on both sides, and \reqq{uvisco} comes into play: on a time scale $ \tau_u \gg \tau_0 $ the $u_i$'s can relax.
As long as the $h_i$ are constant, we have
\begin{align}
u_i(t)=  \nabla^2 h_i+ \lp u_i(t_0) - \nabla^2 h_i \rp e^{-(t-t_0)k_2/\eta_u} , \quad \forall i,
\label{Eq:ui_expo_relax}
\end{align}
where $t_0$ is the time at which the last avalanche occurred. 
The evolution of the $u_i$'s can increase the r.h.s.~of \reqq{hvisco}, so that some blocks may become unstable:  this triggers a secondary avalanche in the system, identified with an aftershock of the seismic context. 
At the end of the aftershock the $h_i$'s are pinned and relaxation resumes, which may trigger an additional aftershock (see Fig.~\ref{Fig:3steps}), itself followed by another one, and so on.
These aftershocks occur without any additional driving: the ensemble of events that occur at a given value of $w$ will be called a \textit{cluster} of events (see Fig.~\ref{Fig:scatter_visco+tau126}).

\paragraph{Driving (time scale $\tau_D$)}
Aside from triggering numerous aftershocks, the effect of relaxation is to suppress the term $k_2 (\nabla^2 h _i- u_i)$ in \reqq{hvisco}.
When finally we have $u_i= \nabla^2 h_i , \forall i$, \reqq{uvisco} cancels on both sides and we say that the system is {\em fully relaxed}.
New instabilities can only be triggered by an increase of $w$, which happens on the slow time scale $\sim \tau_D$.

Note that by definition, when the system is fully relaxed, \req{hvisco} is fulfilled with its last term being exactly zero, i.e.~the configuration is that of an elastic interface with elasticity $k_1$.
We may refer to this abstract elastic interface related to our viscoelastic model as the \textit{flexible} interface.

\paragraph{Summary of the Continuous Dynamics}
Essentially, we have ``main'' avalanches lasting for a time $\sim \tau_0 $ that are triggered by the increase of the drive $w$ (through $k_0$), whereas relaxation (via $k_2, \eta_u$) triggers additional events:
the aftershocks. 
The typical inter-aftershocks time span is expected to be of order $\sim \tau_u$ and the typical inter-main shocks time span of order $\sim \tau_D$.
Since driving is much slower than relaxation, the main avalanches occur only when the interface is fully relaxed, i.e.~when its effective elasticity is $\sim k_1$ (flexible interface).
During any event (main or aftershock), the fast dynamics dominates and the interface evolves essentially as an elastic one with elasticity $k_1+k_2$ (rigid interface).
Thus the viscoelastic interface is expected to evolve between its corresponding rigid elastic interface and the flexible one.

\section{Mean Field: the Fokker-Planck Approach}

The core motivation for studying the mean field dynamics is to get a clear understanding of the phenomenology of our model with the help of analytical and semi-analytical\footnote{By semi-analytical we mean results obtained by the exact numerical integration of some exact equations.} results, which are only available in this simplified case.
Since the mean field is also an approximation for long-range interactions, we may also expect some results to be similar with observations.

The main feature of the mean field model is the existence of periodic oscillations of the average stress that are characteristic of macroscopic stick-slip motion, with large stress drops corresponding to system-size avalanches, independently of finite-size effects.

\subsection{Derivation of the Mean Field Equations}
\label{sec:derivationMF_equations}

Contrary what happened in the case of the purely elastic interface, the ABBM picture does not apply here: since each site has two degrees of freedom $h_i$ and $u_i$ with complex interactions, mapping the complete system state onto a single particle is impossible.
Using a bit more information than a single scalar variable, some things are however possible: a version of the ABBM model for a single particle ``with retardation'' \cite{Zapperi2005} was recently studied in depth \cite{Dobrinevski2013}, displaying aftershocks but no oscillatory behaviour.
In that model, the collective memory of the system is represented via a single kernel interacting with the single particle (representing the interface's center of mass).

Here we want to let each site have its own additional degree of freedom.
To do this, we map the complete system state $\{w, h_i, f^\text{th}_i,u_i, (\forall i)\}$ to the probability distribution $P(\delta^F,\delta^R)$ of a couple of simpler and local variables $(\delta^F_i,\delta^R_i)=\mathcal{F}(w, h_i,u_i, \overline{h}, \overline{u})$.
After writing the dynamics in terms of $\delta$'s, we can derive a simple system of equations for the distribution $P$ (the Fokker-Planck Equations).
We then manage to integrate this system in two relevant limits, which allows to make a few quantitative predictions.

\subsubsection{Definitions}

\paragraph{Fully Connected Model  --  Continuous Equations}
We study the mean field limit via the fully connected approximation: each block position $h_i$ interacts equally with all other blocks $h_j$ via $N-1$ elements of the SLS type.
In Appendix. \ref{App:derivationMF_equations}, we present a pedestrian derivation of the equations directly from this mechanical picture.
As is usually found in mean field models, the equations simply correspond to formally replacing the Laplacian term $\nabla^ 2 h$ with $\overline{h}-h$. 
This reads:
\begin{align}
\eta_0 \partial_t h_i &= f^\text{dis}\eta[h_i,i] + k_0 (w-h_i) +  k_1 (\overline{h}-h_i)+  k_2 (\overline{h}-h_i) - k_2 u_i \notag \\
\eta_u \partial_t u_i &= k_2 (\overline{h}-h_i) -k_2 u_i .
\label{Eq:continuous_MF}
\end{align}
It is worth to notice that both the local and Laplacian relaxation models have exactly the same mean field equations, since $\overline{u}=\text{const.}=0$ (which implies that $\nabla^ 2 u_i$ and $u_i$ both reduce to the simple term $u_i$ in the mean field limit).

\paragraph{Definition of the $\delta$'s  --  Identical Wells}
Using the narrow wells choice of disorder, the expression for $\delta$ (the amount of additional stress that a site can hold before becoming unstable) reads:
$\delta_i = f^\text{th}_i - k_0 (w-h_i) -  k_1 (\overline{h}-h_i) -  k_2 (\overline{h}-h_i) + k_2 u_i$. 
We make a crucial simplification in assuming that all wells are identical, so that we have $f^\text{th}_i  = f^\text{th} = \text{const.}$, so that the definition of the $\delta_i$ reads:
 \begin{align}
\delta_i \equiv f^\text{th} - k_0 (w-h_i) -  k_1 (\overline{h}-h_i) -  k_2 (\overline{h}-h_i) + k_2 u_i
\end{align}
where the randomness remains in the spacings between wells.
Under this assumption, we no longer observe aftershocks in mean field (see Appendix \ref{App:no_AS_in_MF}),
 but otherwise the events statistics is the same.
It is then useful to split $\delta$ in a {\em fast} part, $\delta^F$, and a {\em relaxed} one, $\delta^R$: 
\begin{align}
 \delta^F _i &= f^{\text{th}} - k_0 (w-h_i) - (k_1 + k_2) (\overline{h}-h_i)\notag  \\
  \delta^R _i &=   k_2 u_i,
  \label{Eq:defdeltaF} 
\end{align}
so that $  \delta_i = \delta^F _i + \delta^R_i $.
This splitting of $\delta$ in two variables is crucial in our analysis, it allows to rewrite the entire dynamics solely in terms of the instantaneous values of these variables.

\paragraph{Infinite Size Limit: the $P(\delta^F,\delta^R)$ Distribution}
As for the purely elastic case, we consider the thermodynamic limit $N\to \infty$, where fluctuations vanish and the description of the system via a simple probability distribution becomes exact.
The only difference here is that the sole distribution $P(\delta)$ does not provide enough information to fully characterize the system and its evolution.

Instead, we have to consider the joint probability density distribution $P( \delta^F, \delta^R)$. 
The quantity $P( \delta^F, \delta^R) \d \delta ^ F \d \delta^ R$ represents the probability for a site drawn at random to have a particular set of $(\delta^F, \delta^R)$.
The normalization of $P$ writes:
\begin{align}
\int_\mathbb{R} \d \delta^ F \int_\mathbb{R} \d \delta^R \ P(\delta^F,\delta^R) =1.
\label{Eq:normalizationP2}
\end{align}
We also have the marginal distributions:
\begin{align}
P_F(\delta^ F) &=  \int_\mathbb{R} \d \delta^ R  P(\delta^F,\delta^R) \\
P_R(\delta^ R) &=  \int_\mathbb{R} \d \delta^ F  P(\delta^F,\delta^R) \notag \\
P_\delta (\delta ) &= \int_\mathbb{R} \d \delta^ F \int_\mathbb{R} \d \delta^R  P(\delta^F,\delta^R)  \delta^ {Dirac} (\delta^ F+\delta^R-\delta). \label{Eq:Pdelta_delta}
\end{align}
Our aim is now to translate the evolution equations for $h,u$ or $\delta^F,\delta^R$ into evolution equations for the distribution $P(\delta^F, \delta^R)$ (i.e.~a Master Equation or loosely speaking an equation of the Fokker-Planck type).

\subsubsection{Dynamical Equations for the Distribution $P(\delta^F,\delta^R)$}

Under a small increase of $ \d w $ (on a time scale $\sim \tau_D$), two dynamical regimes are observed: a fast one where avalanches unfold (on a time scale $\sim \tau_0$), and a slow one where the dashpots relax (on a time scale $\sim \tau_u$).

\paragraph{Avalanche Dynamics (time scale $\tau_0$, Fast Part)}

On a short time scale  $(t\simeq \tau_0)$, the dashpots are blocked and only $\delta^ F$ evolves: the avalanche dynamics is very similar to that of the purely elastic interface.
As previously, it is useful to 
 decompose the avalanche in different steps. 
In Appendix \ref{App:MF_fastPart} we provide all the details on how to translate the arguments of sec.~\ref{sec:elastic_depinning_dynamics_Mf_hardcore} 
 to the present case.
Here, we outline the principal results.

The main point of the analysis is to notice that $\int P(-\delta^ R, \delta^ R) \d \delta^R $ plays a role very similar to that of ``$P(0)$'' in the elastic case.
Yet we cannot replace $\int P(-\delta^ R, \delta^ R) \d \delta^R$ with $P_\delta(0)$ everywhere, because the avalanche dynamics reacts differently to each value of $\delta^R$.
Decomposing an avalanche in ``steps'', we find the following evolution of $P(\delta^ F, \delta^ R)$ along the steps:
\begin{align}
\frac{P_{{\rm step} k+1}(\delta^F,\delta^R) - P_{{\rm step} k} (\delta^F,\delta^R)}{\Delta \delta^F_{{\rm step} k}}
&=   \frac{\partial P_{{\rm step} k}}{\partial \delta^ F}(\delta^F,\delta^R)  +  P_{{\rm step} k}(-\delta^ R, \delta^ R) \frac{ g\left( \frac{\delta^ F+\delta^ R}{k_0+k_1+k_2}\right) }{ k_0+k_1+k_2},
 \label{Eq:Pstepk=sVISCO_main}
\end{align}
with driving steps given by the geometrical series:
\begin{align}
\Delta \delta^ F _{{\rm step} k} = k_0 \d w \prod_{j=0}^{k-1} \lp \overline{z}(k_1+k_2) \int P_{{\rm step} j} (-\delta^ R, \delta^ R) \d \delta^R  \rp ,
\label{Eq:DeltadeltakVISCO_main}
\end{align}
where we identify $P_{{\rm step} 0} $ with $P_w$, the distribution before $w$ is increased by $\d w $.
We will discuss the nature of this series (convergent or divergent) in the next subsection.
For now it is enough to assume that it eventually converges to zero.
The total drive occurring during an avalanche is the sum of the initial drive $k_0\d w$ and of the additional drives during the step.
We denote $\Delta^F_{drive}$ this sum: $\Delta^F_{drive} \equiv \sum_k \Delta \delta^ F _{{\rm step} k} $.

To summarize, during an avalanche the interface evolves according to \reqq{Pstepk=sVISCO_main}, \reqq{DeltadeltakVISCO_main} until $\Delta \delta^ F_{{\rm step} j} \approx 0$.
If the r.h.s~of  \req{Pstepk=sVISCO_main} reaches zero everywhere, the distribution $P(\delta^F,\delta^R)$ ceases to evolve, but the interface keeps going forward (until $\Delta \delta^ F_{{\rm step} j} \approx 0$).
This corresponds to the evolution of the corresponding rigid elastic interface (with elasticity $k_1+k_2$). 

Depending on the value of $P_\delta(0)$, the avalanches are expected either to consist in a finite number of steps when the series \req{DeltadeltakVISCO_main} converges, or to be ``infinite'' when the driving steps $\Delta \delta^ F_{{\rm step} k}$ are diverging.

\paragraph{Relaxation (time scale $\tau_u$, Slow Part)}

On longer time scales $(t\simeq \tau_u)$, the dashpots relax: the $\delta^R_i$'s slowly evolve, and if a $\delta_i=\delta^F+\delta^R$ becomes smaller than zero, this triggers a new fast event (aftershock).
The relaxation equation \reqq{uvisco}, $\eta_u \partial_t u_i = k_2 (\nabla^ 2 h_i  -u_i)$, can be rewritten  in terms of $\delta$'s by inverting\footnote{Taking the average of \req{defdeltaF} we get $\overline{\delta^ F} =f^\text{th} -k_0(w-\overline{h}) $. 
Computing the difference  $\overline{\delta^ F}- \delta^ F$ is then easy.} the equations \reqq{defdeltaF}:
\begin{align}
\frac{\eta_u}{k_2} \partial_t \delta^R_i 
&= -\delta^R_i  + k_2  (\overline{h}-h_i)  \notag \\
&= -\delta^R_i  + k_2   \frac{\overline{\delta^F}-\delta_i^F }{k_0+k_1+k_2}.
\label{Eq:relax1}
\end{align}
We note that $\frac{\eta_u}{k_2}  \partial_t \overline{\delta^R} = - \overline{\delta^R}$, so that from any initial condition we end up with $\overline{\delta^R}=0$, and thus $\overline{\delta} \equiv \overline{\delta^F}$.
Assuming that all $\delta^ F_i$'s stay constant (thus $\overline{\delta}$ also is constant), \reqq{relax1} predicts an exponential relaxation of each $\delta^ R_i$ towards:
\begin{align}
\delta^R_{i,\infty} =k_2   \frac{\overline{\delta}-\delta_i^F }{k_0+k_1+k_2},
 \label{Eq:fullyRelaxed}
\end{align} 
where the index $\infty$ denotes the long-time nature of the solution.

This relaxation can decrease $\delta^R$, which might trigger aftershocks if the total $\delta$ is to reach zero.
However this is never the case, due to the simplifying assumption of identical wells ($f^\text{th}_i = f^\text{th} = \text{const.}$).
We prove this result in Appendix \ref{App:no_AS_in_MF}.
The absence of aftershocks allows all the blocks to fully relax after each event, so that the system's state just before any event is always fully relaxed (i.e.~$u_i = \overline h - h_i$, or $\delta^R_{i} =\delta^R_{i,\infty} $).

\subsection{Analytical Integration of the FP Equations}

Using the shorthand $ P_{\delta, {\rm step} j}(0) \equiv  \int P_{{\rm step} j} (-\delta^ R, \delta^ R) \d \delta^R $, we rewrite \reqq{DeltadeltakVISCO_main}:
\begin{align}
\Delta \delta^ F _{{\rm step} k} = k_0 \d w \prod_{j=0}^{k-1} \lp \overline{z}(k_1+k_2) P_{\delta, {\rm step} j}(0) \rp ,
\label{Eq:DeltadeltakVISCO_main2}
\end{align}
and we notice that the convergence of the series to zero is guaranteed if 
\begin{align}
 P_{\delta, {\rm step} j}(0) <  P_{\delta}^c(0) \equiv  \frac{1}{\overline{z}(k_1+k_2)}, \forall j .
 \label{Eq:small_avalanche_condition}
\end{align}
This condition is the \textit{small avalanche condition}.
We now study the two kind of avalanches that arise from this condition: the small ones in the convergent case and the ``infinite'' ones in the divergent case.

\subsubsection{Case of the Convergent Series (Small Avalanches)}

Let's assume that we have $P_{\delta, {\rm step} j}(0) < P_{\delta}^c(0) , \forall j $.
Strictly speaking, the series may not reach zero in any finite number of steps, however we can impose a lower cutoff for the fraction of jumping sites, in which case we have $\Delta \delta^F_{{\rm step} j} \approx 0$ in a finite number of steps (note that in any system of finite size the smallest non zero fraction is $1/N$).
Since we have a prefactor $\d w$, any such cutoff, as small as it is, can be reached in a finite number of steps by choosing a sufficiently small $\d w$. 

Avalanches in this regime involve an infinitesimal total drive $\Delta^F_{drive} \equiv \sum_k \Delta \delta^ F _{{\rm step} k} $, which is proportional to $ \d w$.
The fraction of the sites involved in the avalanche is also infinitesimal.
For a large but finite system ($N<\infty$), this corresponds to a finite avalanche, involving a finite number of blocks, negligible when compared to the system size.

\paragraph{Fully Relaxed State}
Just before the avalanche, the system is fully relaxed because relaxation occurs much faster than driving.
Since the avalanche only involves a finite number of steps and an infinitesimal fraction of the system, the corresponding change in the overall distribution $P$ is also infinitesimal, and we may consider that the system is always fully relaxed during this kind of avalanche.

When the system is fully relaxed, each value of $\delta^F$ is associated to a single value of $\delta^R$, which is $\delta^R_{i,\infty}$: 
thus the distribution $P(\delta^F,\delta^R)$ is non-zero only  on a single line of the $(\delta^F,\delta^R)$ plane and we can re-write it simply in terms of the marginal distribution $P_F(\delta^ F)$:
\begin{align}
P_F(\delta^ F) = P\lp \delta^F, k_2   \frac{\overline{\delta^F}-\delta_i^F }{k_0+k_1+k_2} \rp.
\label{Eq:Pf_suffisant}
\end{align}

In particular, the blocks that jump are those for which $\delta=0$, i.e.~those for which
$
\delta^F_i = -\delta^R_i = -\delta^R_{i,\infty} = -k_2   \frac{\overline{\delta^F}-\delta_i^F }{k_0+k_1+k_2}.
$ 
This corresponds to a single value of $\delta^F$, that we denote\footnote{$\delta^*$ is also the value of $\delta^R$ at which the blocks jump, hence the notation.} $-\delta^ *$:
\begin{align}
\delta^* = \frac{k_2 \overline{\delta^F}}{k_0+k_1}.
\label{Eq:deltaF_jump}
\end{align}
To conclude, at all steps of the avalanche the jumping sites are exactly those with $\delta^F=-\delta^*$, and the variable $\delta^R_i$ can be replaced with $k_2   \frac{\overline{\delta^F}-\delta_i^F }{k_0+k_1+k_2}.$

\paragraph{The Simplified Equation and its Solution}
The distribution $P$ evolves according to \reqq{Pstepk=sVISCO_main}, which drives its r.h.s.~towards zero.
The corresponding (attractive) fixed point is given by:
\begin{align}
 \frac{\partial P_{{\rm step} k}}{\partial \delta^ F}(\delta^F,\delta^R)  +  P_{{\rm step} k}(-\delta^ R, \delta^ R) \frac{ g\left( \frac{\delta^ F+\delta^ R}{k_0+k_1+k_2}\right) }{ k_0+k_1+k_2} 
=0.
\end{align}
Using \reqq{Pf_suffisant}, \reqq{deltaF_jump}, this simplifies into:
\begin{align}
 \frac{\partial P_F}{\partial \delta^F} 
+   \frac{P_F(-\delta^*) }{k_0+k_1+k_2} 
 g\left( \frac{\delta^F + \delta^{*} }{k_0 +k_1+k_2}\right) 
=0.
 \label{Eq:PdeltaF=0?}
\end{align}
Similarly to the elastic case, we use the normalization condition for $P_F$ and find:
\begin{align}
P_F^*(\delta^F) = \frac{1- \displaystyle\int_{-\delta^*}^{\frac{\delta^F+\delta^*}{k_0+k_1+k_2}}  g(z) \d z}{\overline{z}(k_0+k_1+k_2)}.
\end{align}
This exact expression is strongly reminiscent of the fixed point we found in the elastic case. 
Translating this expression into an expression for the more intuitive quantity $P_\delta(\delta)$, we find the same fixed point as for the ``flexible'' elastic interface (with elasticity $k_1$):
\begin{align}
P_\delta^*(\delta) 
= \frac{1- \displaystyle\int_{0}^{\frac{\delta}{k_0+k_1}}  g(z) \d z}{\overline{z}(k_0+k_1)} 
\equiv Q(\delta, k_1)
\label{Eq:Qk1}
\end{align}
The average stress associated to $Q(\delta, k_1)$ can be computed directly with an integration by parts (as in  \req{sigma_mean_general1}):
\begin{align}
\sigma 
&= f^ {th} - (k_0+k_1) \frac{\overline{z^ 2}}{2\overline{z}}.
\label{sigma_mean_general2}
\end{align}
Note that this fixed point $P_\delta^*(\delta) $ is not reached within a single avalanche; instead the distribution $P_\delta(\delta)$ slowly evolves towards it over many cycles of avalanches followed by relaxation.
A direct integration of Eqs. (\ref{Eq:Pstepk=sVISCO_main}, \ref{Eq:relax1}) confirms that the distribution $P$ is indeed driven towards $Q(\delta, k_1)$ (see sec.~\ref{sec:numerical_integration_visco}, \pp{sec:numerical_integration_visco}).

We notice that the fixed point has $P_\delta(0)= 1/\overline{z}(k_0+k_1)$ (for any distribution $g(z)$).
Thus, if $1/\overline{z}(k_0+k_1) \geq P_{\delta}^c(0)  =1/\overline{z}(k_1+k_2)$, we expect that on the way to the fixed point $Q(\delta,k_1)$ of the ``flexible'' interface, the small avalanche condition \reqq{small_avalanche_condition} is violated and  an avalanche with diverging steps $\Delta \delta$ is triggered.

\subsubsection{Case of the Divergent Series}

\paragraph{``Infinite'' Avalanches}

When the small avalanche condition \reqq{small_avalanche_condition} is violated for numerous steps during an avalanche, the magnitude of the driving steps $\Delta \delta^ F_{{\rm step} j}$ becomes larger and larger.
However this growth cannot last forever: the blocks which jump correspond to new $\delta$'s jumping from $0$ to an average value of $\overline{z}(k_0+k_1+k_2)$.
Because the corresponding drive (from $\Delta \delta^F_{{\rm step}  k}$) is  $\sim \overline{z}(k_1+k_2)$, the dissipation\footnote{
Another way to understand this is to notice that sites jump on average from $0$ to $\delta = \overline{z}(k_0+k_1+k_2)$. 
Consider the best scenario for producing an infinite avalanche, when the stationary $P_\delta(\delta)$ is a rectangular function (its the function that decreases the least).
Because of the normalization condition $\int P_\delta = 1$, we cannot hope for anything better than $P_\delta(0) = 1/\overline{z}(k_0+k_1+k_2)$, i.e.~the condition  $P_{\delta, {\rm step} j}(0) \geq  1/(\overline{z}(k_1+k_2))$ cannot be sustained indefinitely.} due to $k_0>0$ prevents the occurrence of any truly infinite avalanche.
After a finite driving from the growing shifts $\Delta \delta^ F_{{\rm step} j}$, these will eventually decrease, converge to zero, and the avalanche will stop.

However, the fact that the total drive ($\sum_j \Delta \delta^ F_{{\rm step} j}$)  is finite (instead of infinitesimal) corresponds to an avalanche involving a finite fraction of the system, or possibly the complete system.
We call this kind of avalanche a \textit{global event}, because it affects the entire system.

\paragraph{Convergence to a Depinning Fixed Point (Fast Part of the Dynamics)}
For a small enough dissipation $k_0$, since there are many steps in this single event, the distribution $P$ actually reaches its fixed point, i.e.~it fulfils: 
\begin{align}
 \frac{\partial P_{{\rm step} k}}{\partial \delta^ F}(\delta^F,\delta^R)  +  P_{{\rm step} k}(-\delta^ R, \delta^ R) \frac{ g\left( \frac{\delta^ F+\delta^ R}{k_0+k_1+k_2}\right) }{ k_0+k_1+k_2} 
=0,
\end{align}
where the $\delta^ R$ are \textit{not} in the fully relaxed state, since they are constant during an avalanche.
We can formally integrate this equation separately for each value of $\delta^ R$, then sum the solutions to get the intuitive distribution $P_\delta(\delta)$. 
We find the fixed point:
\begin{align}
P_* (\delta)   
= \frac{  1 - \displaystyle\int_{0}^{\frac{\delta}{k_0+k_1+k_2}}  g(z) \d z }{\overline{z} (k_0+k_1+k_2)} 
\equiv Q(\delta,k_1+k_2) ,
\label{Eq:Qk1k2}
\end{align}
which is exactly the fixed point of the ``rigid'' elastic interface (with elasticity $k_1+k_2$).
This can also be understood intuitively by remarking that on the short time scale of the avalanche, the dashpots are blocked (they act as rigid bars) so that we just have two springs $k_1, k_2$ acting in parallel.
This corresponds to an elastic interface of stiffness $k_1+k_2$ under a constant ``tilt'' $-k_2 u$.
When proving the STS relation (sec.~\ref{sec:STS}, \pp{sec:STS}), we have seen that the interface, with or without the tilt, had statistically the same evolution equation, so that the convergence to $Q(\delta,k_1+k_2) $ is to be expected.

We note that $P_* (0) = 1/\overline{z} (k_0+k_1+k_2)$ fulfils \reqq{small_avalanche_condition}, which is consistent with our initial hypothesis of a finite avalanche.
The average stress associated to $Q(\delta, k_1+k_2)$ can also be computed directly (integration by parts, as in  \req{sigma_mean_general1}): $\sigma = f^ {th} - (k_0+k_1+k_2) \frac{\overline{z^ 2}}{2\overline{z}}$.

During an event, the $\delta^ R_i$'s are constant.
After this very large event, the $\delta^ R_i$'s are thus very far from their new associated values $\delta^ R_{i,\infty}$.
Thus, the effect of relaxation is a macroscopic change in $P$, that is difficult to compute directly without additional hypotheses.
The qualitative effect of relaxation on this final distribution is presented in the next section, via the numerical integration of Eqs. (\ref{Eq:Pstepk=sVISCO_main}, \ref{Eq:relax1}).

\subsubsection{Comparison of the Two Cases  --  Predictions}

The two cases we presented above may seem contradictory at first.
They key element is to understand that while the fast dynamics of the large events drives the distribution $P_\delta(\delta)$ towards $ Q(\delta,k_1+k_2)$, the mixed dynamics of the small events drives it towards $ Q(\delta,k_1)$.

When the avalanches are ``small'' (i.e.~infinitesimal),
the duet of the fast dynamics and relaxation drives $P_\delta(\delta)$ towards $ Q(\delta,k_1)$.
On the way to this fixed point, the distribution may violate the small avalanche condition \reqq{small_avalanche_condition}, thus triggering a global event.
The condition to obtain a global event is $1/\overline{z}(k_0+k_1) \geq  P_\delta^c(0)=1/\overline{z}(k_1+k_2)$, which simplifies into:
\begin{align}
k_2 \geq k_0,
\end{align}
independently of all other choices (as $g(z), k_1$, etc).

If $k_2 \geq k_0$, when this large event occurs, the fast dynamics works for long enough (at least until the small avalanche condition \reqq{small_avalanche_condition} is again respected), and the fixed point $Q(\delta,k_1+k_2)$ can be reached.
After this, relaxation also produces a large change in $P$, and small avalanches follow.
The function $P$ thus follows a periodic cycle.

If $k_2 < k_0$, there are no global events and the distribution $P_\delta(\delta)$ simply reaches $Q(\delta,k_1)$, irrespective of the precise value of $k_2$.
The dynamics of $P$ is then perfectly stationary.

\subsection{Numerical Integration and Simulations in Mean Field}

\label{sec:numerical_integration_visco}

\subsubsection{{Numerical Integration of the FP Equations}}

\paragraph{The Algorithm}  
Analogously to the purely elastic case, we discretize $P(\delta^F, \delta^R)$ with a bin $\varepsilon$.
The distribution probability is then a matrix $P_{i,j}$ where we identify $P(\delta^F = \varepsilon i, \delta^R = \varepsilon j) \d \delta^F  \d \delta^R \equiv P_{i,j} \varepsilon^ 2$. 
The matrix evolves with the following rules:
\begin{itemize}

\item {\em Driving process}:\\
We shift $P_{i,j}$ of one bin: 
 $P_{i,j} \leftarrow P_{i+1,j} $.\\
\noindent Then we perform the {\em Instability check}.
\item {\em Instability check}: \\
We compute $P_0=\sum_{i=-j} P_{i,j}$.
If $ P_0 \geq 1/ \overline{z} (k_1+k_2) $, we perform the  {\em Driving process}.\\
Else, we compute the total weight of unstable sites:
\begin{align}
P_{\text{inst}} =  \varepsilon  \sum_{(i+j) < 0} P_{i,j}
\end{align}
If $P_{\text{inst}} > \frac{\varepsilon}{\overline{z} (k_1+k_2)} \frac{1}{100} $, then we perform the {\em Avalanche process}. \\
Else we perform the {\em Relaxation process}.

\item {\em Avalanche process}: it is composed by ``jumping sites'' and a ``driving step''.
\begin{itemize}

\item Jumping sites: $\forall (i,j)$,
\begin{align}
&\text{if $i+j\geq0$ :}&
&P_{i,j} \leftarrow  P_{i,j} + \frac{\varepsilon}{\kappa} \left( \sum_{i'|(i'+j<0)} P_{i',j}\right)     g\left(\frac{\varepsilon (i + j)}{\kappa}\right) & \\
&\text{if $i+j<0$ :}&
&P_{i,j} \leftarrow 0,  &
\end{align}
where $\kappa=k_0+k_1+k_2$.

\item Driving step ($\Delta \delta^F_{{\rm step} }$): we shift $P_{i,j}$ of a fraction of bin:
$r = \min (1, \frac{\overline{z} (k_1+k_2) P_{\text{inst}}}{\varepsilon})$,
\begin{align}
P_{i,j} \leftarrow  P_{i,j} +  \left( P_{i+1,j}-P_{i,j} \right) r  
\end{align}
\end{itemize}
\noindent Then we perform the {\em Instability check}.

\item {\em Relaxation process}:\\
We compute $j_{\infty}(i) $, the  single bin associated to $\delta^R_{i,\infty}= j_{\infty}(i) \varepsilon$ as\footnote{It is numerically more stable to associate $\delta^R(i,\infty)$ with two bins, $j_{\infty}(i)$ and $j_{\infty}(i)+1$. The contribution  $\sum _j P_{i,j}$ is split in the two bins using a linear interpolation.}
\begin{align}
j_{\infty}(i)=  \text{Int} \left(  k_2  \frac{- i +  \sum_{i',j} i' P(i',j)  }{\kappa} \right)
\end{align}
so that the relaxation corresponds to:
\begin{align}
P_{i,j_\infty(i)} &\leftarrow  \sum _j P_{i,j}  \notag \\
P_{i,j \ne j_\infty(i)} &\leftarrow 0 
\end{align}
Then we perform the {\em driving process}.

\end{itemize}
This algorithm integrates the fully connected version of the viscoelastic model, and produces the results shown in Figs. \ref{Fig:Pdelta_all}, \ref{Fig:bnw=001_beauGraphe_all}.

We may note that we have simply translated the analysis of sec.~\ref{sec:derivationMF_equations} in an algorithmic format.
In the \textit{Driving Process} we see that the value of $k_0 \d w$ is set to $\varepsilon$, since the initial drive after relaxation is of one bin (of width $\varepsilon$).
In the {\em Instability check} we see that when the steps $\Delta \delta^ F$ are increasing ($ P_0 \overline{z} (k_1+k_2) \geq 1$), we simply drive by one bin ($k_0 \d w$).
The cutoff that we mentioned for the driving steps $\Delta \delta^F_{{\rm step} k}$ is set to $\varepsilon/100$.
In the {\em Avalanche process}, we see that the driving steps $\Delta \delta^ F$ or $r$ saturate to one.
The possibility of driving occurring on length smaller than a bin $\varepsilon$ is accounted for via a smooth shift, reminiscent of the term $\partial P/\partial \delta^ F$.

\subsubsection{Numerical Integrations: Two Cases}

With the help of this numerical scheme, we can directly integrate the dynamics and get exact results, up to the binning precision $\varepsilon$.
This allows us to check our analytical predictions on the behaviour of $P(\delta^ F, \delta^ R)$. 
Since it is difficult to clearly present the evolution over time of functions of two variables, we will focus on the more intuitive distribution $P_\delta(\delta),$ defined by \reqq{Pdelta_delta}.

Depending on the validity of the condition $k_2>k_0$, there are global events, or not.
We comment these two cases below.

\begin{figure}[]
\begin{small}
\begin{center}
\def\svgwidth{\textwidth}
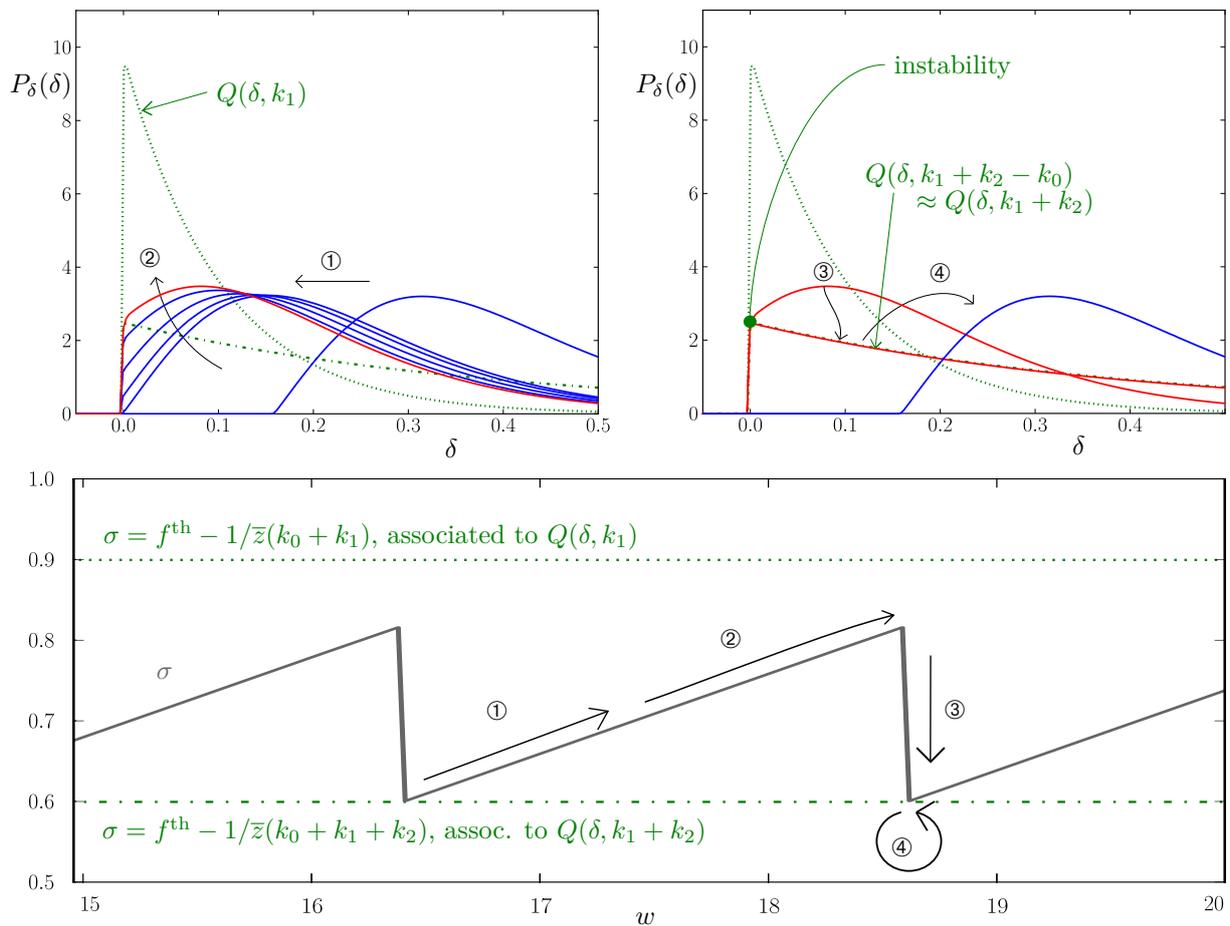
\end{center}
\end{small}
\caption{   {\footnotesize 
Adapted from \cite{Jagla2014a}.
Evolution of $P(\delta)$ (solid line, blue and red) and the stress $\sigma= k_0(w - \overline{h})$ 
(lower panel) computed from direct integration of the evolution equations. 
We used $k_0=0.001, k_1=0.1, k_2=0.3$.
Important curves are highlighted in red.
($1$) driving without any avalanche, linearly increasing stress; 
($2$) driving with elastic-depinning avalanches, slower stress increase.
($3$) global event: $P(\delta)$ collapses to the depinning fixed point $Q(\delta, k_1+k_2)$ (lower dashed curve) and the stress drops to $\sigma(k_1+k_2)$ (lower dashed line). 
($4$) relaxation closes the cycle back to stage ($1$) without altering average stress. 
\label{Fig:Pdelta_all}
}   }
\end{figure}
\paragraph{With Global Events (Periodic Behaviour)}
\begin{figure}[]
\begin{small}
\begin{center}
\def\svgwidth{\textwidth}
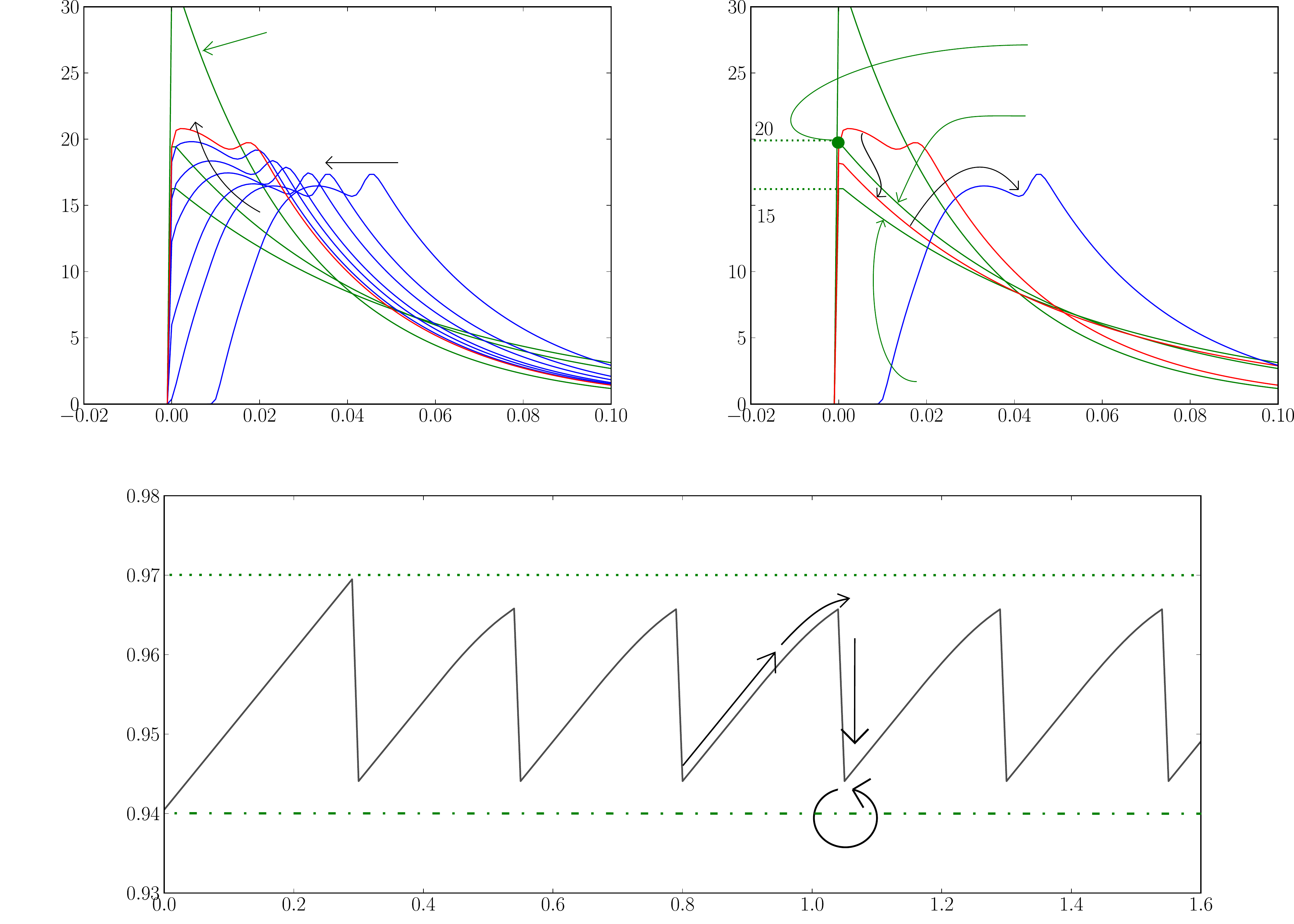
\end{center}
\end{small}
\caption{   {\footnotesize 
Evolution of $P(\delta)$ (solid line, blue and red) and the stress $\sigma= k_0(w - \overline{h})$ 
(lower panel) computed from direct integration of the evolution equations. 
We used a larger value of $k_0$: $k_0=0.1, k_1=0.2, k_2=0.3$.
Important curves are highlighted in red.
The Steps of evolution are essentially the same as in the previous figure.
The point of instability, $P_\delta(0)=1/\overline{z}(k_1+k_2)$ corresponds to the $P_\delta(0)$ of an elastic interface $Q(\delta,k_1+k_2-k_0$ with stiffness  $k_1+k_2-k_0$.
The large value of $k_0$ produces a finite difference between the value of $P_\delta(0)$ at the instability point and for the distribution $Q(\delta, k_1+k_2)$.
\label{Fig:bnw=001_beauGraphe_all}
}   }
\end{figure}
In Figs.~\ref{Fig:Pdelta_all} and \ref{Fig:bnw=001_beauGraphe_all} we show the most interesting case of $k_2>k_0$, where global events happen. 
The first figure is for vanishing $k_0$ (expected physically), the second is the finite $k_0$ case, shown for pedagogical purposes.
In both cases we have $k_0<k_2$ and the evolution of $P_w(\delta)$ is non stationary, with periodic oscillations in time.
The cyclic behaviour can be split in four phases.

In phase \ding{192} the system is driven ($w$ increases), but since $P_\delta(\delta)=0$, there are no avalanches.
The stress $\sigma$ increases linearly with time and the interface does not move ($\overline{h}=\text{const.}$).

In phase \ding{193} the system experiences a few small avalanches, since $0< P_\delta(0) < P_\delta^c(0)$. 
The combination of these small avalanches and relaxation drives $P_\delta$ towards $Q(\delta,k_1)$.
Stress increases (slightly) sub-linearly and the interface slips (infinitesimally).

The ``phase'' \ding{194} corresponds to the instant at which $P_\delta(0)$  reaches the critical point $P_\delta(\delta) =  P_\delta^c(0) = 1/\overline{z }(k_1+k_2)$.
At this point, the infinitesimal increase $\d w$ triggers a global avalanche. 
In Fig.~\ref{Fig:Pdelta_all}, the distribution $P_\delta$ reaches $Q(\delta,k_1+k_2)$ in this single global  avalanche.
In Fig.~\ref{Fig:bnw=001_beauGraphe_all}, the small ratio $k_2/k_0$ (large dissipation $k_0$) is such that at the end of the large event, $P_\delta$ stops between the critical point $P_\delta(\delta) =  P_\delta^c(0)$ and  $Q(\delta,k_1+k_2)$.
There, the instability point and the distribution $Q(\delta,k_1+k_2)$ are more clearly distinguished.
In both cases, the large event corresponds to a large drop of the stress and to a finite slip of the interface.

The ``phase'' \ding{195} corresponds to the relaxation that immediately follows. 
The distribution $P_\delta$ completely changes in this single relaxation operation.
As no event or driving is performed, the stress does not change at all during \ding{195}, and the interface does not move at all either (since only the $u_i$'s evolve via relaxation).
This last phase takes us back to the initial stage: we have an exactly periodic behaviour.

This integration with the choice of parameters $k_2>k_0$ allows to check that global events can actually take us to $Q(\delta,k_1+k_2)$ and that this function is indeed given by our computation \reqq{Qk1k2}.
As phase \ding{193} drives us towards $Q(\delta,k_1)$, we meet the instability point and thus never actually reach  $Q(\delta,k_1)$. 
However in the case $k_2 < k_0$ the convergence to $Q(\delta,k_1)$ is confirmed.

\paragraph{Without Global Events (Stationary Behaviour)}

\begin{figure}[]
\begin{small}
\begin{center}
\def\svgwidth{\textwidth}
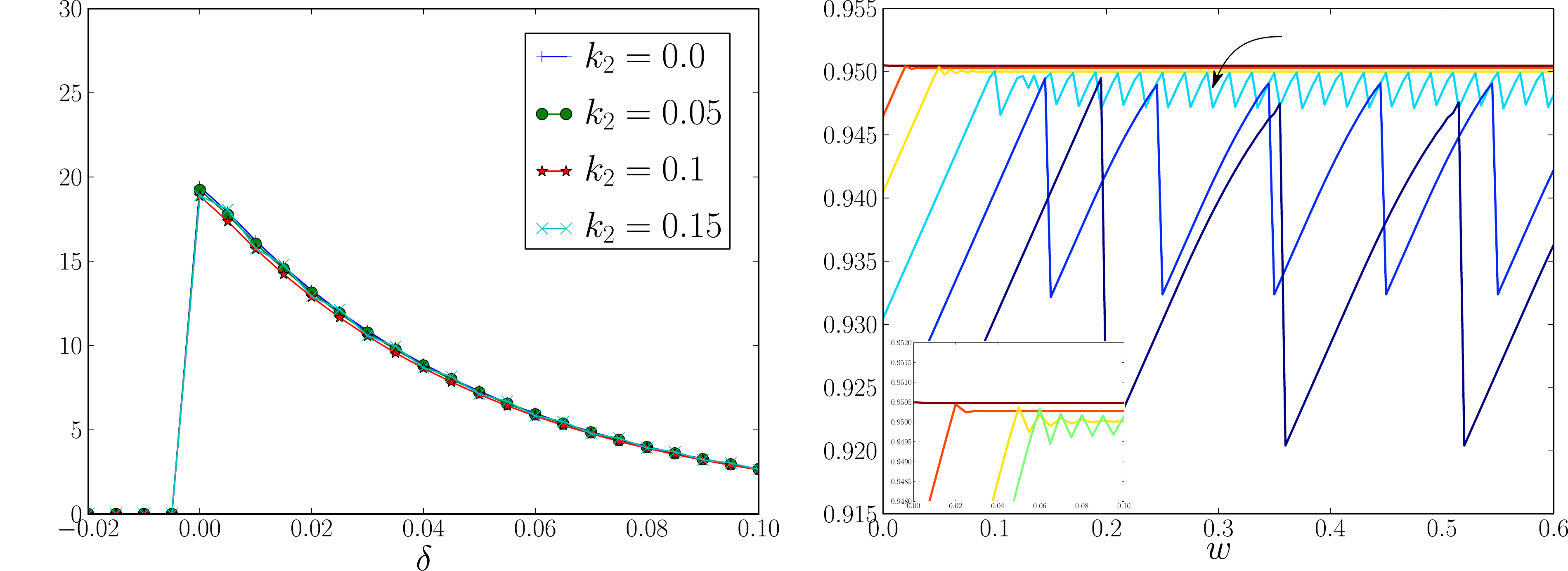
\end{center}
\end{small}
\caption{   {\footnotesize 
Left: 
Collapse of the $P_\delta(\delta)$ distributions for $k_2<k_0$.
We used $k_0=0.2$, $k_1=0.3$ and variable $k_2$'s, 
and we only plot one point every five binning widths for clarity.
\newline
Right: Dependence of the average stress $\sigma(t)$ over time, depending on $k_2$. 
We use $k_2=0, 0.05, 0.1, 0.2, 0.3, 0.4$ (from top to bottom).
For $k_2<0.2=k_0$, there are only very small spurious oscillations due to the finite precision of the numerical integration (finite binning $\varepsilon$).
For $k_2>k_0$, there are large oscillations (with periods larger than a single simulation step).
Inset: focus on the early times dynamics. The larger the $k_2$, the longer it takes to reach the steady state. From left to right, we used $k_2=0, 0.05, 0.10, 0.12$.
\label{Fig:Pdelta_all2_k0_finite}
}   }
\end{figure}

In Fig.~\ref{Fig:Pdelta_all2_k0_finite} (left), we present a few examples of stationary distributions $P_\delta$ obtained using $k_2<k_0$.
In this weakly viscoelastic regime, we observe a convergence of any initial $P_\delta$ to $Q(\delta,k_1)$.
Of course, the critical values $P_\delta^c(0)$ of all these stationary solutions is larger than the (common)  $P_\delta(0)$.
As $k_2$ is increased towards $k_0$, the critical value $P_\delta^c(0)$ gets closer to $P_\delta(0)$, and it takes a longer time for the system to reach a stationary behaviour (see  Fig.~\ref{Fig:Pdelta_all2_k0_finite}, right).

\subsubsection{Relevance of the Viscoelastic ``Perturbation''}

It is important to note that from these analytical results and the corresponding exact integrations, we can conclude that the addition of some ``visco-'' part into the elastic interactions is relevant perturbation, in the macroscopic limit.
Precisely, we see that for any $k_1$  and any $k_2>0$, there will always be a $k_0$ small enough so that the viscoelastic character of the system manifests itself (appearance of global events and periodic oscillations of the average stress).
We can even predict the value of $k_0$ at which this happens, which is simply $k_0^c = k_2$.
In finite dimensions, this feature is also present, although we do not have a precise criterion to predict below which $k_0$ the appearance of the viscoelastic features are expected.

\subsubsection{Monte Carlo Simulations in Mean Field}

\begin{figure}[]
\begin{small}
\begin{center}
\def\svgwidth{9cm}
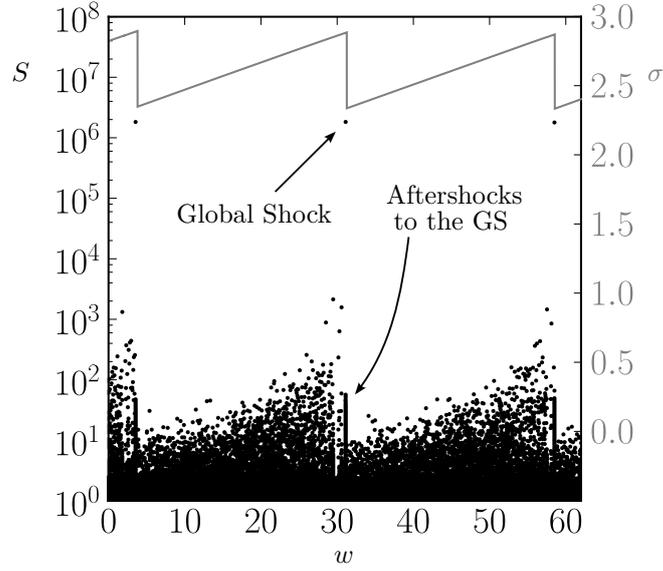
\end{center}
\end{small}
\caption{   {\footnotesize 
Avalanches sizes and stress evolution of the viscoelastic interface in mean field (fully connected).
The threshold forces $f_i^ \text{th}$ are randomly distributed, which allows for numerous aftershocks, especially following the Global Shock (GS).
The fluctuations of the stress (due to finite size effects) during the ``small avalanches'' are negligible compared to the macroscopic stress drops occurring in the GS.
\label{Fig:MonteCarlo_MF}
}   }
\end{figure}
So far, we have always represented the disorder function by narrow wells. 
As long as the associated forces have a random distribution, this choice does not affect the generality of the reasoning, since narrow wells can be chosen to represent e.g.~(discretized) white noise.
However in our mean field calculations we used identical wells: the threshold force associated to each narrow well being unique, randomness remained only in the spacings distribution, $g(z)$.
We have seen that this choice suppresses aftershocks, an important feature of our model.
One may wonder whether this choice also affects the general behaviour of the system.
For instance, does the system still displays periodic oscillations and global events when the threshold force $f^ \text{th}_i$ is random?

To answer this question, we perform Monte-Carlo simulations of the equations \reqq{continuous_MF} using a Gaussian distribution for $f^ \text{th}_i$ and the same algorithm as presented for the 2D case (see \req{deltai_visco_2d} and after).
The results are displayed in Fig.~\ref{Fig:MonteCarlo_MF}, and are very similar to our predictions using the Fokker-Planck formalism.
In particular we observe that the system still displays periodic behaviour and global events, with the fluctuations of the period being due to finite size effects. 
The main difference is the presence of aftershocks, which are especially noticeable following the global shock.

\subsection{Comparison with Experiments}

\subsubsection{Stick-Slip: Friction Forces}

The periodic oscillations we observe are strongly reminiscent of the stick-slip motion expected in friction.  
The oscillations disappear when $k_0$ is large enough compared to the viscoelastic-ness ($\sim k_2$) of the material, as is qualitatively expected in friction.
In terms of stress (or friction force), we predict that the system evolves between two extreme values that are not necessarily reached:
\begin{itemize}
\item
The lower one, $\sigma_{1} = f^ {th} - (k_0+k_1+k_2) \overline{z^ 2} / 2\overline{z}$,  associated to the stationary solution of the rigid interface ($Q(\delta, k_1+k_2)$).
When $k_0\to 0$, this stress is actually reached during the global shocks (i.e.~at the end of the macroscopic slip).

\item
The higher one, $\sigma_{2} = f^ {th} - (k_0+k_1)  \overline{z^ 2} / 2\overline{z} $,  associated to the stationary solution of the flexible interface ($Q(\delta, k_1)$).
When $k_2<k_0$, this stationary stress takes this value.
\end{itemize} 
We can interpret $\sigma_{1}$ as a lower bound for the kinetic friction force, and $\sigma_{2}$ as an upper bound for the kinetic or static friction forces.

In the oscillating regime, the stress at which the macroscopic slip (global event) starts corresponds to the actual static friction force $F_s$, for which we do not have an analytical prediction.
We can only predict that $\sigma_{1}<F_s \lesssim \sigma_{2}$.

The decrease of friction ($\sigma$) with increasing rigidity of the interface ($k_1$ or $k_1+k_2$ depending on $k_0$) is consistent with the intuition that a rigid material will slide more easily than a flexible one.
This feature is also present for the purely elastic interface.
Furthermore, a viscoelastic material that can adapt a lot over a large range of time scales (e.g.~rubber) is expected to adhere much more strongly (e.g.~if $k_2\gg k_1$, $k_2>k_0$), because it can adapt to the local surface profile. 
This latter feature is specific of our viscoelastic model.

\subsubsection{Periodic Events in Other Systems}

\paragraph{``Characteristic'' Earthquakes}

Interpreting our mean field interface as a model for a single fault with an effective long-range elastic interaction, we note a good agreement with the notion of seismic cycle and the observations of characteristic earthquakes (see sec.~\ref{sec:characteristic_EQs}).
On this point, we want to stress out that in our model, the period emerging from the viscoelastic interactions is macroscopic, unrelated to the precise value of the microscopic time scale $\eta_u$ and is not a finite-size effect.

In several earthquake models such as OFC, it is sometimes argued that some almost periodic events reproduce the seimsic cycle.
One needs to be careful when discussing this idea, in particular on estimating the macroscopic character of the cycle.
In many models, the ``cyclic behaviour'' is a purely local effect, which involves only the typical (microscopic) slip length of the block and the corresponding (microscopic) time scale needed to re-load it (we discussed the case of the elastic depinning in sec.~\ref{sec:confusion_stick_slip}).

A different line of argumentation in models such as the OFC is to interpret some system-size events due to finite-size effects as characteristic earthquakes (which occurs in small systems with low dissipation).
The possibility of finite-size effects in seismic faults cannot be discarded entirely, however the absence of a clear correlation between fault size and period of the seismic cycle points against it.

\paragraph{Micro-Crystals Deformation or the ``Avalanche Oscillator''}

The periodic large events we find in mean field are strongly reminiscent of those found in \cite{Papanikolaou2012} (discussed in sec.~\ref{sec:papanikolaou}).
In that paper, using a long-range elastic kernel in two dimensions of space, it was found that system-size events occurred over a large range of parameters (precisely, at sufficiently low strain rates).

The periodic oscillations found in \cite{Papanikolaou2012} were explained through a phenomenological model build on the notion of a susceptibility $\rho$, defined as ``the multiplier giving the net number of local slips triggered by a single slip''.
The equation given in  \cite{Papanikolaou2012} for the evolution of $\rho$ is:
\begin{align}
\rho_{t+1}-\rho_t \propto \lp 1- \frac{S_t}{\overline{S}}\rp,
\label{Eq:phenomeno_papanik}
\end{align}
where the time $t$ used here should be connected to our discrete ``steps''. 

Learning from our analysis in terms of $P_\delta(0)$ and its ``critical'' value $P_\delta^ c(0)=1 /\overline{z}(k_1+k_2)$, we can improve the arguments based on this susceptibility $\rho$.
Essentially, $\rho$ plays the same role as $P_\delta(0) / P_\delta^ c(0)$: when this ratio is larger than $1$, the avalanche involves an increasingly large number of sites, and may involve a finite fraction of the complete system.
A large avalanche decreases $P_\delta(0)$ by a large amount, so that the following avalanches are rather small: this is qualitatively compatible with \reqq{phenomeno_papanik}.
However, by defining $P(\delta^F, \delta^R) $ we have been able to derive the evolution equations for $P_\delta(\delta)$ directly from the dynamical equations, and integrate them in a semi-analytical way, thus giving a clear picture of the origin of the global instability responsible for the global shocks.
Our analysis shows that a full description of such a system necessitates the use of an additional degree of freedom, which encodes the memory of the system.

\subsubsection{Rate and State Friction Laws}
\label{sec:finite_velocity_model}

\includefig{\textwidth}{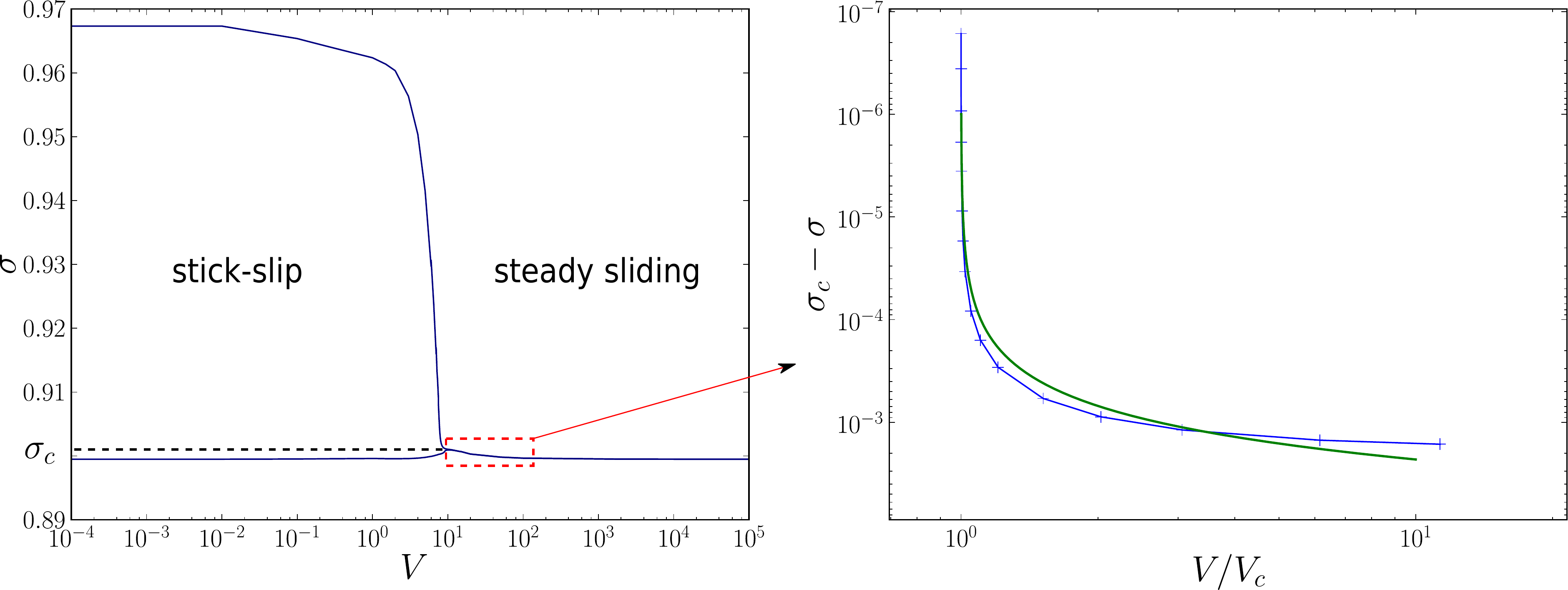}{
Left: Variations of the minimal and maximal stresses during the stick-slip regime (left), or of the stationary stress (right) in the steady sliding regime.
As velocity increases in the steady sliding regime, the friction force ($\propto \sigma$) decreases. 
\newline
Right: Variation of the stationary stress ($\sigma_c-\sigma$) with the adimensionalized velocity $V/V_c$ (blue).
Green: a pure logarithmic behaviour is shown as a guide to the eye.
Notice how we inverted the y-axis to match with the left panel.
\label{Fig:velocity_weakening5}
}
As we discussed in chap.~\ref{chap:friction}, there are phenomenological Rate- and State-dependent Friction (RSF) laws which characterize solid friction rather well.
However these laws lack microscopic foundations, and consequently their adaptation to very small or very large scales is a difficult task.

In the setting of the complete separation of time scales ($\tau_0\ll \tau_u \ll \tau_D$) that we used up to now,
 it is impossible to discuss the effects of the variations of the driving velocity $V_0$ on the friction force, since we assumed quasi-static driving: $V_0=0^+$.
In a paper that should appear soon,
 we discuss these effects in the Mean Field limit, where we can obtain semi-analytical results.

We now relax the constraint on the time scales: $\tau_0\ll \tau_u \simeq  \tau_D$.
In this case, avalanches still unfold infinitely fast (the time of slip is negligible compared to all other times), but the relaxation and driving time scales compete.
In the limit where $\tau_u \gg  \tau_D$, relaxation does not have the time to happen and we recover the purely elastic depinning model, where we can not expect any RSF law.
The other limit, $\tau_0\ll \tau_u \ll \tau_D$, is the one we just studied in the present chapter.
Between these limits, an interesting transition takes place (see Fig.~\ref{Fig:velocity_weakening5}, left panel): at small velocities, one may observe stick-slip behaviour, while at larger velocities we observe steady slip with a friction coefficient that decreases with increasing friction (i.e.~the velocity-weakening effect).

The RSF laws are build mainly on two fundamental observations: the velocity-weakening effect (at small velocities) and the ageing or increase of contact at rest.
Both observations correspond to simple experimental setups, which allow to probe the relevance of our model.
\begin{figure}[]
\begin{small}
\begin{center}
\def\svgwidth{\textwidth}
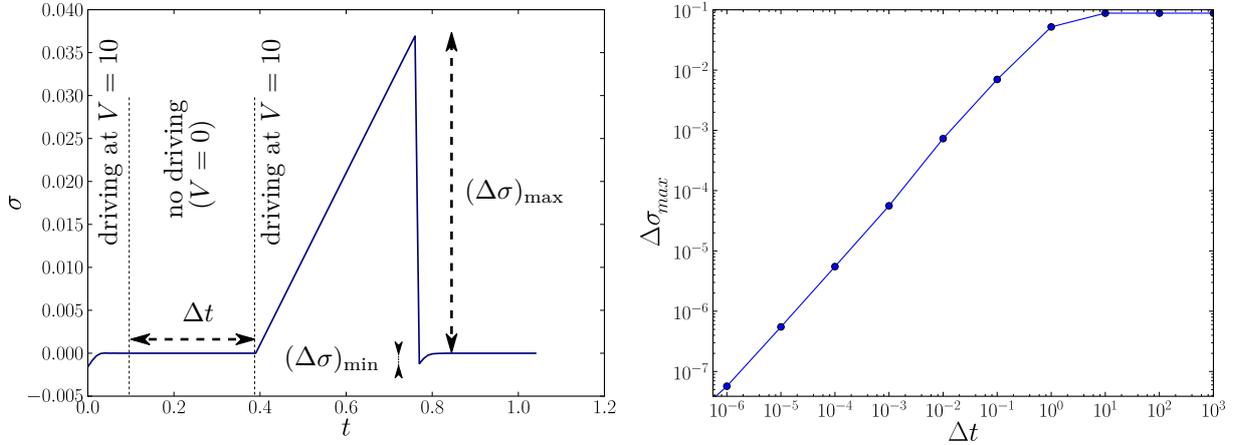
\end{center}
\end{small}
\caption{   {\footnotesize 
Left: Slide-hold-slide experiment in the mean field (fully connected) case.
Starting from some initial configuration, the system quickly reaches a stationary state under steady driving. Pulling is stopped at time $t_1\approx 0.1$, then restarts at time $t_2=t_1+\Delta t\approx 0.4$.
The relaxation occurring when $t\in [t_1,t_2]$ increases the static friction force, which is probed when driving is resumed.
The stress overshoot $(\Delta \sigma)_{max}$ measures this increase of the static friction coefficient.
Due to the large slip during the overshoot, a small decrease in stress quickly follows, after what the system returns to its steady state. 
\newline
Right: Dependence of the stress overshoot on the hold time (in the mean field, fully connected case).
Repeating the slide-hold-slide experiment for many values of the hold time $\Delta t$, we may measure the ageing of contact at rest.
The stress overshoot increases with the hold time, up to a saturation at very long times which corresponds to the complete system being fully relaxed.
\label{Fig:slide_hold_slide_explained}
}   }
\end{figure}

\paragraph{Velocity Weakening}

\textit{On this topic in particular, see our more recent and clearer work: Phys. Rev. E \textbf{92}, 012407 -- Published 16 July 2015.}
In the regime of steady sliding, where the friction force settles to a stationary value, we observe a decrease of friction with increasing velocity (see Fig.~\ref{Fig:velocity_weakening5}).
We study the behaviour close to the transition point ($V_c,\sigma_c$) and compare the decrease of stress $\sigma-\sigma_c$ to the increase in velocity $V/V_c$.
We observe an almost logarithmic decay of the stress with velocity, $\sigma-\sigma_c \sim - \log(V/V_c)$.
This comparison proves that our model captures an important aspect of basic frictional properties.
The complete quantitative comparison has yet to be done.

The decrease of the average stress can be understood in terms of the solutions $P_\delta(\delta)$ of the Fokker-Planck equations, in the case of ``small'' avalanches (steady-state, no system-size instability).
Essentially, the change in velocity lets  $P_\delta(\delta)$ interpolate between $Q(\delta,k_1+k_2)$ and  $Q(\delta,k_1)$.
At very large driving velocity $V_0$, we have $\tau_D \ll \tau_u $ so that the dashpots relax too slowly and thus have no effect: we recover purely elastic depinning, in this case $P_\delta(\delta) = Q(\delta,k_1+k_2)$ (associated to the lower stress $\sigma_1 = f^ {th} - (k_0+k_1+k_2) \overline{z^ 2} / 2\overline{z} $).
At the smallest driving velocities for which we still have a steady state (for lower velocities, there is stick-slip motion), the solution tends  $P_\delta(\delta)$ to  $Q(\delta,k_1)$ (associated to the higher stress $\sigma_{2} = f^ {th} - (k_0+k_1)  \overline{z^ 2} / 2\overline{z} $).

\paragraph{Contact Ageing}
Another experiment we may perform in the regime  $\tau_u\simeq \tau_D$ is the slide-hold-slide experiment, which is a standard test to measure the ageing of contacts at rest.
Consider a solid block pulled on some substrate, sliding steadily.
If at time $t_1$ we stop to pull, the block quickly ceases to move. 
At rest the friction force increases over time, so that when we resume the pulling at time $t_2=t_1 + \Delta t$, the stress (or friction force) overshoots its stationary value: see Fig.~\ref{Fig:CACoulomb_Ageing} (\pp{Fig:CACoulomb_Ageing}) for the experiments and Fig.~\ref{Fig:slide_hold_slide_explained} for our results.

Measuring this overshoot gives a measure of the increase of friction over the time $\Delta t$.
Repeating this experiment many times (as Coulomb did a few centuries ago), we obtain the law of the increase of friction $\Delta\sigma$ over time $\Delta t$. 
We report the results of this ``experiment'' in the right part of Fig.~\ref{Fig:slide_hold_slide_explained}.
Note that this observation should be interpreted using the concept of the joint $P(\delta^F, \delta^R))$ distribution introduced earlier.  
Although we obtain an increase of the static friction over time (as expected), the success of this mean field approach is limited, since we do not obtain a logarithmic increase but a linear one.
This discrepancy can be attributed to the limitation inherent to the mean field, for which it is impossible to obtain an Omori law.
In a finite-dimensional approach with Omori-like laws of the decay of activity over time, we expect to find a logarithmic behaviour.
This is left for future work.

\section{Two Dimensional Results:  Comparison with Seismic Phenomena}
\label{sec:2D_results_visco}

The two dimensional case is expected to be somewhat representative of sliding friction, despite our use of short-range interactions, since sliding surfaces (e.g.~faults) are two dimensional (with surface roughness playing an important role).
Furthermore, this relevance of the two-dimensional case is confirmed a posteriori by qualitative and quantitative agreements of numerical results with several field observations.

We have not fully completed the study of the model, but we can already 
 present several interesting results which contrast strongly with the purely elastic depinning picture and which compare well with experimental observations.

\paragraph{Numerical Scheme}
In two dimensions we must rely on the numerical implementation of Eqs. (\ref{Eq:hvisco}, \ref{Eq:uvisco}), (p.~\pageref{Eq:hvisco}) on a finite system with periodic boundary conditions. 
Implementing a Monte-Carlo integration of the equations, the only approximation we make is to neglect the possibility of backward motion, as explained earlier (sec.~\ref{sec:noBackwardMotion}).
Unlike what we did for the mean field model, here we study the general case of a heterogeneous distribution of pinning wells, using a randomly distributed threshold force $f^\text{th}_i$ (typically a Gaussian distribution with unit mean and variance $3$).

The crucial point that explains the efficiency of our numerical scheme is the use of the narrow wells as representation for the disorder.
In this representation, each block is always in one of the pinning wells and evolves exclusively via finite jumps.
This spares us from computing numerous infinitesimal updates of the interface position under the small increases $\d w $ of the driving.
Inspired by an efficient method originally developed in \cite{Grassberger1994a} (see Appendix \ref{App:Grassberger_efficient_method_two_cases}), we only need to update the sites that participate in an avalanche when they do so, so that we perform the exact dynamics in a time that essentially scales as the total sum of the avalanches area.

For the Laplacian relaxation model, this efficient method does not apply.
 Instead, the solution of the relaxation equation is not local and has to be resolved via an Euler-like method\footnote{Or via Fourier transform, but then we also need to update the whole lattice at each relaxation time step, which is highly inefficient.}.
This is why we favoured the local relaxation model \req{viscoDetaille3} in our presentation.

\subsection{Local Oscillations}

\begin{figure}[]
\begin{small}
\begin{center}
\def\svgwidth{\textwidth}
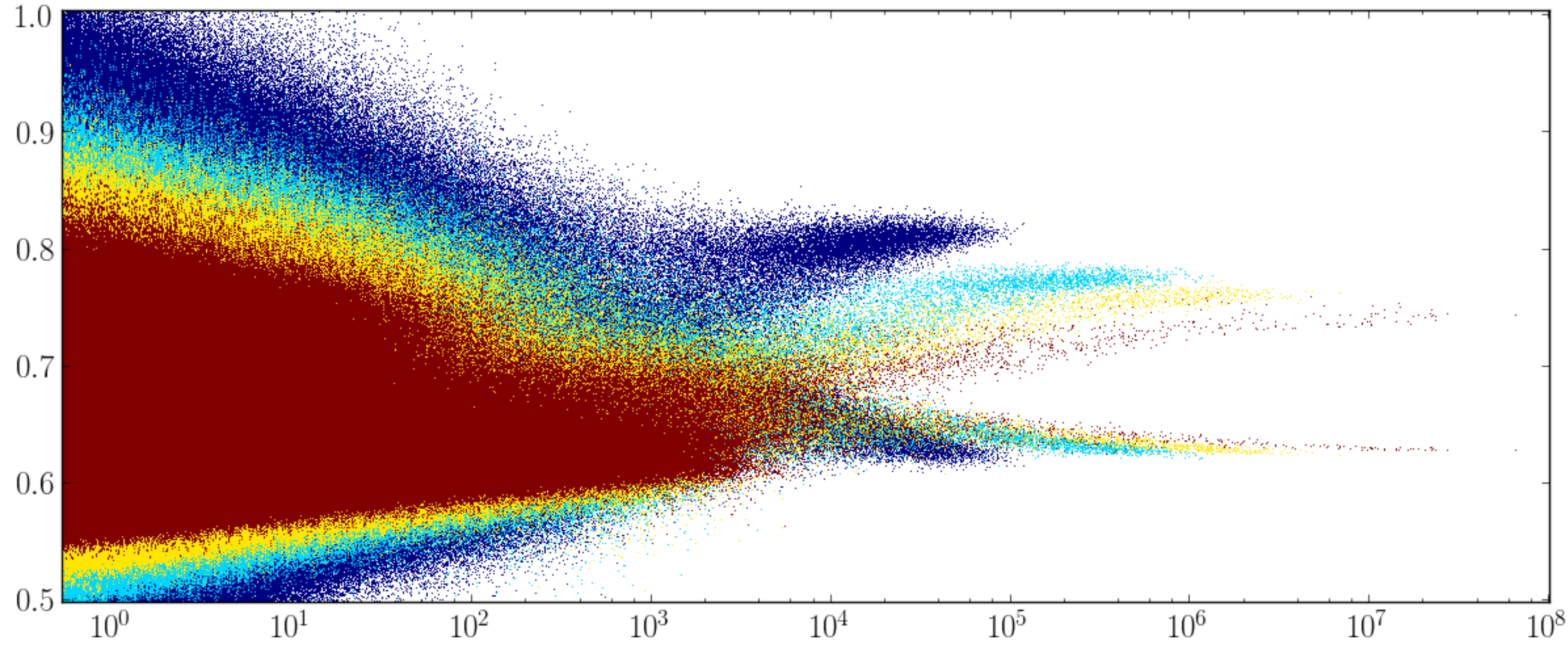
\end{center}
\end{small}
\caption{   {\footnotesize 
The local stress restricted to the cluster area, just before (up, $\sigma_B$) and just after (bottom, $\sigma_A$) it takes place, as a function of the cluster size $S_C$ (the size of a cluster is the sum of the sizes of the events occurring for this $w$).
The local variation of stress vanishes for small avalanches (with fluctuating values of $\sigma_{B,A}$), and saturates to a constant nonzero value for large avalanches (with well defined  values for $\sigma_{B,A}$).
We used $k_1=0.1, k_2=0.9$ and (from left to right): $k_0=0.05, 0.025, 0.018, 0.012$. 
See also the Figure Fig.~\ref{Fig:sasb} in the Appendix 
 for the case with $k_1=0, k_2=1$, which also displays very well defined values of $\sigma_{B,A}$.
\label{Fig:sasb_landes}
}   }
\end{figure}
\begin{figure}[]
\begin{small}
\begin{center}
\def\svgwidth{\textwidth}
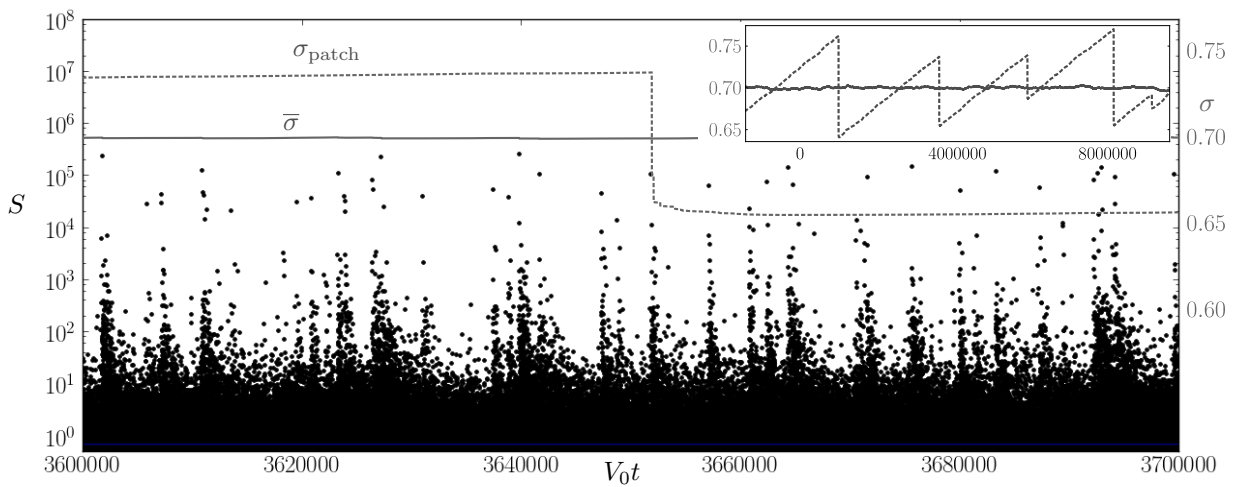
\end{center}
\end{small}
\caption{   {\footnotesize 
Avalanches sizes $S$ and stress evolution of the two-dimensional viscoelastic interface model.
Here we used $\tau_D > \tau_u$ instead of the usual $\tau_D \gg \tau_u$, i.e.~there is some driving occurring together with relaxation.
This makes the qualitative picture more similar with real earthquakes.
In solid grey, the system-average oscillations of the stress.
In dashed grey, the stress averaged over a small patch of the system (patch of  $50\times50$ sites in a $5000 \times 5000$ system).
The stress restricted to a small area has large fluctuations, similarly to the mean field global stress.
In the inset, we show the same quantities but on a longer times, so that the pseudo-periodic oscillations are clearly apparent.
\label{Fig:patches_stress_2D_finiteV}
}   }
\end{figure}

Our mean field prediction (and observation) is that there are periodic\footnote{
The behaviour is exactly periodic for the infinite system and very close to periodic for large systems, the finite size effects being very weak (see Fig.~\ref{Fig:MonteCarlo_MF}).
}
 events which involve the whole system, the so-called global events.
In terms of the average stress, this corresponds to a periodic evolution of the stress with a saw-tooth profile.
In two dimensions on the contrary, the average stress is constant (see Fig.~\ref{Fig:patches_stress_2D_finiteV} or Fig.~\ref{Fig:patches_stress_2D}) and no global events are observed, up to finite size fluctuations.
Nevertheless, a careful analysis of the 2D model shows an interesting reminiscence of the mean field behaviour.
We now provide numerical evidence that periodic stick-slips occur locally, without global synchronization between the different parts of the system.

\begin{figure}[]
\begin{small}
\begin{center}
\def\svgwidth{\textwidth}
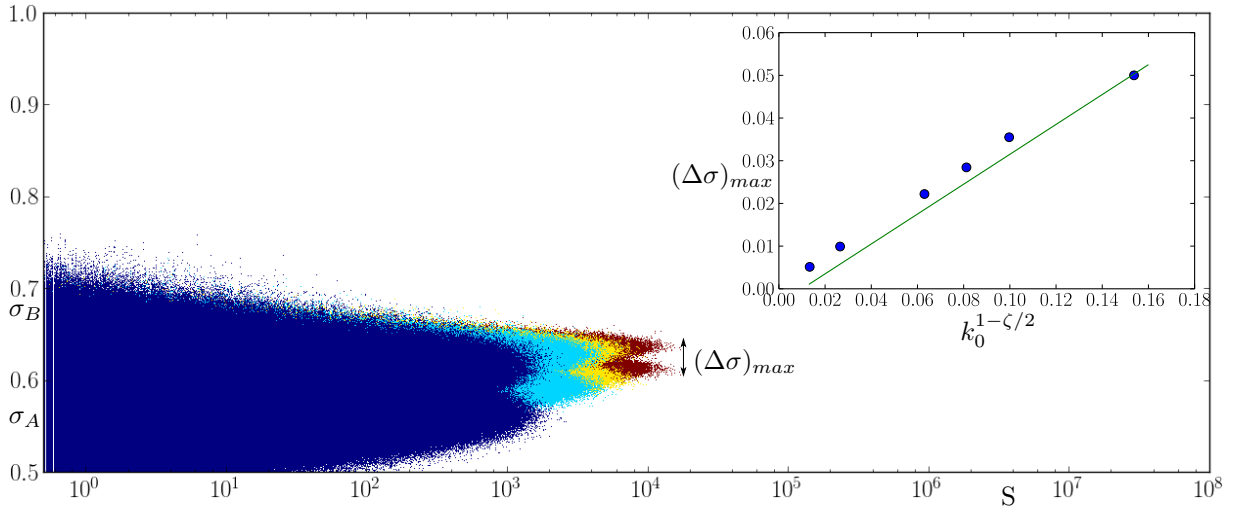
\end{center}
\end{small}
\caption{   {\footnotesize 
Blow-up of Fig.~\ref{Fig:scatter_plot_k2is0_1} to be compared with Fig.~\ref{Fig:sasb_landes}. 
The scale of the plot, the parameters used and the number of avalanches per parameter set ($10^7$) are the same as in  Fig.~\ref{Fig:sasb_landes}, except for the use of  $k_1=1, k_2=0$.
Inset: plot the stress drop $(\Delta \sigma)_{max}$ associated to the largest avalanches against $k_0^{1-\zeta/2}$, with $\zeta =0.75$
The straight line is a guide to the eye. 
When $k_0 \to 0$, the two values $\sigma_B,\sigma_A$ converge to a common value: the stress drop $\Delta \sigma = \sigma_B-\sigma_A$ associated to the avalanches is thus infinitesimal (in particular for the large avalanches).
This can also be seen with even smaller values of $k_0$ in Fig.~\ref{Fig:scatter_plot_k2is0_1}.
\label{Fig:sasb_qEW}
}   }
\end{figure}

In Fig.~\ref{Fig:sasb_landes} we show for each cluster of events the stress average restricted to the cluster area, just before ($\sigma_B$) and just after ($\sigma_A$) it takes place.
Small clusters show broad distributions of $\sigma_B$ and $\sigma_A$, similar to what would be observed for the depinning case.
However, for large clusters both distributions become very narrow: $\sigma_B$ sets to a value that we denote $\sigma_{\max}$, and $\sigma_A$ sets to $\sigma_{\min}$.
This is the fingerprint of the mean field behaviour, suggesting a large scale description of the two-dimensional interface as a terraced structure, with large plateaus of almost constant stress and macroscopic stress differences between plateaus. 
It is remarkable that in the viscoelastic model, the width of the distribution of the local stress ($\sim \sigma_{\max}- \sigma_{\min}$) remains finite  when $k_0\to 0 $,
whereas in the depinning model \cite{Rosso2012} it vanishes as $k_0^{1-\zeta/2}$ when  $k_0\to 0$, since the roughness exponent $\zeta$ is smaller than $2$ in all dimensions (see also Fig.~\ref{Fig:sasb_qEW}).	        

\includefig{\textwidth}{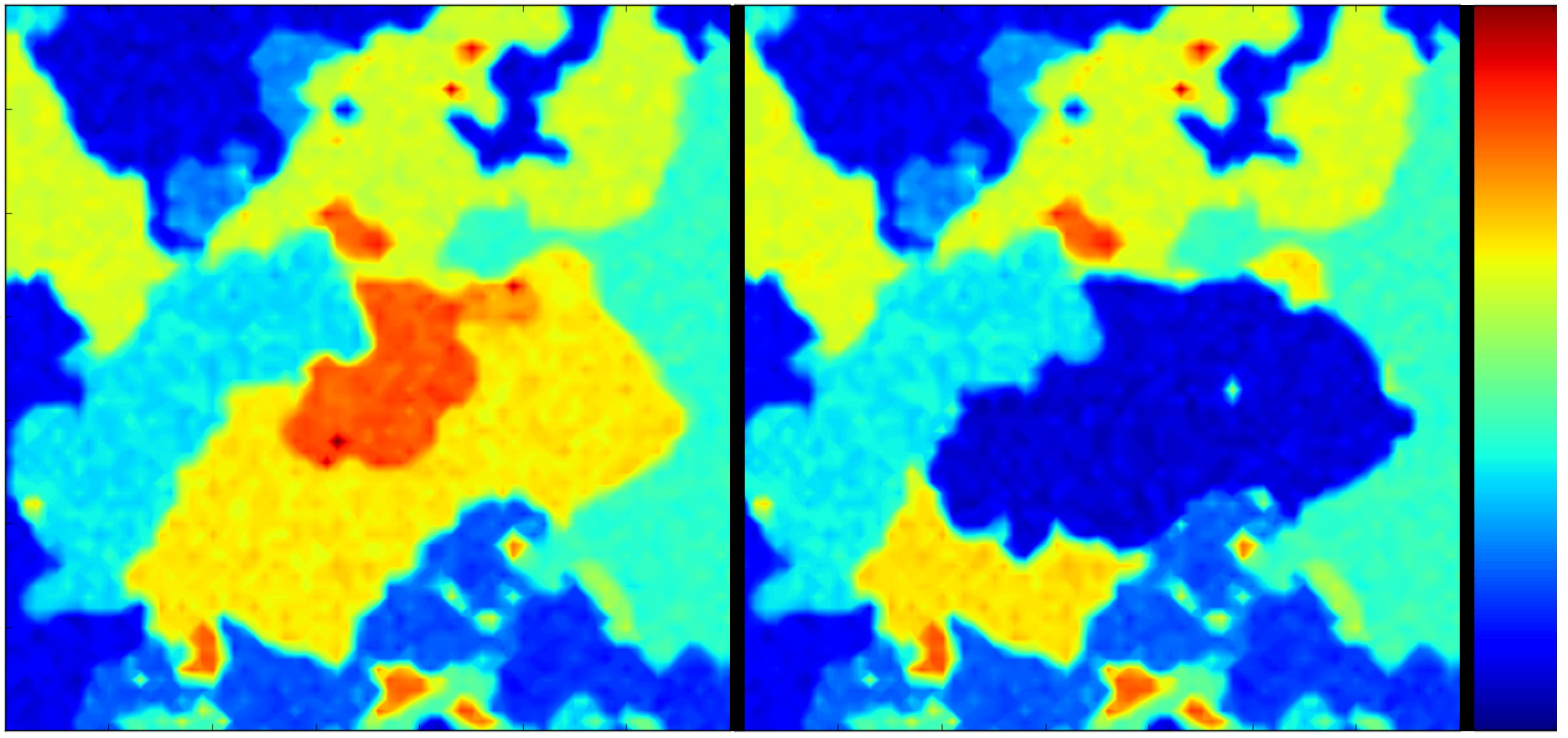}{
Stress map of the viscoelastic interface model in 2D.
A large event is triggered in a region of high stress (left, most red part) , which lowers the stress down to rather homogeneously distributed values $\sim \sigma_{min}$ (left).
The most red spots correspond to $\sigma_{max}=1.95$ and the most blue to $\sigma_{min}= 1.77$.
\label{figBrokenZone}
}
Indeed, we observe (see Fig.~\ref{figBrokenZone}) that different parts of the system have different values of the stress, which range from  $\sigma_{\min}$ to $\sigma_{\max}$ (a range of finite, non-vanishing width).
In analogy with mean field, it is only when the stress of a region reaches a value of $\sim \sigma_{\max}$ that it gets destabilized and that the whole region collapses to $\sigma_{\min}$.

Furthermore, the evolution of the local stress associated to a small patch of the interface is non stationary, and shows an almost periodic oscillation between $\sigma_{\min}$ and $\sigma_{\max}$: see the dashed and dotted lines in Fig.~\ref{Fig:patches_stress_2D}.
Since the oscillations are not synchronized among the different patches, the system does not display any global oscillations (for a large enough system size). 
The evaluation of the characteristic length over which the stress level is strongly correlated has not been performed yet.
We need to assess its dependence on the paramters $k_0, k_1,k_2$ in order to fully describe the model. 
This is left for future work.

Our observation of an almost constant stress drop over a large range of avalanches seems well consistent with the observation of \textit{constant stress drop} for seismic faults (described in sec.~\ref{sec:constant_stress_drop}).
More precisely, if we dismiss our small avalanches (in Fig.~\ref{Fig:sasb_landes}, those with $S\leq 10^4$) as out of the range of interest in earthquakes, then our ``large'' avalanches correspond to the ``small'' earthquakes, which seem to follow a constant stress drop.

\subsection{Aftershocks}

\subsubsection{A Well-Defined Feature}

An important feature of the viscoelastic model is that unlike most avalanches models, it has a very natural definition of aftershocks: aftershocks are the avalanches triggered by the relaxation of the dashpots (the slow evolution of $u_i$'s).
In this sense, for a given increase of $w$ that produces a first avalanche (main shock), all the following ones (occurring at the same value of $w$) are aftershocks.

The aftershocks are not specific of the two-dimensional model: in other dimensions they are also present, including the mean field case.
In our mean field analysis, we simplified the computations by assuming identical wells (i.e.~$f^\text{th}_i = \text{const.}$), a choice that happens to prevent the occurrence of aftershocks.
In the Laplacian relaxation variant of our model, one obtains aftershocks even within this simplifying assumption, in all dimensions.

\begin{figure}[]
\begin{small}
\begin{center}
\def\svgwidth{\textwidth}
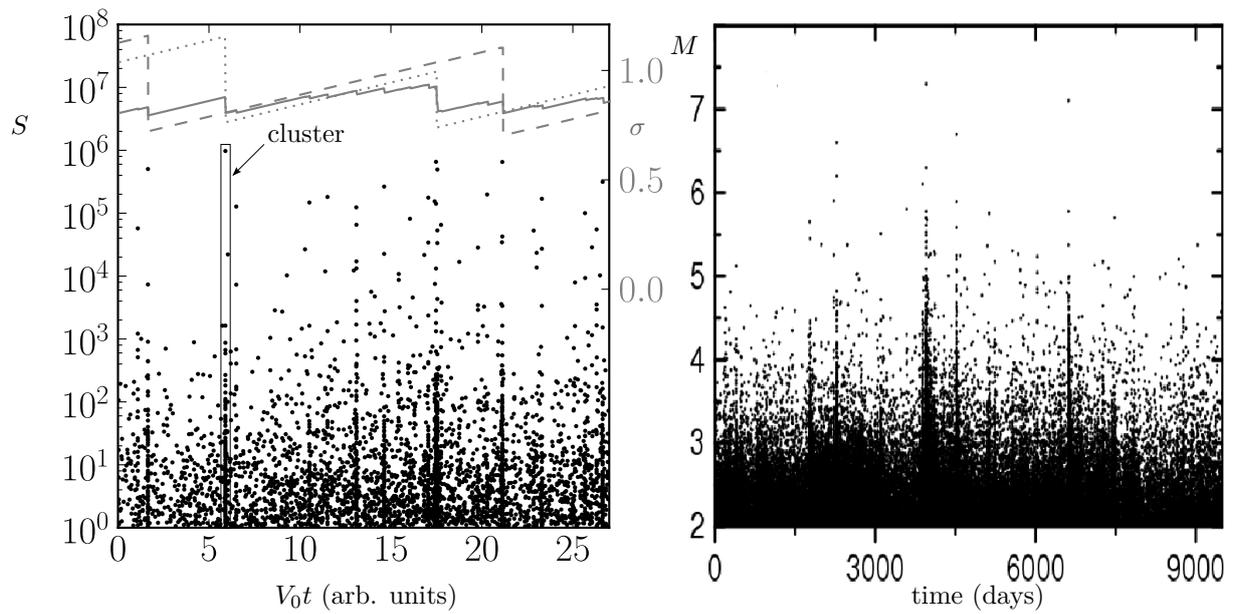
\end{center}
\end{small}
\caption{   {\footnotesize 
Left: Avalanches sizes $S$ and stress evolution of the two-dimensional viscoelastic interface model.
In solid grey, the system-average oscillations of the stress.
In dashed and dotted grey lines, the stress averaged over small patches of the system (patches of  $10\times10$ sites in a $500 \times 500$ system).
The stress restricted to a small area has large fluctuations similar to the mean field global stress.
Avalanches sizes $S$ (indicated by dots) are grouped in clusters, with strong correlations over time inside each cluster.
\label{Fig:patches_stress_2D}
\newline
Right: From \cite{Jagla2010a}. 
Magnitude ($\propto \log(S)$) of earthquakes over the San Andreas area.
Note the strong resemblance between these real events and those in Fig.~\ref{Fig:patches_stress_2D_finiteV}.
\label{Fig:comparison_of_S}
}   }
\end{figure}
In Fig.~\ref{Fig:comparison_of_S}, we compare the synthetic avalanches sizes $S$ over time ($\propto w$) 
 with earthquakes from the San Andreas region.
The possibility of triggering events via two mechanisms with distinct time scales ($\tau_u, \tau_D$) compares well with actual seismic data, as a clear pattern of correlations emerges in both cases.
An important qualitative difference is that a real cluster of events (main shock and its aftershocks) spans a finite time interval, due to the non-complete separation of the relaxation and driving time scales: in reality, $\tau_D/\tau_u$ is different from zero.
This shortcoming can be addressed by using $\tau_D > \tau_u$, i.e.~by allowing some driving to occur while relaxation happens.
The result of these more recent results can be seen in Fig.~\ref{Fig:patches_stress_2D_finiteV}, where the comparison with actual earthquakes is visually excellent.
The similarity of patterns is especially convincing when comparing with the purely elastic depinning result, where events are essentially uncorrelated in time and space (apart from finite size effects, see Fig.~\ref{Fig:scatter_elastic+tau126}).

No matter how convincing this simple kind of comparison may seem, it is insufficient to precisely determine the relevance of our model to seismic phenomena or frictional processes.
In the next subsections, we study the spatial and temporal aftershocks patterns and compare them with seismic data.

\subsubsection{Aftershocks Spatial Evolution: the Aftershock Migration}
\label{sec:migration_AS}
In terms of location and spatial spread over time, our model's aftershocks are qualitatively compatible with an effect observed in seismology, the so-called ``aftershock migration''.

\begin{figure}[]
\begin{small}
\begin{center}
\def\svgwidth{\textwidth}
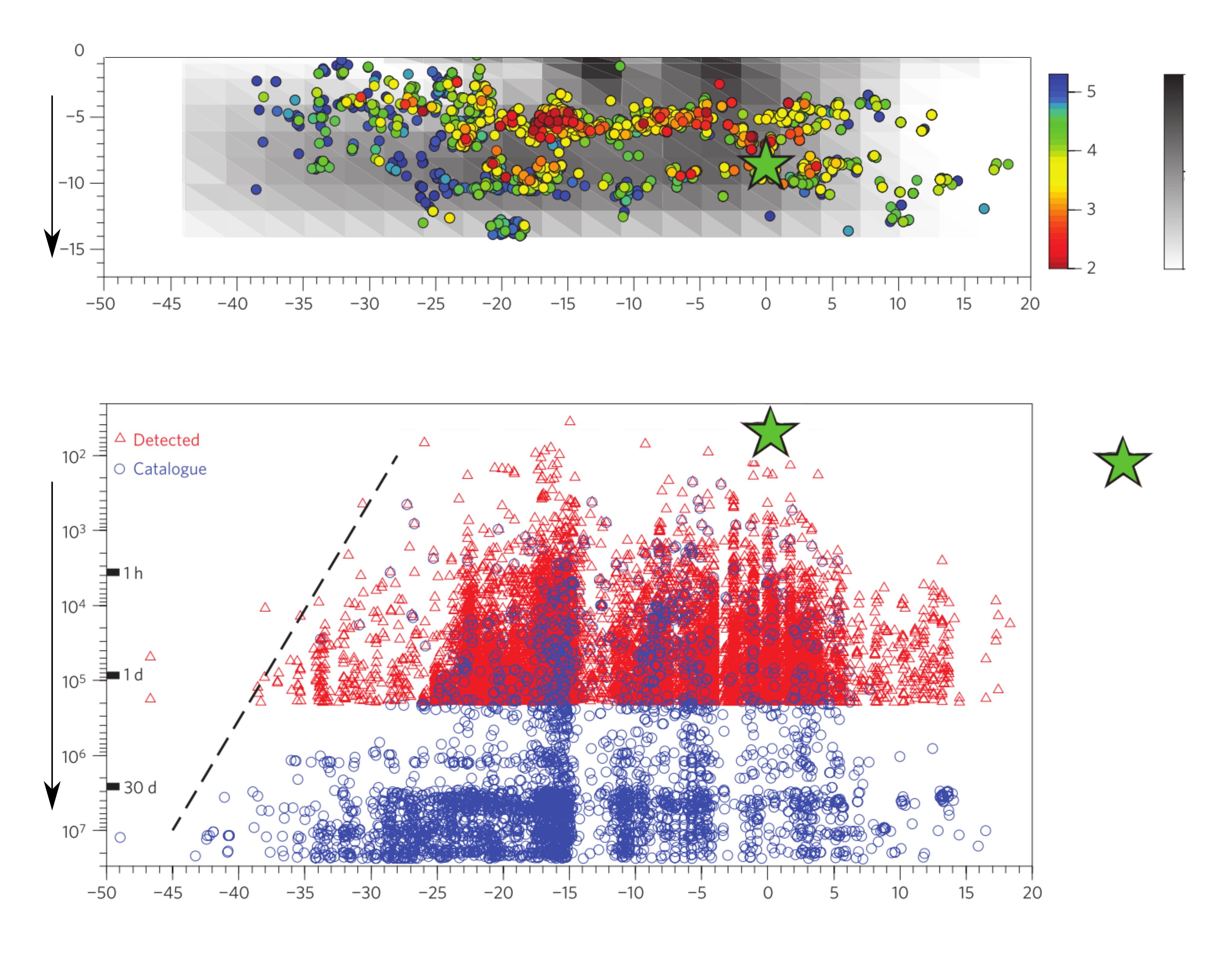
\end{center}
\end{small}
\caption{   {\footnotesize 
Migration of the Parkfield early aftershocks (San Andreas fault), adapted from \cite{Peng2009}. 
Upper panel: location of the events in the fault plane (depth and distance along the fault), with colors indicating the time since main shock.
The grey scale indicates cumulative slip in the first 60 days after the main shock (green star).
\newline
Lower panel: The occurrence times of aftershocks versus the distance along the fault (along-strike distance). 
Blue circles (resp.~red triangles) denote events from a catalogue (resp.~detected via a filtering technique introduced in \cite{Peng2009}).
The black dashed line represents the approximate slope of aftershock migration.
\label{Fig:AS_migration}
}   }
\end{figure}
We report field observations in Fig.~\ref{Fig:AS_migration} (adapted from \cite{Peng2009}) and note that the aftershocks, which correspond to the region of high cumulated slip, spread away from the main shock over time, and more precisely away from the high slip region.
In particular, the boundary of the slip region (which can be measured by cumulated slip or the presence of aftershocks) grows as the logarithm of time, at $\sim 3.4~km$ per time decade.

In a first approximation, we can interpret the aftershock migration effect in terms of our simple viscoelastic interface model. 
Since the region (``slip region'' or region of finite stress drop) where large avalanches occurred has a rather low stress level, ulterior aftershocks are unlikely to be large there, since we need to have $\sigma\sim \sigma_{max}$ to obtain large events.
On the border of the slip region, however, slow relaxation processes trigger aftershocks, which may be large since there are stocks of stress in those neighbourhoods\footnote{
Far away from the affected region, large earthquakes are unlikely: seismic waves can remotely trigger earthquakes, but in a marginal way compared to the aftershock migration effect.
Furthermore, remote triggering via seismic waves can not induce an aftershock spread scaling as $\sim\log t$,  which allows to distinguish it from local effects such as relaxation.
}.
Because the large aftershocks at the border also correspond to further slip and stress drop, they extend the slip area and push the ulterior events further away from the initial shock area.
Thus, the slip area is expected to slowly increase over time, via aftershocks occurring mostly at its border and slowly migrating away from the main shock.

\includefig{\textwidth}{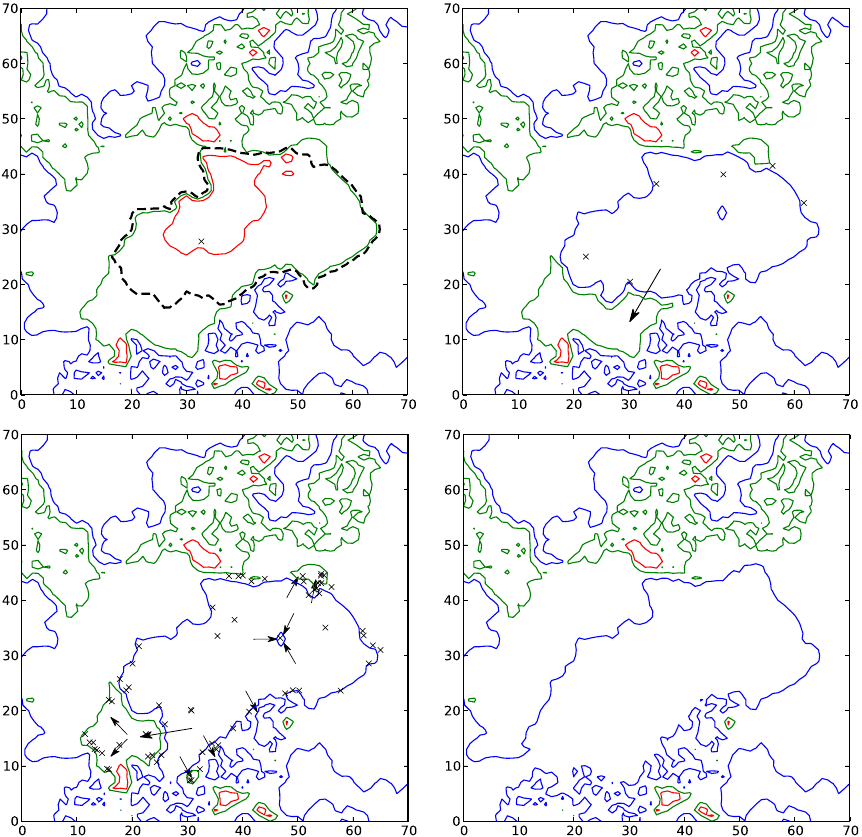}{
Stress map of the viscoelastic interface model in two dimensions. 
Colors indicate stress levels, from high (red, $\sim 2$) to low (blue ,$\sim 1.75$).
A large stress drop corresponds to a large slip.
From left to right and top to bottom: expansion of the affected area is seen to mainly spread (black arrows) around the initial main shock and the subsequent aftershocks (small crosses indicate avalanches' epicentres with $S>5000$). 
The dashed line highlights the initially unstable region (main shock).
Affected regions have low chance to witness new large events, due to the low value of the local stress. 
The simulation was performed using $k_0= 0.012, k_1=0, k_2=1$, and spacings $z$ uniformly distributed, between $0$ and $0.2$. 
The total system size is $15000 \times 15000$: for each elementary surface unit, the local stress was computed by averaging over a square of $100 \times 100$ elemental sites of the discrete system.
\label{Fig:BrokenZone}
}
In Fig.~\ref{Fig:BrokenZone}, we present the whole cluster of aftershocks issued from a single large main shock of our 2D model (see also Fig.~\ref{figBrokenZone} and the Fig.~\ref{Fig:BrokenZone_complementary} of the Appendix for other representations of these data).
We observe that small aftershocks (not indicated) are rather uniformly distributed inside the slip region, while the epicentres of the large ones typically occur at the border, extending the slip region.
As some large areas of high stress can only slip after some small events connect them to the slip region, the growth of the affected area is rather slow.
We conclude that the agreement of the model with experiments is deeper than a simple coincidence, but leave the quantitative comparison for future work\footnote{Actually, since the present model does not reproduce the Omori law of decay of aftershocks over time, we already know that quantitative agreement is out of reach.
However, in the model of viscoelastic interface with Laplacian relaxation, we observe a power-law decay of activity over time, so that quantitative agreement with Omori law is possible.}.

Note that the use of huge system sizes is not a luxury (up to $15000 \times 15000$ sites, running on a single CPU).
As aftershocks spread over the system, the cluster area (area of all the aftershocks belonging to a given cluster) can become several times the size of the largest single event.
Since we do not want to ``feel'' the finite size of the system, we need this cluster area to be much smaller than the system size.
If we want to produce large events and respect this constraint, we typically need very large system sizes.

\subsubsection{Aftershocks Decay Over Time: the Omori Law}
\label{sec:Aftershocks_Law}

A more widely known law about the evolution of aftershocks over time is the Omori law.
Essentially, it states that the number of aftershocks related to a main event decreases as a power-law of time, after a short transient (see sec.~\ref{sec:Omori} for details).
This is compatible with the migration of aftershocks being logarithmic in time (considering aftershock triggering as a local process).

\includefig{12cm}{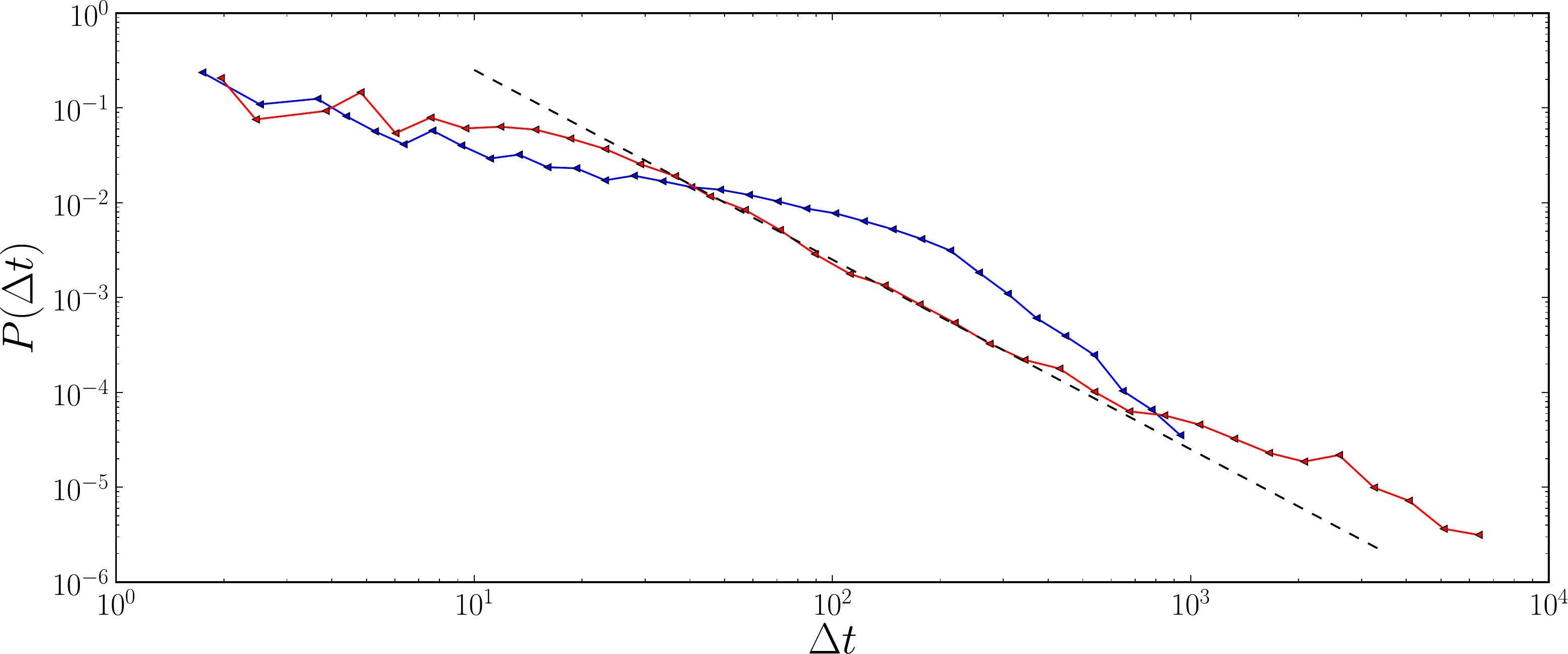}{
Aftershocks decay over time: plot of the density of probability $P(\Delta t)$ for the time since main shock $\Delta t$. 
The model with Laplacian relaxation (red line) has a power-law decay of the aftershocks rate with an exponent $\sim 2$ (dashed line).
The local relaxation model 
 (blue line) has an exponential decay of aftershocks rate, incompatible with the Omori law. 
\label{Fig:Omori}
}
One may have noticed that in the our viscoelastic model, the relaxation of the variable $u_i$ is local and controlled by a single time constant.
This choice yields an unrealistic exponential decay of the aftershocks production rate over time. 
In this respect, it is suitable to consider non-local relaxation mechanisms, as the Laplacian relaxation presented in \req{viscoDetaille3_globalVar3}, which can reproduce the Omori law.
In Fig.~\ref{Fig:Omori}, we compare the decay of the aftershock production rate for the tow models (issued from \reqq{viscoDetaille3} and \reqq{viscoDetaille3_globalVar3}), and observe a  power-law decay for the Laplacian model.
The comparison of the rates of aftershocks production of the two models was recently performed in \cite{Jagla2014-08}. 
For now it is enough to note that simple models for viscoelastic interfaces can reproduce a  power-law decay of aftershocks over time, in qualitative agreement with the observed Omori law.

If one is looking for more complete, detailed models that would reproduce more faithfully the seismic phenomenology, it should prove interesting to consider a variant of our model where the time scale $\tau_u$ would be replaced with a distribution of them, so as to account for the variability in the time scales of relaxation of the various rocks present in the crust. 
This could be done for instance by replacing the SLS (Standard Linear Solid) type of interaction with a Generalized Maxwell model type (also called ``Maxwell-Wiechert model'').
In our work, we focused on building very simple models, in order to be able to extract very general properties, so that this prospect is left for future work.

\subsection{The Gutenberg-Richter Exponent}

As we explained in sec.~\ref{sec:Gutenberg-Richter}, an important feature of earthquakes is the Gutenberg-Richter (GR) law which characterizes the  magnitude-frequency  distribution of seismic events.
The probability for a randomly selected earthquake to be of magnitude $M$ is given by $f(M)\sim 10^ {-bM}$, where $b$ is the GR exponent.
As we discussed in  sec.~\ref{sec:Gutenberg-Richter}, $b$ is found to lie in the range $b \in [0.75,1.25]$.
We note that the total moment (or energy released) in a seismic event corresponds to the size $S$ of an avalanche in our model.
For historical reasons, in seismology the magnitude $M$ of an event is related to the total moment $S$ via $M=6 + (2/3) \log_{10} S $.
This gives an expected exponent for the avalanche size distribution $\tau = 1+ (2/3)b \in[1.5,1.83]$, with the central value $b \simeq 1$ corresponding to $\tau \simeq 1.7$.

\begin{figure}[]
\begin{small}
\begin{center}
\def\svgwidth{\textwidth}
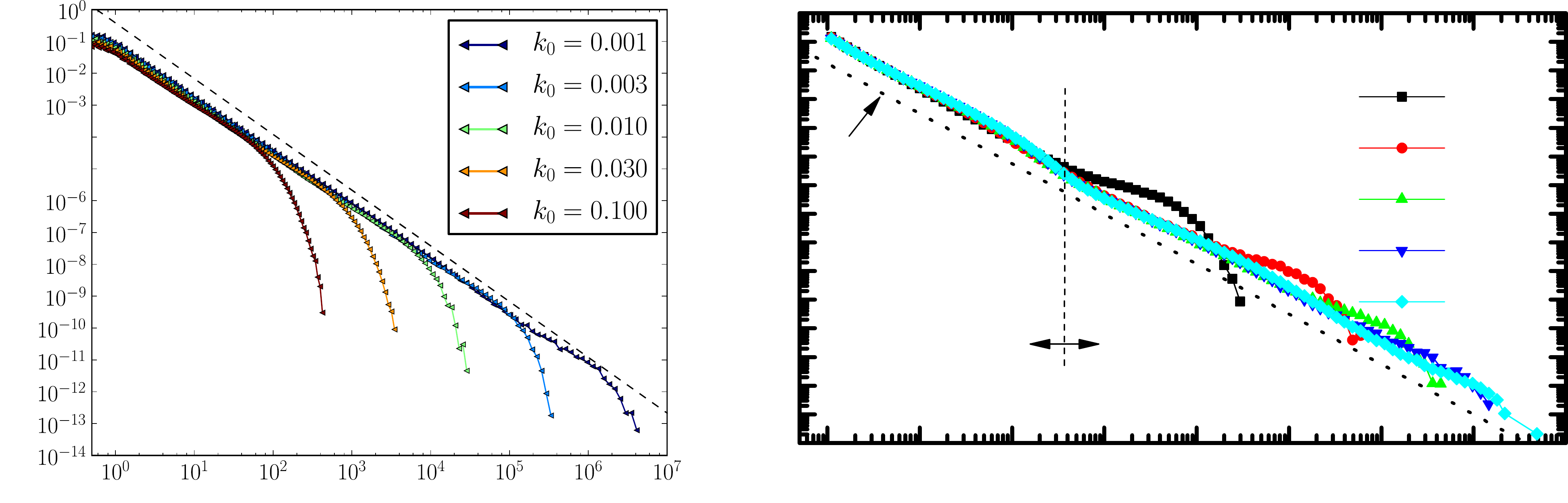
\end{center}
\end{small}
\caption{   {\footnotesize   
Number $N(S)$ of avalanches of size $S$ 
 (not normalized but proportional to $P(S)$) for the two-dimensional viscoelastic interface model.
The dashed lines indicate the pure power-law with exponent $1.75$.
We used $\overline{z}=0.1$ and $f_i^\text{th}$ distributed as a Gaussian with unit variance.
\newline
Left:
We used $k_1=k_2=0.05$. 
The system size is $5000 \times 5000$ and events with $S<0.5$ are not shown.
\newline
Right: 
we use $k_1=0, k_2=1$. The system size is $15000\times 15000$ and events with $S<2$ are not shown.
\label{Fig:exposant_tau_1p75}
}   }
\end{figure}

This expected behaviour is very well compatible with that of the viscoelastic interface model, which displays a power-law decaying distribution $P(S)$ in all the range that we have been able to explore (i.e.~at least over the range $[1, 10^7]$), with an anomalous exponent $\tau \simeq 1.7-1.8$ (see Fig.~\ref{Fig:exposant_tau_1p75}).
This is quite remarkable since in all conventional avalanche models like depinning or directed percolation, this exponent is always smaller than $3/2$, since the mean field value is also the upper bound for the exponent $\tau$ \cite{LeDoussal2009a, Dobrinevski2012}. 
In particular in the 2D depinning case we measured $\tau \simeq 1.27$ (see also \cite{Rosso2009}), which is clearly incompatible with the range observed  experimentally. 
We also note that the depinning mean field behaviour does not convincingly account for earthquakes, as the value $1.5$ is at the edge of the acceptable interval.

The search for ``the right'' exponent may sometimes appear as the ultimate goal, a proof of adequacy of a model with reality.
However, we must remember that the value of $b$ is not very well measured and is subject to intense discussions: whether it is truly universal or subject to regional variations remains an open question.
Given the large variations of $b$, finding a value within the acceptable range is not a conclusive finding. 
Furthermore, since we do not consider realistic long-range elastic interactions, the coincidence has to be taken with caution.

\section{Other Contexts with Viscoelastic-like Effects}
\label{sec:other_contexts}

The generic features of our model are elasticity, disorder, external forcing and plasticity (via the viscoelastic relaxation).
There are actually various situations where extended, slowly driven disordered systems which present some form of memory (either viscoelasticity, relaxation, several degrees of freedom per lattice site, etc.), such as crystal plasticity at slow strain rates \cite{Papanikolaou2012}, amorphous plasticity at slow shear rates \cite{Martens2012}, slowly sheared granular materials \cite{Ben-Zion2003,Dahmen2011}, or seismic faults \cite{Jagla2010}.
We now look for the common points in the definitions of the models describing those examples.

In the following analysis, one should remember that in the depinning transition, the disorder competes with the elastic interactions, so that the decrease of the strength in one is equivalent to the increase of the strength in the other.
For instance, a decrease in the yielding thresholds (pinning force) amounts to the same thing as an increase of the stress, since only the difference $\sigma_i - f_i^ \text{th}$ matters.

In our model \cite{Jagla2014a}, during an avalanche the interface is rigid (with stiffness $k_1+k_2$), while after the avalanche the stiffness decreases to $k_1$, thanks to the relaxation of the viscoelastic elements. 
A primary effect is to enhance the development of avalanches relatively to avalanche triggering, since a stiffer interface corresponds to lower stress thresholds\footnote{
In the model of Marchetti et.~al., in the quasi-static driving limit  -- which is actually not explored in \cite{Marchetti2000} -- during an avalanche the interface would be rigid (with stiffness $\mu$), while after the avalanche its stiffness would decrease to $0$. 
}.
Despite enhancing the growth of already existing avalanches, viscoelasticity is also associated to numerous small avalanches (larger exponent $\tau$).
More generally, the overall dynamics resulting from this primary effect is quite complex, and was the subject of this chapter, so that we now focus only on the origins of this ``primary'' effect.

In the OFCR model for seismic faults \cite{Jagla2010, Jagla2010a}, the relaxation between two events can decrease the local stress $\sigma_i$, so that it is more difficult to trigger an avalanche, but as relaxation of the stress is forgotten during the event (as new stress thresholds are drawn at random), an event which is already started has better chances to be maintained. 
Thus, the effect is qualitatively similar to that in our model.

In the ``avalanche oscillator'' model \cite{Papanikolaou2012}, there is a slow creep (forward motion) of the field $h$ in between events.
This change of $h$ in the inter-avalanches periods corresponds to\footnote{
A change in $h$ also corresponds to some variations in $\sigma$ through the long-range elastic kernel, however the average of that change over the whole system is zero.
} a decrease of $\sigma$, via the term $-k_0 h$.
During the driving phases, this decrease of $\sigma$ tends to inhibit the triggering of new avalanches, whereas during an avalanche the stress is unaffected by the creep and is thus typically higher.
Once again, the effect is qualitatively similar to that in our model.

We now present three other examples of models with similar forms of relaxation.

\paragraph{Elastic Interface Model with Stress Overshoots}

An interesting model of a modified elastic interface embedded in a disordered medium is studied in \cite{Schwarz2003}.
There, the focus is on the effects of inertia and elastic waves, in particular ``stress overshoots'', in which the motion of one region of the interface induces a temporary extra stress on the neighbouring regions, in addition to the static stress.
Precisely, consider the model of the elastic depinning, discretized on a regular lattice: when a site jumps, each neighbour gets an extra stress increase (of value $\propto M$) which lasts for a single time step.
In the memory kernel approach, 
 in the continuum limit, this essentially translates into:
\begin{align}
C(t)=1+ M \delta^D(t),
\end{align}
where $M$ is the amplitude of the overshoots, and $\delta^D $ is the Dirac distribution.

For large enough system sizes, they find a single critical force $F_c(M)$, independent of the previous history of the system, and an hysteresis cycle which seems to vanish in the limit of very large systems.
They also find that for values of $M$ smaller than a critical value $M_c$, the universality class is that of the elastic depinning.
The paper also contains a discussion about earthquakes and some of the features that we observed, as the seismic cycle or the possibility of very large events 
(with sizes much larger than those expected in purely elastic depinning).
The study is however limited to numerical simulations in two dimensions, in the constant force setup, with a focus on the depinned phase ($F>F_c, v>0$).
Furthermore, the largest system sizes used are of $256 \times 256$ (to be compared to $15000 \times 15000$ in our numerical scheme), which limits the accuracy of the quantitative results.

We note that a primary effect of the overshoots is to enhance the development of avalanches relatively to avalanche triggering.
There, this enhancement is directly put in an ad hoc way, and this primary effect is qualitatively similar to that in our model.

\paragraph{Granular Materials with Dynamic Weakening}

\begin{figure}[]
\begin{small}
\begin{center}
\def\svgwidth{8cm}
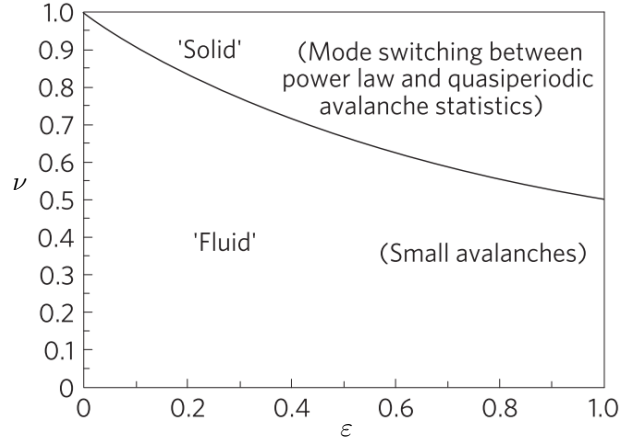
\end{center}
\end{small}
\caption{   {\footnotesize 
From \cite{Dahmen2011}.
Phase diagram of the model for granular materials with relaxation.
At low volume fractions $\nu$ or low relaxation parameter $\varepsilon$, the system is in a fluid phase, with only small avalanches.
At fixed $\varepsilon$, when the volume fraction approaches the critical one, these small avalanches become distributed as power-laws.
In the ``solid'' regime, there are spontaneous switches between a regime with small avalanches distributed as power-laws and a regime with system-size periodic events.
The switching is rare enough so that numerous periodic events can be observed in each periodic sequence.
\label{Fig:Dahmen2011}
}   }
\end{figure}
A model for granular materials with some degree of relaxation (or \textit{Dynamic Weakening}) was presented in \cite{Dahmen2011} (see also \cite{Dahmen2009} for the initial definition). 
In this coarse-grained model, each site (much larger than the grain diameter) can be either fully filled with grains or completely empty.
The fraction of sites occupied by grains is denoted $\nu$, which is proportional to the rescaled volume fraction $\Phi/\Phi_{max}$ of the underlying microscopic granular material.
Quasi-static driving is performed by pulling the grains in the sites at the system boundaries, while sites filled with grains interact elastically with their nearest neighbours (or with all of them in mean field).
The crucial peculiarity of the model lies in the friction law for each site.
Initially, sites have \textit{randomly distributed} static `frictional' failure thresholds $\sigma_{s,i}$ (similar to our $f_i^\text{th}$).
When the shear stress $\sigma_i$ exceeds the local threshold $\sigma_{s,i}$, the grains in the site $i$ slip during one time step.
The stress at which they stop is called the ``arrest stress'' $\sigma_{a,i}$.
The slip of the grains on one site can increase the stress over neighbouring sites, thus triggering an avalanche of numerous slips. 
From the first slip (caused by the stress being larger than $\sigma_{s,i}$) and for all the duration of the avalanche involving this slip, the failure threshold for the site $i$ is set at the ``dynamic failure threshold'' value:
\begin{align}
\sigma_{d,i}= \sigma_{s,i} - \varepsilon (\sigma_{s,i}-\sigma_{a,i}),
\end{align}
 where $\varepsilon$ is a \textit{weakening parameter} that quantifies the difference between effective static and dynamic ``friction'' on meso-scales.
Note that $\sigma_{d,i}$ is automatically larger than $\sigma_{a,i}$, so that the weakening of the failure threshold does not immediately trigger a new slip.
At the end of an avalanche, when all sites have their stress  $\sigma_i$ below $\sigma_{d,i}$, the failure thresholds ``heal'' back to their static value,  $\sigma_{s,i} > \sigma_{d,i}$.

We comment the results obtained for the mean field case of this model in Fig.~\ref{Fig:Dahmen2011}.
A similarity with our model is the observation of periodic events, in the ``solid'' phase (where the relaxation or ``weakening'' plays a significant role).
Another similarity is that 
there are no such large events
when we are far from criticality (small $\nu$, or large $k_0$ in our model), 
In this sense, viscoelasticity is relevant only close to criticality. 
In our model, these features can be explained from first principles. 
An important difference with our model is the observation of an avalanche size exponent of $\tau=1.5$ (as in the mean field theory for the purely elastic depinning model), unlike our model for which we observe a strong deviation from $1.5$ (depending on the values of parameters). 
We note that this model for granular materials is inspired from previous models designed initially for seismic faults \cite{Ben-Zion2003, Ben-Zion1996a}, in which the periodic events naturally identify with \textit{characteristic} earthquakes.
All these results may also be compared to molecular dynamics of disordered solids, as performed in \cite{Salerno2013}, where the inclusion of inertia generates similar features (in $2$ and $3$ dimensions, using long-range interactions).
Part of this comparison is performed in \cite{Dahmen2011} itself, where the model is carefully compared to experiments and simulations of the shearing of granular materials.

Once again, we note that a primary effect of the lowering (``weakening'') of the stress thresholds during an avalanche is to enhance the development of avalanches relatively to avalanche triggering.
The main difference with the previous model is that here it is the disorder (stress threshold) that evolves, not the stress itself.

\paragraph{Amorphous plasticity}

We quickly mentioned the problem of amorphous plasticity earlier, in sec.~\ref{sec:AmorphousPlastoc}. 
The well accepted point of view \cite{Barrat2011,Nicolas2013a} is that flow in disordered (amorphous) media is occurring via local plastic events, corresponding to small size rearrangements, that yield a long-range stress redistribution over the system (Eshelby problem).
The model in \cite{Martens2012} introduces a non-trivial value for  the \textit{restructuring time}, the time needed to regain the original structure after a local rearrangement  (the model is strongly inspired by that originally introduced in \cite{Picard2004a, Picard2005}).

The medium is described by a set of elasto-plastic elements that occupy the sites of a square lattice. 
To model a material with a yield stress, under steady shear at fixed strain rate $\dot{\gamma}$, taking into account the long-range effects of the plastic events via a stress propagator $G(r) = \cos(4\theta)/\pi r^2$, they write the evolution of the stress field $ \sigma(r,t)$ as:
\begin{align}
\partial_t \sigma(r,t) = \mu \dot{\gamma} - 2 \mu \int \d r'~G(r-r') \dot{\varepsilon}^{pl}(r',t),
\end{align}
 where $\mu$ is the shear modulus and $\dot{\varepsilon}^{pl}$ accounts for the change in the strain due to local yielding.
The relaxation of the material to the plastic state is controlled by a Maxwellian viscoelastic mechanism.
\begin{align}
\dot{\varepsilon}^{pl}(r,t) = \frac{1}{2\mu \tau}  n(r,t) \sigma(r,t),
\end{align}
where $\tau$ is a relaxation time scale and $ n(r,t)$ is the local activity, i.e.~$ n(r,t)=0$ in absence of plastic events and  $ n(r,t)=1$ if the local region is in the plastic phase.
The transition from the elastic to the plastic state 
is controlled by the condition $\sigma>\sigma^ {th}$ and by a rate $1/\tau_{pl}$.
Once in the plastic state, the transition back to the elastic state is simply controlled by a rate $1/\tau_{el}$. 
This can be summarized by the following expressions for the transition probabilities:
\begin{align}
P_{01}(n_{ij}(t+dt)=1 | n_{ij}(t)=0; \sigma_{ij}>\sigma_y) &= \frac{dt}{\tau_{pl}} \\
P_{10}(n_{ij}(t+dt)=0 | n_{ij}(t)=1) &= \frac{dt}{\tau_{el}}.
\end{align}
In \cite{Martens2012}, the focus is on the dependence of the dynamics on the ratio of these time scales.

Some results of this model can be compared to ours.
In particular, it is found that shear bands form when the strain rate $\dot{\gamma}$ 
 is low enough and the time scale $\tau_{el}$ needed to restore the elastic state is long enough.
The shear bands correspond to a strongly sheared phase with the geometry of a band (which is as long as the system size),  within a solid non flowing environment\footnote{
This qualitatively echoes with the arrays or channels of flowing vortices surrounded by a pinned crystalline structure that are found in superconductors (sec.~\ref{sec:Marchetti}).
}.
These inhomogeneous patterns disappear when the strain rate ($\dot{\gamma} \leftrightarrow V_0$) is large enough or when the elastic relaxation time scale ($\tau_{el} \leftrightarrow \tau_u$) is small enough (compared to the driving time scale $1/\dot{\gamma} \leftrightarrow \overline{z}/V_0 = \tau_D$).
These shear bands observed using a long-range interaction correspond to a collective organization over the whole system,
as for our global events (observed in mean field).
Some analytical results are found in the mean field version of the model \cite{Martens2012}, that we do not discuss here.

In this last model, the stress can overshoot the yielding value ($\sigma_y \leftrightarrow f^ \text{th}$) for a time $\sim \tau_{pl}$, while it takes a time $\sim \tau_{el}$ to come back to a lower stress.
A long relaxation time $\tau_{el}$ allows the stress to remain high for a longer time, thus enhancing the growth of activity (analogous to enhancing avalanche development). 
Once again, the effect is qualitatively similar to that in our model.
An important difference with our model lies in the use of the quadrupolar stress propagator. 
It would be interesting to consider extensions of our model with this kind of long-range anisotropic interactions in the future. 

We noted recently the existence of another model \cite{Papanikolaou2013} for the shearing of disordered solids which also relies on a mechanism of relaxation (``pin delay'') and in which the phenomenology is reminiscent, with stick-slip like shearing and an increased exponent $\tau$. 
For a general discussion on the role of relaxation mechanisms (or softening mechanisms) in the dynamics of amorphous solids, see \cite{Rodney2011}, or \cite{Daub2010} for a focus on granular materials and the shear transformation zone (STZ) idea.

\subsubsection{Common Features Emerging from models with Relaxation}

It is clear that the idea of accounting for plastic creep or one of the related effects (stress overshoots, dynamics weakening, etc.) has recently gained momentum in the literature.
A recurrent property of these various systems is the presence of system-size events or of some collective organization with a correlation length which equals the system size. 
Our analytical method provides an explanation for the origin of this behaviour, with an unprecedented precision. 
Another robust feature that was reported at least in dynamic weakening and explained in our model is the fact that viscoelastic effects are apparent only close to criticality ($\nu$ large enough, or $k_0$ small enough in our model).
Here  our point is simply to notice that many models in different fields share similar features.
However, the precise universal behaviour of these various model 
 has yet to be determined.

We may already make a distinction between two classes in the aforementioned models. 
In granular materials or more generally in amorphous media, the disorder is at least in part \textit{structural} \cite{Barrat2011}, i.e.~it is generated by the internal organization of the system and it evolves under the dynamics.
On the opposite, our model only includes \textit{quenched} disorder. 
The determination of the precise connection between structural and quenched disorder is currently an open problem.
In friction, we expect both the quenched disorder (heterogeneities in the bulk and the surfaces of each solid) and the structural disorder (self-organization of the asperities and contacts, evolving over time) to be relevant.
In the future, it would be interesting to design a model of friction that would account more carefully for the different forms of randomness characterizing the asperities.

\section{Conclusions}
\label{sec:visco_conclu_2}

As we pointed out in the previous chapter, despite some similarities with the frictional context, the model of a purely elastic interface 
 is unable to provide correct predictions concerning frictional behaviour (sec.~\ref{sec:depinning_output_not_friction}, \pp{sec:depinning_output_not_friction}).

We addressed this problem via the inclusion of viscoelastic interactions, which are a natural way to account for the plastic creep occurring at the contacts.
This addition of viscoelasticity is a relevant change, in the sense that the addition of a very small amount of ``visco-'' to the elastic interactions is enough to affect the behaviour, in the macroscopic limit ($k_0 \to 0$).

In mean field, the relaxation of the viscoelastic elements generates a dynamical instability, which we prove to be responsible for the occurrence of periodic system-size events and macroscopic oscillations of the stress. 
The time scale of these oscillations is distinct from the microscopic time scale associated to the ``visco-'' part of the viscoelastic interactions (which is directly introduced in the equations).
Instead, the oscillations are characterized by a new, emerging time scale. 
The emergence of this cycle results from the competition between the slow viscoelastic relaxation and the fast avalanche dynamics: the slow dynamics drives the system towards a critical point, that we prove to be unstable with respect to the fast avalanche dynamics. 
The ensuing state can be characterized as a Non Equilibrium \textit{Non} Stationary State -- as opposed to the Non Equilibrium Stationary States (NESS).

In two dimensions, we performed simulations on systems of tremendous sizes (up to $15000 \times 15000$, and on a single CPU), which allowed to study regimes otherwise hardly accessible.
The global oscillations found in mean field disappear, but are echoed by coherent oscillations of the local stress on finite regions of large sizes. 
In each region, the oscillations of the stress have roughly the same amplitude and period but a different phase, so that at a given time the stress map has a terraced structure, with large plateaus of almost constant stress and macroscopic stress differences between plateaus.
In this sense, the model displays non-stationarity in its two versions: in mean field it has an exactly periodic behaviour and in two dimensions it oscillates on a local scale.

\paragraph{Comparison of our Results with Friction Experiments}
Our results compare well with the three elementary building blocks of the Rate- and State-dependent Friction laws (RSF laws), in particular in the mean field case, which we have explored further than the 2D case. 
First of all, our mean field model reproduces the existence of stick-slip, with an amplitude of the stress oscillations consistent with experimental observations: it decreases with increasing driving velocity ($V_0$) and with increasing driving spring stiffness ($k_0$).
Second, by studying how the kinetic friction force (in the steady-state regime) depends on the driving velocity, we are able to observe and explain the well-known effect of velocity-weakening (logarithmic decrease of friction with increasing velocity).
Third, the response of our model to intermittent driving allows us to reproduce qualitatively and understand an important aspect of the ageing of contacts: we observe the increase of the static friction force with the time of contact at rest. 

The overall results of this chapter indicate that the small plastic events occurring at the contacts between asperities (responsible for the RSF laws) are well captured by a simple model with viscoelastic interactions.
Our work presents these various macroscopic effects not as resulting from the application of some phenomenological law, but as collective phenomena, emerging from micro- and meso-scopic considerations.
A by-product of our work is to extend the range of applicability of the depinning framework and the related tools to the problem of solid friction.

\paragraph{Comparison of our Results with Earthquakes Statistics}

The avalanches of our model reproduce several important features of earthquakes statistics.
Viscoelastic relaxation produces an increase in the exponent of the avalanche size distribution ($\tau $) which matches the worldwide average value given by the \textit{Gutenberg-Richter} law, a feature usually obtained via a fine-tuning of parameters.
The (synthetic) aftershocks are \textit{naturally defined} as by-products of their corresponding main earthquake, as recognized in geophysics: there, aftershocks are defined as secondary earthquakes triggered by a main one, with a time delay that can range from seconds to years.
In a model slightly different from the main one we presented here, the decay of the aftershocks production rate is qualitatively compatible with the \textit{Omori law} known in geophysics (power-law decay).
In the model that was the main subject of our presentation, the aftershocks production rate follows an unrealistic exponential decay.
The spatial correlations of aftershocks are well consistent with the \textit{migration effect} characterizing real seismicity: the epicentres of large aftershocks are located at the boundary of the slip zone of the preceding ``mother'' quakes.
The linear relationship between area and seismic moment we observe matches with the observation of \textit{constant stress drop} that is often reported in geophysics.
Moreover the oscillations of the stress field 
 are the manifestation of the so-called \textit{seismic cycle}, the quasi-periodic occurrence of large earthquakes in some geographical areas (also referred to as \textit{characteristic earthquakes}).

The fact that our model reproduces the essential features of seismic faults indicates a certain robustness of our description of frictional phenomena, since fault dynamics involves more than solid on solid, dry friction.
We can also conclude that the kind of viscoelastic interaction introduced in our model is essential to capture the basic features of seismic dynamics.

\paragraph{Perspectives}
In the first and last sections (sec.~\ref{sec:previous_litt} and sec.~\ref{sec:other_contexts}) we have seen that in various contexts other than friction or seismic faults (superconductors, granular materials, crystalline and amorphous plasticity, etc.), relaxation effects similar to our viscoelastic relaxation can be relevant. 
We propose a formulation of this kind of problems in terms of a well-defined continuous model (built on two equations, for two fields). 
It may be helpful in further attacking the general problem of relaxation mechanisms in driven disordered systems. 
A singular and advantageous feature of our formulation is the deep connection with the problem of purely elastic depinning: the configurations visited by the viscoelastic interface are also metastable configurations of specific elastic interfaces. 
\\

There are several avenues for future work.
A first is to better characterize the ``viscoelastic depinning'' universality class by extracting all the exponents from the different distributions (distribution of avalanche size, area, duration, etc.) and all their scaling relations.
In the next chapter, we will see that we may expect some scaling relations to hold, despite being in a new, larger universality class.
In particular, in mean field we expect to have a new non trivial exponent, which may be predicted from our analytical calculations.

A second point is that some extensions of our model should prove interesting.
For friction of crystalline solids, one could account for the (visco-)elastic interactions of the solid's bulk using long-range elastic interactions, similar to what was done in \cite{Papanikolaou2012}.
For the case of seismic faults, one could account for the amorphous nature of fault gouge by using a long-range and anisotropic kernel of the Eshelby type (\cite{Martens2012} should be inspiring).
More generally, an open problem is to design a model for friction, which would account for the heterogeneous nature of asperities via a quenched disorder \textit{and} for their displacements via some kind of structural disorder, as what was started in \cite{Dahmen2011}.
Yet another modification of the model would be to account for the fact that in friction, the sliding surface is moving parallel to itself. This may be accounted for using an anisotropic disorder correlator.

\chapter{Directed Percolation: a non-Markovian Variant}
\label{chap:DP}

\vspace{-2cm}
\minitoc
\vspace{2cm}

One of the references of avalanche models studied for their critical properties is Directed Percolation (DP). 
DP is the paradigmatic example of dynamical phase transitions into absorbing states (see \cite{Henkel2008, Hinrichsen2006, Odor2004, Hinrichsen2000} for reviews).
It provides an example of robust universality class with well studied critical behaviour, where power-law distributed avalanches are generated.
One of its most remarkable characteristics is its robustness:
numerous different particular models can effectively be described within the DP scenario.
Simple examples of DP processes are given by cellular automata in which ``active sites'' have some probability to activate their neighbours, possibly propagating activity over large distances, for large periods of time.

It has been shown that the critical properties of the DP transition are however lost if the probability to activate a site for the first time is reduced with respect to the subsequent probabilities \cite{Rousseau1997, Jimenez-Dalmaroni2003}. 
We have shown in \cite{Landes2012} that with an appropriate increase of some of the following activation probabilities, criticality can be restored, in a process that we call ``compensation''. 
Here, we review these results and relate them with the problems of elastic and viscoelastic interfaces.

In this chapter, we first study the conventional DP process and compare it with elastic depinning (sec.~\ref{sec:DP_simple}).
Second, inspired by earthquakes models 
we introduce a non-Markovian variation of the DP process.
We prove that it is critical and study its critical properties (sec.~\ref{sec:nonMarkov_DP}), and discuss its relationship with the viscoelastic interface .
We conclude by discussing the impact of our results and some perspectives in sec.~\ref{sec:conclu_DP}.

\section{Pure Directed Percolation}
\label{sec:DP_simple}

In this section we start by defining a simple cellular automaton that belongs to the DP universality class, and show in what respect it is different from the elastic depinning model (sec.~\ref{sec:DP_and_depinning}).
Then we study the critical behaviour of DP: we define the main critical exponents and show their scaling relations (sec.~\ref{sec:DP_critical_behav}).
This allows us to compare the universality class of DP and depinning via the results obtained at criticality (sec.~\ref{sec:comparasionDepinning}).

\subsection{Link Between Elastic Depinning and Directed Percolation}
\label{sec:DP_and_depinning}

\paragraph{Bond Directed Percolation (Bond DP)}

The Bond DP process is defined as a very simple cellular automaton that produces avalanches: here we present a light variation that highlights the similarity with the avalanches of the elastic depinning model.
The local density of activity $\phi(x,t)$ fully describes the state of the realization of the DP process at time $t$.
Consider an infinite square lattice, for example in two dimensions, with initially no active sites\footnote{In general, the initial condition can consist in any number of active sites.}:
\begin{itemize}
\item[1] Pick a site at random and activate it: $\phi_i(1)=1$.
\item[2] For each active site $\phi_i(t)=1$, de-activate it ($\phi_i(t+1)=0$). 
Each of the neighbours can be activated at time $t+1$, independently, with probability $p$ ($\phi_j(t+1)=1$ in case of success). 
\item[3] If there is one active site or more, then go to Step 2. Else, go to Step 1.
\end{itemize}
Note that a site which is activated by more than one neighbour ends up in the same state as a site activated only once: a site can not be ``doubly activated''. 
Usually, DP is defined as above except for the Step 3 which does not send back to Step 1, so that only one avalanche (or \textit{cluster} of connected sites) is produced.
Here we simply let the algorithm repeat indefinitely from the initial condition with no active sites.
A striking difference between this model and depinning is the absence of any interface or field that would evolve during the avalanche and remember its total progression: in DP, only the instantaneous activity matters.
The link between Directed Percolation and (Isotropic) Percolation is explained in Fig.~\ref{Fig:iso_VS_directed}
\includefig{14cm}{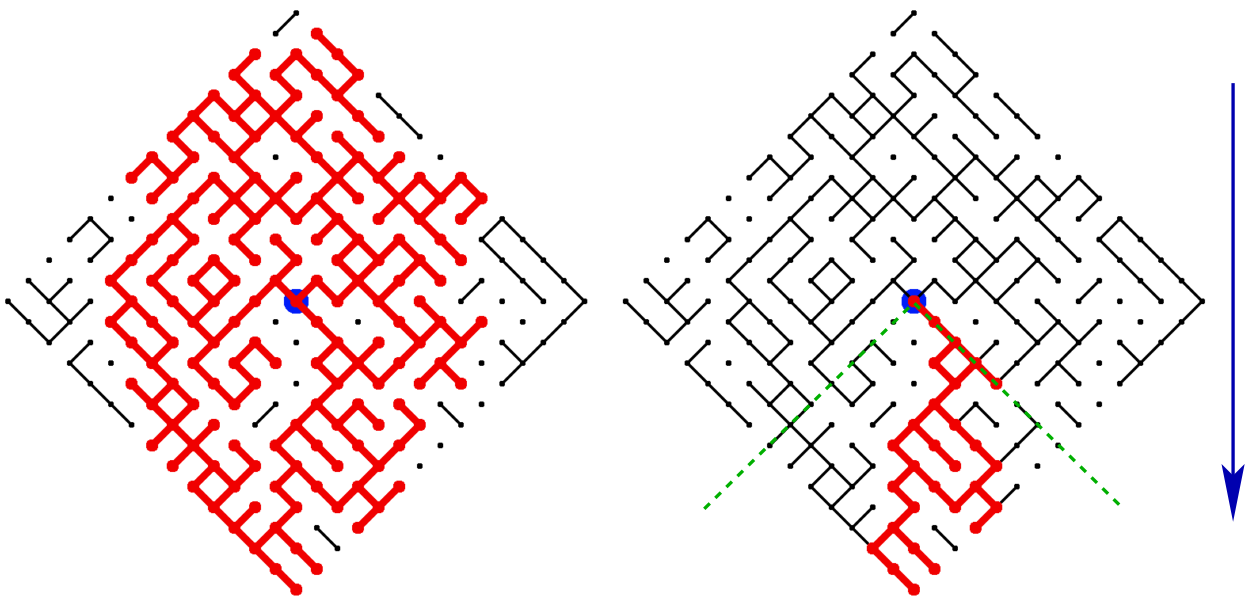}{
From \cite{Hinrichsen2006}.
Isotropic and Directed Percolation. 
Bonds connecting two sites are highlighted in bold black (drawn at random).
The site initially active (seed) is highlighted in blue.
Sites which belong to the connected cluster of active sites are highlighted in red.
\newline
Left: Isotropic Percolation in $d=2$ dimensions.
\newline
Right: Directed Percolation in $d=1+1$ dimensions.
Direction of time is downwards, following the arrow.
Note that the DP cluster is restricted to the ``light-cone'' represented with dashed green lines.
\label{Fig:iso_VS_directed}
}

\paragraph{A Special Case of Elastic Depinning}

Let us recall the continuous equation of motion of the elastic interface using the narrow wells disorder:
\begin{align}
\eta_0 \partial_t h_i = k_0( V_0 t - h_i) + k_1 (\nabla^2 h)_i - f_i^ \text{th}.
\end{align}
Using the narrow wells disorder, the quasi-static dynamics stipulates that a site is active if the local stress $\sigma_i \equiv  k_0( V_0 t - h_i) + k_1 (\nabla^2 h)_i$ is larger than its random threshold $ f_i^ \text{th}$.

We want to give a mapping between depinning and a \textit{modified} DP model.
Let us assume that the narrow wells are equally spaced (by one unit length) and that the threshold forces $f_i^ \text{th}$ are exponentially distributed:
\begin{align}
\mathbb{P}(f_i^ \text{th}=X)\d X =\lambda  e^{-\lambda X} \d X, \qquad g(z)=\delta^{Dirac}(z-1),
\end{align}
where $g(z)$ is the spacings distribution of the ``narrow wells'' introduced in sec.~\ref{sec:narrow_wells}. 
We will see that the choice of the exponential distribution is crucial.
Let us suppose that the site $i$ is inactive and receives some additional stress $\varepsilon$, so that its stress goes from $\sigma_i$ to $\sigma_i+\varepsilon$.
Since the site is inactive, we already know that $f_i^ \text{th} > \sigma_i $.
The probability that it remains inactive is:
\begin{align}
\mathbb{P}(f_i^ \text{th}>\sigma_i+\varepsilon| f_i^ \text{th} > \sigma_i) 
&= \frac{ \mathbb{P}(f_i^ \text{th}>\sigma_i+\varepsilon) }{\mathbb{P}( f_i^ \text{th} > \sigma_i)}, \qquad \sigma_i+\varepsilon>\sigma_i \\
&= \frac{ \int_{\sigma_i+\varepsilon}^\infty \lambda e^{-\lambda s} \d s}{\int_{\sigma_i}^\infty \lambda e^{-\lambda s} \d s} \\
&=  e^{-\lambda \varepsilon} \\
&= \mathbb{P}(f_i^ \text{th}> \varepsilon ).
\end{align}
This result is independent of the value of $\sigma_i$, i.e.~the site does not remember the increases of stress it witnessed.
This memory-less property is characteristic of the exponential distribution.
We can verify the consistency of the scheme via an example: the probability of not being activated under an increase $2\varepsilon$ of the stress should be the product of not being activated by two consecutive increases by $\varepsilon$.
This is indeed the case, since $e^{-2\lambda\varepsilon}=e^{-\lambda\varepsilon}e^{-\lambda\varepsilon}$.

When a site is activated, its stress decreases to a new value, $\sigma_i$, and its threshold is drawn at random again.
If the new threshold $f_i^ \text{th}$ is lower than the new stress, the sites is active again (it self-activates).
This happens with a probability $p_\text{self}=\mathbb{P}(f_i^ \text{th} < \sigma_i) = \int_0^{\sigma_i} \lambda e^{-\lambda s}\d s = 1-e^{-\lambda \sigma_i}$.
Note that $p_\text{self}$ does depend on $\sigma_i$.

\paragraph{Dynamical Rules of the Special Model}
We can now describe the interface dynamics in a probabilistic setup, i.e.~without drawing the thresholds $f_i^ \text{th} $ in advance.
Starting from an initially flat\footnote{
We can also start from any random distribution with finite first and second moments, the two dynamics quickly collapse onto the same one, for a finite system.
}  $d$-dimensional interface discretized on a square lattice with $N$ sites, $ \{ h_i(0)=\sigma_i(0)=0, \forall i \in[1,N] \} $, the quasi-static dynamics ($V_0=0^ +$) reads:
\begin{itemize}
\item[1] Increase time until one site is active.
We do not know the position of the thresholds in advance, so the stress increase $(\Delta \sigma) = k_0 V_0 \Delta t$ needed to activate one site is a random variable distributed as:
\begin{align}
 \mathbb{P}(\Delta \sigma) 
 &= 1- \mathbb{P}(\text{no site is activated up to time }\Delta t) \\
 &= 1- (e^{-\lambda (\Delta \sigma)})^N.
\end{align}
Since all sites are equivalent, the activated site can be picked at random.
Declare the site active and increase all $\sigma_i$'s by $(\Delta \sigma)$.
\item[2] For all active sites, increase $h_i$ by one (decrease $\sigma_i$ by $k_0+2d k_1$) and draw a new threshold $f_i^ \text{th}$. 
The $2d$ neighbouring $\sigma_j$'s are all increased by $k_1$ (due to the term $k_1 \nabla^2 h$). 
Each of them is thus activated independently\footnote{If a site has several active neighbours, each one of them successively tries to activate it, each with the same probability.} with probability $p=1 - e^{-\lambda k_1}$.
The site $i$ re-activates itself with probability $p_\text{self}=1-e^{-\lambda \sigma_i}$. 
\item[3] If there is one active site or more, then go to Step 2. Else, go to Step 1. 
\end{itemize}
This algorithm is actually independent from the field $h$: it can be followed by simply following the update rules for the $\sigma_i$'s, as was the case in the OFC* model (which is equivalent to elastic depinning, see sec.~\ref{sec:OFC*}, \pp{sec:OFC*}).

\paragraph{Link with a Modified DP Process}
The link with directed percolation is twofold.
First, if we artificially set the probability of self-activation $p_\text{self}$ to zero and identify the constant $p=1 - e^{-\lambda k_1}$ with the activation probability of DP, we exactly obtain the bond DP process.
If we artificially set $p_\text{self}$ to a constant non-zero value, 
we also obtain the DP universality class: the critical value $p_c$ to obtain criticality will change, but not the exponents.

Second, the above algorithm defines a modified DP process, associated to a field $\sigma$ (it is DP-like in the sense that it evolves only by random activations). 
An interface $h$ can be associated to this process by demanding that $h_i$ advances by $1$ whenever the site $i$ becomes active: the link with the density of activity $\phi$ is then given by  $\phi = \partial_t h$.

Note that if we associate a field $h$ (using $\phi = \partial_t h$) to the usual DP process, the Langevin equation for this $h$ will \textit{not} be that of the elastic depinning.
For completeness, we now provide the Langevin equation for the DP universality class.

\paragraph{Langevin Equation for the DP}
\label{sec:langevin_4_DP}

The DP process can be described by a Langevin equation for the activity density $\phi$ \cite{Jimenez-Dalmaroni2003, Hinrichsen2000}:
\begin{align}
\partial_t \phi = D \nabla^2 \phi + r \phi - \frac{1}{2} u \phi^2 +  \xi(x,t), 
\label{Eq:DP_langevin}
\end{align}
where the noise $\xi$ is Gaussian, has zero average $ \overline{\xi(x,t)} = 0$ and is delta correlated: $ \overline{\xi(x,t)\xi(x',t')} = \Gamma \phi~\delta(x-x') \delta(t-t')$.
Note that this is a multiplicative noise, since activity can not be generated from an inactive region.
The distance to criticality is controlled by $r \propto p_c-p$.
The key ingredient in the DP process lies in the non-linear term $- \frac{1}{2} u   \phi^2$, which  makes the local density saturate: it is the continuous translation of the prescription that a single site cannot be ``doubly activated''.

The Langevin equation \reqq{DP_langevin} can be reformulated into the so-called \textit{Reggeon field theory}, which is the most practical way to perform a renormalization group analysis of the DP universality class.
For references on the Reggeon field theory, see the reviews \cite{Hinrichsen2000,Hinrichsen2006}.
For an interesting application of this field theory to a modified DP model, see the excellent \cite{Jimenez-Dalmaroni2003}.

\paragraph{Link with the NDCF Universality Class}
\label{NDCF}

Here, our argument focused mostly on showing that DP and elastic depinning are in different universality classes. 
In \cite{Alava2001}, it is shown heuristically that the minimal Langevin equation for the NDCF class (Non Diffusive Conserved Field class), given by:
\begin{align}
\partial_t \phi &= D \nabla^2 \phi + r \phi - \frac{1}{2} u \phi^2 +  \xi(x,t)  - \mu \psi \phi \\
\partial_t \psi &= D' \nabla^2 \phi
\end{align}
can be mapped onto a model of interface depinning (and vice-versa).
The NDCF class corresponds to the DP model with an additional field $\psi$, and corresponds to the ``modified DP''  process with a non-trivial $p_\text{self}$ that we defined above.
However, the path followed in \cite{Alava2001} is more rigorous than ours, as two mappings are provided: one from a microscopic model in the depinning class to the Langevin equation for the NDCF class, and one from a microscopic model in the NDCF class to the Langevin equation for the depinning class.
Furthermore, a table of exponents for these classes and related models is provided, along with a detailed physical interpretation of all the terms of the Langevin equations.

In this chapter we are interested in a different modification of the pure DP process, which does not map to depinning.
 For additional details on the relation between NDCF and depinning, we refer to \cite{Alava2001}.

\subsection{Critical Behaviour of Directed Percolation}\label{DP}
\label{sec:DP_critical_behav}

In its discrete version, DP is a dynamical model defined on a lattice, where each site is associated with a state (active or inactive, $\phi_i=0$ or $1$) that evolves in time. 
We have presented the bond DP process up to here.
A commonly considered variant is site DP,
in which 
a site on the lattice will be active at time $t+1$ with probability $p$ if at least one of its neighbours is active at time $t$.
In bond DP, a site will be active at time $t+1$ with probability $1-(1-p)^k$, $k$ being the number of its active neighbours at time $t$. 
The configuration with no active sites is called an \textit{absorbing state} because once it is reached, the dynamics stops.
In DP, the absorbing state is unique, it is $\{ \phi_i=0, \forall i\} $.

\includefig{15cm}{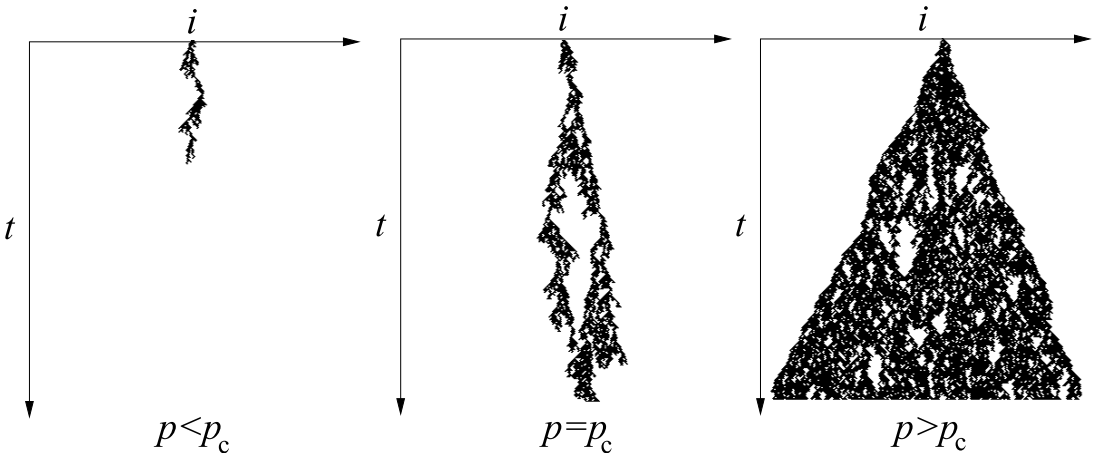}{
From \cite{Hinrichsen2006}.
Realizations of the $1+1$-dimensional DP process below, at, and above threshold.
\label{Fig:oneD_DP_simu}
}
For a small $p$ the system is trapped in the absorbing state exponentially fast, while for large $p$ the system has a finite probability to remain active indefinitely (see Fig.~\ref{Fig:oneD_DP_simu}).
There exists a threshold $p_c$ at which the system is critical, and in which the surviving probability decays algebraically with time. 
Around the threshold $p_c$ the system displays a non equilibrium phase transition from a fluctuating phase to the absorbing state.
As for standard equilibrium phase transitions, universal behaviour and critical exponents are expected.
It was found that both site and bond DP belong to the same universality class:
here, we focus on bond DP on a two dimensional square lattice, for which $p_c \simeq 0.287338$ \cite{Dickman1999}.

\paragraph{Order Parameter}
As $p$ is the control parameter of the transition, we denote the distance from criticality as $\Delta \equiv |p-p_c|$.  
Two different order parameters can be defined, depending on the initial condition.
When  the initial condition corresponds to a fully active lattice the relevant question is to determine the density of active sites when $t \to \infty$ (the stationary state), namely $\rho_{\text{st}}$. For $p<p_c$,  $\rho_{\text{st}} =0$, for $p>p_c$, $\rho_{\text{st}} =\Delta^\beta$. 
When the initial condition corresponds to a lattice with a single active site (the ``seed''), a \textit{cluster} of active sites spreads from it. 
Here the relevant question is to determine  the probability to remain out of the absorbing state  when $t \to \infty$, namely $Q_{\text{st}}$. For $p<p_c$,  $Q_{\text{st}} =0$, for $p>p_c$, $Q_{\text{st}} =\Delta^{\beta^\prime}$.

\paragraph{Correlation Length}
Similarly to the case of equilibrium phase transitions, when approaching criticality,  a diverging length $\xi_\perp \sim \Delta^ {-\nu_\perp}$ describes the spatial correlations. 
In dynamical phase transitions, there is also a characteristic scale for time correlations, $\xi_\parallel \sim \Delta^ {-\nu_\parallel}$. These scales are independent of the observable and thus of the initial condition, while one expects the two distinct order parameters $\rho_{\text{st}}$ and $Q_{\text{st}}$ to be characterized by different\footnote{
More precisely the field theory of absorbing phase transitions shows \cite{Grassberger1983c} that the density exponent 
$\beta$ is associated with the annihilation operator while the survival exponent $\beta^ \prime$ is associated with the creation operator.
} 
exponents $\beta$ and $\beta^ \prime$.
We will see that other quantities display power-law behaviour with different critical exponents, however it is possible to write scaling relations that constrain the set of critical exponents to only four  independent quantities.

\subsubsection{Observables and Scaling Relations}

In practice, in numerical simulations it  is convenient to start from the single seed initial condition and let the cluster evolve up to time $t$. 
To characterize the growth of spreading clusters, one measures the survival probability $Q(t)$ and the average\footnote{
Decent statistics are obtained via the averaging of numerous realizations of the process, but we do not explicitly write $\langle Q \rangle, \langle N \rangle,$ etc., as we ought to.
} 
number of active sites at time $t$, $N(t)$. 
These two quantities obey the scaling forms:
\begin{eqnarray}
Q(t) & \propto &t^{-\delta}g_1(t/\xi_\parallel)  \label{Q1}\\
N(t)& \propto &t^{\eta}g_2(t/\xi_\parallel)
\end{eqnarray}
where $g_1$ and $g_2$ are $1$ at  $t=0$, and 
$g_i(x)\to 0$ for $x \to \infty$,
below threshold. 
When we consider \textit{surviving clusters only}, we can measure the average spatial extension of the cluster at time $t$, namely $L^d(t)$, and the average density $\rho(t)$ of active sites at time $t$ inside this region. These two quantities obey the scaling forms:
\begin{eqnarray}
\rho(t) & \propto &t^{-\theta}g_3(t/\xi_\parallel)  \\
L(t)& \propto &t^{1/z}g_4(t/\xi_\parallel)
\end{eqnarray}
where $g_3$ and $g_4$ behave similarly to  $g_1$ and $g_2$ below threshold.

\paragraph{Scaling Relations From Above the Threshold}
Above threshold, both $Q(t)$ and $\rho(t)$  approach their asymptotic stationary state, $Q_{\text{st}}$ and  $\rho_{\text{st}}$, at a characteristic time $\sim \xi_\parallel$, so that two scaling relations can be written:
\begin{eqnarray}
\beta= \theta \nu_\parallel  \\
\beta'= \delta \nu_\parallel 
\end{eqnarray}
At the  critical point the scale invariance predicts that if time is rescaled by a factor $b$, space should be rescaled by a factor 
$b^{\nu_\perp / \nu_\parallel}$. Thus the size of a cluster grows as $L(t) \sim t ^ {\nu_\perp / \nu_\parallel}$ and a third scaling relation can be written:
\begin{equation}
 z=\frac{\nu_\parallel }{ \nu_\perp}
\end{equation}

A \textit{generalized hyperscaling relation} \cite{Mendes1994a} valid below the upper critical dimension \cite{Hinrichsen2000} relates the four quantities previously defined.
Namely $N(t)$ can be expressed as the sum of two contributions: the active sites of surviving clusters ($\sim \rho(t) L^d(t)$) which have probability $Q(t)$, and the contribution of dead clusters.
 This reads:
\begin{eqnarray}
N(t) &=& L^d(t) \rho(t) \cdot Q(t) + 0 \cdot (1-Q(t))   \nonumber\\
\eta &=& \frac{d}{z} -\theta -\delta. \label{hypers}
\end{eqnarray}

\paragraph{Scaling Relations From Below the Threshold}

Below threshold, each cluster can be identified with an avalanche and dies in a finite time $T$.  
We define the size $S$ of an avalanche as the total number of activations that occurred, and are mainly interested in its statistics,
 $P(S)$, which is expected to follow a power-law at criticality: $P(S) \sim S^{-\tau}$. 
The characteristic size of an avalanche is related to $T$ through  \cite{Dickman1999}:
\begin{equation}
\label{LL}
S(T) \sim \int_0^T \frac{N(t)}{Q(t)} \d t \sim T^{1+ \eta + \delta }
\end{equation}
Assuming that fluctuations around this characteristic value are small, we can write $P(S) \, \d S \sim - Q^\prime(T)  \,\d T$
where $- Q^\prime(T) \sim T^{-\delta-1}$  stands for the rate of death. Combining the latter relation with Eq.(\ref{LL}) we have
$P(S) \sim T^{-(1+\eta+2\delta)}
\sim S^{- \left( \frac{1+\eta+2\delta}{1+\eta+\delta}  \right) }$, and a scaling relation for the exponent $\tau$ can thus be written:
\begin{equation}
 \tau = \frac{1+\eta+2\delta}{1+\eta+\delta} = 1 + \frac{\delta}{1+\eta+\delta}. \label{tau}
\end{equation}

\includefig{12cm}{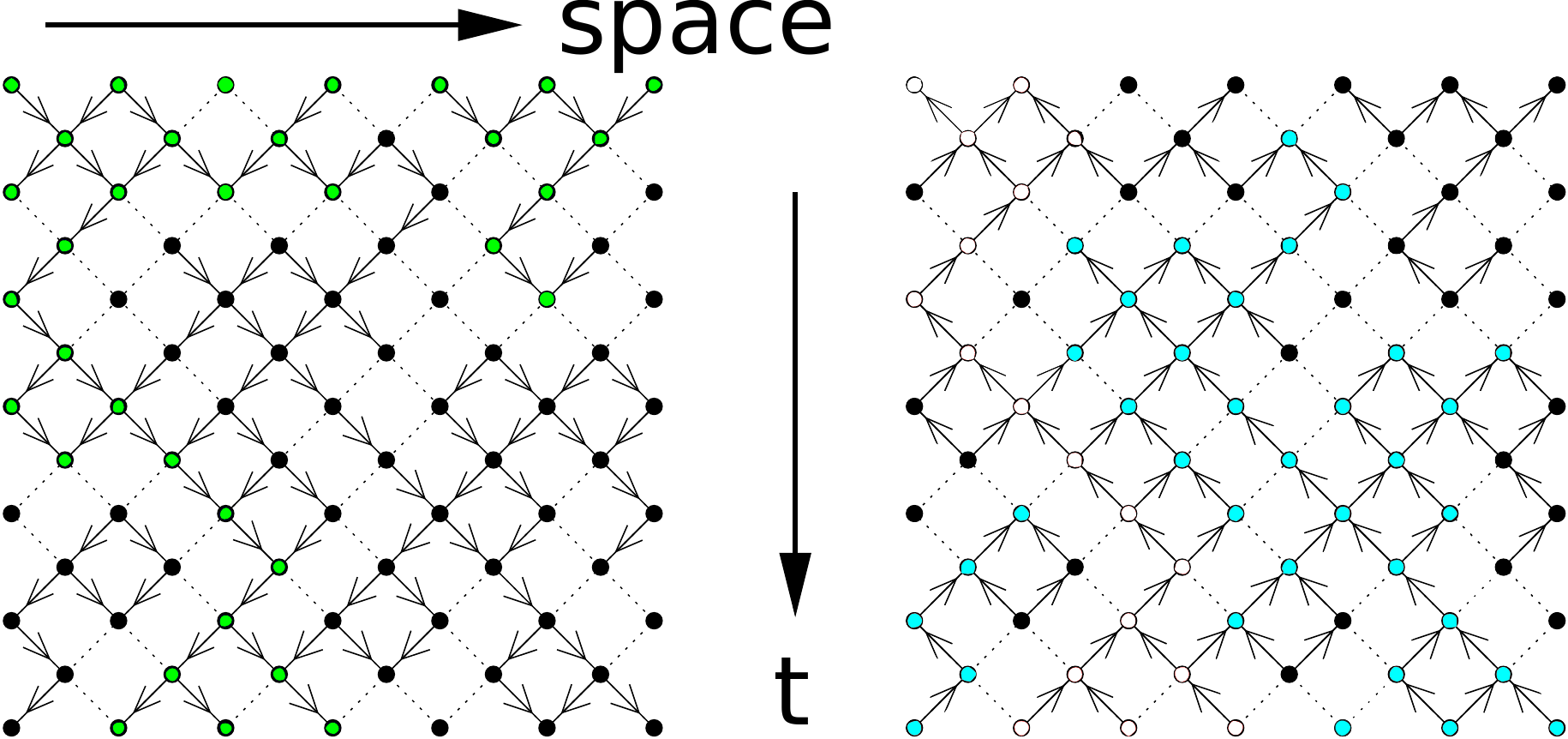}{
$1$-dimensional bond DP. Normal direction of time is downwards. The arrows are given once and for all and are the same for both panels. Final time is $t=12$.
Left: DP starting with a fully active lattice: $M=7$ occupied green seeds, with final density $\rho(t) = \frac{3}{7}$. 
\newline
Right: DP with time reversed, starting from a single seed (seven times). 
In light blue, the paths which die before the end. In open circles, the paths that survive until $t$. There are exactly $m=3$ seeds that participate in surviving walks: $Q(t) = \frac{3}{7}$.
\label{timeR1}
}

\paragraph{Bulk and Spreading Exponents}

The exponents and relations that we introduced here are general features of all absorbing phase transitions, which are characterized by only four independent exponents: $\delta, \theta, z$ and $\nu_\parallel$. 
The exponents $\beta, \theta, \nu_\parallel,\nu_\perp $ and $z$ are called ``bulk exponents'' because they can be measured both from the fully active initial condition and from the single seed initial condition (with averages performed over surviving runs exclusively).
The exponents $\beta^{\prime}, \delta, \eta$ and $\tau$  are called ``spreading exponents'' because they are measured  starting from a single seed, with averages performed over all runs.

\paragraph{Time Reversal Symmetry: an Additional Scaling Relation}

DP has an additional symmetry associated with time reversal, which implies that $\theta=\delta$ \cite{Grassberger1979, Hinrichsen2000}.
This is schematically proved in Fig.~\ref{timeR1} for $1$-dimensional bond DP, where arrows, drawn with probability $p$, connect neighbouring sites.
An activated site at the start of an arrow activates the site at the end of the arrow.
The key observation is that the direction of time is arbitrary: starting from the top 
is equivalent to starting from the bottom with reversed arrows. The survival probability $Q(t)$ with fully active initial condition (with normal direction of time) is exactly equal to the density $\rho(t)$ with single seed initial condition in reversed time.
This relation thus reads:
\begin{eqnarray}
Q(t) &=&\rho(t) \\
\delta &=& \theta.
\end{eqnarray}
The relation is exact for bond DP, while in general $Q(t) \sim \rho(t)$, thanks to the universality of DP.
A necessary condition for this time-reversal symmetry is the uniqueness of the absorbing state: in a process with multiple absorbing states or ageing, one cannot freely reverse the arrow of time.

\paragraph{Exponents: Numerical Values in Two Dimensions}

We recall two-dimensional DP exponents precisely measured in numerical simulations  \cite{Dickman1999}:
\begin{eqnarray}
\label{pureDP}
\delta=\theta= 0.4505 \pm 0.001  &\;& z=1.766 \pm 0.002  \nonumber\\
\nu_\parallel=   1.295\pm 0.006      &\;&   \eta= 0.2295 \pm 0.001.
\end{eqnarray}
From Eq.~(\ref{tau}), we compute $\tau=1.268$, very close to the depinning 2D value, $\tau_{depinning}=1.27\pm0.01$, but different because the two models belong to different universality classes.

\subsection{Comparison with Depinning}
\label{sec:comparasionDepinning}

\paragraph{Successful Identifications}
We first recall two scaling relations and provide a few additional ones that can easily be deduced: 
\begin{align}
\text{DP: }\quad \xi_\perp= \vert\Delta \vert ^{-\nu_\perp},\qquad
  \xi_\parallel =\vert \Delta \vert ^{-\nu_\parallel},\qquad
  \xi_\parallel \sim \xi_\perp^z,\qquad
  S_m^{DP} = \xi_\perp^d \xi_\parallel \rho \sim \Delta^{-(d\nu_\perp + \nu_\parallel -\beta)},
\end{align}
where $\Delta=\vert p-p_c\vert$ and $S_m^{DP}$ is the cutoff for the large avalanches, defined only when $p<p_c$.

We quickly recall a few scaling relations from the elastic depinning model.
We denote $\Delta=F-F_c$ the distance to criticality.
\begin{align}
\text{Depinning: }\quad \xi=\vert \Delta \vert^{-\nu}, \qquad
&T_m =\vert \Delta \vert^{-\nu z}, \qquad
T_m = \xi^ z, \qquad 
S_m = \xi^ {d+\zeta}, \qquad \\
v &\sim \Delta^{\beta_{el}},\qquad
\beta_{el} = \nu (z-\zeta), 
\end{align}
where $\beta_{el}$ is the $\beta$ exponent of the \textit{el}astic depinning, and $T_m, S_m$ are the cutoffs for the distributions of the avalanches duration and size, respectively.

There are certain quantities that are straightforwardly identified.
For instance, the correlation length $\xi$ of the depinning can be associated to the correlation length of  DP: $\xi \rightarrow \xi_\perp$.
We write down all these associations:
\begin{align}
\nu \rightarrow \nu_\perp, \qquad
\xi \rightarrow \xi_\perp, \qquad
T_m \rightarrow \xi_\parallel,\qquad
z \rightarrow z,\qquad
S_m \rightarrow S_m^{DP}, \qquad
\zeta \rightarrow \frac{\nu_\parallel - \beta }{\nu_\perp},
\label{Eq:zeta_DP}
\end{align}
where the last one comes from the identifications of the $S_m$'s.
If we now inject this $\zeta$ in the depinning relation $\beta_{el} = \nu (z-\zeta)$ and use the previous associations, we get:
\begin{align}
\beta_{el} 
= \nu (z-\zeta) 
= \nu\lp z- \frac{\nu_\parallel - \beta }{\nu_\perp} \rp 
= \nu \lp z -z - \frac{\beta}{\nu_\perp} \rp
=\beta.
\end{align}
In this sense, we may identify $v(t) \rightarrow \rho(t)$, which is perfectly consistent with the analysis provided in sec.~\ref{sec:langevin_4_DP}, \pp{sec:langevin_4_DP}, where we identified $\partial_t h_i$ with the local activity $\phi_i$.

We can derive more associations by considering observables derived from the ones introduced above.
For instance the expression for the avalanche duration exponent $\tau_T = 1 + (d+\zeta-2)/z$ of the depinning is consistent with the expression for the survival exponent $\delta$ that can be found using the associations above.
However providing an extensive list is not our aim here.
See \cite{Paczuski1996a} for an early review of avalanches model and a comparison of their universality classes.

\subsubsection{Differences Between the Models}

\paragraph{The Statistical Tilt Symmetry (STS) }
A first important difference is the violation of the STS, $\nu = 1/(2-\zeta)$,  in the DP model.
We can compute $\zeta$ ``for the DP'' from \reqq{zeta_DP}: in two dimensions, $\zeta_{DP} \approx 0.97$. 
As $\nu_\perp \approx 0.733 \neq 0.97 \approx 1/(2-\zeta_{DP})$.
This is to be expected, since the STS relation lies on the assumption of a quadratic Hamiltonian, i.e.~an interaction linear in $h$. 
The non linear term $-\frac{1}{2}u\phi^2$ of the DP clearly violates this assumption.

\paragraph{Susceptibility: a Scaling Relation for $\tau$}
In the elastic depinning as in DP, by definition $P(S)\sim S^{-\tau} g(S/S_m)$, so that  $\langle S \rangle \sim  S_m^{2-\tau}$. 
However, the relation to the susceptibility, $\langle S \rangle = \chi \sim \Delta^{-(1+\nu \zeta)}$,  holds only for the elastic depinning, and yields:
\begin{align}
\tau = 2 - \frac{\zeta +1/\nu}{d +\zeta}.
\label{Eq:tau_DP_XX}
\end{align}
As there is no notion of driving in DP, there is no susceptibility either.
Thus, $\langle S \rangle$ is not controlled as some average response would be, and \reqq{tau_DP_XX} is violated.
Injecting the appropriate numbers from the associations made above into \reqq{tau_DP_XX}, we find $\tau \approx 1.21$, quite far from its actual value $\tau \approx 1.265$.

We must conclude that although depinning and DP are two avalanche models with a few similarities, they are \textit{not} in the same universality class.
Furthermore, re-interpretations of scaling arguments from one model to the other proves unsuccessful.

\paragraph{Time Reversal Symmetry}

Conversely, the equivalent of time reversal symmetry, valid in DP, is violated in depinning.
This is due to the presence of the field $h$, which acts as a memory kernel for the interface (or for the ``modified DP'' defined earlier). 
Since the depinning is defined by more than its instantaneous velocity (or activity, $\phi \leftrightarrow \partial_t h)$), its dynamics is not symmetric by time reversal.

\section{A Non-Markovian Variation of DP}
\label{sec:nonMarkov_DP}

\includefig{9cm}{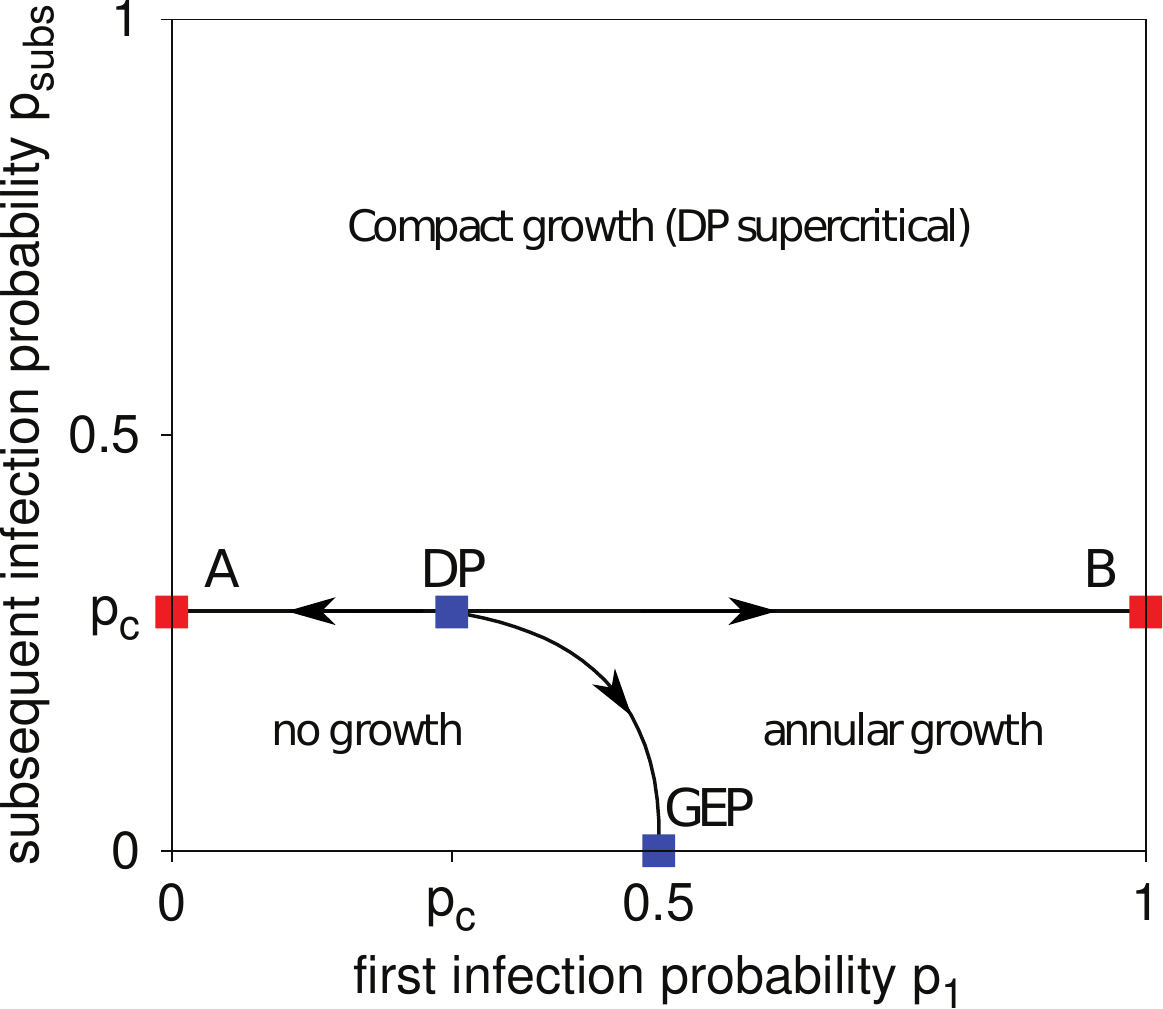}{
Adapted from \cite{Rousseau1997}.
Phase diagram of the model with a first infection probability $p_1$ different from the subsequent reinfection probability $p_{\text{subs}}$ (the IM).
Arrows indicate the RG-flow. 
\label{phased}
}
A generalization of the bond DP process is the modified first Infection Model (IM) \cite{Rousseau1997,Jimenez-Dalmaroni2003, VanWijland2002, Dammer2003}. In this variant,
the probability to activate a site for the first time is given a value $p_1$ different from the value of the subsequent activations (that we call $p_{\text{subs}}$). 
This has been considered as a model to describe epidemic processes with partial immunization.
In this context, the activation of a site is called {\em infection}, and it is understood that the possibility of the subsequent reinfection probability $p_{\text{subs}}$ can differ from the first infection probability $p_1$ due to ``immunization'' effects. 
The question of the relevance of this change was a debate for some time, but was finally settled  \cite{Rousseau1997,Jimenez-Dalmaroni2003} and the phase diagram of this problem in $d=2$ is the one reproduced in Fig.~\ref{phased}.

DP critical behaviour occurs at $p_1=p_{\text{subs}}=p_c$: at this point, $Q(t)$, $N(t)$, $L^2(t)$ and $\rho(t)$ have power-law distributions corresponding to pure DP.
In the curved line terminated in the points ``DP'' and ``GEP'', the system experiences a phase transition corresponding to the so called General Epidemic Process (GEP). 
The fixed point of (bond) GEP is located at $p_1=1/2$, $p_{\text{subs}}=0$ and corresponds exactly to the problem of bond Isotropic Percolation.

Along the AB line (Fig.~\ref{phased}), except for the unstable DP fixed point, the system is not critical. 
In particular, the surviving probability $Q(t)$ 
and the size distribution of the avalanches $P(S)$ 
decays faster than a power-law.
The instability of the DP fixed point was shown in \cite{Jimenez-Dalmaroni2003}: the renormalization flow takes one from any point in AB (outside the DP point) to either A or B.

\subsection{First Attempt Model}

Instead of the case in which there are different probabilities for the first infections, here we focus on the case in which different probabilities occur for successive {\em attempts}, namely irrespective if the activation of the site actually occurred or not. 
The state is defined by the number of \textit{trials} of activation, not the number of infections.
We will refer to this variant as the Attempt Model (AM), to distinguish it from the  Infection Model (IM) previously described.
The AM is a sort of milder modification of the original DP problem, compared to the IM. 
We expect the phase diagram of the AM to be qualitatively similar to that of the IM.

In particular, the DP fixed point is clearly located at the same position,
while the GEP point is slightly different. As we stated before, for the IM the GEP point corresponds to two-dimensional \textit{bond} Isotropic Percolation ($p_1=0.5$, $p_{\text{subs}}=0$). Instead, for the AM it corresponds to two-dimensional \textit{site} Isotropic Percolation ($p_1\simeq 0.592746$, $p_{\text{subs}}=0$).
Indeed, we observe that for the AM, when $p_{\text{subs}}=0$, a site can be activated only at the very first attempt, with probability $p_1$ (no matter if we consider site or bond DP), 
thus the sites that are activated once with this rule are exactly the sites activated in $d$-dimensional site Isotropic Percolation.

The main difference between AM and IM is that the AM has a non-singular limiting behaviour as $p_1\to 0$, leading in particular to a finite mean event size $\langle S \rangle$ in this limit, whereas 
for the IM $\langle S \rangle$ goes to $0$ as $p_1\to 0$.

\paragraph{Motivation for the AM: link with the Viscoelastic Interface Model}

We start by referring to the simpler case of the OFCR model.
During an avalanche 
of the OFCR model, the bulk dynamics is unaffected by the relaxation process: once a block jumped during an avalanche, its behaviour is fully controlled by its new stress and stress threshold.

However, the spreading of an avalanche is modified by relaxation.
Under relaxation ($\partial_t \sigma_i = k_0 V_0 + R \nabla^ 2 \sigma _i$), a block $i$ can see its stress either decrease or increase.
When the stress increases under relaxation, the block becomes closer to its activation threshold and the probability for a neighbouring active site to activate it is increased.
Reciprocally, a decrease in the stress due to relaxation decreases the activation probability for this site.
When a spreading avalanche encounters for the first time a site that relaxed for some time, the probability to activate it is thus different from the bulk one.

This probability evolves under the activation attempts:
when a relaxed site has an active neighbour, it receives an additional amount of stress $k_1$ (or $\alpha$) which takes it closer to its threshold.
Under a few attempts of activation by neighbours, the relaxed site should thus either be activated, or have the same probability of activation as any other site.
Note that in this respect, the simple fact that a neighbouring site attempts to activate a site increases its activation probability, independently of the success of this activation. 
This explains our motivation to study the Attempt Model rather than the IM.

The parallel between the AM and the viscoelastic model is the same, the only difficulty is to understand that the bulk activation probability is independent of the precise value of the auxiliary field $u$.

\subsection{Criticality Recovered with Compensation}
\label{results}

Our main point here is to show that for the AM the lack of criticality generated by a value of the first attempt $p_1$ smaller (resp.~larger) than $p_c$ 
can be  ``compensated''
by a second attempt probability $p_2$ larger (resp.~smaller) than $p_c$.  
We will present strong numerical evidence showing how this compensation occurs, restoring critical behaviour in the system\footnote{
We will not discuss the possibility of compensation in the IM since we cannot be conclusive at present. 
Although it seems that compensation can be obtained, numerical evidence is not enough for a discussion on the variation or not of the obtained critical exponents.
}.

In addition, a remarkable result is that at compensation, several critical exponents of the problem, 
in particular the bulk exponents $\theta, z$ and $\nu_\parallel$, take their normal DP values, 
while the spreading exponents ($\delta, \eta, \tau$) depend on the precise values of $p_1$ and $p_2$.

We consider the case in which the first two \textit{attempts} $p_1$ and $p_2$ differ from the subsequent ones, 
that from now on we consider to be equal to the critical DP value: $p_{i>2} \equiv p_{\text{subs}}=p_c=0.287338$. 

\paragraph{Heuristic Argument for the Compensation}

A heuristic argument suggesting that such a compensation can result in criticality is the following.
As a perturbation, the relevant character of a change in  $p_1$ was demonstrated in \cite{Jimenez-Dalmaroni2003} for the IM.
The analysis presented there indicates that a change in $p_2$ generates qualitatively the same kind of perturbation 
(to leading order) 
than a different $p_1$. 
Therefore, it is not surprising that there are particular combinations of $p_1$ and $p_2$ at which the leading term of both perturbations cancel each other. 
These particular combinations will be the compensating pairs of values $(p_1,p_2)$. However, the fact that we do not recover the pure DP exponents indicates that higher order terms do not vanish, but result in a marginal perturbation.

\subsubsection{Numerical Evidence for Compensation and Variable Critical Exponents}

We present first the numerical evidence of the compensation effect. 
In all simulations, we start from a single active site (seed) a time $t=0$, which is in a state of being attempted twice, and let the clusters grow until time $t=10^6$, or their natural death. 
The lattice is large enough so that the boundaries are never reached by the cluster.
To be very precise about our choices: a site that has been successfully infected at the first attempt is still in a state of being attempted just once.

\includefigtwo{0.48\textwidth}{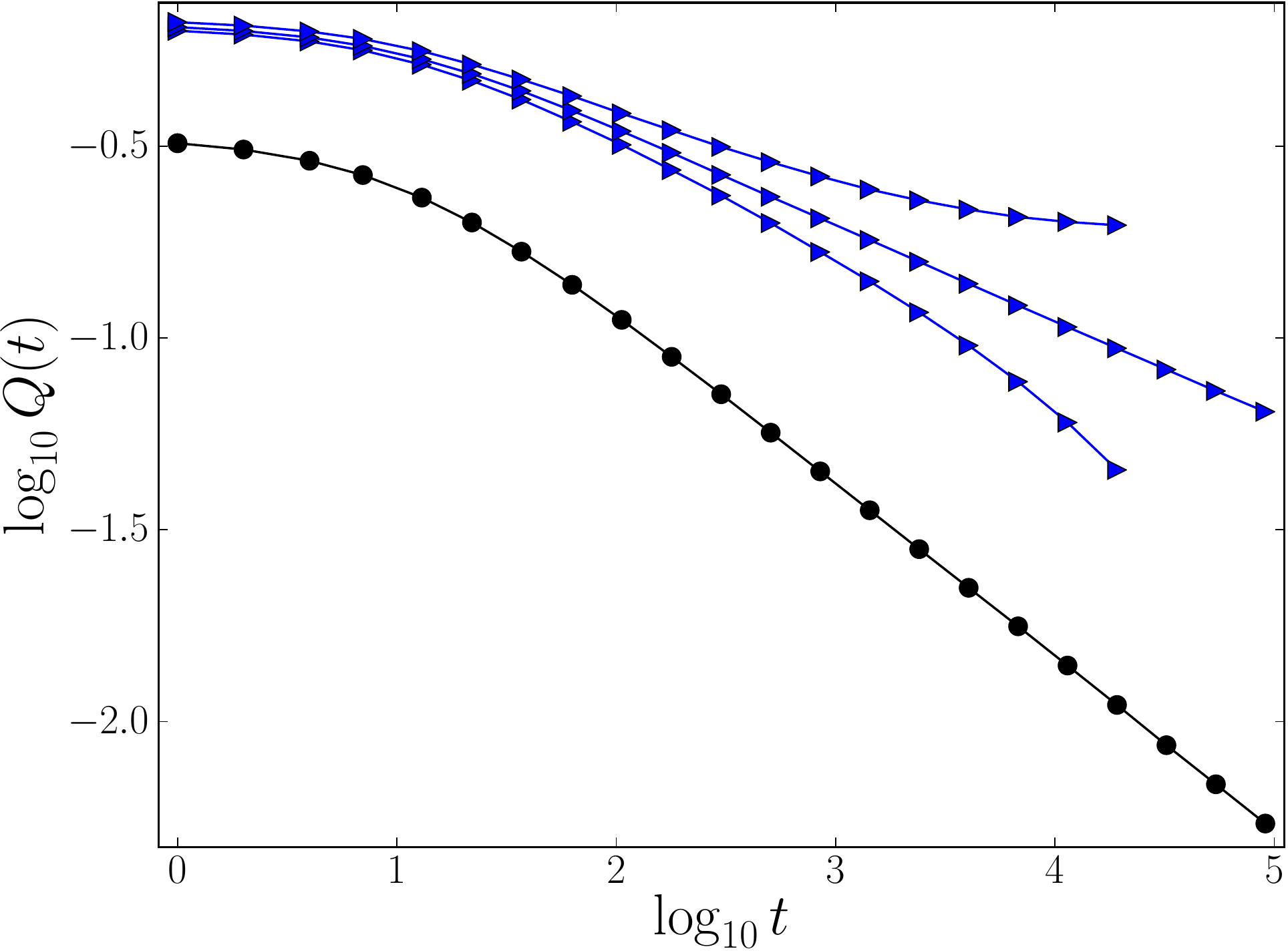}{0.48\textwidth}{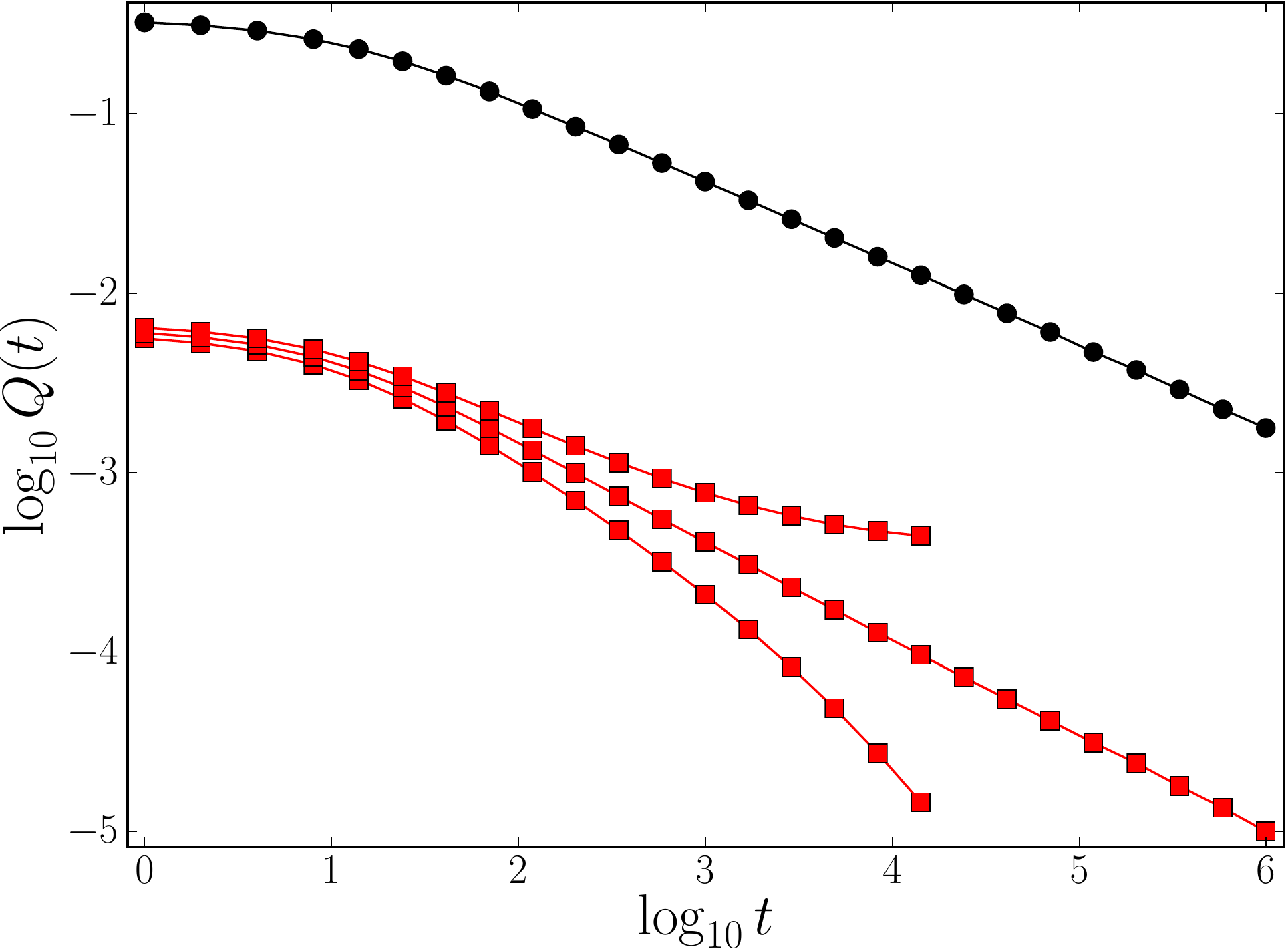}{
$Q(t)$ for different choices of $(p_1,p_2)$. 
Black circles represent the pure DP at $p_1=p_2=p_{\text{subs}}=p_c$ with a power-law exponent $\delta=0.4505$ (averages performed over $10^6$ samples). 
\newline
Left: Triangles represent the AM with $p_2=0, p_{\text{subs}}=p_c$. 
From top to bottom, we used $p_1 = 0.494, 0.4888, 0.485$. 
For $p_1=0.4888$, $Q(t)$ displays a clear power-law with $\delta=0.25 \pm 0.01$ (averages performed over $10^5$ samples). 
\newline
Right: Squares represent the AM with $p_1=0.01, p_{\text{subs}}=p_c$. 
From top to bottom, we used $p_2 =0.62, 0.60, 0.58 $. 
For $p_1=0.600$, $Q(t)$ displays a clear power-law with $\delta=0.53 \pm 0.01$ (averages performed over $10^8$ samples). 
\label{qt2}
}

We investigated two pairs of compensating points $(p_1,p_2)$ and compared with usual DP (in which $p_1=p_2=p_{\text{subs}}=p_c$).  
For the first one, we set $p_2=0$ and varied $p_1$ in order to find the critical point. 
In Fig.~\ref{qt2} (left), we show a few results for different values of $p_1$. A careful study around the point $p_1=0.4888$ shows that we recover the critical character of the surviving probability at ($p_1=0.4888 \pm 0.0005, p_2=0$).
The critical exponent $\delta$ measured at the compensation point ($\delta = 0.25 \pm 0.01$) is different from that at DP. 

The second compensation point is searched by setting $p_1=0.01$ and varying $p_2$.
In Fig.~\ref{qt2} (right), we show the critical character of the point $(p_1=0.01, p_2=0.6000 \pm 0.0005)$. As for the previous point, this level of precision on the location of the critical point was obtained from a careful numerical study.
Similarly we find a new value for $\delta$: $0.53\pm 0.01$.

Given the width of the range of times explored, we rule out the possibility of a simple crossover between a pseudo-critical behaviour and a non-critical behaviour that might exist at long times. 
As mentioned above, a Renormalization Group analysis (such as that performed in \cite{Jimenez-Dalmaroni2003}) would allow to solve this issue once and for all.

\paragraph{Critical Behaviour of the Bulk Observables}

Let us present the critical behaviour of the quantities related to the bulk exponents, $ \theta, z, \nu_\parallel$. 
$L(t)$ corresponds to the mean cluster width averaged over runs that survive until time $t$.
In Fig.~\ref{rhot} (left) we compare our data at two compensation points and at the  DP point:
we notice that the $z$ exponent does not change, unlike the coefficient before the power-law.

In Fig.~\ref{rhot} (right), $\rho(t)$ corresponds to the mean density averaged over runs that survived until $t$. 
The density  of a single run is measured as the ratio of the number of active sites at $t$ over the number of sites that were activated at least once.
Again, one may notice in Fig.~\ref{rhot} (right) that the exponent $\theta$ remains unchanged between the different critical points.

\includefigtwo{0.46\textwidth}{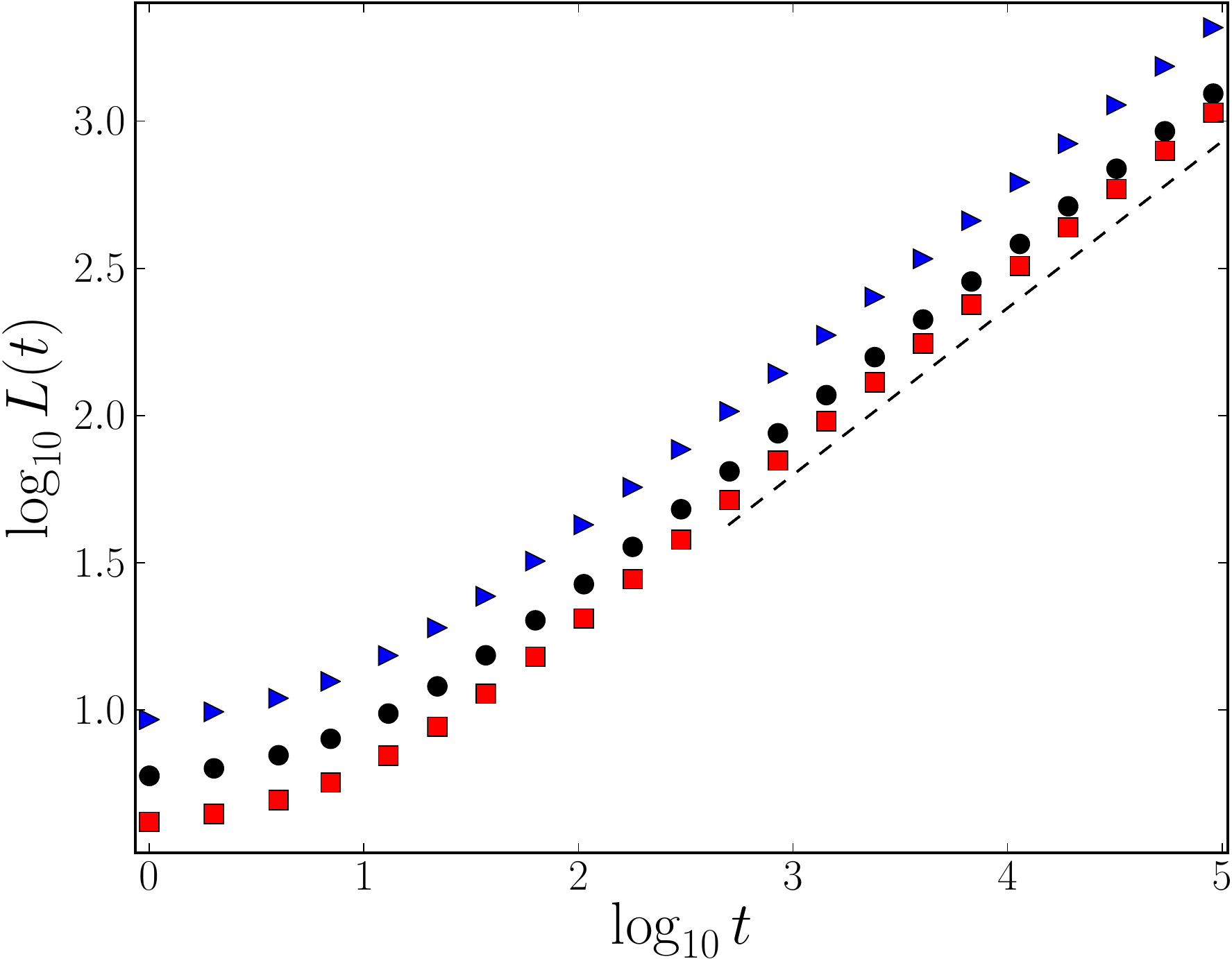}{0.485\textwidth}{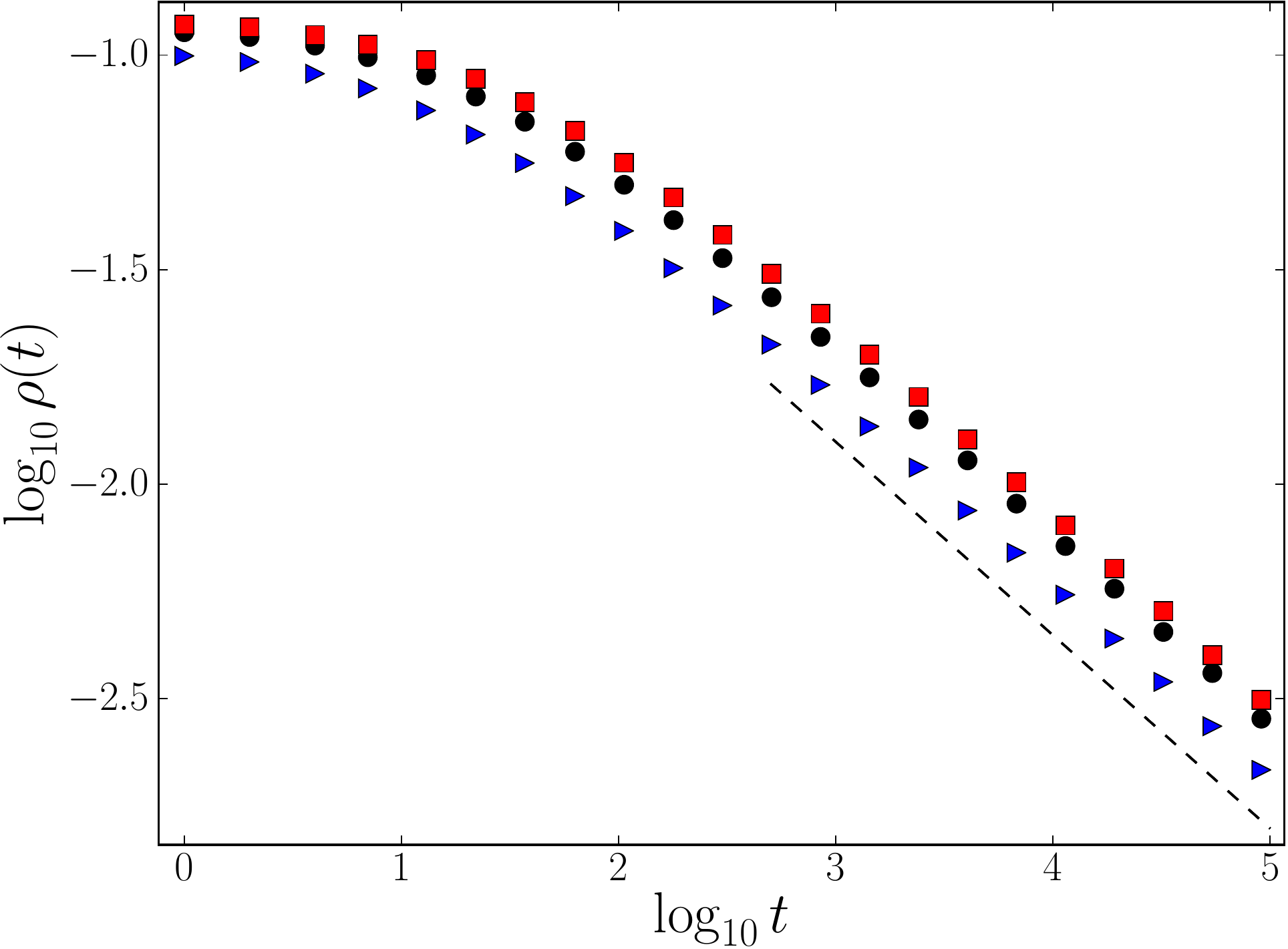}{
Left: $L(t)$ for the AM at criticality. 
Circles represent the pure DP at $p_1=p_2=p_{\text{subs}}=p_c$. 
Squares represent the AM with $p_1=0.01, p_2=0.600, p_{\text{subs}}=p_c$. 
Triangles represent the AM with $p_1=0.4888, p_2=0, p_{\text{subs}}=p_c$.  
The dashed line corresponds to a slope $1/z=0.566$, using the exponent $z$ measured in pure  DP  (\ref{pureDP}). 
Averages are performed over $10^5-10^8$ samples. 
\newline
Right: $\rho(t)$ for the AM at criticality. 
Circles represent the pure DP. 
Squares represent the AM with $p_1=0.01$, $p_2=0.600$, $p_{\text{subs}}=p_c$. 
Triangles represent the AM with $p_1=0.4888, p_2=0, p_{\text{subs}}=p_c$.  
The dashed line corresponds to the exponent  measured in pure DP, $\theta=0.4505$ (\ref{pureDP}).
Averages are performed over $10^5-10^8$ samples.
\label{rhot}
}

We want to check if $\nu_\parallel$ changes with $p_1$ and $p_2$.
To do this we set $(p_1=0.01, p_2=0.600)$ and use different values of $p_{\text{subs}} < p_c$ , thus varying $\Delta$, and observe the deviation from power-law behaviour in Fig.~\ref{collapse} (left). 
We consider the scaling law in Eq.(\ref{Q1}):
using the value of $\delta= 0.53$ extracted from Fig.~\ref{qt2} (right) and the DP value given in Eq.(\ref{pureDP}), 
we obtain a perfect collapse for the survival probability.
This shows that $\nu_\parallel$ does not change between compensation and DP. 
\includefigtwo{0.48\textwidth}{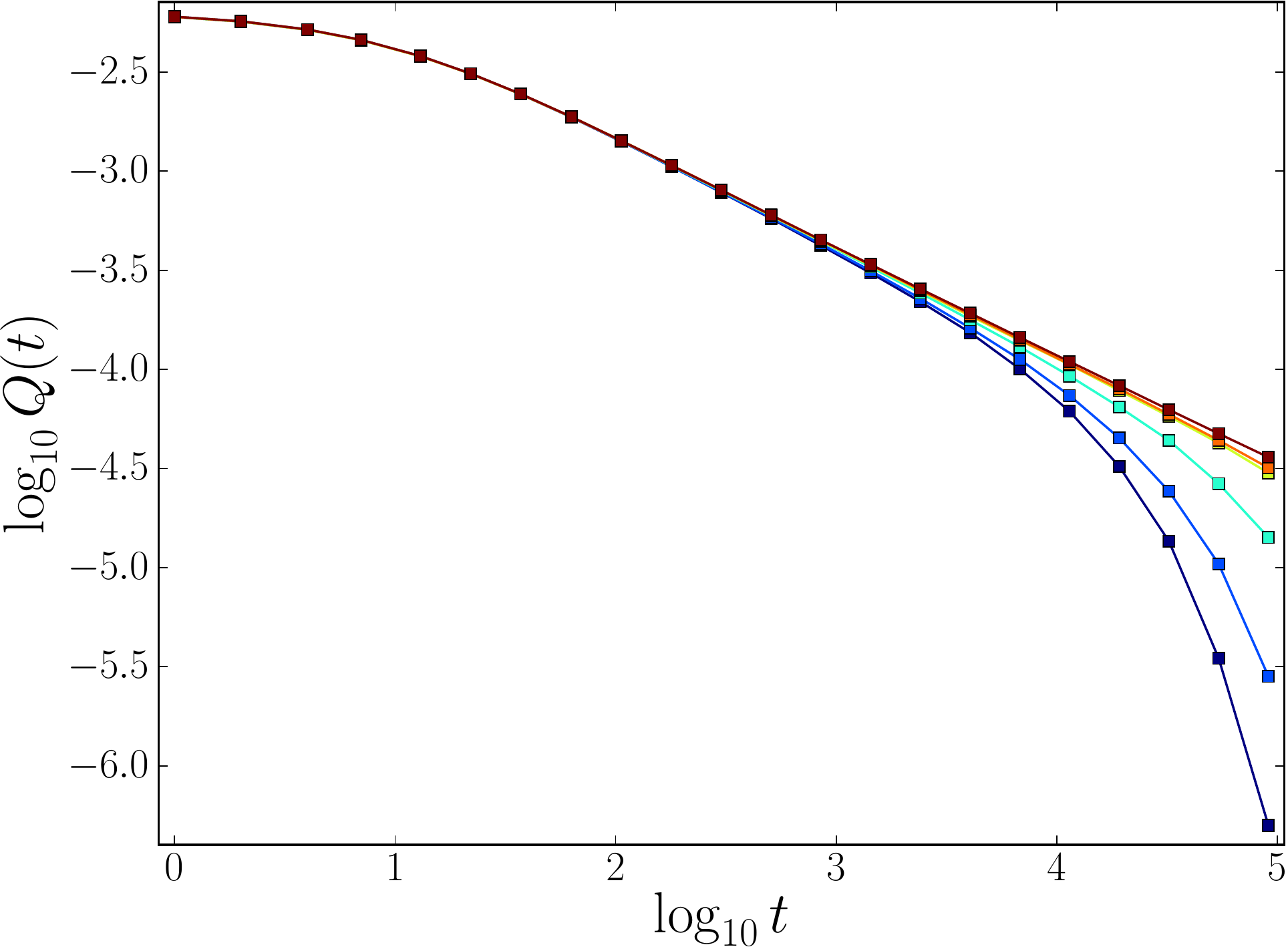}{0.47\textwidth}{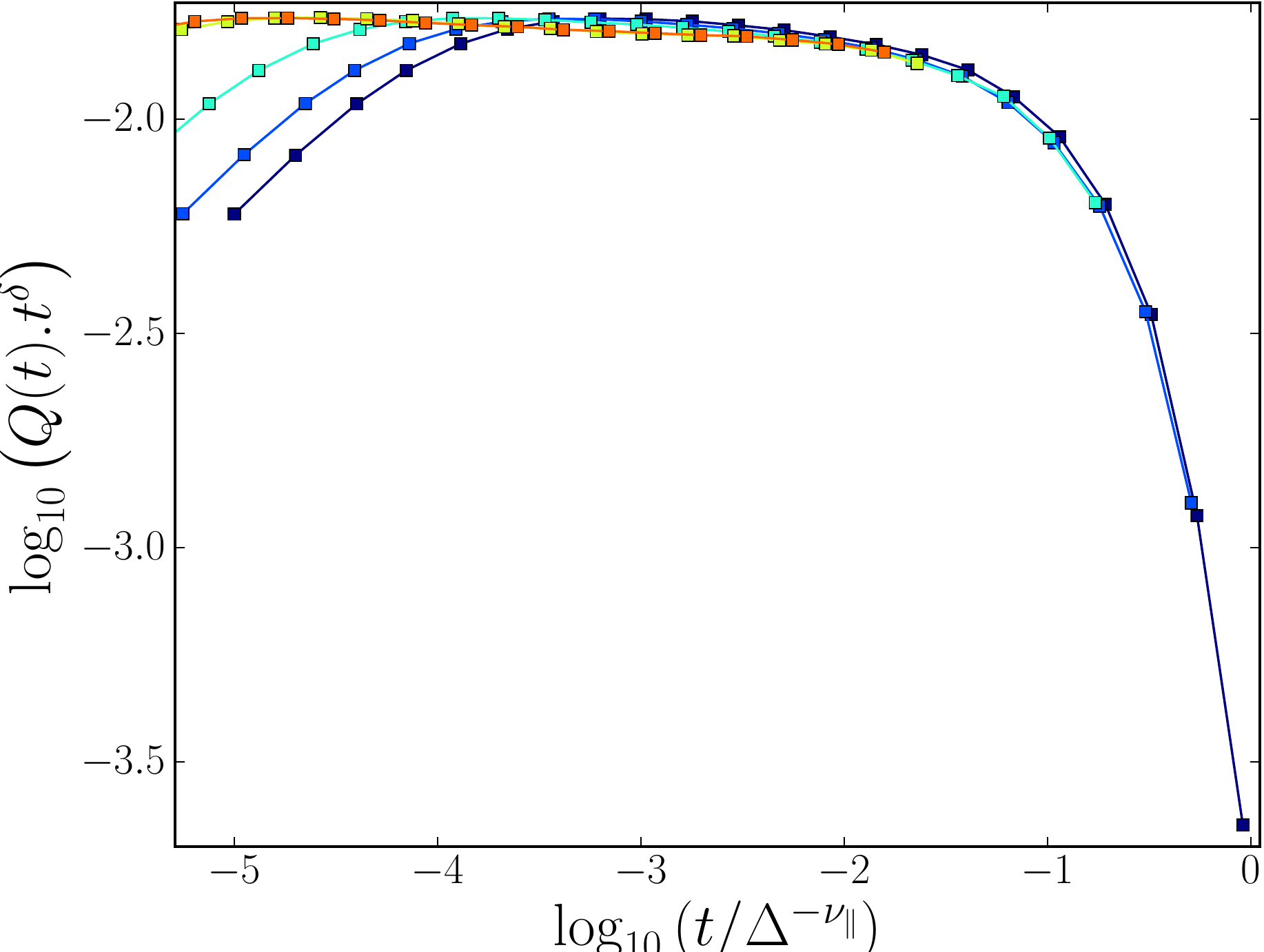}{
Left: $Q(t)$ for the AM for $p_1=0.01$, $p_2=0.600$ and different $p_{\text{subs}}$. 
From top to bottom, $p_{\text{subs}}=0.287338$, $0.28733$, $0.28732$, $0.2873$, $0.2872$, $0.287$.
Averages are performed over $8\, 10^6$ samples. 
\newline
Right:
We collapse these data, plotting $Q(t) \cdot t^\delta$ against $t/ \Delta ^ {-\nu_\parallel}$.
We used $\delta=0.53$ as measured in figure \ref{qt2} and the pure DP value  $\nu_\parallel= 1.295$ as in (\ref{pureDP}).
\label{collapse}
}

\paragraph{Critical Behaviour of the Spreading Observables}

The scenario is different for the spreading exponents $\delta, \eta$ and $ \tau$. 
We already saw that $\delta$ changes at compensation. 
In addition, in Fig.~\ref{P(S)} (left),
the number of active sites averaged over all runs, $N(t)$, is seen to depend on the compensation pair $(p_1,p_2)$. For the compensated point $(p_1=0.4888, p_2=0)$ we measure $\eta= 0.44 \pm 0.01$ and for $(p_1=0.01, p_2=0.600)$ we measure $\eta= 0.15 \pm 0.01$.
At compensation, we expect the hyperscaling relation (\ref{hypers}) to hold. 
As $z$ and $\theta$ are found to be constant, the only way to preserve this relation is to have $\delta + \eta = d/z -\theta = \text{const}$. This constant is $0.680 \pm 0.002$, if we refer to \cite{Dickman1999}. 
For the point $(p_1=0.4888, p_2=0, p_{\text{subs}}=p_c)$, we find that $\delta+\eta = 0.69 \pm 0.02$.
For the other compensation point $(p_1=0.01, p_2=0.600, p_{\text{subs}}=p_c)$, we find $\delta+\eta = 0.68 \pm 0.02$.
These results are consistent with the expected value, for both compensation points.

\includefigtwo{0.475\textwidth}{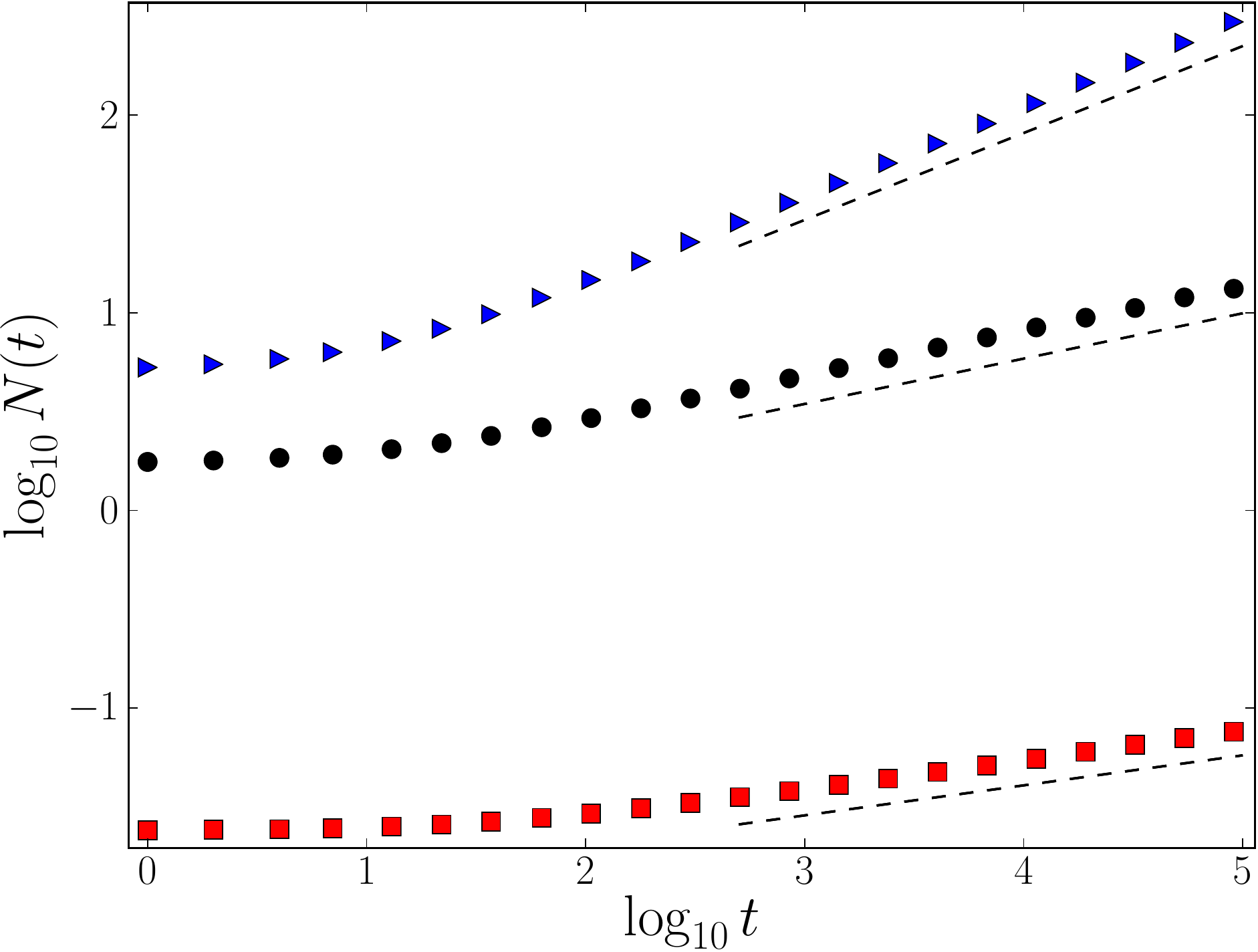}{0.48\textwidth}{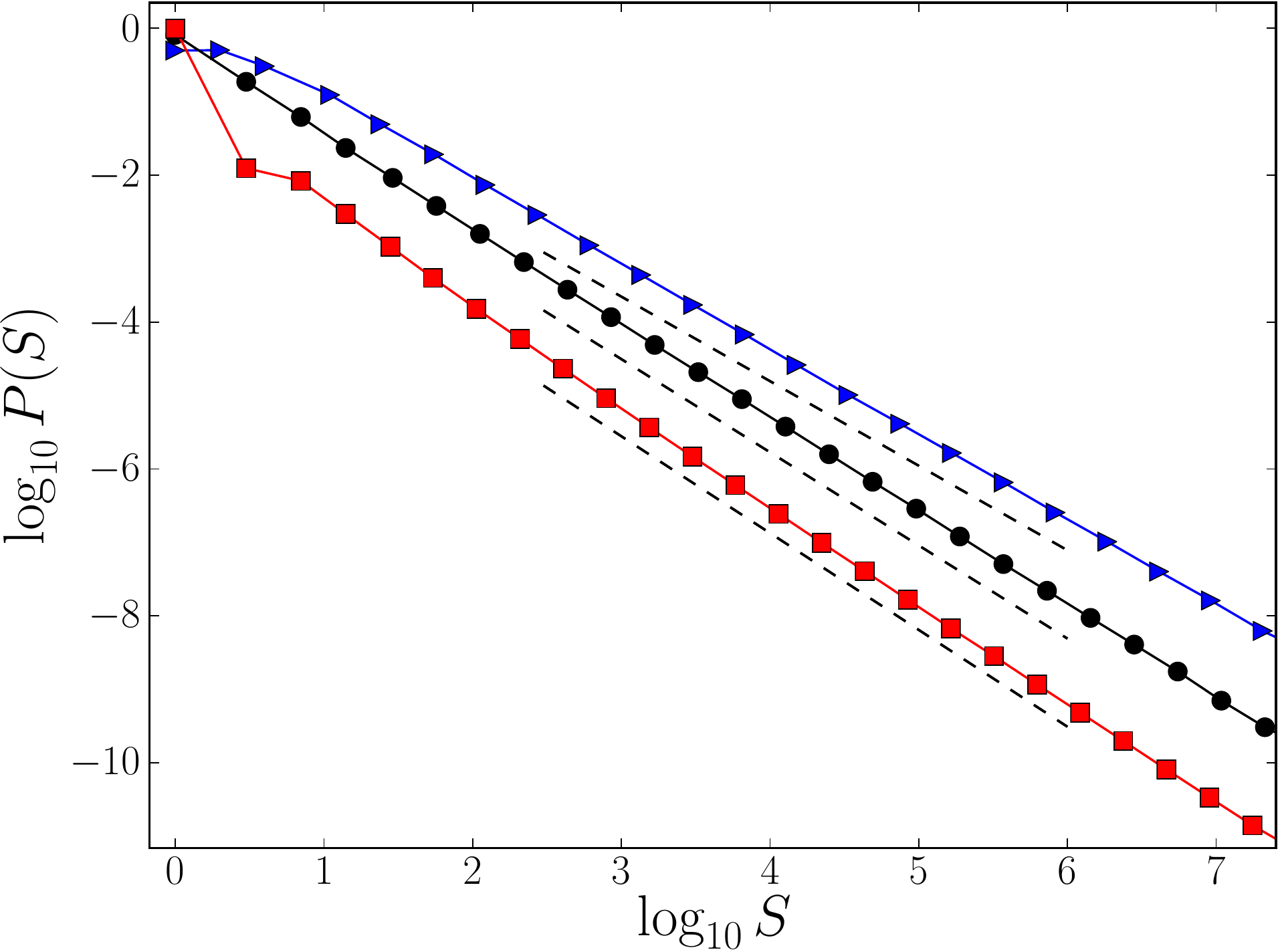}{
Left: $N(t)$ for the AM at criticality. 
Triangles represent the AM with $p_1=0.4888, p_2=0, p_{\text{subs}}=p_c$.  
There we measure $\eta=0.44 \pm 0.01$ (dashed line). 
Circles represent the pure DP at $p_1=p_2=p_{\text{subs}}=p_c$. 
There the dashed line corresponds  to the exponent  measured in pure  DP, $\eta= 0.2295$  (\ref{pureDP}).
Squares represent the AM with $p_1=0.01, p_2=0.600, p_{\text{subs}}=p_c$. 
We measure $\eta=0.15 \pm 0.01$ (dashed line).
Averages are performed over $10^5-10^8$ samples.
\newline
Right:
$P(S)$ for the AM at criticality. 
Circles represent the pure DP at $p_1=p_2=p_{\text{subs}}=p_c$. 
We check that $\tau=1.268 \pm 0.005$.
Squares represent the AM with $p_1=0.01$, $p_2=0.600$, $p_{\text{subs}}=p_c$. 
We measure $\tau= 1.318 \pm 0.005$.
Triangles represent the AM with $p_1=0.4888$, $p_2=0$, $p_{\text{subs}}=p_c$.  
There we measure $\tau=1.151 \pm 0.005$. 
Averages are performed over $10^5-10^8$ samples.
\label{P(S)}
}

In Fig.~\ref{P(S)} (right), we present the probability density function $P(S)$.
The scaling relation (\ref{tau})  holds for the compensation process. 
In particular for the first compensation point $(p_1=0.4888, p_2=0)$, using $\delta= 0.25 \pm 0.01$, the equation (\ref{tau}), and $\delta+\eta = 0.69 \pm 0.02$, we expect $\tau=1.148 \pm 0.006$. We measure $\tau = 1.151 \pm 0.005$.
For the other compensation point we expect $\tau = 1.315 \pm 0.006 $ and measure $\tau = 1.318 \pm 0.005$. 
These results are all consistent with the  expected values, within our numerical precision.
We see that the relations derived in the first section are still valid, except for the time-reversal symmetry which is violated, since $\delta \neq \theta$.

\subsubsection{Interpretation of the Results}

\paragraph{Existence of Compensation, Criticality}
A generic description of the behaviour of the model can be presented in the  $(p_1, p_2)$  parameter plane (Fig.~\ref{p1p2}).
In this plane there is a line along which the behaviour of the system is critical.
This line passes through the DP point $(p_1=p_c, p_2=p_c)$.
The values of the bulk exponents  $z$, $\theta$ and $\nu_\parallel$ are constant all along the line.
The three spreading exponents $\delta$, $\eta$ and $\tau$ change continuously when we move along the line, but always respect the relations (\ref{hypers}) and (\ref{tau}), 
so that there is only one independent exponent that changes.
The value of $\delta$ passes from lower-than-DP values when $p_1>p_c>p_2$, to larger-than-DP values
when $p_1<p_c<p_2$.
Out of this line, there is in general a stretched exponential contribution to the distribution of the relevant quantities of the problem.

Although we do not have an analytical proof of our main claim, i.e.~the existence of a critical line in the $(p_1, p_2)$ plane, we can simply demonstrate that there is a singular line in some respect. Along the diagonal of the $(p_1, p_2)$ plane, the DP point separates a long term survival probability $Q_{st}$ of zero (towards the origin, $p_1=p_2=0$) and a finite value of $Q_{st}$ (towards larger values of $p_1$ and $p_2$).
The values of $Q_{st}$ in other parts of the $(p_1, p_2)$ plane must smoothly match this known behaviour.
In particular, we will have a singular line separating a region with $Q_{st}=0$, towards the origin and along this line, from another region with $Q_{st}\neq 0$, to the right and above this line. This proves that there is a singular line with respect to $Q_{st}$ in the $(p_1,p_2)$ plane.
Our expectation is that this singular line is also a critical line in which quantities are power-law distributed.

A complete proof would require to adapt and extend the renormalization-group analysis of \cite{Jimenez-Dalmaroni2003} to the present case. 
To confirm our numerical findings, we would need to prove that our the Attempt Model, at compensation, corresponds to a marginal perturbation of the DP equations.
As we explained earlier, a quick analysis of the perturbation introduced by a modified first attempt probability shows that its nature is very similar to that introduced by a modified second attempt probability.
Thus, it seems probable that the leading orders of these two perturbations can cancel each other by appropriately selecting the pair ($p_1,p_2$), thus resulting in a marginal perturbation.

\includefig{8cm}{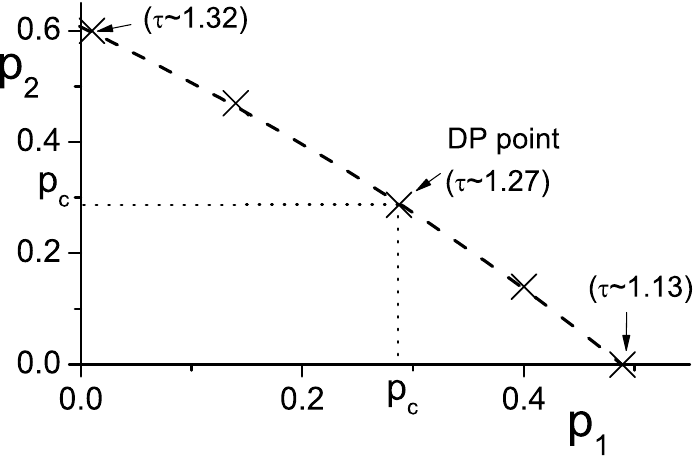}{
Phase diagram of the system in the $(p_1, p_2)$ parameter space with $p_{\text{subs}}=p_c=0.287338$.
The dashed line (schematic) is a critical line on which quantities in the system are power-law distributed. 
Above the line there is annular growth, and below there is sub-critical growth.
The bulk exponents $\theta, z, \nu_\parallel$ are equal to DP values all along this line, 
whereas the spreading exponents $\delta, \eta, \tau$ vary continuously along the line (representative values of $\tau$ are indicated).
The crosses correspond to those points along the critical line that were numerically determined.
\label{p1p2}
}

\paragraph{Bulk and Spreading Exponents}

\includefig{8cm}{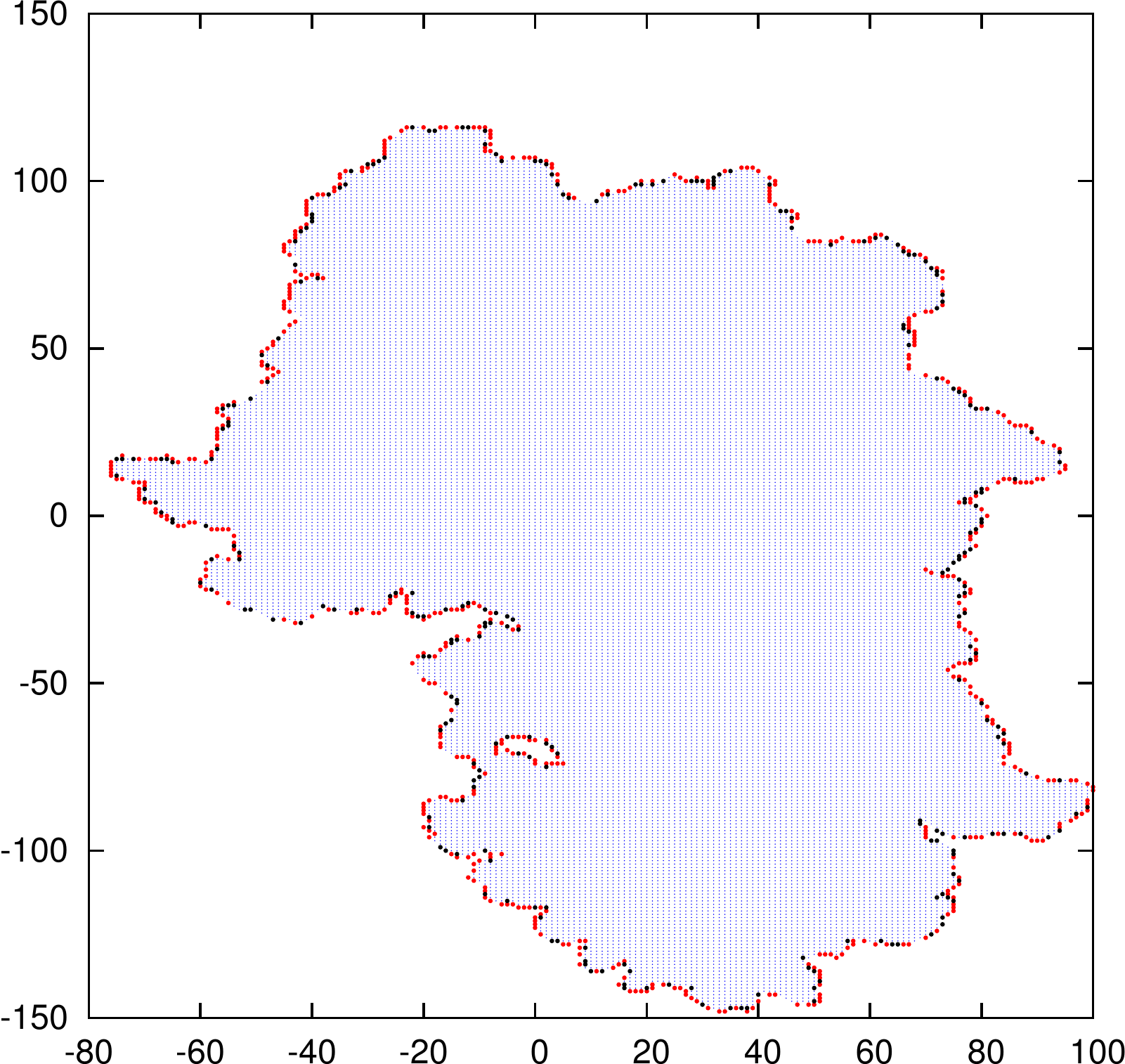}{
Snapshot of a growing cluster in the AM at compensation ($p_1=0.01$, $p_2=0.600$). Red (resp.~black) points at the border are those sites that have been attempted once (resp.~twice). 
The whole interior is formed by sites that have been attempted more than twice (blue)
\label{cluster}
}
We can understand the behaviour of the bulk exponents if we think that these exponents can 
be measured starting from a fully active lattice.
In this case the evolution of the system coincides with that of pure DP after a few time steps. 
However, bulk exponents can also be measured on the surviving runs started from a single active site.
In this case space-time is divided in two regions: the active one, and the outer, inactive one. 
In Fig.~\ref{cluster} we show a snapshot of a AM growing cluster at a given $t$. 
We see that sites that make the difference with usual DP  are mostly located at the boundary of the active region.
We consider a large box  of size $\ell_\perp \ll \xi_\perp$ in space and $\ell_\parallel \ll \xi_\parallel$ in time, sufficiently far away from the boundary with  the inactive region. 
Its statistical properties will be completely independent of its precise location and 
are indistinguishable from those of a box with the same size, with the fully active initial condition.
Since the role of the boundaries is asymptotically small, this shows that the bulk exponents $\theta, \nu_\perp, \nu_\parallel$ and thus $z$ are unchanged by the compensation process also if we use the single seed  initial condition.

On the contrary, the spreading exponents $\delta, \eta, \tau$ are naturally defined only in the seed initial condition, and involve averages over all runs.  
They are strongly affected by the spreading properties of the active cluster, it is thus not so surprising to see them depend continuously on $p_1$ and $p_2$.

\subsubsection{Generality of the Results}

Here we have focused on the case of two spatial dimensions, but
qualitatively the same behaviour is obtained for one dimension. However in one dimension the deviations from criticality when $p_1$ is changed are much weaker than in two dimensions, making the determination of the compensation condition much more difficult numerically.

In this respect we want to mention that we have found other realizations of the DP process 
where the effect is quantitatively much more important. 
For instance, the process in which we try to activate neighbours with probability $p$, 
having in addition a self-activation probability $p_\text{self}$ of the same site, also belongs to the DP universality class. 
In this case we have observed that a lower probability to activate neighbours for the first time can be compensated by larger self-activation probabilities during the next steps,
and in this case the quantitative effect is much more important. 
In particular we have obtained avalanche size distributions with $\tau$ as large as $\simeq 1.7$.

\subsection{Discussion of Related Models}\label{discussion}

A similar scenario happens in $1$-dimensional models which display critical behavior, despite their breaking of the time-reversal symmetry. 
In these models \cite{Mendes1994a, Jensen1993a, Jensen1993, Marques1999, Dickman1999, Odor1998, Park2007, Munoz1996, Munoz1998} each site is active or inactive, as in DP, but is equipped with an additional auxiliary field $\phi$ which allows for a large degeneracy of the absorbing state.
On this point, let us remark that the AM can formally be described by a Markovian dynamics on three fields $\phi, \phi_1$, with $\phi$ the activity field at time $t$ and $\phi_1$ a record of the number of local attempts: $\phi_{att}=0$ for a virgin site, $\phi_{att}=1$ for a site attempted once and $\phi_{att}=2$ for a site attempted twice or more.
Using these two fields, it is easy to describe the AM in a purely Markovian way.

\paragraph{Threshold Transfer Process (TTP)}
We discuss  DP with auxiliary fields using the example of the Threshold Transfer Process (TTP) \cite{Mendes1994a, Odor1998}.
In the $1$-dimensional TTP, a site may be vacant, singly or doubly occupied, 
corresponding respectively to states $\sigma_i = 0, 1$ or $2$.
The auxiliary field $\phi$ denotes the density of singly occupied sites. 
A doubly occupied site corresponds to the active state. 
Initially, only the site at the origin is doubly occupied,
while the state $\sigma_i$ of each other site is $1$ with probability $\phi_{\text{init}}$, and $0$ otherwise. 
At each time step, a site $i$ is selected at random.
If $\sigma_i(t) < 2$, then $\sigma_i(t+1) = 1$ with probability $r$ and $0$ with probability $1-r$, irrespective of the precise initial value. 
If $\sigma_i(t) = 2$, the site releases one particle to all neighbours with $\sigma(t) < 2$. 
Contrary to the DP case, there are infinitely many absorbing states since any configuration with no doubly occupied site is absorbing.

In the TTP, $r$ plays the role of control parameter, and in $d=1$, $r_c = 0.6894$ \cite{Odor1998}. At criticality the bulk exponents and the hyperscaling relation behave as in DP, independently of the initial condition. 
However the spreading exponents continuously depend on the initial condition $\phi_{\text{init}}$.
Setting the initial density of singly-occupied sites to its stationary value $\phi_{st}=r_c$, one recovers the full set of DP critical exponents \cite{Mendes1994a}.
As far as we know, a theoretical explanation for the continuous change in the spreading exponents $\delta, \eta$ is still an open question.

\paragraph{DP in a Box}

It is worth mentioning a second class of models with similar behaviour, which corresponds to DP with special absorbing boundary conditions.
In particular DP with  absorbing walls at positions $x(t) = \pm C \cdot t^{1/z}$ shows spreading exponents that continuously depend on $C$ \cite{Kaiser1995, Kaiser1994c}. Analogous results with a moving active wall are presented in \cite{Chen1999}.
Moreover, one dimensional models with soft or hard walls conditions can be studied analytically  in the case of Compact DP.
They can be mapped onto compact first attempt (for soft walls) and compact first infection (for hard walls). 
Dickman  showed \cite{Dickman2001c} that in this case the critical behaviour is maintained when $p_1$ is reduced, i.e.~in this case we do not have a stretched exponential contribution.

\paragraph{Conclusion: (Number of) Fields Counting}
Memory effects in immunization problems, or the presence of auxiliary fields in TTP-like models, introduce high degeneracy of the absorbing state and thus break the time reversal symmetry. 
In these systems, at criticality, the bulk DP exponents are recovered. 
However, if the initial condition is sufficiently far from its stationary value (which is $\phi_{\text{st}}$ for TTP-like models, and the fully twice-attempted lattice for the compensation model $\phi_{att}=2$, or the fully once-infected lattice in the modified first Infection Model), the spreading exponents depend continuously on the initial condition. 
Non-stationarity seems to play a key role in the observed anomaly of the spreading exponents.

At the beginning of this chapter, we discussed the link between the microscopic dynamics of the elastic interface model and a Modified DP Process.
We saw that for this ``modified DP'' to follow the same dynamics as an interface, it needed an additional field $h$, that controls the self-activation probability: $p_\text{self}= 1-e^{-\lambda h}$.
The link between the activity $\phi$ and $h$ is simply given by $\phi=\partial_t h$.
In the case of the viscoelastic interface, described by two fields $h$ and $u$, the scenario is similar.
We do not give the full derivation here, but simply mention that a ``very modified DP'' process can be defined, in which the fields $\phi, h$ and $u$ are necessary to fully describe the system and its evolution.
This draws a parallel with the AM:
starting from the ``modified DP'' with fields $\phi, h$, if we then add the Attempt rule (using the field $\phi_{att}$), we obtain a model with three fields $\phi, h, \phi_{att}$, reminiscent of the case of the viscoelastic interface model.

\section{Conclusion: An Extended Universality Class}
\label{sec:conclu_DP}

We have shown that DP universal behaviour is strongly affected by changes in the first probabilities to activate sites. 
This modification corresponds to a special case of ``long term memory'', where each site remembers 
exactly how many times it has been activated (or attempted) before. Our main result is that, although the change of the very first attempt probability takes the model out of criticality, by changing the second attempt probability in the opposite direction, we can restore critical behaviour, in a process we called compensation. 
Several critical exponents found at the point of compensation do not coincide with those of pure DP: in particular, a time-reversal symmetry known to be valid for DP is violated at compensation, thus changing the value of one of DP's fundamental exponents, while the other three conserve their values. 
A remarkable feature of the criticality with compensation is that the exponents depend continuously on the precise choice of the activation probabilities, while almost all the scaling relations of DP are preserved.

An issue well known in the DP literature is the absence of experimental realizations of the DP universality class \cite{Hinrichsen1999}.
Here, we have shown that by including a simple memory effect (which can also be expressed in terms of additional fields) into the DP dynamics, we obtain an extended version of the DP universality class.
The pure DP is contained in this extended model, with some exponents varying continuously with the parameters $p_1, p_2$ while others do not change; similarly several scaling relations are maintained, while one is violated.
In this respect, our model offers the hope to find universality classes larger than DP, but sharing some crucial features with it (like the non linearity of the bulk of DP).
Finding an experimental realization of this extended-DP class should be easier than for the more restrictive pure DP class.

Aside from any application to a concrete situation, we want to stress the fact that the present model provides a link between two classes of models with very different behaviour: the models with an auxiliary field and the Infection Model. Although we obtained the same results as in models with an auxiliary field (criticality  with  scaling relations preserved, time-reversal symmetry broken), our microscopical description fits in the framework of modified First Infection Models, for which analytical computations have been successful \cite{Jimenez-Dalmaroni2003}. This may be an interesting approach to the open problem of initial-condition-dependent exponents in absorbing phase transitions.

\chapter{General Perspectives}
\label{chap:Conclusion}

Here, we do not intend to summarize the results obtained during this thesis, which are already summarized in the conclusions of each chapter.
Instead, let us explain the path that was followed in this thesis, and try to put our results into perspective.
\\

Seismic phenomena represent a striking natural realization of the kind of scale-free statistics expected in out-of-equilibrium phase transitions.
However, they also represent a physical situation in which various areas of the natural sciences are involved (from chemistry to planetary science, through geology).
The approach of the statistical physicist is to cut into this broad diversity of mechanisms, trying to sort out the relevant ones.
This angle has already proven successful: for example, over the last 30 years the universality behind the notion of disorder has been clarified, so that we know that many variations in the disorder distribution are irrelevant at the macroscopic scale.
This allows to consider the precise nature and distribution of rocks as an essentially irrelevant parameter. 
Similar arguments apply to other variable parameters, which turn out to be irrelevant in the macroscopic limit (e.g.~the precise value of the strength of the interactions).

Relying on these powerful simplifications, simple statistical physics models show that the competition between elasticity, disorder and driving force is already enough to reproduce some of the main features characterizing faults dynamics.
However, a closer inspection of the experimental evidence reveals important discrepancies between these models and real seismic faults.
These failures point out the need for including at least one new element into the statistical physics models.
\\

In order to choose adequately this new ``ingredient'', we considered the simpler case of the frictional behaviour of dry materials in the well-controlled environment of the laboratory.
There, it is clear that some mechanisms at the level of contacts are responsible for the various effects observed in friction at low sliding velocity (velocity-weakening, ageing of contacts at rest).
As we want to keep our models simple, we decided to account for these mechanisms by considering the interactions within each surface to be viscoelastic rather than purely elastic.

This simple choice proves to be a very good one, since our analytical and numerical results compare very well with seismological observations and friction experiments (see sec.~\ref{sec:visco_conclu_2} for a summary).
This success is a proof that, in order to understand friction at small sliding velocities, one must take into account both disorder and viscoelasticity. 
A side effect of our study is to clearly set the problem of dry friction into the field of disordered systems.
A more detailed account of the ability of our model to reproduce frictional behaviour would be interesting, and further comparison with experiments is an interesting lead for future work.

Independently from the precise performance of the model at matching with experiments, it is interesting to question the generality of the approach. 
We answer to this interrogation by noting that some common features seem to emerge in various models including both microscopic disorder and some relaxation mechanisms.
In particular, in our model, we have proven that the addition of viscoelasticity is a relevant change, in the sense that the addition of a very small amount of ``visco-'' to the elastic interactions affects the macroscopic behaviour.
The separate observations of universality in disordered systems and in viscoelastic materials date from a long time, but for the combination of both the consensus is just starting to emerge.

Our model is a good candidate to study the extended universality class of ``viscoelastic depinning'', as its simple formulation allows for analytic treatment in mean field (that we performed) and extensive numerical simulations (as we did in two dimensions).
In this respect, an avenue for future work is to perform the full analysis of the model in finite dimensions, extracting all the exponents and discussing the scaling relations.
In particular, in two dimensions, we need to evaluate the characteristic length over which the stress level is strongly correlated.

In these future works, we may be guided by our results on the modified Directed Percolation (DP) model, for which the new universality class displays many common features with the particular case of pure DP, indicating a relative robustness of the ``stationary features''.
Natural extensions of our model that should be considered are the study of the finite dimensional case with long-range interactions (expected to be similar to \cite{Papanikolaou2012}) and with quadrupolar, long-range stress redistribution (Eshelby problem).

\appendix
\begin{appendices}

\chapter{Additional Proofs and Heuristics}

\vspace{-2cm}
\minitoc
\vspace{2cm}

\section{Appendix to chapter \ref{chap:friction}}

\subsection{Why Lubricants May be Irrelevant}
\label{App:lubricants_irrelevant}
In this thesis, we are mainly interested in dry friction, as opposed to lubricated friction.
However, in real contact mechanics, there are always some impurities, gas molecules or even liquids between the substrate (lower, motionless solid) and the (upper) sliding solid. 
As friction is controlled by the properties of the surfaces in presence, the molecules adsorbed on the substrate (and on the upper solid) are expected to play a major role, even if they are present in very small quantities.

Consider a perfectly flat substrate, i.e.~without a single one-atom imperfection on length scales of several micrometers (this can be accomplished using mica surfaces, that are rather easily produced with such flatness).
On this substrate, molecules of gas (or oil lubricants, etc.) can be adsorbed, allowing for ``lubricated'' friction (in the broad sense).

If the adsorbate is in small enough quantity, only a single layer of molecules (or less) will be present on the substrate. Increasing the quantity of adsorbate, one can obtain several ``layers'' of them. 
For a thickness of a few layers, the first layer is generally adsorbed on the surface, with possibly a regular crystalline structure, while the remaining ones are either in the same crystalline order, or in a \textit{fluidized} state, which can be an un-jammed state for granular matter, fluid, or other ``flowing'' states of matter.
In this case, friction is controlled mainly by the solidification or \textit{fluidization} of this adsorbate layer.
See \cite{Aranson2002} for a detailed study of this case.

If a liquid is present in large quantity ($\sim 10~\mu m$ or more)  between the substrate and the sliding solid, then the hydrodynamic approach becomes relevant, and friction is controlled by the bulk hydrodynamics of this intermediate liquid, along with the adsorption properties on each of the two surfaces (which control the boundary conditions of the hydrodynamic problem).
The nature of the interactions between the two bare surfaces is then completely irrelevant.\\

However we have seen that most surfaces 
are not at all flat, and even those we call ``smooth'' in everyday life are actually quite rough at small   --  and not so small  --   length scales. 
This diminishes the a priori crucial role of adsorbates, since the ``true'' contact area is much smaller than what one would naively expect. 
At the rare contact points that are relevant for the friction of rough surfaces, the hydrodynamics of adsorbates is often irrelevant.

\subsection{Self-Similarity (and related definitions)} 
\label{App:self_similarity}
Numerous objects have the property that they ``look the same'' at various length scales. 
Here we make this idea more precise by defining various mathematical properties related to this idea.

Let us first define the property of self-similarity (and other related properties) in the general case.
A function $g(x)$ is said to be self-similar if an only if (iff) it satisfies:
\begin{align}
g(x) = \Lambda g(\Lambda^{-1} x), \qquad \forall \Lambda>0,~\forall x.
\end{align}
This is a \textit{re-scaling}, and it correspond intuitively (e.g.~for $\Lambda>1$) to do two things at the same time: ``zoom in'' in the $x$-direction and magnify (or also ``zoom in'') in the $g$-direction.
This can be extended in $2$ or more dimensions, where the condition becomes  (e.g.~in $d=2$ dimensions):
\begin{align}
g(x,y) = \Lambda_1 \Lambda_2 g(\Lambda_1^{-1} x, \Lambda_2^{-1} y) , \qquad \forall \Lambda_{1,2}>0,~\forall (x,y).
\end{align}
Self-similarity is a very stringent constraint, since the re-scaling in the $x$- and $g$-directions  (and the $y$ direction in 2D) has to be exactly the same.

A more general property defining objects with ``similar'' appearance at different length scales is \textit{self-affinity}.
A function $g(x)$ is said to be self-affine iff it satisfies:
\begin{align}
g(x) = \Lambda^b g(\Lambda^{-1} x), \qquad \forall \Lambda>0,~\forall x,
\end{align}
where $b$ is the self-affinity or scaling exponent.
We see that self-affinity is an anisotropic transformation which contains self-similarity as a special case ($b=1$).
The anisotropy can be stronger in two or more dimensions. 
In $d=2$ dimensions, the condition for self-affinity becomes 
\begin{align}
g(x,y) = \Lambda_1^{b_1} \Lambda_2^{b_2} g(\Lambda_1^{-1} x, \Lambda_2^{-1} y) , \qquad \forall \Lambda_{1,2}>0,~\forall (x,y),
\end{align}
where $b_1,b_2$ are the self-affinity or scaling exponents related to the affine transformation. 
This may be referred to as ``anisotropic'' self-affinity, but this wording is misleading, because even for $b_1=b_2 \neq1$, we already have an affine transformation (and not a similarity transformation)\footnote{
Please note that in part of the literature, these two concepts are sometimes mistaken for one another, or simply melted and seen as equivalent. 
When considering functions, it seems quite natural that the ordinate and abscissa do not share the same scaling exponent, so that considering self-affinity seems very natural. 
However, when considering geometrical objects such as self-similar or self-affine objects, the distinction becomes important. Not all \textit{fractals} are \textit{self-similar fractals}.
}.

Self-affinity is a rather general property, however it is interesting to note that it only allows to compare fully deterministic objects. 
This is already quite general, since even for stochastic systems one may consider e.g.~the correlation function $g(x_1,x_2)$, which despite being a deterministic object, helps in characterizing the fluctuations of the system.
For instance, if the system state is described by $\sigma(x)$, one may define $g(x_1,x_2) \equiv \langle \sigma(x_1)\sigma(x_2) - \sigma(x_1)^2 \rangle$. 
As $g$ is deterministic, it may be self-affine. 
However, if we are interested in many moments of some random distribution, or even in its full distribution, then we need an additional definition: \textit{statistical self-affinity}.
A stochastic process $g$ is said to be statistically self-affine iff:
\begin{align}
g(x) \overset{\text{Law}}{=} \Lambda^b g(\Lambda^{-1} x), \qquad \forall \Lambda>0,~\forall x,
\end{align}
where the equality is ``in Law''. 
This definition allows to analyse the properties of random processes. 
The definition of statistical self-similarity is obvious (just take $b=1$).

\subsection{Fraction Brownian Motion}
\label{App:fBm}

Here, we use Fraction Brownian Motion (fBm) as a non-trivial and statistically self-affine process, to exemplify the notion.
Besides, it can be useful by itself. 

As its name suggests, fBm is a generalization of Brownian Motion (BM). 
The fBm process is defined as a continuous-time Gaussian process which is self-affine and has a specific covariance function. 
The covariance function is:
\begin{align}
\langle h(x_1)h(x_2)\rangle  = \frac{1}{2	} \lp x_1^{2H} + x_2^{2H} - |x_1-x_2|^{2H}  \rp ,
\end{align}
where $H$ is the \textit{Hurst exponent};
the Gaussianity hypothesis means that 
\begin{align}
\mathbb{P}(h(x)) \d x = \frac{1}{\sigma_x \sqrt{2\pi} } e^{-\frac{h(x)^2}{2\sigma_x^2}} \d x,
\end{align}
(where the variance $\sigma_x$ can be computed, $\sigma_x^2 = \langle h(x)^2 \rangle = x^{2H}$);
and the self-affinity or scaling exponent is $H\in [0,1]$:
\begin{align}
h(\Lambda x)  \overset{\text{Law}}{=} \Lambda^H h(x).
\end{align}
The case $H=1/2$ reduces to BM, since $  x_1 + x_2 - |x_1-x_2| = \min(x_1,x_2)$, which is the covariance of the BM.
The definition of the corresponding discrete process can be done rather naturally. 
To stay concise, we do not give it here.
\\

We are going to see that intuitively, the fBm represents sub- or super-diffusive processes. 
We start with simple Brownian motion for simplicity.

Originally, Brownian motion is understood as representing the position $X(t)$ of a particle diffusing over time.
For a large number $N$ of independent random walkers (each following a different realization $X_i(t)$ of the same law), the average density $\rho(x,t)$ of particles is supposed to follow the \textit{equation of diffusion} or \textit{heat equation}, i.e.~$\partial_t \rho =  \Delta \rho$.
The initial condition with all particles at $x=0$ at time $t=0$ is $\{X_i(0)=0, \forall i\}$ and corresponds to a Dirac\footnote{For reasons of coherence of notations which will be clearer later, we denote $\delta^D$ the Dirac distribution.}
 distribution $\rho(x,0)=\delta^ D(x)$ for the density.
The solution to the heat equation with this boundary condition is $\rho(x,t)= (2\pi t)^{-1/2} e^{-x^2/2t}$, which is exactly the probability density of the Brownian Motion (this result also holds for any boundary condition).
This means that BM is a good candidate to represent the microscopical motion of diffusing particles, since in the limit of large enough time (larger than the typical collision length) and large number of particles (in order to give a meaning to the notion of density), it gives the same result as the continuous equation of diffusion. 
A well-known side-product of this result is that the typical distance $\sqrt{\langle x^2(t)\rangle}$ from the origin of a random walker (understand BM process) at time $t$ is typically $\sim t^{1/2}$. 
The exponent $1/2$ characterizes the spread of the Brownian walker.

Let us see what the corresponding exponent is for fBm.
The \textit{increments} of the fBm are said to be \textit{stationary}, i.e.~any function of the difference $h(x)-h(x+s)$ does \textit{not} depend on $x$. 
In this sense, despite the fact that the fBm is non-Markovian\footnote{A Markovian process is a (random) process of which the future evolution only depends on the present state (it has no memory of the past). Mathematically, $X(t>t_1)$ only depends on $X(t_1)$, not on $X(t<t_1)$. A non-Markovian process is a process for which $X(t)$ also depends on the values of $X$ at times earlier than $t$.}
 (except for $H=1/2$), it does not properly speaking display \textit{ageing}, because its evolution (encoded in the increments) does not explicitly depend on time.
  More precisely, using the covariance formula we compute the second moment\footnote{Thanks to the hypothesis of Gaussianity, all the other moments also depend only on $s$.}:
\begin{align}
\left \langle \big( h(x)-h(x+s) \big )^2 \right \rangle = s^{2H}, \label{Eq:second_moment}
\end{align}
which is independent of $x$.
So with $h(0)=0$, we have  $\sqrt{\langle h^2(x) \rangle} = x^H$, which is often written as $h(x)\sim x^H$ in the physics community.
This means that a particle following a fractional Brownian motion has $X(t)\sim t^H$: depending on $H$, the fBm is either positively correlated ($H>1/2$, \textit{super-diffusive}), or negatively correlated ($H<1/2$, \textit{sub-diffusive). }
Although this is not properly speaking \textit{ageing}, we say that the fBm with $H>1/2$ displays a long-term memory or has long-range\footnote{When physicists talk about \textit{long-range} (resp.~\textit{long term}), they usually mean that correlations in space (resp.~time) are power-law decaying functions of the distance (resp.~time interval). 
Short-range usually means an exponential decay (or step function with $0$ value at infinity).
} correlations, since the decay of its correlation function occurs with a power-law.

A few particular values of $H$ are especially interesting. 
\includefig{\textwidth}{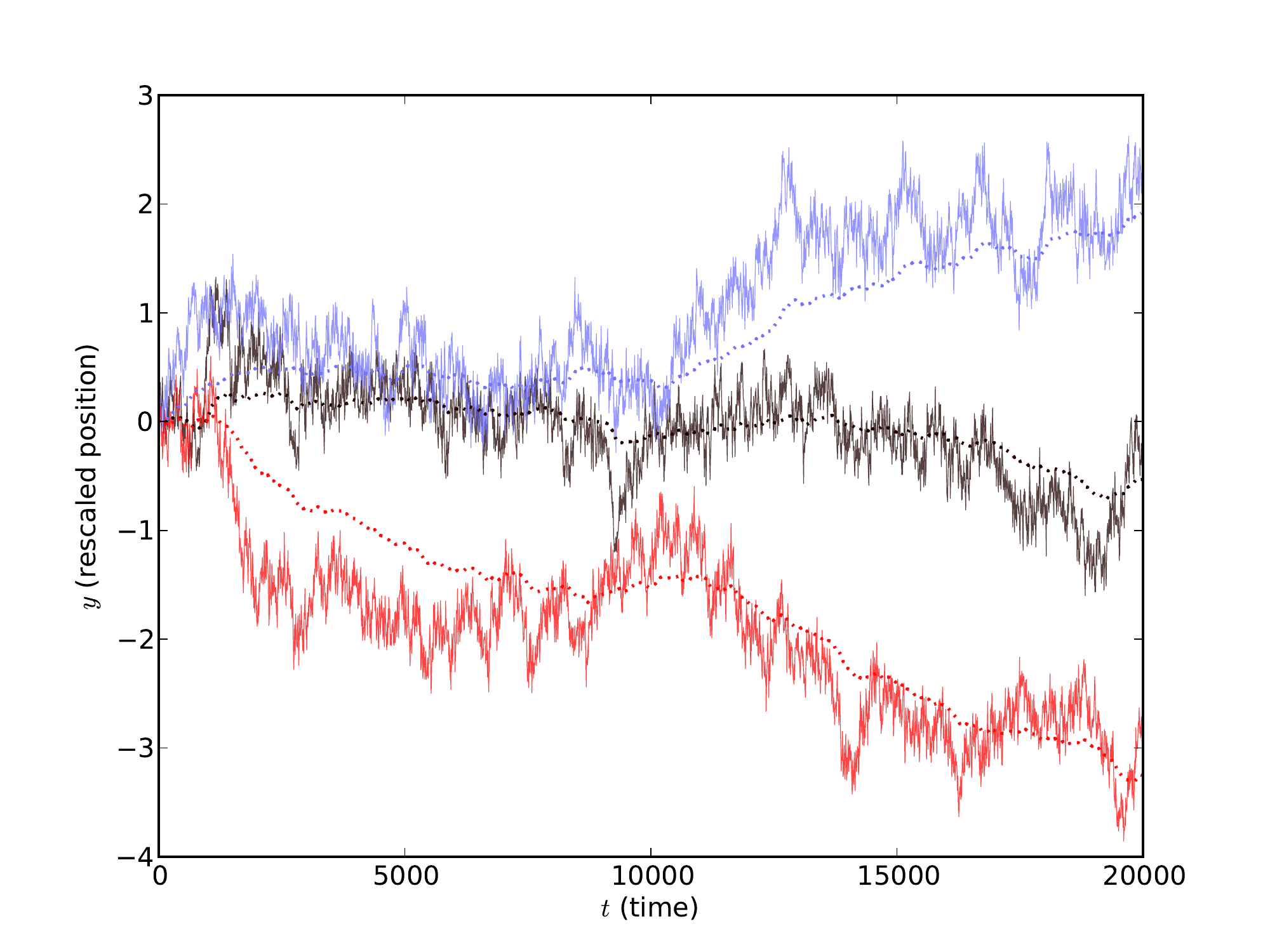}{
Illustration of a statistically self-affine process: the rescaled fractional Brownian motion $y(t)$.
Each color indicates a different realization of the underlying generating  process. 
For each realization, the dotted line corresponds to the fBm with $H=2/3$ and the plain line $H=1/4$.
We see how positive ($H>1/2$) or negative ($H<1/2$) auto-correlation translates in terms of local variations (ignoring the bulk variation $t^H$, thanks to the rescaling).
\label{Fig:fractional_example}
}
\begin{itemize}
\item As said above, with $H=1/2$ we recover Brownian motion (memoryless, simple diffusion, etc.). 
\item For $H=1$, the process becomes very smooth: the function $h(x)$ becomes differentiable! (As opposed to BM which is continuous everywhere but nowhere differentiable). The motion is essentially ``ballistic''.
\item For $H\approx0$, the process is extremely anti-correlated. There, the discrete picture is clearer: if at time $n$ the process increased in value, at the next time step ($n+1$), it will decrease below this value, to keep the balance. This is expected, since $X(t)\sim t^H\sim t^0 \sim O(1) $. 
The intuitive picture is less easy for the continuous process.
\item For $H=-1/2$ (!), we have white noise. This negative value needs an explanation. 
Remember that the BM is not differentiable in the common sense. 
However, the differential elements $dB(s)$ of a Brownian motion are known to be a simple white noise (i.e.~just a sequence of independent Gaussians variables).
A usual construct of the discrete BM $B(t)$ is to simply compute the integrand of a white noise $\eta(s)$: $B(t) = \int_0^t \eta(s) \d s$.
In this sense, the derivative of BM is white noise, and formally, one may say that the derivative of an fBM with some $H$ is an fBm with $H-1$. The case $H=-1/2$ is especially meaningful. 
\item For $H>1$, the same remark as previous item holds: formally, fBm only exists for $H\in[0,1]$, but procedures to generate fBm can be extended to other values of $H$, producing processes with many similarities with fBm. For $H\in[1,2[$, the process is very smooth (differentiable everywhere).
\end{itemize}

To understand how $h(x)$ depends on $H$ (besides the bulk variation that goes as $\sim x^H$), it is practical to define the rescaled fBm:
\begin{align}
y(t) \equiv \frac{h(t)}{\sqrt{ \langle h^2(t)\rangle}} = \frac{h(t)}{t^H}
\end{align}
Which has $\mathbb{P}(y(t))\d t = \frac{1}{ \sqrt{2\pi} } e^{-\frac{y^2}{2}} \d t$.
This process still has the same qualitative correlations as fBm, except for the bulk drift.
We show the behaviour for three different realizations, each time for two values of $H$, in Fig.~\ref{Fig:fractional_example}.

\paragraph{Link with the fractal dimension} 
It is difficult to discuss self-affine or self-similar processes without mentioning fractals. 
We do not go into much detail here, but just mention that for a (statistically) self-affine process $h(\mathbf{x})$ embedded in $d$-dimensional space with Hurst exponent $H$, the associated fractal dimension $D$ (Hausdorff dimension) is given by the relation:
\begin{align}
D+H=d+1.
\end{align}
However this is \textit{not} true for all processes with a behaviour in $h(x)\sim x^ H$ ! It is indeed a very specific property.
For more on the fBm and the link with fractals, one may consult (with some caution with the vocabulary, that has since changed) the seminal  works of Mandelbrot: \cite{Mandelbrot1982}, and more specifically \cite{Mandelbrot1968}.
A more recent review which deals in particular with the subtleties concerning the difference between fractals and self-affine processes can be found in \cite{Schlather2004}.

\section{Appendix to chapter \ref{chap:visco}}

\subsection{Remark on Terminology}
\label{App:important_remark}
We have presented numerous variations based on the elastic depinning model, some of them being in different universality classes, some of them being more intimately connected to the ``basic'' depinning problem.

However, in this thesis we focus on a particular case of depinning, and we use ``depinning'' as short hand for the problem under specific assumptions:
\begin{itemize}
\item Overdamped limit (no inertial term).
\item Drive is elastic , i.e.~$F_\text{drive} = k_0 (w -h)$, with steady driving ($w=V_0 t$).
\item Quasi-static limit: $V_0 \ll h_0/ \eta_0$ , or $V_0=0^+$.
\item We are interested in the dynamics close to the transition, i.e.~$k_0 \ll k_1$.
\item Short-range correlated disorder for the pinning \textit{force} (Random Field), not periodic.
\item Short-range elastic kernel: $F_\text{elastic} = k_1 \Delta h$.
\item The dimensionality will usually be $d=2$, but we may also study $d=1$ and the mean field cases.
\item Temperature is Zero: there is no noise term in the Langevin equation of motion.
\end{itemize}

\subsection{Derivation of the Mean Field Equations}
\label{App:derivationMF_equations}
We study the mean field limit via the fully connected approximation.
In practice, each block position $h_i$ interacts with the positions of all other blocks via $N-1$ springs of elastic constant $k_1/N$ ($N$ being the number of blocks in the system) and via $N-1$ Maxwell elements (i.e.~spring in series with a dashpot). 
As usual for fully connected systems, the final equation for the site $i$ is obtained by replacing any occurrence of $\Delta h_i$ with $\overline{h}-h_i$.
\begin{figure}[]
\begin{small}
\begin{center}
\def\svgwidth{9cm}
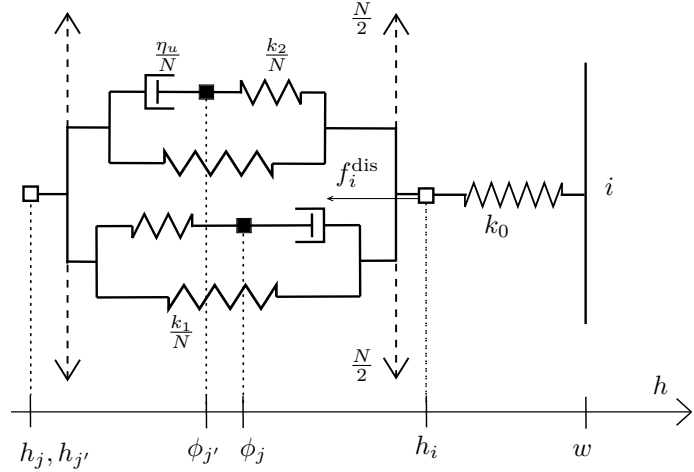
\end{center}
\end{small}
\caption{   {\footnotesize 
Mechanical circuit of the mean field viscoelastic interface.
\label{Fig:schemaV3mf}
}   }
\end{figure}

Here  we give a precise derivation of this result, directly from the mechanical circuit associated to the fully connected model pictured in Fig.~\ref{Fig:schemaV3mf}. In order to have a simple symmetry in the equations, $(N-1)/2$ Maxwell elements  (numbered $j$) are in the order $h_i$-dashpot-$\phi_j$-spring-$h_j$ and the remaining $(N-1)/2$ (numbered $j'$) are in the inverse order ($h_i$-spring-$\phi_{j'}$-dashpot-$h_{j'}$), as pictured in Fig.~\ref{Fig:schemaV3mf}.
The force balance on $h_i$, $\phi_j$ and $\phi_{j'}$ gives:
\begin{align}
 \eta \partial_t h_i =& f^\text{dis}_i(h_i) + k_0 (w-h_i) + \frac{k_1}{N}\sum_{j,j'\neq i}(h_{j} - h_i) \notag  \\
 &+ \frac{\eta_u \partial_t}{N}\sum_{j\neq i} (\phi_j - h_i) + \frac{k_2}{N}\sum_{j'\neq i} (\phi_j-h_i) \\ 
0&=\frac{k_2}{N} \sum_{j\neq i} (\phi_j - h_j) + \frac{\eta_u \partial_t}{N} \sum_{j\neq i} (\phi_j - h_i)  \\
0&=\frac{k_2}{N} \sum_{j'\neq i} (\phi_{j'}-h_i)      +  \frac{\eta_u \partial_t}{N} \sum_{j'\neq i} (\phi_{j'} - h_{j'})
\end{align}
Defining  $\overline{h} = \frac{1}{N} \sum_{i=1}^N h_i $ and $u_i =   \sum_{j\neq i} (h_i - \phi_j) +  \sum_{j'\neq i} (\phi_{j'} - h_{j'})$, 
this simplifies into:
\begin{align}
\eta \partial_t h_i &= f^\text{dis}_i(h_i) + k_0 (w-h_i) + (k_1 + k_2) (\overline{h}-h_i) - k_2 u_i \notag \\
\eta_u \partial_t u_i &= k_2 (\overline{h}-h_i) -k_2 u_i \label{relax2}
\end{align}
which is just the $d$-dimensional result, after replacement of $\Delta h$ by $\overline{h}-h$.
One may notice that formally, the $d$-dimensional expression (\req{definitionGeneraleDe_u}) taken at $d=(N-1)/2$ gives the exact same definition for $u_i$ as the one found here. Thus another way of defining the mean field is to take a large $N$ and formally set $d$ to $(N-1)/2$.

\subsection{Mean Field Dynamics: Fast Part }
\label{App:MF_fastPart}

As for the purely elastic interface (see sec.~\ref{sec:elastic_depinning_dynamics_Mf_hardcore}), it is 
 useful to artificially decompose the dynamical evolution in different steps. 

In a first step, the center of the parabolic potential moves from $w$ to $w + \d w$ and all $\delta^ F_i$'s decrease by $\Delta \delta_{{\rm step} 0} = k_0 \d w$: $P(\delta^ F, \delta^ R) \d \delta^ F \d \delta^ R$ is increased by $ (\partial P_w / \partial \delta^ F) \d \delta^ F  \d \delta^ R k_0 \d w $.
Still in this first step, a fraction  $P_w(\delta^ F=-\delta^ R, \delta^ R) k_0  \d w $  of the sites \textit{with a given } $\delta^ R$ becomes unstable and moves to the next wells: $P(\delta^ F, \delta^ R) \d \delta^ F \d \delta^ R$ is increased by $P_w(-\delta^ R, \delta^ R) k_0  \d w g_1(\delta^ F+\delta^ R) \d \delta^ F$, where $g_1(\delta^ F+ \delta^ R) \d \delta^ F$ is the probability for a block to fall in the range $[\delta^ F, \delta^ F +\d \delta^ F]$ after a jump\footnote{By definition, $g_1(\delta^F+\delta^R) \d \delta^ F = g(z) \d z$.} from some $\delta^ F_{initial}=-\delta^ R$.
The new $\delta=\delta^ F+\delta^ R $ is given by $z (k_0+k_1+k_2)$,  with $z$'s drawn from the distribution $g(z)$. 
This  writes:
\begin{align}
\frac{P_{{\rm step} 1} (\delta^ F, \delta^ R) - P_w(\delta^ F, \delta^ R)}{ k_0  \d w }
&=  \frac{\partial P_w}{\partial \delta^ F} (\delta^ F, \delta^ R)  +  P_w(-\delta^ R, \delta^ R) \frac{ g\left( \frac{\delta^ F+ \delta^ R}{k_0+k_1+k_2}\right) }{ k_0+k_1+k_2}.  \label{Eq:PdeltaVISCO}
\end{align}
In this expression we have not accounted for the increase of $\overline{h}$ due to the numerous jumps.
This increase is given by the fraction of jumping sites multiplied by their average jumping distance, i.e.~it is worth\footnote{
The average jump size of any finite number of jumps is not $\overline{z}$, so this expression should be puzzling.
However, we work with $P(\delta^ F, \delta^ R)$, i.e.~we work in the infinite system size limit. 
In this limit an infinitesimal fraction of sites that jump corresponds to infinitely many sites, so that the average jump is exactly $\overline{z}$.
}
 $\overline{z} P_w(-\delta^ R, \delta^ R) k_0 \d w$.
The corresponding change in the $\delta^F$'s is a uniform decrease by $\overline{z} ( k_1+k_2) P_w(-\delta^ R, \delta^ R) k_0 \d w $ (see \req{deltas_depinning_MF}).

This shift in the $\delta^ F$'s is accounted for in a second step, which acts on $P_{{\rm step} 1}(\delta^ F, \delta^ R)$ exactly as the first did on $P_w(\delta^ F, \delta^ R)$, but with an initial drive given by the shift $\Delta \delta^ F_{{\rm step} 1}= \overline{z} (k_1+k_2) P_w (-\delta^ R, \delta^ R) k_0 \d w $:
\begin{align}
\frac{P_{{\rm step} 2}  (\delta^ F, \delta^ R) - P_{{\rm step} 1}(\delta^ F, \delta^ R)}{\Delta \delta_{{\rm step} 1}}
&=  \frac{\partial P_{{\rm step} 1}}{\partial \delta^ F} (\delta^ F, \delta^ R)  +  P_{{\rm step} 1}(-\delta^ R, \delta^ R) \frac{ g\left( \frac{\delta^ F+ \delta^ R}{k_0+k_1+k_2}\right) }{ k_0+k_1+k_2}.  \label{Eq:Pdelta2VISCO}
\end{align}
In turn, this second step does not account for the increase of $\overline{h}$ due to the ``driving'' by $\Delta \delta^ F_{{\rm step} 1}$: this is accounted for in a third step, and so on.

As these steps go on, the drive from the increase in $\overline{h}$ is given by the geometrical series:
\begin{align}
\Delta \delta^ F _{{\rm step} k} = k_0 \d w \prod_{j=0}^{k-1} (\overline{z}(k_1+k_2) P_{{\rm step} j} (-\delta^ R, \delta^ R) ) , 
\label{Eq:DeltadeltakVISCO}
\end{align}
where we identify $P_{{\rm step} 0} \equiv P_w$.
The convergence of the series to zero is guaranteed if $P_{{\rm step} j} (-\delta^ R, \delta^ R) < 1/(\overline{z}(k_1+k_2)), \forall j $.
Strictly speaking we may not reach zero in any finite number of steps, however it is natural to impose a lower cutoff for the fraction of jumping sites (in any finite size system the minimal non zero value is $1/N$), so that we may have $\Delta \delta_{{\rm step} j} \approx 0$ in a finite number of steps.
On the other hand, if we have $P_{{\rm step} j} (-\delta^ R, \delta^ R) > 1/(\overline{z}(k_1+k_2))$ for numerous consecutive steps, the magnitude of the avalanche increases, the shifts $\Delta \delta_{{\rm step} j}$ may become finite (instead of infinitesimal), and the relevance of this artificial decomposition for analytical aims becomes dubious.

The general set of equations for the $P_{{\rm step} k}$'s is 
a closed form since  $P_{{\rm step} k}$ only depends on the previous $P_{{\rm step} j}, (j<k)$.
Denoting $s\equiv \text{step} k$ the internal time of the avalanche in terms of steps, we can write the evolution as:
\begin{align}
\frac{ \partial P_s}{\partial s} \frac{1}{\Delta \delta _{{\rm step} k} }
&=  \frac{\partial P_s}{\partial \delta^ F}  +  P_s(-\delta^ R, \delta^ R) \frac{ g\left( \frac{\delta^ F+\delta^ R}{k_0+k_1+k_2}\right) }{ k_0+k_1+k_2}. \label{Eq:Pstepk=sVISCO}
\end{align}
This evolution stops either when $P_{{\rm step} k}(-\delta^ R, \delta^ R)=0$ (hence $\Delta \delta^ F_{{\rm step} k+1}=0$), or when the r.h.s of \req{Pstepk=sVISCO} is zero.
The latter case corresponds to a convergence to the fixed point of the corresponding elastic interface with elasticity $k_1+k_2$.
In the former case, some additional driving (increase in $w$) will eventually lead to $P_w(-\delta^ R, \delta^ R)>0$. 
Upon successive increases of $w$, \req{Pstepk=sVISCO} will be iterated again and again, each time with a renewed initial drive $k_0 \d w$: this drives the distribution $P(\delta^ F,\delta^ R)$ towards a fixed point where the r.h.s of \req{Pstepk=sVISCO} cancels.

\subsection{Relaxation Does Not Trigger Aftershocks in Mean Field}
\label{App:no_AS_in_MF}
Consider a block that participates in an avalanche: it jumps from $\delta(0)=0$ at time $0$ to some $\delta(1)=z >0$ at time $1$.
This can be decomposed as a jump from $\delta^F(0) = -\delta^R(0)$ to $\delta^F(1) = -\delta^R(0) +z$.
At the beginning of the avalanche the dashpots are fully relaxed, so that $\delta^R(0)=\delta^R_{i,\infty}(0) =k_2   \frac{\overline{\delta}-\delta_i^F(0) }{k_0+k_1+k_2}$.
After the avalanche, $\delta^R$ relaxes to a new value of 
\begin{align}
\delta^R_{i,\infty}(1) 
&= k_2   \frac{\overline{\delta}-\delta_i^F(1) }{k_0+k_1+k_2} \\
&= \delta^R_{i,\infty}(0) - \frac{k_2}{k_0+k_1+k_2} z.
\end{align}
In terms of the variable $\delta$, this means a shift from $z$ to $z(1-\frac{k_2}{k_0+k_1+k_2} )$, i.e.~the overall shift due to the avalanche and the relaxation is still positive, and there is no aftershock.

The meticulous reader may also consider the evolution under driving: as $w$ increases by $\d w$, $\delta^F$ decreases by the same amount.
Assuming that no avalanche occurs upon driving (otherwise we refer to the case above), this corresponds to a shift $\delta^R_{i,\infty}(1) = \delta^R_{i,\infty}(0) + \frac{k_2}{k_0+k_1+k_2} \d w$, i.e.~an increase of $\delta^R$ after driving.

We conclude that thanks to the approximation of $f^\text{th}_i=f^\text{th}=\text{const.}$, the relaxation does not trigger aftershocks in the mean field regime.

Note that within the approximation $f^\text{th} = \text{const}$, even in finite dimensions there are no aftershocks.
This is true only for the particular model studied in this chapter, which has a ``local'' relaxation kernel. 
For the Laplacian relaxation kernel of the model, in finite dimension we still have aftershocks, even within this approximation.

\subsection{Additional Figures to sec.~\ref{sec:2D_results_visco}}

The figures \ref{Fig:sasb} and \ref{Fig:BrokenZone_complementary} are complementary material to section \ref{sec:2D_results_visco}.

\includefig{\textwidth}{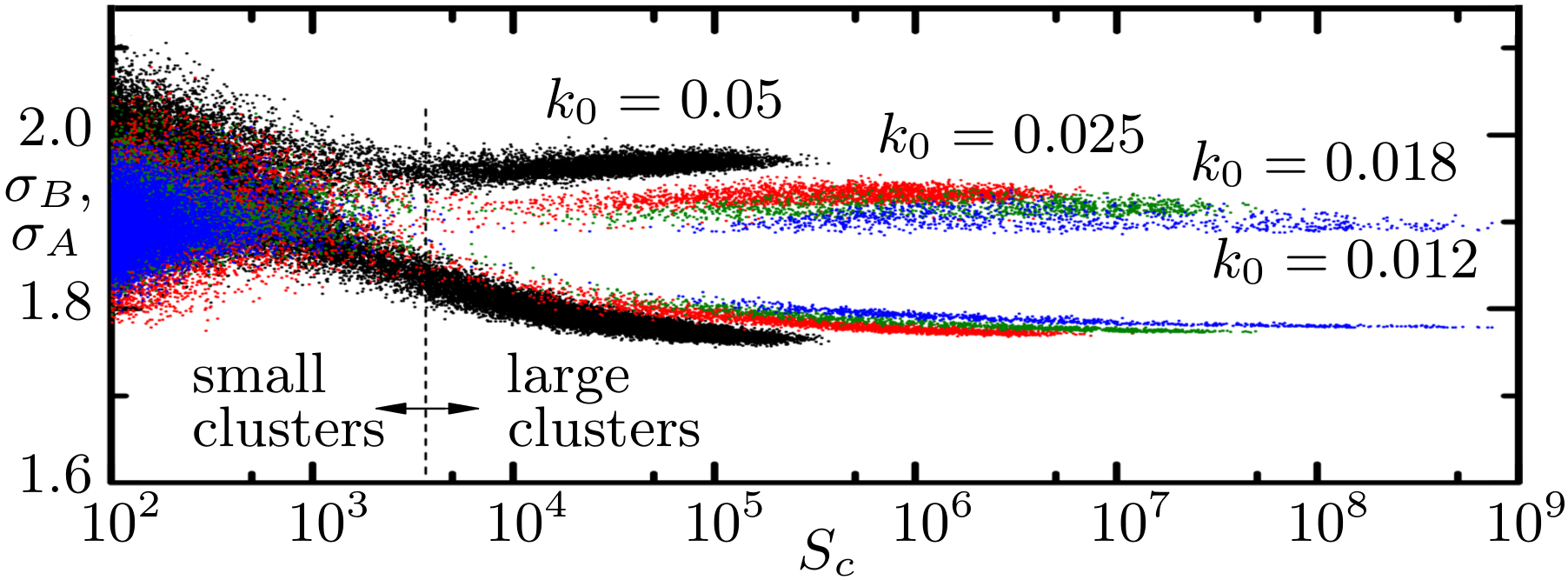}{
The local stress restricted to the event area, just before (up, $\sigma_B$) and just after (bottom, $\sigma_A$) it takes place, as a function of the cluster size $S$ (the size of a cluster is the sum of the sizes of the events occurring for this $w$).
The local variation of stress vanishes for small avalanches (with fluctuating values of $\sigma_{B,A}$), and saturates to a constant nonzero value for large avalanches (with well defined  values for $\sigma_{B,A}$).
We used $k_1=0, k_2=1$.
\label{Fig:sasb}
}

\includefig{15cm}{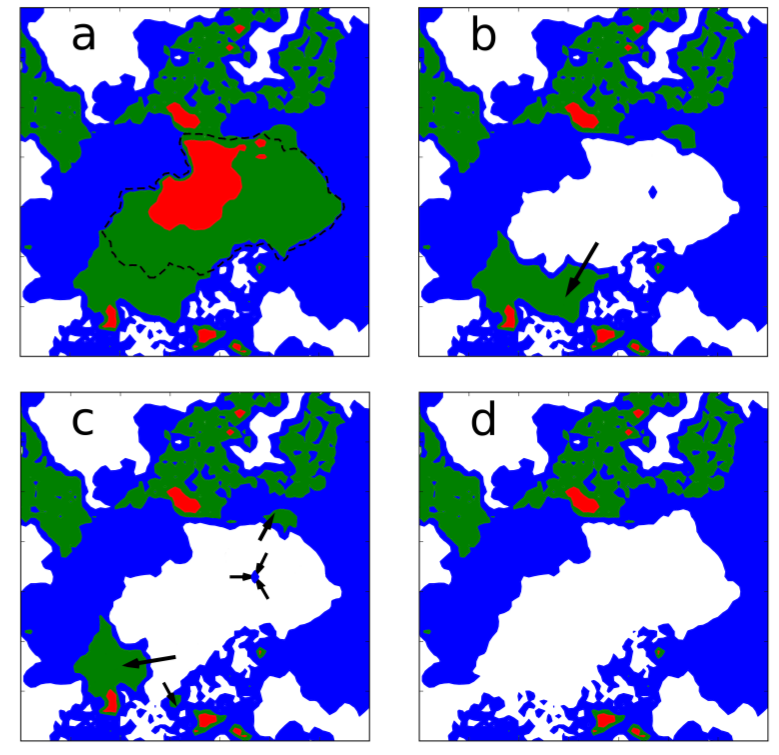}{
Complementary figure to Fig.~\ref{Fig:BrokenZone}
\label{Fig:BrokenZone_complementary}
}

\section{Numerical Methods}

One can (naively) integrate the dynamics using Euler steps, but this is highly inefficient and should not be avoided whenever it is possible. 
We do not actually use Euler steps, but a better scheme that we explain in the next subsection.

\subsection{Viscoelastic Two dimensional case: Details on the Numerical Integration Procedure}
\label{App:Grassberger_efficient_method_two_cases}

We provide here details on the integration of the dynamic equations of the viscoelastic model. Our starting point is the set of equations (2) of the main text:
\begin{align}
\eta \partial_t h_i
&=  k_0 (w-h_i) + f_i^\text{dis}(h_i)+k_1 \Delta h _i + k_2 (\Delta h_i - u_i) \label{1}\\
\eta_u \partial_t u_i
&= k_2 (\Delta h_i - u_i)\label{2}
\end{align}
with $w=V_0t$.
For the numerical work, it is convenient to introduce variables $F_i$ and $G_i$, defined as:
\begin{align}
F_i&\equiv k_2 (\Delta h_i - u_i), \\
G_i&\equiv k_1 (\Delta h_i)+k_0(w-h_i).
\end{align}
Using $F_i$ and $G_i$, the model equations can be written as
\begin{align}
\eta \partial_t h_i
&=  f_i^\text{dis}(h_i)+G_i + F_i\label{hdot}\\
\eta_u \partial_t F_i +k_2 F_i&= \eta_u k_2 (\Delta \partial_t h)_i.
\label{fdot}  
\end{align}
It is thus clear that $G_i$ represents the force onto $h_i$ exerted through $k_1$ and $k_0$ springs, whereas $F_i$ is the force coming from the branches that contain the dashpots and $k_2$ springs.

We work in the case in which temporal scales are well separated: $\tau \ll \tau_u \ll \tau_D$. 
This corresponds to $\eta \ll \eta_u \ll \overline z k_0 /V_0$. 
As discussed in the main text, within the narrow well approximation the actual integration of Eqs. [\ref{hdot}] and [\ref{fdot}] does not need a continuous time algorithm, but can be presented in the form of a discrete set of rules.
From a relaxed configuration with $F_i=0$ at time $t$, the load increase triggers a new instability of Eq. [\ref{hdot}] when the total force from the springs, here $G_i$, reaches $f_i^{th}$, and this occurs after a time interval:
\begin{equation}
\delta t=\min_i \left (\frac{f_i^{th}-G_i}{k_0 V_0}\right )
\label{dtext}
\end{equation}
Thus at time $t+\delta t$ an avalanche starts at position $i$, producing the advance of $h_i$ to the next potential well $h_i \leftarrow h_i +z$, and a corresponding rearrangement of the forces according to (in two dimensions):
\begin{align}
F_i&\leftarrow F_i-4k_2z\\
G_i&\leftarrow G_i-(4k_1+k_0)z\\
F_j&\leftarrow F_j+k_2 z\\
G_j&\leftarrow G_j+k_1 z
\end{align}
where $j$ are the four neighbour sites to $i$, and the value of $f_i^{th}$ is renewed from its probability distribution. All successive unstable sites are 
treated in the same way until there are no more unstable sites. This defines the primary avalanche.
At this point the relaxation dynamics [\ref{fdot}] begins to act,  until some site eventually becomes unstable. Note that due to the discrete pinning potential, in this stage $h$ remains constant, namely the relaxation dynamics is simply:
\begin{equation}
\eta_u \partial_t F_i =-k_2 F_i,
\label{fdot2}  
\end{equation}
This means that a given site $i$ will trigger an avalanche due to relaxation if for some increase in time $\delta t$ the total force from the springs on this site, here $F_i+G_i$, reaches $f_i^{th}$, i.e., if
\begin{equation}
F_ie^{\frac{-k_2 \delta t}{\eta_u}} +G_i=f_i^{th},
\label{dt}  
\end{equation}
(note that in order to have a solution, $F_i$ must be negative, as the l.h.s.~is lower that the r.h.s.~at the starting time).
This leads to the determination of $\delta t$ as
\begin{equation}
\delta t =- \frac{\eta_u}{k_2}\min_i \left [\ln \left(\frac{f_i^{th}-G_i}{F_i}   \right )  \right ]
\label{dtint}
\end{equation}
Once all the secondary avalanches generated by relaxation have been produced and  $F_i$ has relaxed to zero, the external driving is increased again, according to [\ref{dtext}].

This is the main scheme of the simulation. 
We should mention however, that its efficient implementation relies on a classification scheme of all sites, in such a way that the determination of the next unstable site in [\ref{dtext}] and [\ref{dtint}] does not require a time consuming sweep over the whole lattice. 
In fact, following Grassberger \cite{Grassberger1994a} we classify the sites according to their value of the r.h.s.~of [\ref{dtext}] and [\ref{dtint}], and bin them, in such a way that the determination of the next unstable site can be limited to the bin corresponding to the lowest values of these quantities. 
When sites change their $h$ values along the simulation, they are reaccommodated in the bins using a doubly linked list algorithm (via a matrix of fixed size, which contains the next and previous site of each site).\\

My codes will be available on bitBucket soon, at: \cite{LandesCodeAll}.

\end{appendices}

\phantomsection
\addcontentsline{toc}{chapter}{References}
\footnotesize{

\newcommand{\etalchar}[1]{$^{#1}$}

}

\newpage
\phantomsection
\addcontentsline{toc}{chapter}{Abstract}
\section*{Abstract}

\noindent
\textbf{Viscoelastic Interfaces Driven in Disordered Media and Applications to Friction}\\

\noindent
\textbf{Abstract:}\\
Many complex systems respond to a continuous input of energy by an accumulation of stress over time, interrupted by sudden energy releases called avalanches. Recently, it has been pointed out that several basic features of avalanche dynamics are induced at the microscopic level by relaxation processes, which are neglected by most models. During my thesis, I studied two well-known models of avalanche dynamics, modified minimally by the inclusion of some forms of relaxation.

The first system is that of a viscoelastic interface driven in a disordered medium. In mean-field, we prove that the interface has a periodic behaviour (with a new, emerging time scale), with avalanche events that span the whole system. We compute semi-analytically the friction force acting on this surface, and find that it is compatible with classical friction experiments. In finite dimensions (2D), the mean-field system-sized events become local, and numerical simulations give qualitative and quantitative results in good agreement with several important features of real earthquakes.

The second system including a minimal form of relaxation consists in a toy model of avalanches: the Directed Percolation process. In our study of a non-Markovian variant of Directed Percolation, we observed that the universality class was modified but not completely. In particular, in the non-Markov case an exponent changes of value while several scaling relations still hold. This picture of an extended universality class obtained by the addition of a non-Markovian perturbation to the dynamics provides promising prospects for our first system.\\

\noindent
\textbf{Keywords:} \\
quenched disorder, out-of-equilibrium, dynamical phase transition, friction, earthquakes, depinning transition, visco-elastic, Directed Percolation, multiple absorbing state transition.

\newpage
\phantomsection
\addcontentsline{toc}{chapter}{R\'esum\'e}
\section*{R\'esum\'e}

\noindent
\textbf{Interfaces visco\'elastiques sous for\c cage en milieu al\'eatoire et applications \`a la friction.}\\

\noindent
\textbf{R\'esum\'e:}\\
De nombreux syst\`emes complexes soumis \`a un ajout continu d'\'energie r\'eagissent \`a cet ajout par une accumulation de tension au cours du temps, interrompue par de soudaines lib\'erations d'\'energie appel\'ees avalanches. R\'ecemment, il a \'et\'e remarqu\'e que plusieurs propri\'et\'es \'el\'ementaires de la dynamique d'avalanche sont issues de processus de relaxation ayant lieu \`a une \'echelle microscopique, processus qui sont n\'eglig\'es dans la plupart des mod\`eles. Lors de ma th\`ese, j'ai \'etudi\'e deux mod\`eles classiques d'avalanches, modifi\'es par l'ajout d'une forme de relaxation la plus simple possible.

Le premier syst\`eme est une interface visco\'elastique tir\'ee \`a travers un milieu d\'esordonn\'e. En champ moyen, nous prouvons que l'interface a un comportement p\'eriodique caract\'eris\'e par une nouvelle \'echelle temporelle (\'emergente), avec des avalanches qui touchent l'ensemble du syst\`eme. Le calcul semi-analytique de la force de friction agissant sur la surface donne des r\'esultats compatibles avec les exp\'eriences de friction classique. En dimension finie (2D), les \'ev\'enements touchant l'ensemble du syst\`eme (trouv\'es en champ moyen) deviennent localis\'es, et les simulations num\'eriques donnent des r\'esultats en bon accord avec plusieurs caract\'eristiques importantes des tremblements de terre, tant qualitativement que quantitativement.

Le second syst\`eme incluant \'egalement une forme tr\`es simple de relaxation est un mod\`ele jouet d'avalanche: c'est la percolation dirig\'ee. Dans notre \'etude d'une variante non-markovienne de la percolation dirig\'ee, nous avons observ\'e que la classe d'universalit\'e \'etait modifi\'ee mais seulement partiellement. En particulier, un exposant change de valeur tandis que plusieurs relations d'\'echelle sont pr\'eserv\'ees. Cette id\'ee d'une classe d'universalit\'e \'etendue, obtenue par l'ajout d'une perturbation non-markovienne offre des perspectives prometteuses pour notre premier syst\`eme.\\

\noindent
\textbf{Mots-clefs:} \\
d\'esordre gel\'e, hors d'\'equilibre, transition de phase dynamique, friction, s\'eismes, transition de d\'epiegeage, visco-\'elastique, Percolation Dirig\'ee, transition de phase à \'etats absorbants multiples.

\newpage

\phantomsection
\addcontentsline{toc}{chapter}{Remerciements}
\section*{Remerciements}

Cette th\`ese n'aurait pas \'et\'e possible sans le soutien, les encouragements, les discussions, les enseignements et l'aide en g\'en\'eral de mes proches et de certains de mes moins-proches. 
Je tiens ici \`a rendre hommage \`a ceux qui ont \`a mes yeux jou\'e un r\^ole important dans le d\'eveloppement de mon go\^ut pour la science, puis du d\'eveloppement de cette carri\`ere scientifique naissante, qui se concr\'etise aujourd'hui en ce manuscrit.

Tout d'abord, je remercie les membres de mon jury de th\`ese, tout simplement pour avoir accept\'e d'en faire partie, mais aussi et surtout pour l'int\'er\^et port\'e \`a ma th\`ese, pour leurs remarques, tr\`es utiles, ainsi que pour leur gentillesse et leur sympathie lors de nos \'echanges.
En particulier, je remercie mes deux rapporteurs pour leur relecture attentive : Jean-Louis Barrat pour sa critique tr\`es encourageante et Stefano Zapperi pour ses remarques qui m'ont \'et\'e d'une granfe utilit\'e, en particulier dans le domaine de la friction, o\`u mon expertise est encore assez limit\'ee.
Je salue \'egalement l'attention de Leticia Cugliandolo, pour sa lecture d\'etaill\'ee de mon manuscrit.

Les ``stars'' de ma th\`ese sont, bien entendu, Alberto Rosso, mon directeur, mais aussi Eduardo Jagla, qui a jou\'e lui aussi un r\^ole important dans le d\'eveloppement de mes comp\'etences scientifiques, et un r\^ole crucial dans les choix scientifiques. Apr\`es deux s\'ejours de un mois chacun en Argentine, \`a Bariloche, il est pour moi comme un second directeur de th\`ese.
Cela ne doit pas obscurcir les m\'erites d'Alberto : il a \'et\'e un directeur de th\`ese formidablement d\'evou\'e, tr\`es attach\'e \`a ma r\'eussite, et qui m'a toujours pouss\'e \`a donner le meilleur de moi m\^eme. Son expertise dans le domaine du d\'epi\'egeage, qu'il m'a transmise de la fa\c con la plus douce (\`a l'oral, au tableau), m'a \'et\'e tr\`es pr\'ecieuse au cours de ces trois ans. Le probl\`eme du d\'epi\'egeage, un sujet riche et tr\`es explor\'e par la communaut\'e depuis 30 ans a \'et\'e le pivot de ma th\`ese, et Alberto m'a fourni les fondations permettant de planter fermement ce pivot en terre.
De son enseignement, je retiendrai tout particuli\`erement son souci du d\'etail, dans la r\'edaction comme dans la pr\'eparation d'expos\'es : la minutie \'etait toujours mise au service de la recherche de la plus grande clart\'e.

Au sujet de l'Argentine, je remercie le partenariat ECOS-Sud d'avoir rendu possible ma collaboration avec Eduardo, et de m'avoir permis de visiter ce beau pays qu'est la Patagonie, deux fois ! Cela m'a aussi permis de faire na\^itre des amiti\'es, avec Ezekiel Ferrero, que je pense avoir l'occasion de revoir souvent maintenant qu'il est en France, avec Alejandro Kolton, qui m'a chaleuresement accueilli et avec Luis Arag\'on, qui m'a introduit aupr\`es de ses amis, m'a fait d\'ecouvrir beaucoup de l'Argentine : les bi\`eres maison, l'asado, bref la joie de vivre patagonienne; et avec qui j'esp\`ere pouvoir collaborer dans le futur. 

Je remercie au passage mes relecteurs de th\`ese : Alberto (qui a tout lu), mais aussi Luis Arag\'on et Eiji Kawasaki, qui m'ont bien aid\'e a peaufiner mon premier chapitre.
Sur ce point, je suis \'egalement endett\'e aupr\`es de Pierpaolo Vivo, qui m'a grandement aid\'e \`a corriger mon ``Research Statement'', une \'etape difficile et bien angoissante dans la p\'eriode de recherche de post-docs. 
Je remercie \'egalement Shamik Gupta pour son aide lors de la r\'edaction de certains r\'esum\'es importants (et Mikhail Zvonarev, pour cette m\^eme aide), et pour m'avoir fait r\'ep\'eter (avec Haggai Landa) ma soutenance de th\`ese, avec des commentaires tr\`es constructifs.

Pour les discussions scientifiques au cours de cette th\`ese, je remercie Kabir Ramola, vite passionn\'e par un probl\`eme qui n'\'etait pas le sien; Martin Trulsson pour une discussion particuli\`erement enrichissante, et plus g\'en\'eralement tous les gens du LPTMS avec qui j'ai pu \'echanger au cours de ces trois ann\'ees.
Au del\`a de l'aspect scientifique, le labo a \'et\'e pour moi un espace convivial pour travailler, et parfois aussi pour d'autres choses ! Je retiendrai nos discussions philosophiques et politiques avec Arthur, Matthieu, et Andrey; les discussions sur la morale et Dieu avec Yasar, les encouragements (m\^eme \`a distance) de Pierre-\'Elie, les curiosit\'es trouv\'ees par Ricardo, et en g\'en\'eral les discussions avec les anciens puis avec les nouveaux, au d\'etour d'un repas ou d'une cigarette. Et bien s\^ur, pour les bons souvenirs de M2, scientifiques ou non, je pense \`a Thomas ainsi qu'\`a Tanguy le Bot.
Je remercie Claudine pour sa disponibilit\'e, son efficacit\'e, et sa connaissance (impressionnante) des ``r\`egles'' \'ecrites et non-\'ecrites, qui g\`erent le labo, la fac, le CNRS, enfin bref, le monde.
C'est gr\^ace \`a Robin Masurel que je me suis vraiment pench\'e sur l'origine microscopique de la plasticit\'e lors de la r\'edaction de cette th\`ese, merci \`a lui.
Merci \`a Didier pour ``l'harmonie'' parfois trouv\'ee pendant la th\`ese, gr\^ace \`a son art.

Je veux maintenant rendre hommage \`a ceux qui ont permis de stimuler et aiguiller mon go\^ut pour la science dans des directions int\'eressantes.
J\'er\^ome Perez m'a aid\'e \`a me diriger vers la recherche fondamentale, comme c'\'etait mon souhait avant m\^eme de rentrer \`a l'Ensta ParisTech.
Mikko Alava a \'et\'e le premier \`a me faire confiance, il m'a fourni ma premi\`ere exp\'erience de recherche, qui est plus est dans un domaine qui me passionnait d\'ej\`a, bien que de fa\c con diffuse (les syst\`emes ``complexes''). 
Merci \`a Matti Pelt\"omaki, qui a \'et\'e un excellent tuteur \`a l'occasion de ces quatre mois pass\'es en Finlande.
En termes de recherche th\'eorique, ma formation doit beaucoup \`a mon stage avec ``Fred'' (Fr\'ed\'eric van Wijland), toujours de bon conseil pour m'aider \`a trouver une bonne direction (scientifiquement parlant).

Certains de mes profs de pr\'epa, et aussi du Lyc\'ee, m'ont permis d'affirmer mon go\^ut pour la physique : Mme Bonzon et Mme Gros m'ont permis d'\'evoluer vers le haut et de commencer \`a prendre le chemin de la rigueur, et Mnm's (Marie-No\"elle Sanz) m'a fait courir dessus.
Ma ma\^itrise de l'anglais ne doit rien \`a mes qualit\'es naturelles de linguiste : m\'ediocre jusqu'en 1\`ere, je n'ai pu d\'ecoller que gr\^ace \`a mes s\'ejours chez des amis d'amis bienveillants : chez les Gulliford dans le Kent (Londres, merci \`a Vincent), puis chez les Dong dans le Connecticut (merci \`a Sophie).

Il est important de remonter plus loin, et de remercier mes parents, qui m'ont certes fourni un patrimoine g\'en\'etique correct, mais surtout un environnement ad\'equat au d\'eveloppement de l'esprit scientifique : de la philo, des lettres, un go\^ut pour la d\'ecouverte, mais aussi et surtout pour la recherche de la v\'erit\'e, que malheureusement tous n'ont pas la chance de recevoir si jeune. 
Tout cela, accompagn\'e de beaucoup de Lego, m'a je crois pouss\'e \`a ``devenir scientifique'' (au lieu de ``ing\'enieur Lego'').

Certes, Alberto m'a support\'e pendant ces trois ans, mais il n'est pas le seul : je dois beaucoup de mon \'equilibre et de ma pers\'ev\'erance \`a mes amis -- ils se reconna\^itront --, mais aussi et surtout \`a Anne, qui m'accompagne depuis maintenant 6 ann\'ees.
\vspace{2cm}

Et pour finir, je citerai quelques extraits d'un des webcomics qui m'ont fait ``perdre'' du temps durant ces trois ann\'ees :\\
\url{http://xkcd.com/169/} : 
``Communicating badly and then acting smug when you're misunderstood is not cleverness.''\\
\url{http://xkcd.com/1349/} : ``Computers are just carefully organized sand. \textit{Everything} is hard until someone makes it easy.''

\newpage

\end{document}